\documentclass[twocolumn, preprint]{aastex631}

\newcommand\tauVeff{$\hat{ \tau }_V \ $}
\newcommand\tauVeffNoSpace{$\hat{ \tau }_V$}
\newcommand\Av{$\textrm{A}_{\textrm{V}} \ $}
\newcommand\AvNoSpace{$\textrm{A}_{\textrm{V}}$}

\usepackage{rotating}
\usepackage{tabularx}
\usepackage{tikz}
\usepackage{lipsum}
\usepackage{amsmath}
\usepackage{soul} 

\usepackage{xcolor}

\shortauthors{de la Vega et al.}

\graphicspath{{./figures/}{}}

\usepackage{afterpage}

\begin{document}

\title{Improved SED-Fitting Assumptions Result in Inside-Out Quenching at $z\sim0.5$ and Quenching at All Radii Simultaneously at $z\sim1$}

\correspondingauthor{Alexander de la Vega}
\email{alexandd@ucr.edu}

\author[0000-0002-6219-5558]{Alexander de la Vega}
\altaffiliation{Current address: Department of Physics \& Astronomy,
University of California, 900 University Ave, Riverside, CA 92521, USA.}
\affiliation{Department of Physics \& Astronomy,
Johns Hopkins University, 3400 N. Charles Street, Baltimore, MD, 21218, USA}

\author[0000-0002-3838-8093]{Susan A. Kassin}
\affiliation{Space Telescope Science Institute, 3700 San Martin Dr., 
Baltimore, MD 21218, USA}

\affiliation{Department of Physics \& Astronomy,
Johns Hopkins University, 3400 N. Charles Street, Baltimore, MD, 21218, USA}

\author[0000-0003-4196-0617]{Camilla Pacifici}
\affiliation{Space Telescope Science Institute, 3700 San Martin Dr., 
Baltimore, MD 21218, USA}

\author[0000-0003-3458-2275]{St\'{e}phane Charlot}
\affiliation{Sorbonne Universit\'{e}, CNRS, UMR7095, Institut d’Astrophysique de Paris, 98 bis bd Arago, 75014 Paris, France}

\author[0000-0002-9551-0534]{Emma Curtis-Lake}
\affiliation{Centre for Astrophysics Research, Department of Physics, Astronomy and Mathematics, University of Hertfordshire, Hatfield
AL10 9AB, UK}

\author[0000-0002-7636-0534]{Jacopo Chevallard}
\affiliation{Sub-department of Astrophysics, Department of Physics, University of Oxford, Denys Wilkinson Building, Keble Road, Oxford OX1 3RH, UK}

\author[0000-0001-6670-6370]{Timothy M. Heckman}
\affiliation{Department of Physics \& Astronomy,
Johns Hopkins University, 3400 N. Charles Street, Baltimore, MD, 21218, USA}

\author[0000-0002-6610-2048]{Anton M. Koekemoer}
\affiliation{Space Telescope Science Institute, 3700 San Martin Dr., 
Baltimore, MD 21218, USA}

\author[0000-0002-9593-8274]{Weichen Wang}
\affiliation{Dipartimento di Fisica G. Occhialini, Università degli Studi di Milano Bicocca, Piazza della Scienza 3, 20126 Milano, Italy}
\affiliation{Department of Physics \& Astronomy,
Johns Hopkins University, 3400 N. Charles Street, Baltimore, MD, 21218, USA}

\defcitealias{Calzetti94}{C94}
\defcitealias{Chevallard13}{CCWW13}
\defcitealias{Greener20}{G20}
\defcitealias{Medling18}{M18}

\begin{abstract}

Many studies conclude that galaxies quench from the inside-out by examining profiles of specific star-formation rate (sSFR). These are usually measured by fitting spectral energy distributions (SEDs) assuming a fixed dust law and uniform priors on all parameters. Here, we examine the effects of more physically motivated priors: a flexible dust law, an exponential prior on the dust attenuation \AvNoSpace, and Gaussian priors that favor extended star-formation histories. This results in model colors that better trace observations. We then perform radial SED fits to multiband flux profiles measured from Hubble Space Telescope images for 1,440 galaxies at $0.4<z<1.5$ of stellar masses $10^{10}-10^{11.5}\ M_{\odot}$ \textcolor{black}{using both the traditional and the more physically motivated assumptions}. The latter results in \textcolor{black}{star formation rate and \Av profiles that agree with measurements from spectroscopy and \Av profiles that behave correctly as a function of inclination}. Since green valley galaxies at $z\sim1.3$ are expected to evolve into quiescent galaxies at $z\sim0.9$, we compare their sSFR profiles \textcolor{black}{using the more physically motivated assumptions}. Their slopes are similar at all masses ($0.06 - 0.08~\textrm{dex}~\textrm{kpc}^{-1}$), and the normalizations for the quiescent galaxies are lower. \textcolor{black}{Therefore}, the sSFR profiles decline with time as quenching occurs at all radii simultaneously. We compare profiles of green valley galaxies at $z\sim0.9$ and quiescent galaxies at $z\sim0.5$. The former are shallower at all masses by $\sim0.1~\textrm{dex}~\textrm{kpc}^{-1}$. The sSFR profiles steepen with time as galaxies quench from the inside-out. In summary, at $z\sim0.9-1.3$, galaxies quench at all radii simultaneously, and at $z\sim0.5-0.9$, they quench from the inside-out. 

\end{abstract}


\section{Introduction} 
\label{sec:intro}

A major unsolved question in galaxy evolution concerns the end of star formation in galaxies \citep[e.g.,][]{Bell04, Faber07, YPeng10}. While numerous mechanisms to terminate star-formation, or ``quench'' have been proposed, conclusive evidence for the frequency with which galaxies quench under each mechanism has proven difficult to determine (see, e.g., \citealt{Man18}). 

One way to learn about the quenching of galaxies is \textcolor{black}{to study} their spatially resolved spectra. \textcolor{black}{Such} data permit the study of the dust-corrected specific star-formation rate (sSFR; SFR/stellar mass) as a function of radius and indicate which regions within galaxies were the first to quench. For nearby galaxies, studies that have examined these spectra generally find that galaxies have the lowest sSFRs in their centers and the highest sSFRs in their outskirts \citep[e.g.,][]{Belfiore18, Ellison18, Lin19}. In other words, nearby galaxies quench from the inside-out.  

An alternative approach to understanding quenching is to examine galaxies at low and high redshifts. Determining how galaxies evolve with time can be done by connecting galaxy populations at different redshifts. At $z\sim1$, quiescent galaxies are rarer than at $z\sim0$, but they begin to become more common with time \citep{Bell04, Faber07, Brammer11}. At these intermediate redshifts, many studies have examined how galaxies quench. Most have done this by performing radial or spatially resolved spectral energy distribution (SED) fitting of galaxies using Hubble Space Telescope (HST) images \citep[e.g.,][]{Morishita15, Nelson16, Nelson21, Abdurrouf18, Liu18, Morselli19, Hemmati20}. Nearly all of these studies find that galaxies quenched from the inside-out at $z\sim1$, in qualitative agreement with results for nearby galaxies, though these results are obtained in different ways. \textcolor{black}{The rest of this Section will summarize studies that have examined galaxies at similar redshifts to this work. The studies below make similar SED-fitting assumptions, namely assuming uniform priors on all SED-fitting parameters and a single attenuation curve.}

\citet{Morishita15} examined the radial distribution of stellar mass and rest-frame colors of 3,785 massive galaxies (\textcolor{black}{$\log \left(\textrm{M}_{\star}/\textrm{M}_{\odot}\right) \gtrsim 10$}) at $0.5 < z < 3.0$ by performing SED fits to HST images from the Cosmic Assembly Near-infrared Deep Extragalactic Legacy Survey (CANDELS; \citealt{Grogin11, Koekemoer11}). This paper finds that massive galaxies at these redshifts build up dense centers and sufficiently red rest-frame U-V colors to quench from the inside-out at $z\sim2$. 

\citet{Nelson16} used spatially resolved grism spectra from the 3D-HST survey \citep{Skelton14, Momcheva16} to construct maps of the H$\alpha$ emission line by stacking spectra of 3,200 individual galaxies at $0.7 < z < 1.5$. This work constructs these maps as a function of stellar mass and location relative to the relation between stellar mass and SFR, known as the star-formation main sequence \citep[SFMS;][]{Noeske07, Whitaker14, Lee15, Schreiber15, Tomczak16}. They obtain stellar mass surface density profiles by measuring surface brightness profiles in F140W and multiplying by spatially integrated mass-to-light ratios. For SFR surface density profiles, H$\alpha$ surface brightness profiles are multiplied by the ratio of SFR to H$\alpha$ luminosity \citep[e.g.,][]{Kennicutt98}. The H$\alpha$ luminosities were not corrected for dust. This work finds that star-formation is enhanced \textcolor{black}{at all radii for galaxies above the SFMS and suppressed at all radii for galaxies below the SFMS}. For high-mass galaxies (\textcolor{black}{$10.5 < \log \left(\textrm{M}_{\star}/\textrm{M}_{\odot}\right) < 11$}), the largest difference in sSFR between galaxies on the SFMS and below it is found in the centers of galaxies, \textcolor{black}{which, they write,} suggests inside-out quenching \citep[see also][]{Tacchella18}.

\citet{Nelson21} performed dust corrections to the grism data presented by \citet{Nelson16} and compared their corrected sSFR profiles with those from the Illustris-TNG50 simulation \citep{Nelson18, Pillepich18}. Dust attenuation is measured via spatially resolved SED fitting of images from CANDELS. This paper finds that observed and simulated sSFR profiles are similar in both normalization and shape. \textcolor{black}{It is found that} galaxies at stellar masses $\log \ \textrm{M}_{\star} > 10.5$ have centrally suppressed sSFR profiles, indicative of inside-out quenching, which is attributed to feedback from central supermassive black holes. At lower stellar masses, \textcolor{black}{it is found that} galaxies on, above, and below the SFMS have approximately flat sSFR profiles. 

\citet{Abdurrouf18, Liu18, Morselli19} and \citet{Hemmati20} all performed spatially resolved or radial SED fitting on CANDELS images for galaxies at intermediate redshifts ($0.2 < z < 1.8$) and concluded that galaxies quench from the inside-out at high \textcolor{black}{mass ($\log \left(\textrm{M}_{\star}/\textrm{M}_{\odot}\right) > 10.5$)} and potentially from the outside-in at lower \textcolor{black}{mass ($\log \left(\textrm{M}_{\star}/\textrm{M}_{\odot}\right) < 10.5$)}. \citet{Abdurrouf18} examined sSFR profiles for 152 galaxies at $0.8 < z < 1.8$ and stellar masses \textcolor{black}{$\log \left(\textrm{M}_{\star}/\textrm{M}_{\odot}\right) > 10.5$} by constructing maps of stellar mass and SFR surface densities. \citet{Liu18} inferred sSFR profiles for 4,377 galaxies at stellar masses \textcolor{black}{$9 < \log \left(\textrm{M}_{\star}/\textrm{M}_{\odot}\right) < 11$} and at redshifts $0.5 < z < 1.0$ using radial profiles of the rest-frame $UVI$ colors. \citet{Morselli19} constructed maps of stellar mass and surface densities for 712 galaxies at $0.2 < z < 1.2$ and stellar masses $\log \ \textrm{M}_{\star} \geq 9.5$. They measured sSFR profiles using these maps and found that galaxies below the SFMS show the greatest suppression in sSFR in their centers at all masses considered. Lastly, \citet{Hemmati20} also measured maps of stellar mass and SFR surface densities for 545 galaxies at $0.8 < z < 1.0$. For each galaxy in their sample, they fit Schechter functions to the distributions of SFR surface density and stellar mass surface density. They found that the knees of these Schechter functions lie at larger radii for galaxies that lie farther below the SFMS, which, \textcolor{black}{they write,} suggests that galaxies quench from the inside-out. 

Many studies have shown that results from SED fits are sensitive to \textcolor{black}{modeling assumptions}, and particular attention has been paid to the treatment of dust attenuation and star-formation histories \citep[SFHs; e.g.,][]{Pacifici12, Pacifici15, Pforr12, Salmon16, Iyer17, LoFaro17, Carnall18, Leja19, Lower20, Wang23}. Recent work has built upon earlier studies that make conventional SED-fitting assumptions. For example, \citet{Pacifici15} demonstrated that using a single dust law and uniform priors results in \textcolor{black}{stellar masses that are systematically overestimated by $\sim0.1$ dex and SFRs that are underestimated by $\sim0.6$ dex}, when compared to a more sophisticated approach wherein a flexible attenuation curve and nonparametric SFHs are adopted (SFHs in this work are derived from a post-treatment of cosmological simulations). \citet{Leja19} showed that using a flexible attenuation curve and nonparametric SFHs yields \textcolor{black}{stellar masses that are $\sim0.1 - 0.3$ dex higher and SFRs that are $\sim0.1 - 1.0$ dex lower} than those obtained by SED fits that assume a single attenuation curve and parametric SFHs (their nonparametric SFHs assign weights to specified bins in lookback time). 

\textcolor{black}{In this paper, we examine the effects of varying SED-fitting assumptions on the radial properties of a large sample of massive galaxies at intermediate redshift ($0.4 < z < 1.5$). The paper is laid out as follows. In Section \ref{sec:data}, we describe the HST imaging and other datasets used in this paper. In Section \ref{sec:general_cuts}, we select our sample. We measure radial flux profiles and their uncertainties in Section \ref{sec:flux_profiles} and describe how SED fitting is performed in Section \ref{sec:sed_model}. We construct five sets of assumptions to understand which changes to which priors affect the profiles. These are introduced in Section \ref{sec:five_sets_assumptions}, where model colors are compared against observed colors. We find that changes to the dust law and star-formation histories have the most significant effects. In Section \ref{sec:define_galaxies}, we describe how we define star-forming, green valley, and quiescent galaxies. We then create radial profiles of the stellar mass surface density, SFR surface density, sSFR, \AvNoSpace, and mass-weighted age in Section \ref{sec:science_profiles}. In the same Section, we also examine how these profiles change as a result of changing the SED-fitting assumptions. This is done by examining how the models behave in color-color space with respect to observations. In Section \ref{sec:choosing_assumption}, we compare our radial profiles with those measured from spectroscopy and determine that the fifth set of assumptions is most appropriate for our sample. We also compare our profiles with expected behavior as a function of inclination. In Section \ref{sec:ssfr_optimal}, we examine sSFR profiles under the fifth set of assumptions and provide a physical interpretation of these profiles in Section \ref{sec:evolution_optimal}. We summarize and conclude in Section \ref{sec:conclusions}. } 

All magnitudes in this paper are in the AB system \citep{Oke83} and we adopt the cosmological parameters measured by \citet{Planck16}.

\section{Data and Data Products}
\label{sec:data}

\subsection{Hubble Images and Photometry}
\label{sec:images_phot}

Publicly available multiband imaging and integrated photometry from the CANDELS \citep{Grogin11, Koekemoer11} in the Great Observatories Origins Deep Survey \citep{Giavalisco04} North and South fields,  hereafter GOODS-N  and  GOODS-S, are used.  The images were taken with the {\it Hubble Space Telescope} (HST) Advanced Camera for Surveys (ACS) in the F435W, F606W, F775W, and F850LP bandpasses, and the Wide Field Camera 3 (WFC3) in the F125W, F140W, and F160W bandpasses. Images in the F140W bandpass were taken as part of the 3D-HST survey \citep{Skelton14}. 

Mosaicked images of each bandpass are from \citet{Koekemoer11}. All are aligned to the same reference frame and drizzled to a common pixel scale of 0.\arcsec06 pixel$^{-1}$. \textcolor{black}{We refer to these as images in the rest of the paper.} The point spread function (PSF) of each mosaic is then degraded to the resolution of the F160W bandpass using PSF-matching kernels from \citet{Skelton14}. The PSF of the F160W mosaic has an FWHM of 0.\arcsec18. 

Integrated photometry is adopted for all bandpasses from \citet{Guo13} for GOODS-S and \citet{Barro19} for GOODS-N. These measurements use HST \textcolor{black}{images} that are PSF-matched to the resolution of the F160W image. Sources are selected in the F160W bandpass. 

We use Galactic reddening maps measured by \citet{Schlegel98}, later recalibrated by \citet{Schlafly11}, to correct for Galactic extinction for all filters. The $E(B-V)$ values measured in GOODS-N and GOODS-S are converted to \Av assuming $\textrm{R}_{\textrm{V}} = 3.1$. Then, the extinction curve from \citet{Fitzpatrick99} is used to derive an extinction for each filter. These corrections are small and are \textcolor{black}{as follows: 0.0419, 0.0276, 0.0183, 0.0138, 0.0081, 0.0070, and 0.0063 magnitude in GOODS-N and 0.0288, 0.0190, 0.0125, 0.0094, 0.0055, 0.0049, and 0.0043 magnitude in GOODS-S in the F435W, F606W, F775W, F850LP, F125W, F140W, and F160W filters, respectively}. 

\subsection{Axis Ratios and Position Angles}

From these images, axis ratios of galaxies, defined as the ratio of the minor to major axes, and position angles of the major axes were measured by \citet{Guo13} and \citet{Barro19} for GOODS-S and GOODS-N, respectively. \textcolor{black}{We adopt these measurements for our analysis.}  These parameters are measured by applying the Source Extractor software \citep{Bertin96} to the F160W \textcolor{black}{images}. \citet{Guo13} and \citet{Barro19} used the same Source Extractor parameters to detect sources in each field.  Typical fractional errors in the axis ratio are 2\%, and typical errors in the position angle are $<5^{\circ}$.  

\subsection{Redshifts}

Redshifts are adopted from a collection of spectroscopic and photometric redshifts compiled by \citet{Kodra23}. To measure photometric redshifts, they used five different codes, each of which uses different techniques to estimate the probability distribution function (PDF) of a galaxy's redshift. Each photometric redshift we adopt for this paper is the probability-weighted expectation value of the combined redshift PDF from the five codes for a given galaxy. Cuts on the photometric redshift uncertainties are discussed in detail in Section \ref{sec:general_cuts}. 

\subsection{X-Ray Data}
\label{sec:xray}

Active galactic nuclei (AGNs) are identified and removed using Chandra X-ray catalogs in GOODS-N \citep{Xue16} and GOODS-S \citep{Luo17}. These surveys have exposure times of $\sim$2 Ms and $\sim7$ Ms, and reach flux limits of $3.5 \times 10^{-17} \ \textrm{erg cm}^{-2} \ \textrm{s}^{-1}$ and $1.9 \times 10^{-17} \ \textrm{erg cm}^{-2} \ \textrm{s}^{-1}$ in the full band (0.5-7 keV) in GOODS-N and GOODS-S, respectively. 

\section{Sample Selection}
\label{sec:general_cuts}

\subsection{The Full Sample}
\label{sec:full_sample}

We begin with all sources in the integrated flux catalogs for GOODS-N and GOODS-S. These sources are then cross-matched with catalogs of spectroscopic and photometric redshifts. A maximum separation of 0.\arcsec2 was used in the cross-matching. All sources are matched. There are a total of 70,375 sources: 35,445 in GOODS-N and 34,930 in GOODS-S. Of these, 64,117 have photometric redshifts and 6,258 have spectroscopic redshifts. 

\textcolor{black}{We remove galaxies contaminated by bright foreground stars and the foreground stars themselves}. We require the SExtractor stellarity index CLASS\_STAR to be $<0.78$ \citep{Boada15}. To remove \textcolor{black}{galaxies} that fall on diffraction spikes or are on the edges of the \textcolor{black}{images}, we require the flag PhotFlag to be zero.  These cuts remove 3,348 sources, leaving 67,027.

Galaxies are selected to span redshifts $0.4 < z < 1.5$, so that the HST bandpasses span the rest-frame ultraviolet to near-infrared (NIR). This wavelength range is optimal for measuring stellar masses and star-formation rates \citep{Pacifici12, Pforr12, Buat14}. At $z=0.4$, the rest-frame wavelength range is $\lambda \sim3100 - 11000$ \AA, and at $z=1.5$ it is $\lambda \sim1750 - 6200$ \AA. This redshift cut removes 40,134 galaxies and leaves 26,893. 

For our analysis, galaxies are selected to have an F160W magnitude brighter than 25 to ensure high signal-to-noise ratios (S/Ns) in the photometry. This cut removes low-mass galaxies (often less than $10^9 \ \textrm{M}_{\odot}$) and decreases the sample by 15,119, with 11,774 galaxies left.

We require galaxies to be in the following \textcolor{black}{images}: F435W, F606W, F775W, F850LP, and F125W. 
\textcolor{black}{We do not require galaxies to be in the F140W mosaic,} because it is not necessary for our analysis: it has significant wavelength overlap with F125W and F160W, and it \textcolor{black}{is the shallowest of all of the images}. This restriction reduces the sample by 1,034, leaving 10,740 galaxies. Of these \textcolor{black}{10,740 galaxies,} 8,889 are observed in F140W.

To remove X-ray AGN, we then remove all galaxies detected in the X-ray catalogs (Section \ref{sec:xray}). This removes 185 galaxies and leaves 10,555.

We remove galaxies that are not resolved in the F160W bandpass, i.e. those with half-light radii smaller than the FWHM of the PSF, which is 0.\arcsec18 \citep{Koekemoer11, Wuyts12}. The galaxies that are excluded are mostly low-mass blue galaxies at low redshift (i.e. stellar mass less than $10^9 \ \textrm{M}_{\odot}$, redshift $z < 1$, and rest-frame $U-V < 1.0$ mag). The cut is achieved by selecting galaxies with half-light radii in F160W (column name FLUX\_RADIUS\_2\_F160W in the catalogs) to be $>3$ pixels. This removes 225 galaxies and leaves 10,330. 

\begin{figure}
    \centering
    \includegraphics[width=\columnwidth]{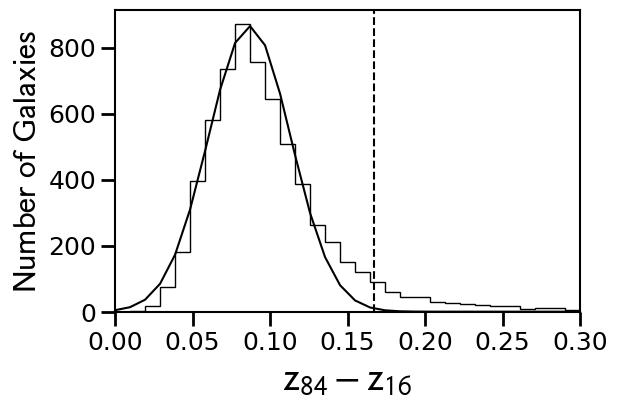}
    \caption{\textcolor{black}{Only photometric redshifts with reasonable uncertainties are used in this work.  The distribution of the widths of the posteriors of the photometric redshifts of \citet{Kodra23} are shown. The width of the posterior is the difference between the 84th percentile of the posterior, $z_{84}$, and the 16th percentile, $z_{16}$}. The solid line is a Gaussian fit to the histogram, and the dashed line is three standard deviations above the mean. Galaxies with photometric redshift posterior widths beyond the dashed line \textcolor{black}{at $z_{84} - z_{16} = 0.168$ are considered outliers and} are excluded.}
    \label{fig:photo_z_width}
\end{figure}

Galaxies are further selected to have spectroscopic redshifts or photometric redshifts with small uncertainties. Of the 10,330 galaxies that remain, 3,804 have spectroscopic redshifts (543 are measured using the HST grism) and 6,526 have photometric redshifts. Here we discuss the photometric redshifts. \citet{Kodra23} measured Bayesian photometric redshift posterior distributions following \citet{Dahlen13}. Figure \ref{fig:photo_z_width} shows the distribution of the photometric redshift posterior widths, defined as the difference between the 16th and 84th percentiles of each photometric redshift posterior. The distribution peaks at a width of $\sim0.08$ and has a long tail toward higher widths. We fit a Gaussian to the bulk of the distribution, where the widths are $< 0.15$. The fit is shown as the solid line in Figure \ref{fig:photo_z_width} and has a mean of $\mu = 0.087$ and standard deviation of $\sigma = 0.027$. We exclude galaxies with widths $>3\sigma$ above the mean, 0.168, indicated by the dashed line in Figure \ref{fig:photo_z_width}. This cut removes 634 galaxies and leaves 9,696. Galaxies that are removed by this cut have similar distributions in stellar mass and SFR (which come from the physical parameter catalogs by \citealt{Santini15}) compared to the galaxies that are retained. Therefore, removing them does not bias our sample.

The remaining 9,696 galaxies are referred to in this paper as the ``full sample." Their distributions in redshift and color-color space are shown in the top row of Figure \ref{fig:sample}.

\subsection{The Mass-selected Sample: \textcolor{black}{$\log \left(\textrm{M}_{\star} / \textrm{M}_{\odot}\right)\geq 10$}}
\label{sec:mass_selected_sample}
This work relies on measurements of radial profiles of galaxies. To ensure a high S/N in the radial profiles, we select galaxies with stellar masses greater than \textcolor{black}{$\log \left(\textrm{M}_{\star} / \textrm{M}_{\odot}\right) \geq 10$}. The stellar masses are measured using SED-fitting (see Section \ref{sec:sed_model}). 

\begin{figure*}
    \centering
    \includegraphics[width=\textwidth]{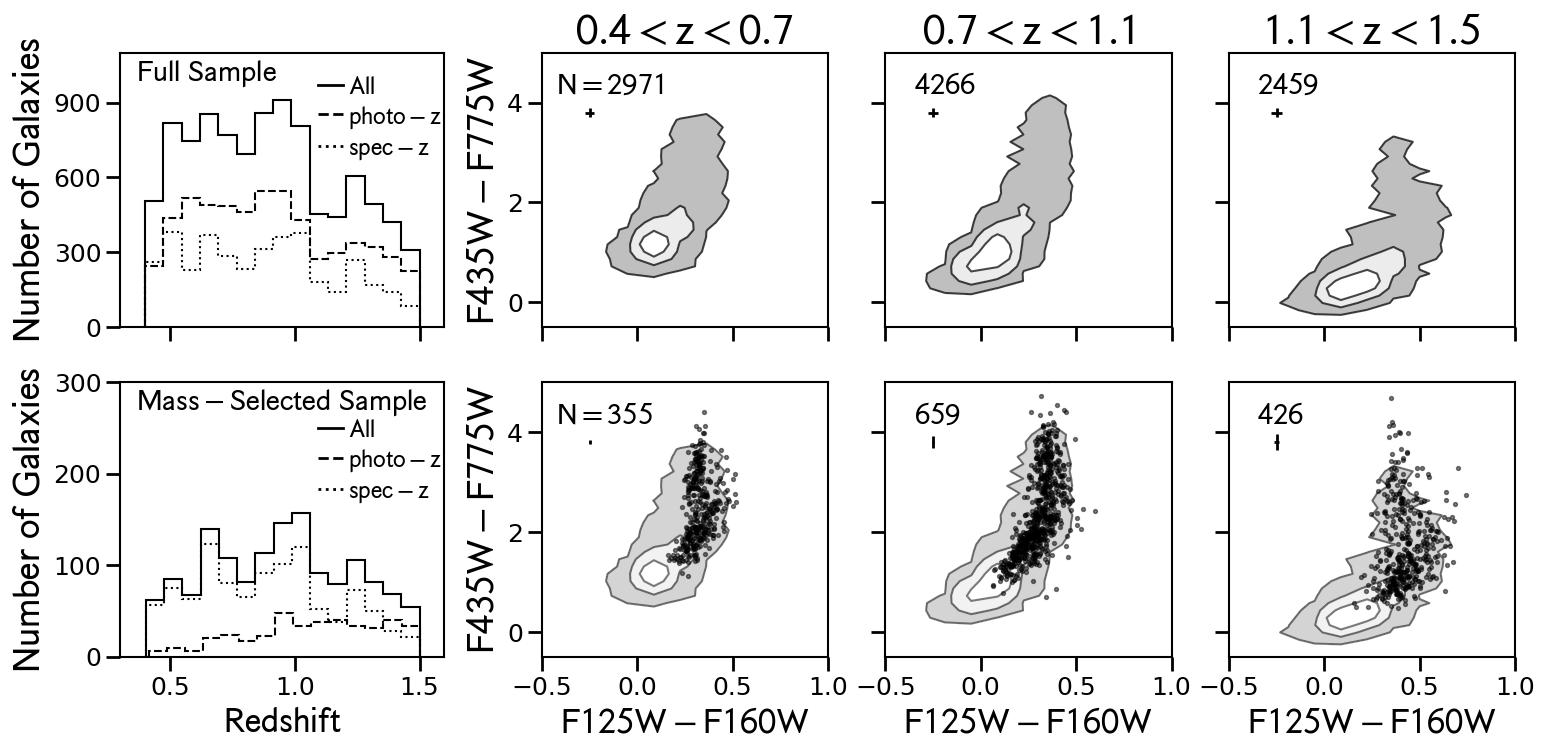}
    \caption{\textcolor{black}{The} distributions of redshifts and integrated optical-to-near-infrared (NIR) colors are shown for the full sample in the top row, and for the mass-selected sample in the bottom row. \textcolor{black}{The mass-selected sample consists of the reddest galaxies in the full sample, as expected. Higher-mass galaxies are observed to be older, dustier, more metal-rich, and more likely to be quiescent than lower-mass ones. Radial profiles are measured for galaxies in the mass-selected sample.}
    Distributions of all, photometric, and spectroscopic redshifts are shown as solid, dashed, and dotted lines, respectively. In the color-color plots for three redshift bins that we will use throughout the paper, contours represent the distributions of the full sample. White, light-gray, and dark-gray contours enclose 50\%, 16\%, and 2\%, respectively, of the maximum number density of the color-color distribution. Colors of the mass-selected sample are shown as black dots. The number of galaxies in each redshift bin and median \textcolor{black}{uncertainty} in each color are indicated in each panel.}
    \label{fig:sample}
\end{figure*}

In this paper, we examine five sets of SED-fitting assumptions. We select galaxies with stellar masses greater than or equal to $10^{10} \ \textrm{M}_{\odot}$ under any set of assumptions. These galaxies encompass the ``mass-selected sample.'' Of the 9,696 galaxies in the full sample, 1,909 are in the mass-selected sample. Its distributions in redshift and color-color space are shown in the bottom row of Figure \ref{fig:sample}. 
\textcolor{black}{The mass-selected sample makes up the reddest galaxies in the full sample. This is because galaxies are more likely to be older, dustier, more metal-rich, and quiescent at higher mass \citep[e.g.,][]{Gallazzi05, GarnBest10, Peng10}. }

\subsubsection{Visual Inspection \textcolor{black}{to Remove Contaminated and Merging Galaxies}}
\label{sec:visual_inspection}
\textcolor{black}{Lead author} A.dlV. visually inspected the galaxies in the mass-selected sample to remove major mergers, galaxies contaminated by neighbors, and \textcolor{black}{galaxies} whose images are affected by artifacts and diffraction spikes from bright stars. \textcolor{black}{A total of 32 major mergers and 426 galaxies that are too close to neighbors are removed. The 32 major mergers and 426 contaminated galaxies are evenly spread out in stellar mass and redshift; removing them does not affect our results}. Artifacts affect 11 galaxies, which are removed. A total of 1,440 galaxies remain.

Only galaxies that are obviously undergoing mergers (those that display features such as, e.g., multiple nuclei, off-center nuclei, or disturbed morphologies in close proximity to projected neighbors) or are close enough to neighboring galaxies such that their light profiles become difficult to distinguish, are removed. This is done because of the following. Merging and contaminated galaxies contain multiple generations of stellar populations whose SFHs are likely complex and cannot be modeled under the SED-fitting assumptions used in this work (Section \ref{sec:five_sets_assumptions} and Table \ref{tab:priors}). Consequently, inferences from fitting their SEDs are unreliable. 

\section{Construction of Flux Profiles}
\label{sec:flux_profiles}

\subsection{Measuring Radial Flux Profiles}
\label{sec:measure_flux_profiles}

\textcolor{black}{
Radial flux profiles are created for the 1,440 galaxies in the mass-selected sample (Section \ref{sec:mass_selected_sample}) in each of the HST bandpasses.  In Figure~\ref{fig:profile_example}, these radial flux profiles are shown for four example galaxies. Flux profiles are created in elliptical annuli starting in the center of each galaxy using the {\tt photutils} \citep{Bradley20} task {\tt EllipticalAnnulus}, the shapes of which do not change with radius. For the four example galaxies in Figure \ref{fig:profile_example}, they are shown as white ellipses. The galaxy centers and ellipse shapes are adopted from \citet{Guo13} and \citet{Barro19}, for GOODS-S and GOODS-N, respectively, for the F160W images.  We choose to use annuli widths of 2 pixels because it is greater than the PSF half-width at half maximum (1.5 pixels) and small enough to provide high spatial resolution. Over the redshift range studied, 2 pixels span 0.65 -- 1.03 kpc.  The radius of each annulus is defined to be the middle of each annulus.  The flux profile of each galaxy is measured out to a radius where the S/N in an annulus is $\geq3$ in two bandpasses. We restrict the profiles to even higher S/Ns in the reddest bandpass, as we explain below. This restriction is indicated by red circles and red vertical lines in Figure \ref{fig:profile_example}.}

\textcolor{black}{Given the radial flux profiles that reach an S/N of 3, we apply an additional restriction on the S/N of the reddest rest-frame bandpass. We chose to do this in the reddest bands to ensure the mass-to-light ratio is constrained at all radii, as it is most easily measured using the reddest bands \citep{Bell01, Taylor11, Ge21}. Specifically, we require an S/N of at least 5 in an annulus in F125W for $z \leq 0.8$ and F160W for $z > 0.8$. These wave bands correspond to the rest-frame R and I wave bands. Examples of these cuts are shown as red ellipses and red vertical lines in the bottom row of each panel in Figure \ref{fig:profile_example}. }

\textcolor{black}{For galaxies whose light profiles are contaminated by neighbors, we visually inspect the surface brightness profiles in F125W and F160W for galaxies at $z \leq 0.8$ and $z > 0.8$, respectively.  The presence of the neighbor is indicated by a ``bump'' in the profiles, and the flux profile is cut at the first local minimum we find before the bump. }

\begin{figure*}[t!]
    \centering
    \includegraphics[width=\textwidth]{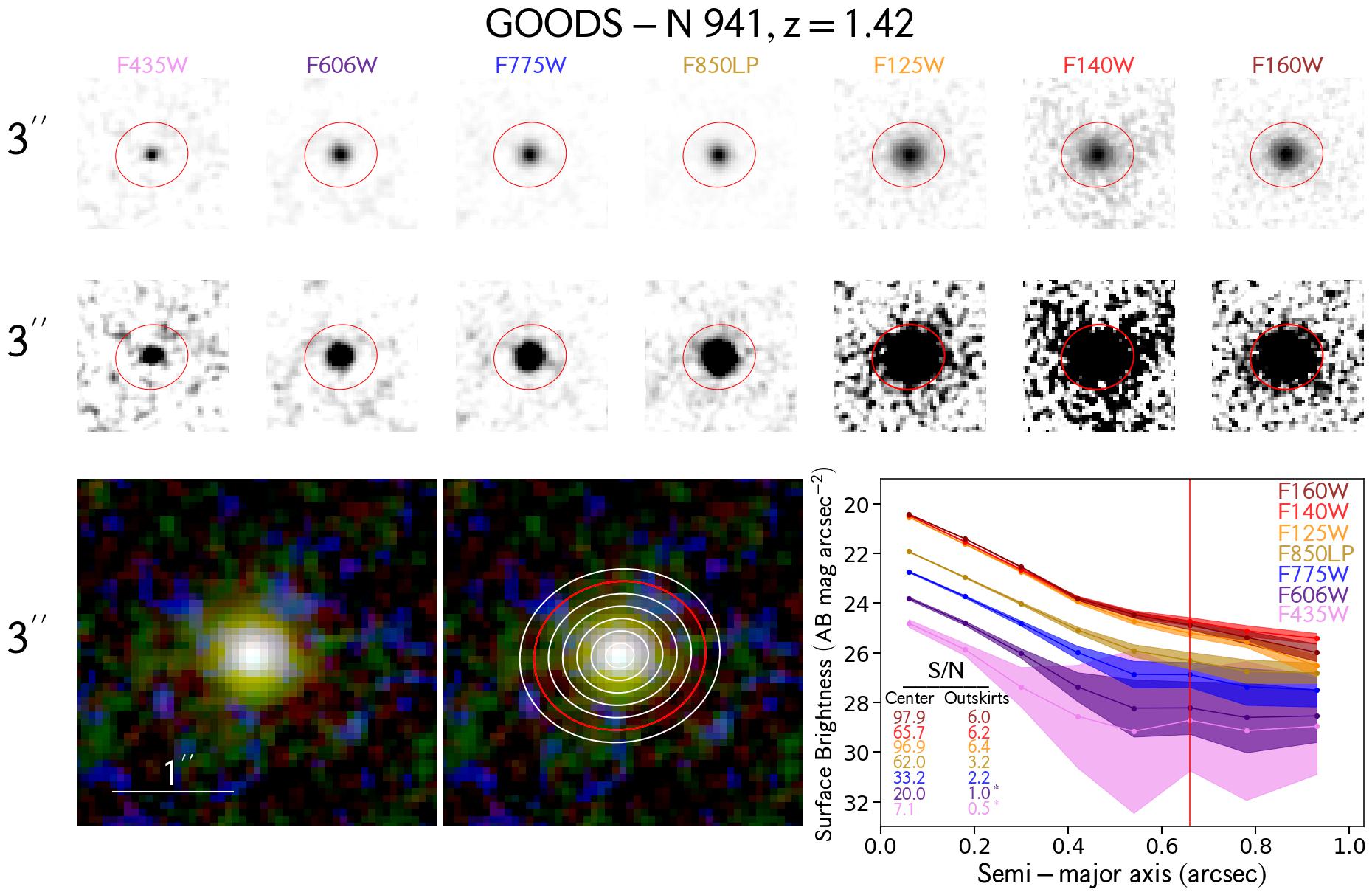}
    
    \caption{HST postage stamps, RGB images, and radial flux profiles of four example galaxies in our sample. Top and middle rows show postage stamps for all seven HST bandpasses in a low- and high-contrast stretch, respectively. The bottom row shows RGB images with and without annuli used to create the profiles drawn on top in white, and the resulting surface brightness profiles of the galaxies in all seven HST bandpasses. For GOODS-S 17058, we plot every other annulus for clarity. The RGB image is made from the F435W, F850LP, and F160W bandpasses. The red annuli in all images, and the red vertical lines in the surface brightness profile plots indicate the cuts in radius based on the S/N in F125W or F160W, depending on the redshift (Section \ref{sec:measure_flux_profiles}). For each galaxy, in the lower-left corner of the surface brightness profile plots, we list the S/N in the center and at the cut in radius above in all bandpasses. These share the same colors as the profiles. S/N values with asterisks indicate fluxes that are treated as upper limits in the SED fits (see Section \ref{sec:upper_limits}). }
    \label{fig:profile_example}
\end{figure*}

\addtocounter{figure}{-1}

\begin{figure*}
    \centering
    \includegraphics[width=\textwidth]{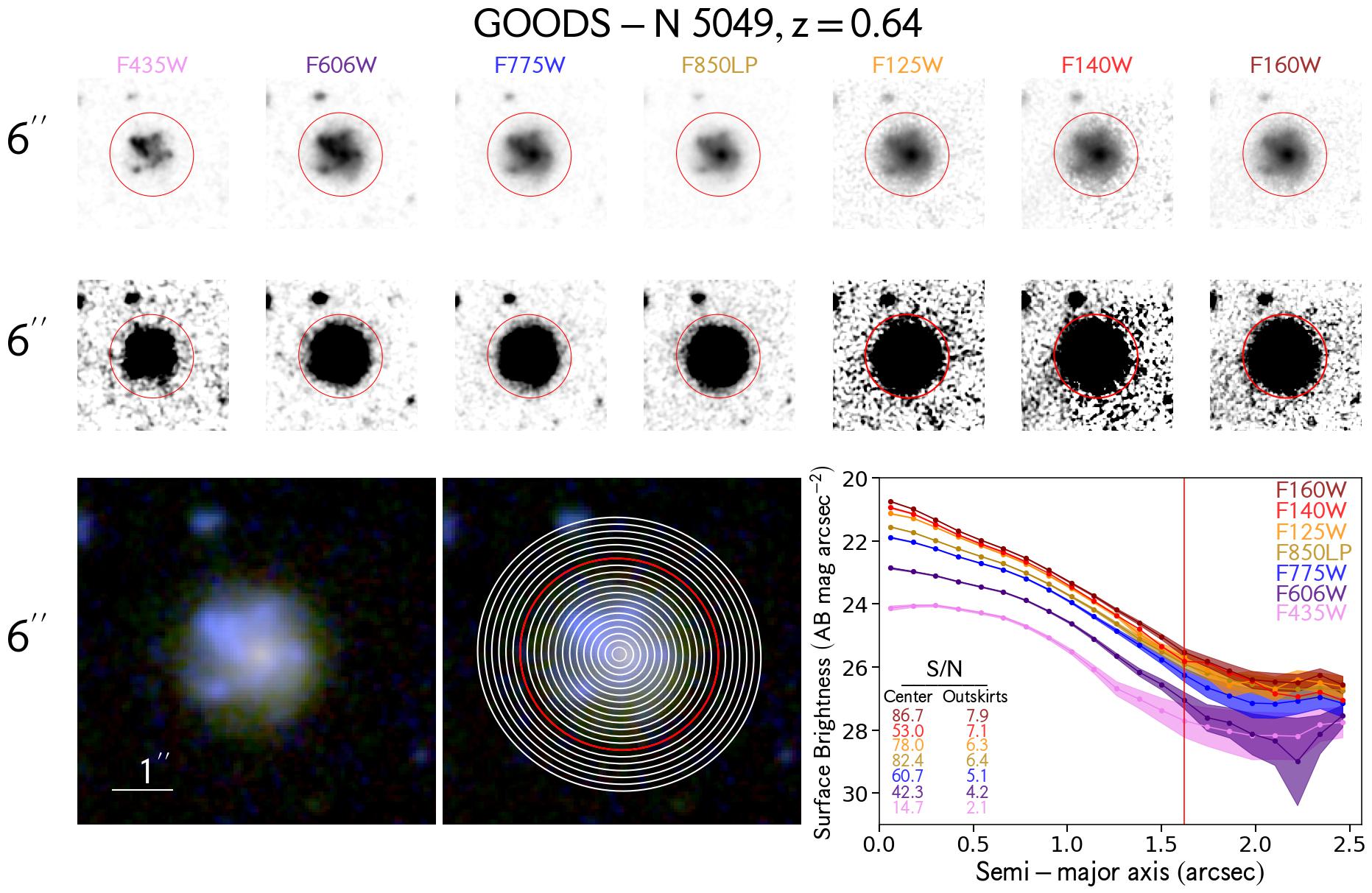}
    
    \caption{-- continued.}
    \label{fig:profile_example2}
\end{figure*}

\addtocounter{figure}{-1}

\begin{figure*}
    \centering
    \includegraphics[width=\textwidth]{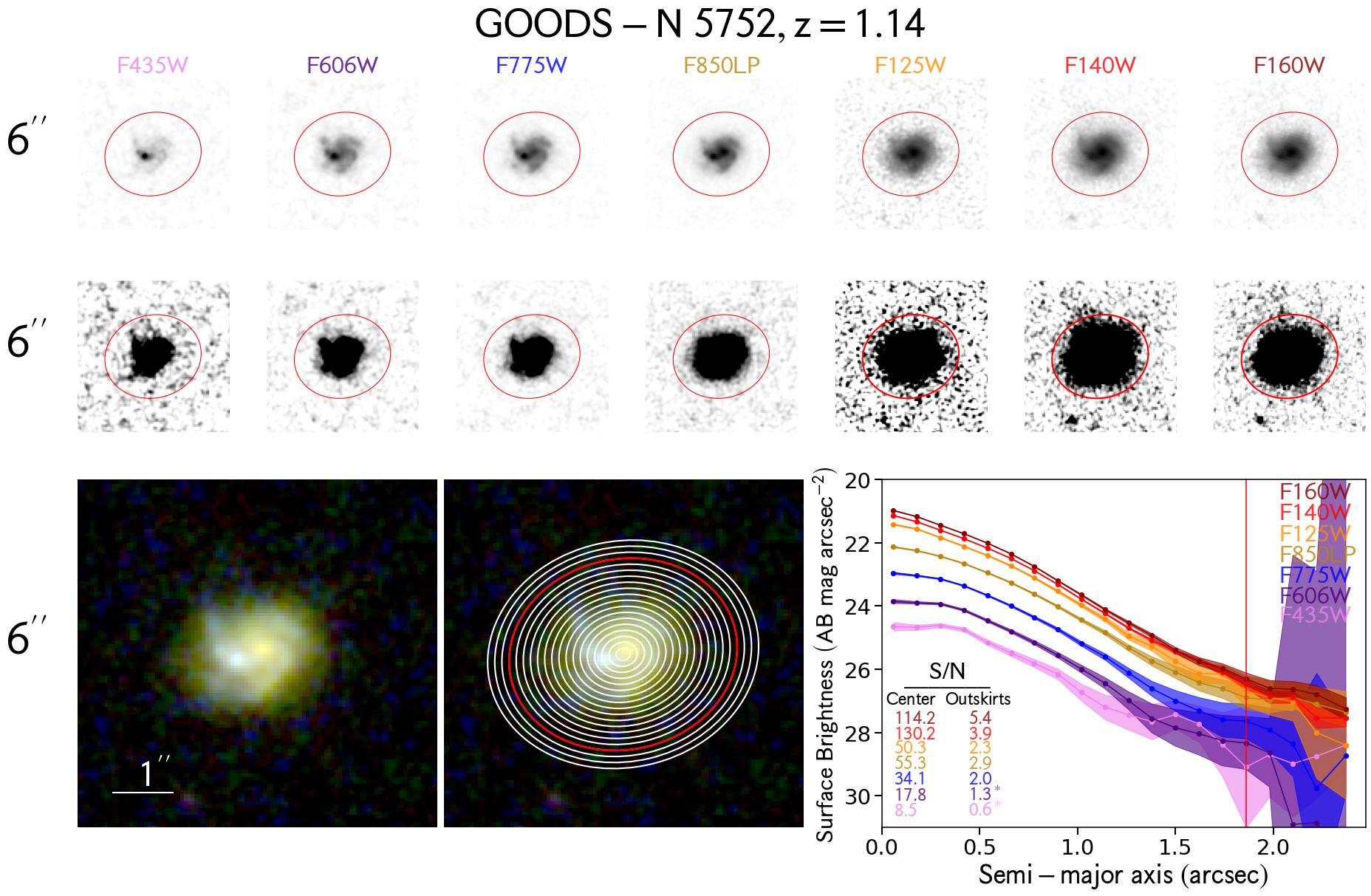}
    
    \caption{-- continued.}
    \label{fig:profile_example3}
\end{figure*}

\addtocounter{figure}{-1}

\begin{figure*}
    \centering
    \includegraphics[width=\textwidth]{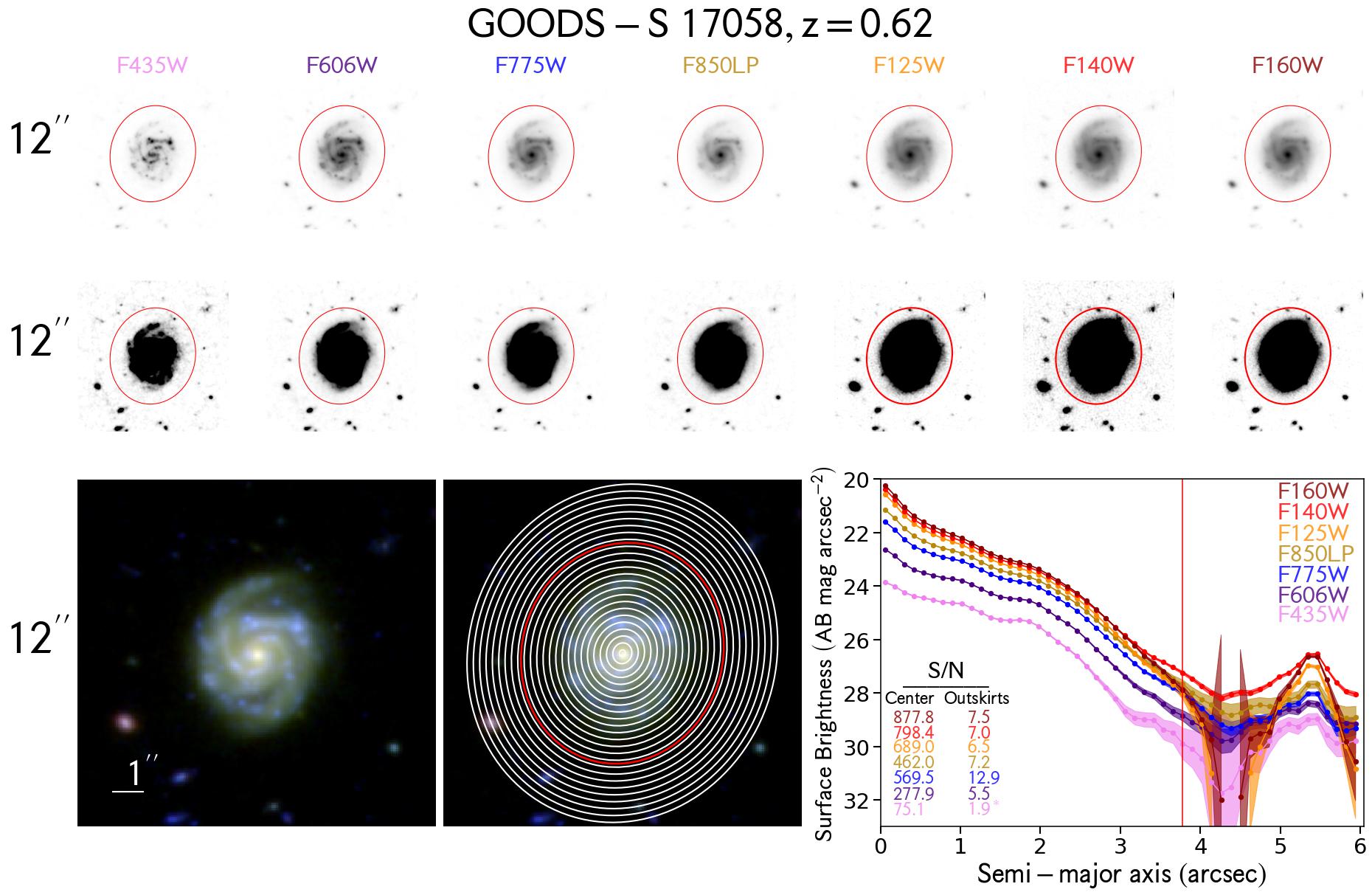}
    
    \caption{-- continued.}
    \label{fig:profile_example4}
\end{figure*}

\subsection{Uncertainties on the Flux Profiles}
\label{sec:prof_uncertainty}

\textcolor{black}{We account for three sources of uncertainty in the radial flux profiles: the uncertainty due to background subtraction, the uncertainty from the inverse variance weight images, and the Poisson noise from galaxies. The uncertainties on the flux profiles are dominated by those due to background subtraction (30-90\% of the uncertainty in each annulus) at most radii beyond the centers of galaxies, especially in the outer parts. Below we describe how these are calculated as they play an important role in determining how far out in radius we can accurately measure galaxy flux profiles. Another contribution to the uncertainty is from the HST inverse variance weight images (30-60\% of the uncertainty). These take into account noise from flat-fielding, dark current, and readout, as well as the sky background Poisson noise \citep{Casertano00, Koekemoer11}. Finally, although dominant (20-60\% of the uncertainty) only in the centers of galaxies and nearly negligible ($\lesssim 10$\%) at larger radii, Poisson noise from the galaxies is taken into account.}

Each of these \textcolor{black}{sources of uncertainty} is added in quadrature to yield the total uncertainty in flux for a given aperture. In Figure \ref{fig:profile_example}, the shading around each radial flux profile \textcolor{black}{indicates} the total flux uncertainty. In general, the uncertainty in any bandpass is small wherever the galaxy is bright. \textcolor{black}{In the centers, where the S/N is the highest, the Poisson noise is $\lesssim 0.01~\textrm{mag~arcsec}^{-2}$ and other sources of uncertainty are of similar size or smaller.} At radii where the galaxy is faint, the uncertainty grows. The uncertainty is largest at the farthest distances \textcolor{black}{from the centers of galaxies} because \textcolor{black}{the surface brightness there is the lowest}. At these distances, the errors due to sky subtraction and \textcolor{black}{those quantified in the} weight \textcolor{black}{images} dominate. \textcolor{black}{In these regions, errors due to sky subtraction are often $\sim0.2 - 0.3~\textrm{mag~arcsec}^{-2}$ and those quantified in the weight images are $\sim0.1 - 0.2~\textrm{mag~arcsec}^{-2}$.}

\textcolor{black}{To obtain the uncertainty from the weight image, first the weight image is inverted to get a variance image. Then, for a given annulus, the square of each pixel value is summed for all pixels in the annulus. The error from the weight image is taken as the square root of this sum.}

\textcolor{black}{The Poisson noise is measured as follows. The science image is first multiplied by the exposure time image to get a Poisson variance image. Then, the square root of this image is divided by the exposure time image. For a given annulus, the square of each pixel value is summed. The Poisson noise is taken as the square root of this sum.}

\textcolor{black}{A final} uncertainty is added to all fluxes (for each bandpass and radial bin) to account for zero-point errors and mismatches between the template SEDs and observed SEDs. This additional uncertainty is \textcolor{black}{set to be} 5\% of the flux (see, e.g., Table 4 of \citealt{Dahlen13} and Table 11 of \citealt{Skelton14}).

\subsubsection{Uncertainties due to Background Subtraction}

\begin{figure}
    \centering
    \includegraphics[width=\columnwidth]{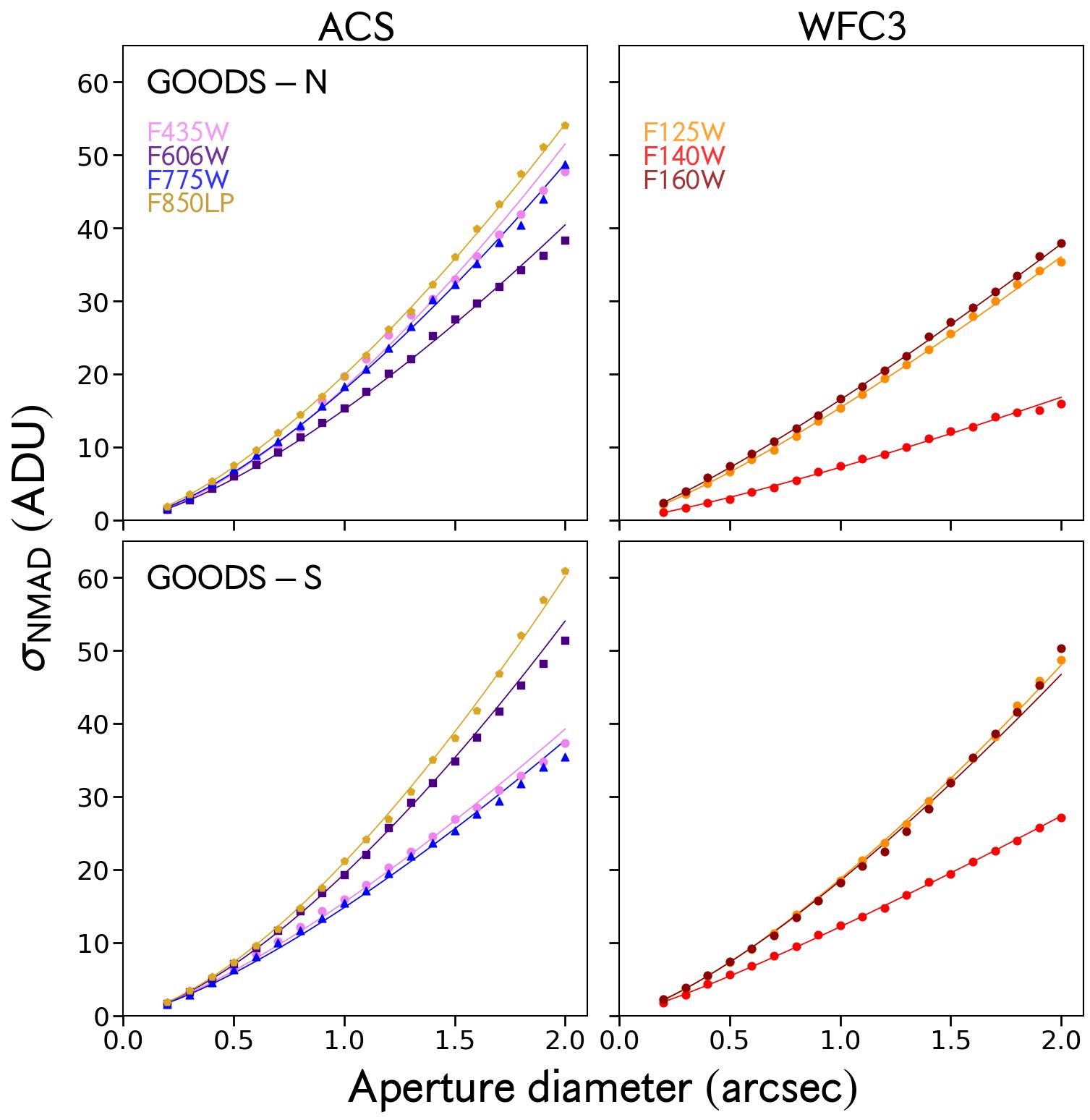}
    \caption{The uncertainty due to background subtraction \textcolor{black}{is calculated by first performing aperture photometry on noise-equalized images in empty regions and in circular apertures of various radii. Then, the normalized median absolute deviation ($\sigma_{\textrm{NMAD}}$) of the flux distribution for apertures of a given radius is measured, and a power law is fit to $\sigma_{\textrm{NMAD}}$ as a function of aperture area in pixels. Lastly, $\sigma_{\textrm{NMAD}}$ is divided by the average of the square root of the weight map within an aperture. For an aperture of radius equal to 0.\arcsec25, the uncertainty due to background subtraction in F160W is $\sim1$ nJy in the Hubble Ultra Deep Field and $\sim10$ nJy in the Wide portion of GOODS-S. Here, we show $\sigma_{\textrm{NMAD}}$ as a function of the aperture diameter} in the four ACS bands (left) and the three WFC3 bands (right) for GOODS-N in the top row and GOODS-S in the bottom row. The \textcolor{black}{$\sigma_{\textrm{NMAD}}$ measured} in the F435W, F606W, F775W, and F850LP bandpasses \textcolor{black}{is} plotted as dots, squares, triangles, and pentagons, respectively. The power-law fits to each set of $\sigma_{\textrm{NMAD}}$ values in each bandpass are shown as thin lines. Colors \textcolor{black}{indicate the bandpass and} are the same as in Figure \ref{fig:profile_example}. The uncertainty due to background subtraction rises quickly with increasing aperture size and is steeper for the ACS bands than for the WFC3 bands, corresponding to differences such as PSF, detector pixel scale, readnoise, and background level, which combine to produce somewhat different scalings in each case. \textcolor{black}{This uncertainty is taken into account in the radial profiles.}}
    \label{fig:bckg_error}
\end{figure}

Uncertainties due to background subtraction are measured using the 
empty aperture technique \citep{Labbe03, Wuyts08, Whitaker11, Skelton14}. \textcolor{black}{In this technique}, the normalized median absolute deviation (NMAD) of the background is measured for a range of aperture sizes. \textcolor{black}{In practice}, apertures are randomly placed in empty regions of the \textcolor{black}{images} such that they do not overlap each other, any galaxies, or the edges of the \textcolor{black}{images}. \textcolor{black}{A power law is then fit} to the relation between NMAD and aperture size. \textcolor{black}{In Figure \ref{fig:bckg_error}, the NMAD of the background is shown as a function of aperture diameter in all bandpasses as dots. Power-law fits to the NMAD as a function of aperture diameter are shown as lines.} Full details are as follows. 

First, the \textcolor{black}{science image} is multiplied by the square root of the \textcolor{black}{weight image} to produce a noise-equalized image. Noise equalization is crucial, as the HST \textcolor{black}{images} in certain bandpasses vary widely in depth. Next, 2,000 circular apertures of radius 1\arcsec\ are randomly placed \textcolor{black}{on} the noise-equalized image. The apertures are placed such that no aperture overlaps with another aperture, a segmentation map of a galaxy as identified by Source Extractor (from \citealt{Guo13} and \citealt{Barro19}), or with the edge of the image. Aperture photometry is performed in the noise-equalized image, and the NMAD of the resulting flux distribution is computed. This process is repeated for various aperture sizes, ranging in radius from 0.\arcsec1 to 1.\arcsec0, as shown by the points in Figure \ref{fig:bckg_error} and \textcolor{black}{is} performed \textcolor{black}{for} all HST bands using the same apertures.

To calculate the error due to background subtraction for any aperture size, we fit a power law of the form $\sigma_{\textrm{NMAD}} = \alpha ( \sqrt{A} )^{\beta}$, indicated by the lines in Figure \ref{fig:bckg_error}, where $\sigma_{\textrm{NMAD}}$ \textcolor{black}{indicates} the NMAD values measured in \textcolor{black}{various} aperture sizes, and $A$ is the area \textcolor{black}{of the aperture in pixels}. The fit parameters \textcolor{black}{for the GOODS-N and GOODS-S fields are listed separately} in Table \ref{tab:bckg_sub_err}. The uncertainty due to background subtraction in \textcolor{black}{an} aperture is computed by calculating $\sigma_{\textrm{NMAD}}$ using the power-law fits \textcolor{black}{to $\sigma_{\textrm{NMAD}}$ as a function of aperture area} and dividing by the average value of the square root of the weight \textcolor{black}{mosaic} within the aperture. This last step ensures the uncertainty has the same units as the science \textcolor{black}{mosaic}. \textcolor{black}{For a galaxy of brightness $\sim10~\mu\textrm{Jy}$ in F160W with outskirts that have a surface brightness of $\sim26~\textrm{mag~arcsec}^{-2}$, the uncertainty due to background subtraction in an annulus of area $\sim1$ square arcsec is $\sim0.1~\textrm{mag~arcsec}^{-2}$.}

Less than half a percent of galaxies in the mass-selected sample (six \textcolor{black}{out of} 1,440) possess apertures \textcolor{black}{with} areas larger than the largest aperture used to measure the uncertainty due to sky subtraction. For these galaxies, $\sigma_{\textrm{NMAD}}$ is extrapolated from the power-law fits. \textcolor{black}{The largest annulus has an area that is 39\% larger and an error that is 29\% larger than the largest aperture used to measure the error due to background subtraction.}

\begin{table*}
    \caption{Power-Law Parameters for Uncertainties due to Background Subtraction\\For each parameter and bandpass combination, the first line gives the value for GOODS-N and the second line gives the value for GOODS-S.}\label{tab:bckg_sub_err}
    
    \begin{center}
    \begin{tabular}{l c c c c c c c}
    \hline \hline
    Parameter & F435W & F606W & F775W & F850LP & F125W & F140W & F160W \\
    \hline
        
    $\alpha$ & 0.31 & 0.33 & 0.36 & 0.40 & 0.56 & 0.27 & 0.65 \\
    & 0.42 & 0.36 & 0.39 & 0.35 & 0.48 & 0.52 & 0.50 \\
        
    $\beta$ & 1.51 & 1.42 & 1.45 & 1.45 & 1.23 & 1.22 & 1.20 \\
    & 1.34 & 1.48 & 1.35 & 1.52 & 1.36 & 1.17 & 1.34 \\
    \hline

    \end{tabular}
    \end{center}
\end{table*}

\section{SED Modeling}
\label{sec:sed_model}

\textcolor{black}{In this Section, we fit the SED at each radius in the flux profiles with the {\sc beagle} SED-fitting tool to obtain profiles of star-formation rate, stellar mass, age, and dust \citep[][version 0.20.2]{Chevallard16}. Additionally, we fit the integrated fluxes of the galaxies to obtain integrated stellar masses and star-formation rates.}

\subsection{The BEAGLE SED-fitting Tool and Treatment of Its Output}
\label{sec:beagle}
 
{\sc beagle} has multiple options for star-formation and chemical enrichment histories and dust attenuation. We assume delayed-$\tau$ star-formation histories and that metallicity is the same for all stars at a given radius. We use the empirical starburst attenuation curve measured by \citet{Calzetti94} and the attenuation curve relations derived by \citet{Chevallard13}. {\sc beagle} can fit SEDs using stellar and nebular emission templates simultaneously, and we use this option. For stellar emission, it uses the updated version of the \citet{BC03} stellar population synthesis (SPS) models described by \citet[][see also \citealt{VidalGarcia2017}]{Gutkin16}. For nebular emission, it adopts the photoionization models developed by \citet{Gutkin16}, which combined the updated \citet{BC03} models with models of photoionized interstellar gas from {\sc cloudy} \citep{Ferland13}. {\sc beagle} adopts a Bayesian framework. It uses the {\sc MultiNest} tool \citep{Feroz09} to efficiently explore parameter space and accurately quantify uncertainties on fit parameters.

For each physical property of interest (e.g., stellar mass, SFR, \AvNoSpace, and \textcolor{black}{mass-weighted age}), {\sc beagle} returns a posterior distribution. We use the median of the posterior as the most representative value of the physical property. 
The majority of the sample shows little to no difference between the median and the maximum a posteriori, the peak value of the posterior PDFs, suggesting that the distributions are not heavily skewed or multimodal.
We use SFRs averaged over the last 100 Myr in this work. Stellar mass and SFR surface densities are measured by taking the median stellar mass and SFR from their respective posteriors and dividing by the area of the annulus that these properties were measured in. We measure sSFRs by dividing the SFR surface density by the stellar mass surface density. With this process, we obtain radial profiles of physical properties (hereafter ``science profiles'') for the physical parameters.

\subsection{Upper Limits on Flux Profiles}
\label{sec:upper_limits}

When fitting the SED at each galaxy radius, if the flux in a given bandpass \textcolor{black}{has} S/N $\leq 2$, it is customary to remove it from the SED fit. However, we \textcolor{black}{choose to treat it as an upper limit. In practice, this is done by replacing the flux value with a number that is close to zero and its uncertainty with the value of the flux. This was done in a similar manner by} \citet{Shanks21}. 
\textcolor{black}{The flux is replaced with an arbitrarily low value, $10^{-20}$ Jy, which is} well below the sensitivity limits of the images. 
\textcolor{black}{When these replacements are made, in the SED fits, the model fluxes are forced to be lower than the observed ones. This functions as an upper limit.}
    
\subsection{Priors in SED Fitting}
\label{sec:priors}

\begin{figure*}[t!]
    \centering
    \includegraphics[width=\textwidth]{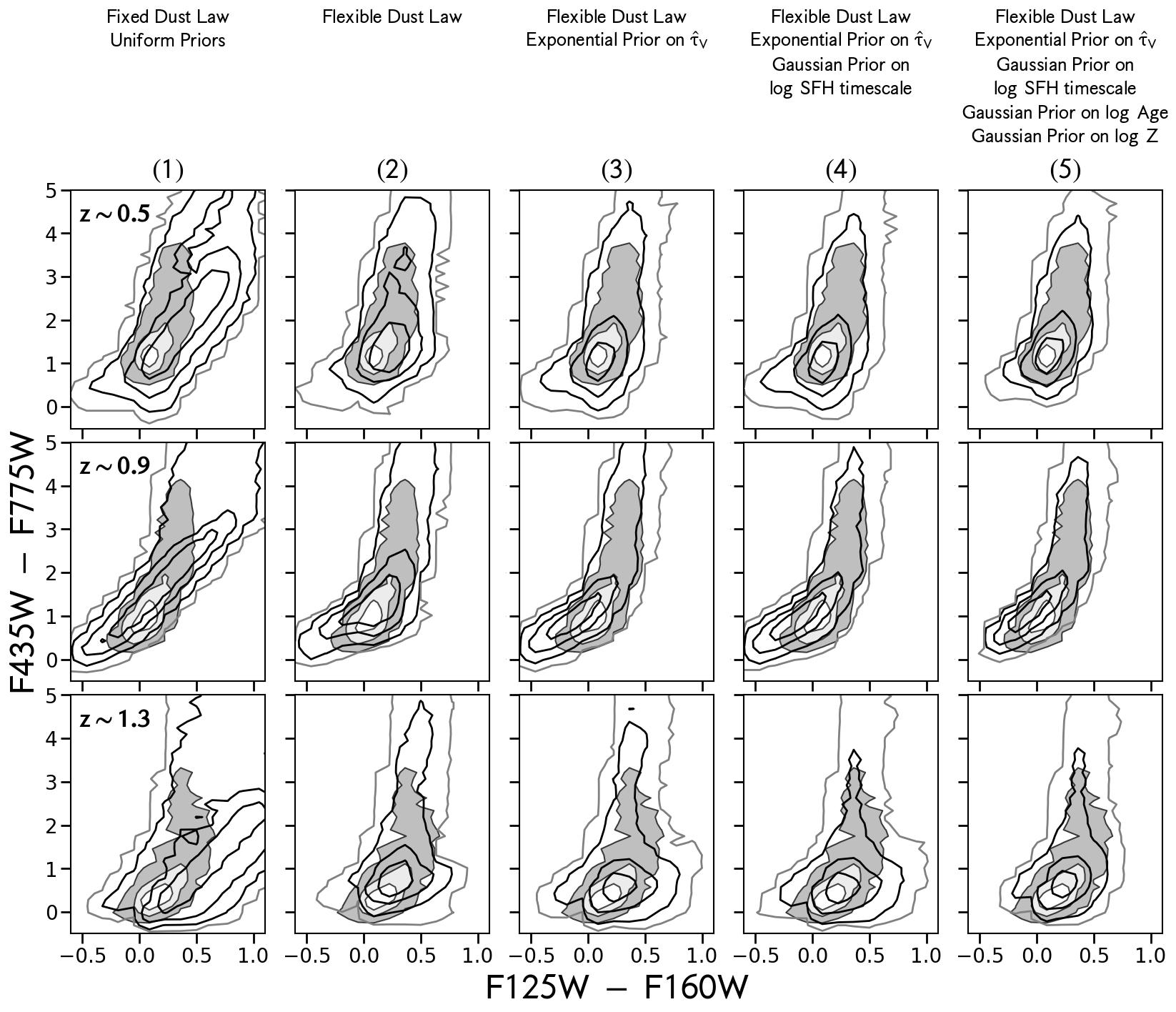}
    \caption{Observed and model color-color distributions for the full galaxy sample \textcolor{black}{for} the five sets of assumptions \textcolor{black}{are shown as open and filled contours, respectively. Models under the first set of assumptions do not trace the observations. As assumptions are modified, there is progressively greater overlap until the fifth set is reached, where the overlap is highest. A different set of model assumptions is shown in each column. Three redshift bins are shown from top to bottom. Contour levels are 50\%, 16\%, and 2\% of the maximum number density of models or data. For the models, an additional contour at 0.1\% is shown. Note that the offset in the bulk colors of the models at $z\sim0.9$ relative to the observations results from emission lines contaminating the WFC3 colors and the inclusion of extreme values in the priors on the nebular emission (Section \ref{sec:considerations_for_priors}).}
    }
    \label{fig:model_obs_colors}
\end{figure*}

One of the main aspects of this paper is an examination of five different sets of priors in our {\sc beagle} SED fits, which we detail in the following Section.  Priors, or existing knowledge about the expected distributions of parameters being fit for (e.g., stellar mass, star-formation rate) should be appropriate for the data being fit. For example, star-forming galaxies generally contain dust, so priors on the amount of dust should allow for high attenuation values. Whether priors are appropriate for the galaxies being fit can be tested by comparing the colors of models with those of the galaxies, as we do in Figure \ref{fig:model_obs_colors}. The models should trace the data -- the empty contours in Figure \ref{fig:model_obs_colors} should trace filled contours.  If the models are too narrow for the data, as in the leftmost panels in \textcolor{black}{the Figure}, the SED fits will lead to artificially narrow posterior distributions \citep[e.g.,][]{CurtisLake13, Pacifici15}. On the other hand, if the models are distributed too broadly relative to the data, degeneracies between model parameters and uncertainties in the data may lead to unrealistic fits \citep[e.g.,][]{Li21}. Furthermore, models that are overly complex (e.g., too many parameters relative to \textcolor{black}{the constraining power of} the data or model parameters that are highly degenerate with each other) may result in poorly constrained results or overfitting \citep[e.g.,][]{Lower20, CurtisLake21}.

\section{Five Sets of Assumptions in SED Fitting}
\label{sec:five_sets_assumptions}

We examine the effects of adopting five different sets of SED-fitting assumptions on the output radial properties of galaxies. These sets of assumptions incorporate different dust laws in addition to different priors on the parameters that we fit. They are described in Table \ref{tab:priors}. We find that the most important priors are those on dust attenuation and star-formation histories, as discussed in Section \ref{sec:effects_assumptions}. From left to right in Table \ref{tab:priors}, for each set of assumptions, we list the priors assumed for each parameter. We refer to these sets \textcolor{black}{as} first to fifth. They start with commonly used assumptions in the first set, and progressively incorporate modifications until we reach the fifth set, whereby all SED-fitting parameters have been modified. In the second and third sets, changes are in the priors for the treatment of dust in the models. Specifically, the priors on the attenuation curve and the amount of dust in the models are varied. For the fourth set, the prior on $\tau$, the star-formation history timescale, is varied. Finally, for the fifth set, we additionally change the priors on the stellar ages and metallicity. The effects of changing the priors on the star-formation histories are discussed in Appendix \ref{sec:priors_sfh}.

We note that the priors on the stellar mass and the initial mass function are kept the same for all assumptions. \textcolor{black}{The prior on the log of the stellar mass (units of solar masses) is uniform over the range [5, 13], and we adopt the initial mass function developed by \citet{Chabrier03}.}
\textcolor{black}{Also}, for all sets of assumptions, the priors on the nebular emission are the same. \textcolor{black}{We assume that the nebular and stellar metallicities are equal, the dust-to-metal mass ratio is fixed to 0.3 \citep[e.g.,][]{Brinchmann13}, and the ionization parameter behaves according to eq. 25 in \citet{Chevallard16}}.

\begin{deluxetable*}{l c c c c c}[t!]
\tablecaption{Priors used in SED-fitting \label{tab:priors}}

\tablewidth{\textwidth}
\tablehead{
& Fixed Dust Law & Flexible Dust Law & Flexible Dust Law & Flexible Dust Law & Flexible Dust Law \\
& Uniform Priors & & Exponential Prior & Exponential Prior & Exponential Prior \\
& & & on \tauVeff & on \tauVeff & on \tauVeff \\
& & & & Gaussian Prior on & Gaussian Prior on \\
& & & & log SFH timescale & log SFH timescale \\
& & & & & Gaussian Prior \\
& & & & & on log Z \\
& (1) & (2) & (3) & (4) & (5) \\
\hline
Parameter & functional form & functional form & functional form & functional form & functional form \\
& range & range & range & range & range \\}
\startdata
SFH & delayed-tau & $\ldots$ & $\ldots$ & $\ldots$ & $\ldots$ \\
& -- & -- & -- & -- & -- \\
log Age\tablenotemark{a} & uniform & $\ldots$ & $\ldots$ & $\ldots$ & $\mathcal{N}$(9.0, 1.0)\tablenotemark{i} \\
\vspace{0.1cm}
& [8.0, age of Universe] & $\ldots$ & $\ldots$ & $\ldots$ & [8.5, age of Universe] \\
log $\tau$\tablenotemark{b} & uniform & $\ldots$ & \ldots & $\mathcal{N}$(9.0, 1.0)\tablenotemark{h} & $\ldots$  \\
\vspace{0.1cm}
& [8.0, 10.0] & $\ldots$ & $\ldots$ & [8.3, 10.0] & $\ldots$  \\
$\log\left(\textrm{Z}_{\star} \ / \ \textrm{Z}_{\odot}\right)$\tablenotemark{c} & uniform & $\ldots$ & $\ldots$ & $\ldots$ & $\mathcal{N}$(-0.6, 0.8)\tablenotemark{i} \\
\vspace{0.1cm}
& [-1.0, {\bf 0.3}] & $\ldots$ & $\ldots$ & $\ldots$ & $\ldots$ \\
\tauVeffNoSpace\tablenotemark{d} & uniform & $\ldots$ & exponential\tablenotemark{e} & $\ldots$ & $\ldots$ \\
\vspace{0.1cm}
& [0.0, 4.0] & [0.0, 6.0] & $\ldots$ & $\ldots$ & $\ldots$ \\
Attenuation & eq. 25  & eqs. 7-11 & $\ldots$ & $\ldots$ & $\ldots$ \\
\vspace{0.1cm}
Curve & \citetalias{Calzetti94} & \citetalias{Chevallard13} \\
$\mu$\tablenotemark{f} & -- & uniform & $\ldots$ & $\ldots$ & $\ldots$ \\
& -- & [0.1, 0.7] & [0.1, 1.0]\tablenotemark{g} & $\ldots$ & $\ldots$ \\
\hline
\enddata
         

\tablecomments{Ellipses ($\ldots$) indicate that priors are the same as the last column to the left containing text. References in table: \citet[][text]{Calzetti94} (\citetalias{Calzetti94}) and \citet{Chevallard13} (\citetalias{Chevallard13}). $\mathcal{N}$($x,y$) denotes a Gaussian distribution with mean $x$ and standard deviation $y$. \\
$^a$ Log of the maximum stellar age in yr. \\
$^b$ Log of the SFH timescale in yr, assuming a delayed-tau star-formation history ($\textrm{SFR}(t) \propto t e^{-t / \tau}$). \\
$^c$ Log of the stellar metallicity in solar units. We assume $\textrm{Z}_{\odot} = 0.0152$ \citep{Caffau11, Bressan12}. \\ 
$^d$ Effective $V$-band optical depth; see \citet{CF00, daCunha08, Chevallard13}. \\ $^e$ The exponential prior is $P(\hat{ \tau }_V) = e^{ -\hat{ \tau }_V }$. This prior is assumed because the distribution of \Av of nearby galaxies in SDSS is approximately exponential \citep{Brinchmann04}. We expect that the distribution of \Av is similar to that in SDSS at higher redshifts. \\
$^f$Fraction of optical depth arising from the ISM, see \citet{CF00, daCunha08, Chevallard13}. \\
$^g$ We adjust the prior on $\mu$ as well in this set as expanding the range of $\mu$ is a small effect. \\
$^h$ A broad Gaussian prior on log $\tau$ is assumed as massive quiescent galaxies at $z\sim1$ have SFHs  that peak $\sim1-2$ Gyr after they begin forming stars \citep{Pacifici16}. \\
$^i$ The broad Gaussian prior on log Age is assumed as massive galaxies at $z\sim1$ have light-weighted ages of $\sim1$ Gyr \citep{Hathi09, Onodera15, EstradaCarpenter19, Costantin22}. A broad Gaussian prior on log $\textrm{Z}_{\star}$ is assumed to account for the range in metallicity typically found for massive galaxies at $z\sim1$ \citep{Maiolino08, Henry13, Guo16a, Maiolino19}. }
\end{deluxetable*}

\subsection{The First Set: Fixed Dust Law and Uniform Priors}
\label{sec:first_set}

The main features of the first set of assumptions are a fixed attenuation curve and uniform priors \textcolor{black}{on} all SED-fitting parameters. These features are informed by the priors commonly adopted in SED fitting (e.g., \citealt{Wuyts12, Hemmati14, Hemmati20, Cibinel15, Morishita15, Mosleh17, Mosleh20, Abdurrouf18, Guo18, Sorba18, Morselli19, Suess19a}. The attenuation curve adopted is that of \citet{Calzetti94} and the optical depth of dust in the rest-frame $V$ band, \tauVeffNoSpace{}, spans the range \textcolor{black}{0--4}. 

\textcolor{black}{Model colors under this set of assumptions do not trace observed colors, as shown by the poorly overlapping sets of contours in Figure \ref{fig:model_obs_colors}. The overlap is quantified by measuring the fraction of model galaxies that lie within a rectangular region in color-color space that encompasses 90\% of the observations: $-0.25 < \textrm{(F125W - F160W)} < 0.5$ for $z \sim0.5$ and $z\sim 0.9$ and 
$0 < \textrm{F435W - F775W} < 2.7$ for $z\sim1.3$. Only 51\%, 65\%, and 29\% of the model colors in the first set lie in the same region in color-color space in the lowest, middle, and highest redshift bin, respectively.  We conclude from this that the first set of assumptions may not be the best choice for this data set.}

\textcolor{black}{This mismatch between data and models can be attributed mainly to the priors on dust attenuation and stellar age.} As only one attenuation curve is assumed, the reddening due to dust in color-color space only varies along one direction; hence, the models concentrate in a swath. A uniform prior on the attenuation means that dust-free and very dusty models are equally likely, so a large fraction of the models lie at colors much redder than the data \citep[see also][]{Harvey25}. Uniform priors on the log maximum age and log SFH timescale yield star-formation histories that are rising or roughly constant, producing an abundance of blue colors. 

\subsection{The Second Set: Flexible Dust Law}
\label{sec:second_set}

For the second set of assumptions, the relation between the slope of the attenuation curve and the rest-frame $V$-band optical depth in the interstellar medium (ISM) developed by \citet{Chevallard13} is adopted. This is because multiple theoretical and observational studies have found that galaxies possess diverse attenuation curves (see \citealt{Salim20} for a review on the diversity of attenuation curves). 

The \citet{Chevallard13} relation we adopt is (their Equation 10)\textcolor{black}{:}
$$n_V^{\textrm{ISM}}(\hat{\tau}_{V}^{\textrm{ISM}}) = \frac{2.8}{1+3\sqrt{\hat{\tau}_{V}^{\textrm{ISM}}}},$$
where $n_V^{\textrm{ISM}}$ is the power-law slope of the attenuation curve in the rest-frame $V$ band ($\hat{\tau}(\lambda) \propto \lambda^{-n}$), $\hat{\tau}_{V}^{\textrm{ISM}} = \mu$\tauVeff{} is the optical depth in the rest-frame $V$ band due to dust in the ISM, and 
$$\mu = \frac{\hat{\tau}_{V}^{\textrm{ISM}}}{\hat{\tau}_{V}^{\textrm{BC}} + \hat{\tau}_{V}^{\textrm{ISM}}},$$
where $\hat{\tau}_{V}^{\textrm{BC}}$ is the optical depth of the dust originating in the birth clouds surrounding young stars \citep[see][]{CF00}. 
Taken together, these equations imply myriad attenuation curves can be fit under this set of assumptions with varying contributions from young stars and the diffuse ISM. Qualitatively, \textcolor{black}{the resulting attenuation curves are steep at low optical depths and shallow at high optical depths.}
\textcolor{black}{Because} dust from young stars and the diffuse ISM can contribute to \tauVeff{} in varying amounts, which does not occur under the first set of assumptions, the range on \tauVeff{} is expanded to 0--6 \textcolor{black}{and a uniform prior is assumed}. We adopt this range from \citet{Walcher08}.
We set priors on both parameters of the \citet{Chevallard13} model, \tauVeff and $\mu$. The slope of the attenuation curve, $n_V^{\textrm{ISM}}$, is analytically linked to the optical depth and is not an independent parameter.

The \citet{Chevallard13} relation is based on the results of  radiative transfer studies by \citet{Silva98}, \citet{Pierini04}, \citet{Tuffs04}, and \citet{Jonsson10}. Observationally, several studies find qualitative agreement between observed variations in attenuation curves and what is predicted by the \citet{Chevallard13} attenuation curve relation at low \textcolor{black}{redshift} (\citealt{Wild11, Battisti17, Leja17, Salim18}; though see \citealt{Qin22}) and high redshift \citep{Kriek13, Salmon16, Tress18, Tress19, Barisic20, Nagaraj22}. 

Under the second set of assumptions, in the column second-from-the-left in Figure \ref{fig:model_obs_colors}, the model colors trace the data much better than under the first set of assumptions. In the lowest, middle, and highest redshift bins, 83\%, 83\%, and 84\% of the models in the second set lie in the color-color box defined in Section \ref{sec:first_set}, respectively. The improved overlap between observations and models is a consequence of the attenuation curve being allowed to vary. The model colors do not extend to colors as red as those under the first set of assumptions because the prior on $\mu$ does not permit many models to be dominated by dust from the ISM. However, \textcolor{black}{in Figure \ref{fig:model_obs_colors},} the innermost empty contours, i.e., the bulk of the model colors, are redder than the observations. This is due to the uniform prior on \tauVeffNoSpace. \textcolor{black}{The median ISM optical depth for this prior is 1.2.  Therefore, many models have optically thick ISM dust and therefore redder colors than observed.}

\subsection{The Third Set: Flexible Dust Law and Exponential Prior on \tauVeffNoSpace}
\label{sec:third_set}

\textcolor{black}{For the third set of assumptions, we adopt an exponentially declining (referred to as ``exponential'') prior on \tauVeff instead of a uniform prior, and the upper limit on $\mu$ is extended from 0.7 to 1.0. The exponential prior is based on spectroscopic observations of nearby galaxies, which find that the distribution of attenuations is approximately exponential (Figure 21 in \citealt{Brinchmann04}). This finding also applies at the higher redshifts studied here because, just as in the local Universe, galaxies with little dust are more common than those with a lot. This is because both the correlation between mass and dust and the shape of the mass function do not significantly change over the redshifts studied here (see \citealt{GarnBest10, Ly12, Dominguez13, Kashino13, Momcheva13, Price14, Puglisi16, Ramraj17, Shapley22, Shapley23} and \citealt{Moustakas13, Muzzin13, Davidzon17, Leja20}, respectively)}. 
The upper bound on $\mu$ is increased because quiescent galaxies lack young stars, so their dust will mostly arise from the diffuse ISM, implying $\mu \approx 1$ \citep{daCunha08}.

Under the third set of assumptions, in Figure \ref{fig:model_obs_colors}, model colors trace the observed colors to a similar degree as under the second: 87\%, 74\%, and 82\% of the models lie in the same region \textcolor{black}{as} 90\% of the observations at $z\sim0.5$, $z\sim0.9$, and $z\sim1.3$, respectively. This is because under the exponential prior, there are more models with little attenuation than there are models with high attenuation. This yields bluer model colors that align \textcolor{black}{better} with the observed colors. However, \textcolor{black}{in the lowest-redshift bin,} model colors extend too blue in \textcolor{black}{F435W-F775W} and \textcolor{black}{at both $z\sim0.5$ and $z\sim1.3$, they are too blue and too red in F125W-F160W}. 

\subsection{The Fourth Set: Flexible Dust Law, Exponential Prior on \tauVeffNoSpace, and Gaussian Prior on log SFH Timescale}
\label{sec:fourth_set}

For the fourth set of assumptions, a Gaussian prior on the log of the SFH timescale is adopted. Specifically, we adopt a wide Gaussian prior centered on $\log(\tau) = 9.0$ with a lower limit of $\log(\tau) = 8.3$\textcolor{black}{, which was found to better model galaxies than a lower limit of $\log(\tau) = 7.5$} \citep{Wuyts11, Pacifici16}. 

While Gaussian priors may draw the posterior toward the median (see, e.g., \citealt{Li18}), we adopt a broad Gaussian prior with a standard deviation of 1 dex. In Figure \ref{fig:all_param_one_bin}, we show that posteriors are not significantly influenced by their Gaussian priors.

For \textcolor{black}{this} set of assumptions, in Figure \ref{fig:model_obs_colors}, the model colors trace the observed colors as well as those under the third set. For the lowest-, middle-, and highest-redshift bins, 90\%, 77\%, and 85\% of the model colors in the fourth set lie in the \textcolor{black}{same region as 90\% of the observations}, respectively.
For galaxies with the reddest F435W-F775W colors, model\textcolor{black}{s} in the fourth set trace \textcolor{black}{them} better than in the third as a result of the changes in the prior on the SFH timescale. Under the third set, SFHs peak quickly after the onset of star-formation, which produces colors that are much redder than the observations. Under the fourth set, SFHs are \textcolor{black}{also} allowed to extend to longer timescales. Therefore, the model colors  are bluer, and \textcolor{black}{trace} the observations better.

\subsection{The Fifth Set: Flexible Dust Law, Exponential Prior on \tauVeffNoSpace, Gaussian Prior on log SFH Timescale, Gaussian Prior on log Age, and Gaussian Prior on log Z}
\label{sec:fifth_set}

\begin{figure*}[t!]
    \centering
    \includegraphics[width=\textwidth]{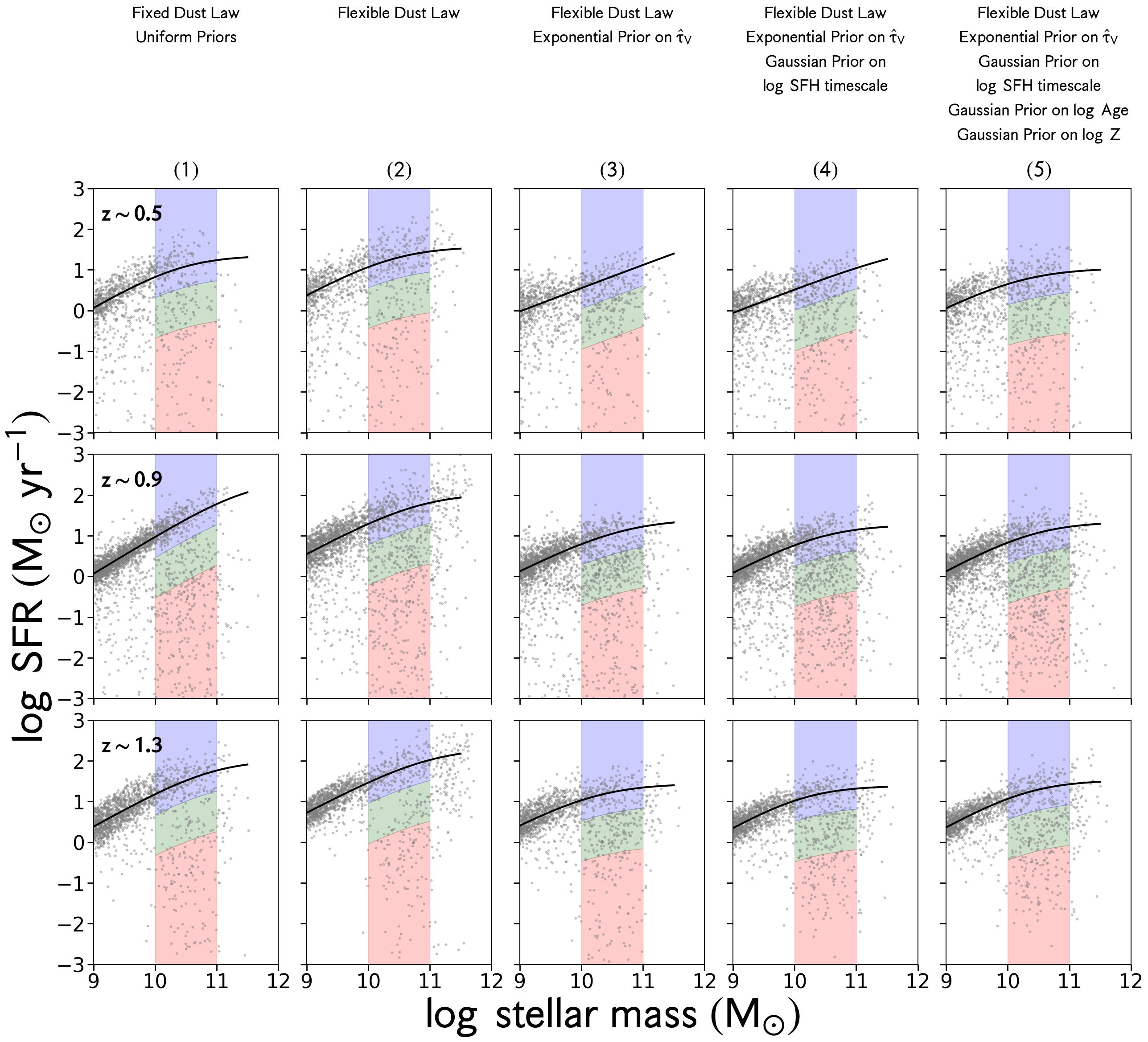}
    \caption{\textcolor{black}{}
    The stellar mass-SFR diagram \textcolor{black}{for five sets of SED-fitting} assumptions and all redshift bins. \textcolor{black}{Galaxies are selected to be star-forming, green valley, or quiescent according to their location with respect to the median star-formation main sequence (SFMS). The measurement of the SFMS and selection of galaxies is performed for each set independently. Therefore, the SFMS and selection of galaxies differ set by set. 
    Galaxies are shown as gray points.} 
    \textcolor{black}{A different set of assumptions is in each column. Three redshift bins are shown in each of the rows.}
    Median SFMS relations are shown as thick black lines. Star-forming, green valley, and quiescent galaxies are denoted by the blue, green, and red swaths, respectively. }
    \label{fig:sfms_all_redshift_bins}
\end{figure*}

Finally, under the fifth set of assumptions, Gaussian priors on the log of the maximum age (i.e., the age of the oldest stars) and log of the metallicity are adopted instead of uniform priors. This is because the distributions of the logarithm of the ages of massive galaxies at low and intermediate redshift more closely resemble Gaussians than uniform distributions \citep{Gallazzi05, Gallazzi14}. A broad Gaussian prior centered on log(Age) = 9.0 with a lower limit of log(Age) = 8.5 is adopted based on estimates of light- and mass-weighted ages of massive galaxies at $z\sim1$ \citep[e.g.,][]{Hathi09, Onodera15, Wu18, EstradaCarpenter19, Costantin22}. 
This prior has a standard deviation of 1 dex, which is much larger than the relative change in the age of the Universe from $z=1.5$ to $z=0.4$, 0.32 dex. Therefore, this prior is broad enough to cover changes in the ages of galaxies with redshift. 
The prior on the log of the metallicity is assumed to be a broad Gaussian centered at $\log\left(\textrm{Z}_{\star} \ / \ \textrm{Z}_{\odot}\right) = -0.6$, which is the typical metallicity of low-mass ($\log \ \textrm{M}_{\star} \sim8$) star-forming galaxies \citep[e.g.,][]{Maiolino08, Henry13, Gallazzi14, Guo16a}. 

Under the fifth set of assumptions, in the right column of Figure \ref{fig:model_obs_colors}, the models trace the observations the best. The color-color region \textcolor{black}{containing} 90\% of the \textcolor{black}{observations} (see Section \ref{sec:first_set}) also contains 89\%, 89\%, and 93\% of the models under the fifth set at $z\sim0.5, z\sim0.9$, and $z\sim1.3$, respectively. 

Strong emission lines in the models result in minor contamination of the colors at $z\sim0.9$ and $z\sim1.3$ for all sets of assumptions. This is due to H$\alpha$ being redshifted into F125W at $z\sim0.9$ and into F160W at $z\sim1.3$. Additionally, [O {\sc iii}] is redshifted into F125W at $z\sim1.3$. As a result, there are small offsets ($\sim0.1$ mag) between the innermost contours of the model and observed colors, as seen in the middle and bottom rows in Figure \ref{fig:model_obs_colors}. However, these minor offsets do not bias the fits to the radial flux profiles, which are the focus of this work (see Section \ref{sec:effects_assumptions}). This is because the outskirts of the radial color-color profiles of star-forming galaxies, where the S/N is the lowest, coincide with or lie at the edges of the regions in color-color space where the density of the models is the highest.

\subsection{Considerations for the Priors}
\label{sec:considerations_for_priors}

\textcolor{black}{If one wants to use these priors, here are some considerations that should be made. 
We focus on the treatment of nebular emission and what may need to change when adapting the priors to other datasets.}

\textcolor{black}{Our priors on nebular emission are based on models by \citet{Gutkin16}, which explore large variations in metallicity, ionization parameter, and dust-to-metal mass ratio. For certain populations of galaxies, e.g., lower-mass galaxies (stellar masses $\lesssim 10^9 ~\textrm{M}_{\odot}$) or at higher redshifts (e.g., $z \gtrsim 2$), more specific variations would be ideal. Setting wide priors on the ionization parameter and nebular metallicity for these populations would be useful.}

The priors of the fifth set should still hold under different SPS models, and the results should fall within the modeling uncertainties related to stellar models and stellar tracks because that set is informed by spectroscopic results.

Priors on dust attenuation and the maximum stellar age should still hold when changing the form of the SFH. We assume smooth, parametric star-formation histories as our dataset comprises several broadband bandpasses. Further, rising and falling SFHs are expected for this redshift and mass range even when using nonparametric SFHs and metallicity histories \citep[see][]{Pacifici16}. Inevitably, when switching to these, priors on other parameters will not be applicable anymore (e.g., $\tau$, the SFH timescale). However, the priors on dust attenuation and the prior on the maximum stellar age should still hold because these parameters are independent of the choice of SFH and are informed by spectroscopic results. For datasets that contain spectra, especially those with absorption features, adding bursts to SFHs would be beneficial, regardless of the form of the SFH. 

We did not assume redshift-dependent priors as our sample is restricted to a small redshift range: $0.4 < z < 1.5$. However, for a dataset that spans a larger range in redshift (e.g., $0<z<6$), redshift-dependent priors would be useful.

A parameter that could have been varied is the fraction of the NIR luminosity that is due to thermally pulsing asymptotic giant branch (TP-AGB) stars. This was not varied in this work, as our dataset lacks rest-frame NIR and mid-infrared data, which are most sensitive to changes in the TP-AGB fraction \citep[e.g.,][]{Maraston05, Kriek10, Zibetti13, Villaume15, Lu24}. 

Finally, we note that the fifth set of priors is based on having constraints from the rest-frame UV to NIR. In general, these will be applicable to similar samples with only a portion of the wavelength range available. If, for example, longer wavelengths were to be included and some galaxies showed unexpectedly strong IR fluxes, we might have to reconsider the restrictions we put on the dust and/or age priors and make them wider.

\section{Defining Star-forming, green valley, and quiescent galaxies}
\label{sec:define_galaxies}

\textcolor{black}{We bin the observations (i.e., the full sample) in stellar mass and redshift. Then we break galaxies up into three types depending on their distance from the SFMS \citep[i.e. the relation between stellar mass and SFR; e.g.,][]{Daddi07, Elbaz07, Noeske07, Rodighiero10, Karim11, Whitaker12, Whitaker14, Speagle14, Lee15, Schreiber15}: star-forming, green valley, and quiescent. We measure integrated measurements of stellar mass and star-formation rate by fitting the integrated fluxes under each set of assumptions. These quantities are shown in Figure \ref{fig:sfms_all_redshift_bins} as gray dots. Each redshift bin is shown in a row, and each set of assumptions is shown in a column. The median SFMS in a given redshift bin and set is shown as a thick black line. Details on the measurement of the median SFMS are provided below. Star-forming galaxies are defined as those with SFRs higher than 0.5 dex below the median SFMS relation and are represented by the blue swaths in the Figure. Green valley galaxies are defined to be those galaxies with SFRs from 0.5--1.5 dex below the SFMS (green swaths). Quiescent galaxies are defined to be those with SFRs lower than 1.5 dex below the SFMS and are denoted by red swaths.}

We note that the definition of star-forming, green valley, and quiescent varies according to the set of assumptions used. Reassuringly, however, the median color-color tracks of star-forming, green valley, and quiescent galaxies defined under each set of assumptions are consistent with each other and discrepant by at most 0.3 mag. We discuss the color-color tracks in more detail in Appendix \ref{sec:color_color_tracks_appendix}. 

\begin{figure*}[t!]
\begin{center}
    \includegraphics[width=0.98\textwidth]{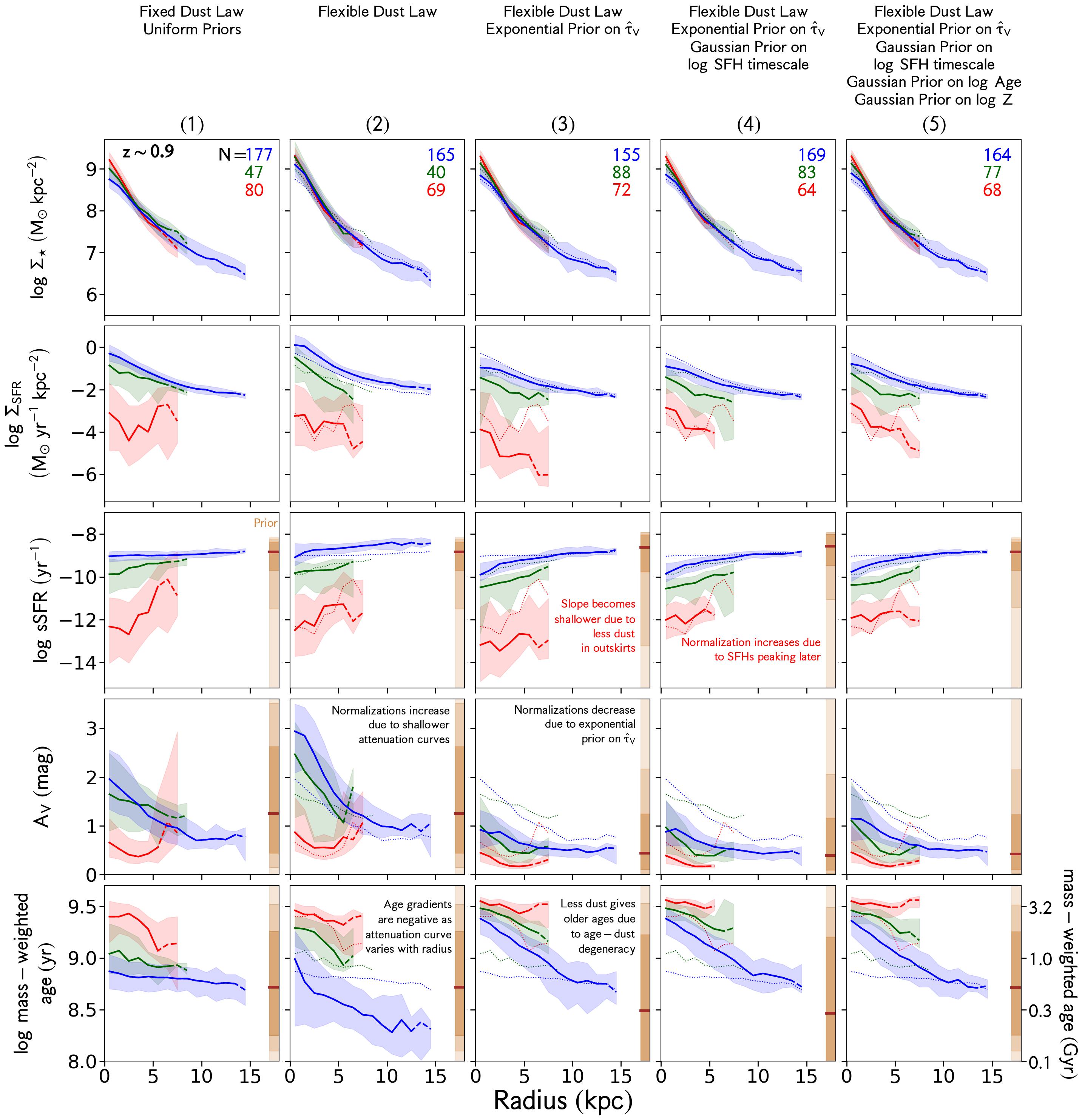}
    
    
    \caption{\textcolor{black}{Science profiles for one stellar mass ($\log \ \left( \textrm{M}_{\star} \ / \ \textrm{M}_{\odot} \right) \sim10.3$) and redshift bin ($z\sim0.9$) created under five sets of SED-fitting assumptions from left to right. From top to bottom, we show profiles of the $\Sigma_{\star}, \Sigma_{\textrm{SFR}}$, sSFR, \AvNoSpace, and mass-weighted age. Star-forming, green valley, and quiescent galaxies are shown in blue, green, and red, respectively. Median profiles are plotted as thick solid lines, where at least 10 galaxies contribute, and as dashed lines where at least five contribute. The 16th--84th percentile ranges are shown as shaded regions. The numbers of galaxies that contribute to the profiles are given in the top row. Profiles for the first set of assumptions are shown in the other panels as dotted lines. Priors for the five sets of assumptions are shown in brown. Their medians are shown as horizontal solid lines. The 16th--84th percentile ranges, the 5th--95th percentile ranges, and the full extents of the priors are shown as increasingly lighter brown shaded regions.}}
    \label{fig:all_param_one_bin}
\end{center}

\end{figure*}

\begin{figure*}[t!]
    \centering
    \includegraphics[width=\textwidth]{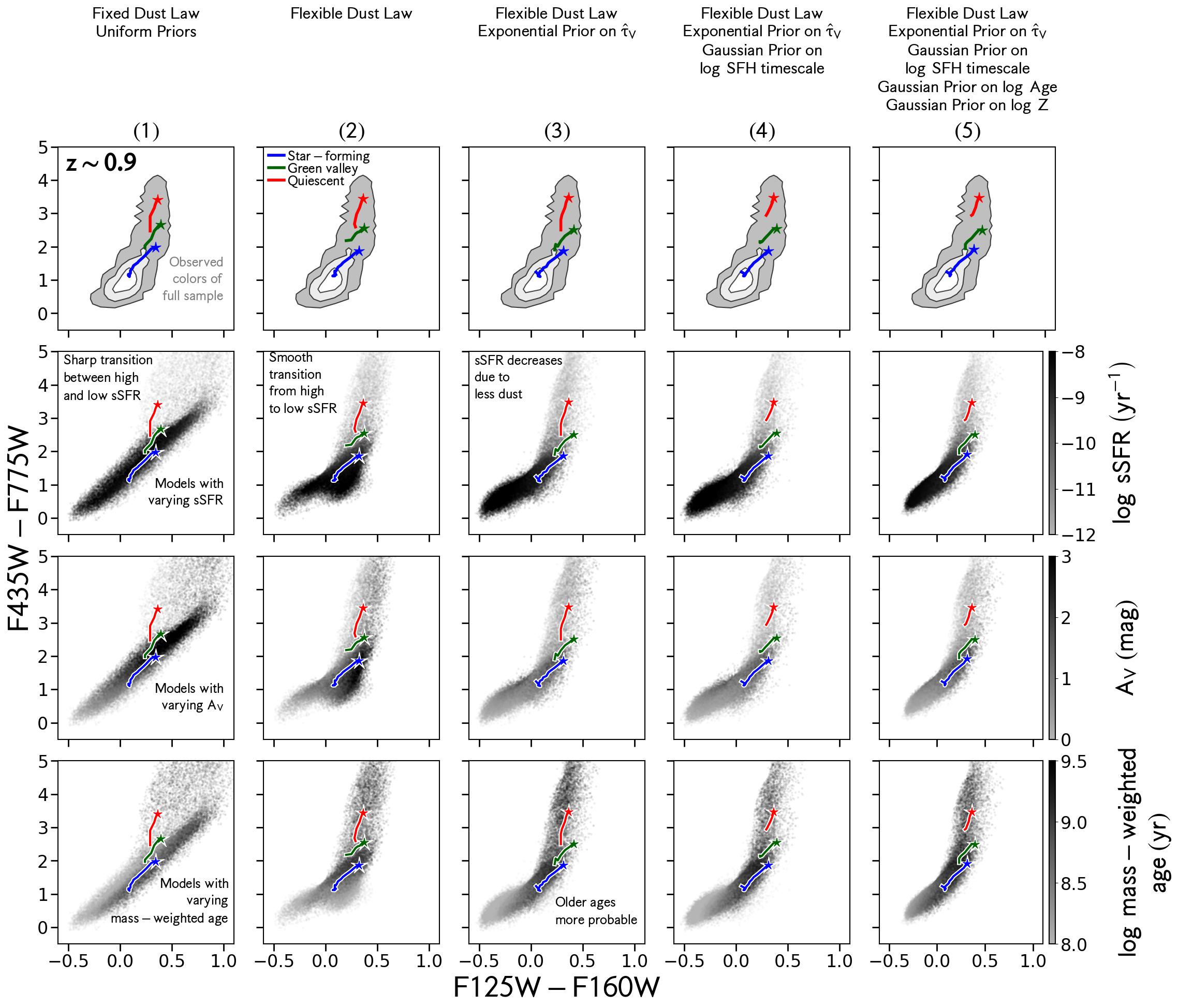}
    
    \caption{\textcolor{black}{Color--color tracks for one stellar mass ($\log \ \left( \textrm{M}_{\star} \ / \ \textrm{M}_{\odot} \right) \sim10.3$) and redshift  ($z\sim0.9$) bin for the five sets of SED-fitting assumptions are shown from left to right. For comparison, the middle row in Figure \ref{fig:model_obs_colors} is reproduced in the top row. Star-forming, green valley, and quiescent galaxies are shown in blue, green, and red, respectively.} Median colors of the centers of galaxies are shown as stars and color-color tracks become bluer with increasing radius.
    Model colors \textcolor{black}{for each set of assumptions are shown in different columns and are colored by log sSFR, \AvNoSpace, and log mass-weighted age from the second to fourth rows. Under the first set of assumptions, there is a sharp transition in the F435W-F775W color between models with high and low sSFRs, a lot and little dust, and young and old ages. When a flexible dust law is assumed, there is a smooth transition in F435W-F775W color. When adopting an exponential prior on the dust, normalizations in the model sSFRs and ages change due to the age--dust degeneracy.}}
    
    \label{fig:color_color_tracks}
\end{figure*}

\textcolor{black}{For each set of assumptions and for each redshift bin,} the median SFMS relation is measured by first excluding galaxies with sSFRs (i.e., SFR / stellar mass) lower than half the inverse of the age of the Universe \citep[e.g.,][]{Wuyts12, Pacifici16} at the median redshift of the redshift bin. This cut excludes quiescent galaxies such that the median SFMS relation is determined predominantly using star-forming galaxies. \textcolor{black}{All quiescent galaxies have sSFRs lower than this cut}.
The remaining galaxies are binned in stellar mass in bins 0.25 dex wide such that each bin has at least 10 galaxies. 
When fitting the SFMS, we fit the median SFR in bins of stellar mass and account for the dispersion around the median, which we take to be half the difference between the 16th and 84th percentiles in SFR.
The median SFR as a function of stellar mass is fit using Equation 2 from \citet{Lee15}. This function behaves like a power law at low stellar masses and asymptotically approaches a maximum SFR at masses higher than a transitional mass. The fit relation is taken to be the SFMS in a given redshift bin and is shown under each set of assumptions as thick black lines in Figure \ref{fig:sfms_all_redshift_bins} for all mass and redshift bins we consider.

\section{Creation of the Science Profiles}
\label{sec:science_profiles}

{\textcolor{black}{In this Section we describe how radial profiles of stellar mass surface density, SFR surface density, sSFR, \AvNoSpace, and mass-weighted age are created.  Collectively, we refer to these as ``science profiles." We describe how these profiles are created for each of the five sets of SED-fitting assumptions in Section \ref{sec:construct_median_profile}. We examine and explain the differences among them by studying how the models for the different SED-fitting assumptions behave in color-color space with respect to observations in Section \ref{sec:effects_assumptions}. Additionally, in Section \ref{sec:effects_assumptions}, we show that the science profiles are not dominated by the adopted priors. Finally, we compare the science profiles fit under the first set of assumptions with those from studies in the literature, which make similar assumptions, in Section \ref{sec:compare_literature}. }

\subsection{Calculating the Median Science Profiles}
\label{sec:construct_median_profile}

\textcolor{black}{First, radial science profiles are created for individual galaxies in the mass-selected sample (Section \ref{sec:mass_selected_sample}).  Given the radial flux profiles of a galaxy, the observed SED at each radius is fit for each of the five sets of assumptions. This results in science profiles for the galaxy. Since the radial flux profiles are measured along the semi-major axes of galaxies, the science profiles are circularized by multiplying by the square root of the axis ratios of the galaxies.}

\textcolor{black}{Galaxies are then divided into three redshift bins, $0.4< z <0.7$, $0.7 < z < 1.1$, and $1.1 < z < 1.5$ (see Figure \ref{fig:sample}), which are each further divided into two mass bins, $10 \leq \log \left(\textrm{M}_{\star} / \textrm{M}_{\odot}\right) < 10.5$ and $10.5 \leq \log \left(\textrm{M}_{\star} / \textrm{M}_{\odot}\right) < 11.0$. Medians of the science profiles are then created in these bins in 1 kpc sized radial bins. These profiles are shown in Appendix \ref{sec:science_profiles_all} for all sets of assumptions. To ensure sufficient statistics, at least five galaxies are required to contribute to a median profile at all radii. To characterize the dispersion about the median, the 16th and 84th percentiles are measured.}

\textcolor{black}{Individual science profiles \textcolor{black}{incorporated} into the median are required have adequate S/N at a given radius in key wave bands. The rest-frame UV is important for accurate SED fitting because it is particularly sensitive to SFR, age, and dust \citep{Pforr12, Buat14}. Therefore, an S/N of at least 5 is required at a given radius in the waveband which spans the rest-NUV, namely F435W for $z \leq 0.8$ and F606W for $z > 0.8$. Further, this requirement must be met in at least three radial bins. As a note to the reader, this cut is only for incorporation into the median, unlike the cuts described in Section \ref{sec:flux_profiles}.}

\subsection{Explaining the Behavior of the Science Profiles Using Color--Color Diagrams}
\label{sec:effects_assumptions}

\textcolor{black}{To showcase the most important changes in the profiles among the five sets of assumptions, one example bin in mass and redshift is discussed: namely $10.0 < \log\left(\textrm{M}_{\star}/\textrm{M}_{\odot}\right) < 10.5$ and $z\sim0.9$.  The profiles for this bin are shown in Figure \ref{fig:all_param_one_bin} for each of the sets of assumptions in the columns.   The differences are explained with the behavior of the models in color-color space in Figure \ref{fig:color_color_tracks}.}

\textcolor{black}{The most important differences are listed below and indicated with annotations in Figures \ref{fig:all_param_one_bin} and \ref{fig:color_color_tracks}. Most of these arise when a flexible dust law and an exponential prior on the optical depth in the rest-frame $V$-band are assumed.
\begin{enumerate}
    \item In the first set, models with lower sSFRs have redder F435W-F775W colors, become older with redder F125W-F160W color, and dustier with redder F435W-F775W and F125W-F160W colors (left column in Figure \ref{fig:color_color_tracks}). There is a sharp transition in the F435W-F775W color between models with high and low sSFRs, a lot and a little dust, and young and old ages. The color-color tracks of quiescent galaxies cross this transition. These galaxies have steeply rising sSFR and \Av profiles and steeply declining age profiles (left column in Figure \ref{fig:all_param_one_bin}). Color--color tracks of star-forming galaxies have bluer colors than this transition. These galaxies have flat sSFR profiles and \Av and age profiles that decline mildly with radius. Green valley galaxies lie between these extremes. 
    \item In the second set, when assuming a flexible dust law, models also have lower sSFRs with redder F435W-F775W colors, but the trends between age and dust in the models and F435W-F775W and F125W-F160W colors are more complex than those in the first set (second column in Figure \ref{fig:color_color_tracks}). There is no longer a sharp transition, but a smooth one in F435W-F775W color. Consequently, quiescent galaxies have sSFR and \Av profiles that rise mildly with radius and age profiles that decline mildly with radius (second column in Figure \ref{fig:all_param_one_bin}). Age profiles of star-forming galaxies decline with radius because the attenuation curve varies with radius. The normalizations of the \Av profiles for all galaxies increase because attenuation curves are shallower than that assumed in the first set. 
    \item In the third set, assuming an exponential prior on optical depth affects the \AvNoSpace, age, and sSFR profiles (third column in Figure \ref{fig:all_param_one_bin}). Normalizations of the \Av profiles decrease because models with little to no dust are more common than those with a lot. Less dust at all radii also results in older ages at all radii due to the age--dust degeneracy. Quiescent galaxies have shallower sSFR profiles because there is less dust, and therefore lower sSFR, in the outskirts of these galaxies. 
    \item In the fourth set, the normalizations of the sSFR profiles of quiescent galaxies increase when assuming a Gaussian prior on the log of the SFH timescale (fourth column in Figure \ref{fig:all_param_one_bin}). This is because the SFHs of these galaxies peak later in time, which raises the SFR and sSFR at later times. 
    \item There are no significant changes between profiles in the fifth set and those in the fourth (right column in Figure \ref{fig:all_param_one_bin}). 
\end{enumerate}}

\textcolor{black}{There are several features of the median profiles that are the same among all the sets of assumptions. It is reassuring to see these qualitative similarities, as these trends have been found by multiple studies. Namely, star-forming galaxies have the lowest stellar mass surface densities, the highest SFR surface densities and sSFRs, the most dust, and the youngest ages. Quiescent galaxies have the highest stellar mass surface densities, the lowest SFR surface densities and sSFRs, the least dust, and the oldest ages. Green valley galaxies lie in between these extremes in all cases.}

We also examine whether priors in each set dominate the posteriors. This happens when their medians and percentile ranges are similar. In this work, neither the medians nor the 16th--84th percentile ranges of the priors are similar to those of the profiles. This can be seen by comparing the brown shaded regions in Figure \ref{fig:all_param_one_bin}, which indicate the priors, with the profiles. This is true even in the outskirts of galaxies, where the S/N is the lowest. This is because we have a strict limit on the S/N. In the third set, we purposely down-weight the models with large \AvNoSpace; hence, the posteriors become narrow and more similar to the shape of the prior. This is what we expect to see based on the studies that informed the priors adopted in the third set (see Section \ref{sec:third_set}) and the validation against profiles measured from spectroscopic datasets (see Section \ref{sec:comparison_sami_manga}).

\begin{figure*}[t!]
    \centering
    \includegraphics[width=\textwidth]{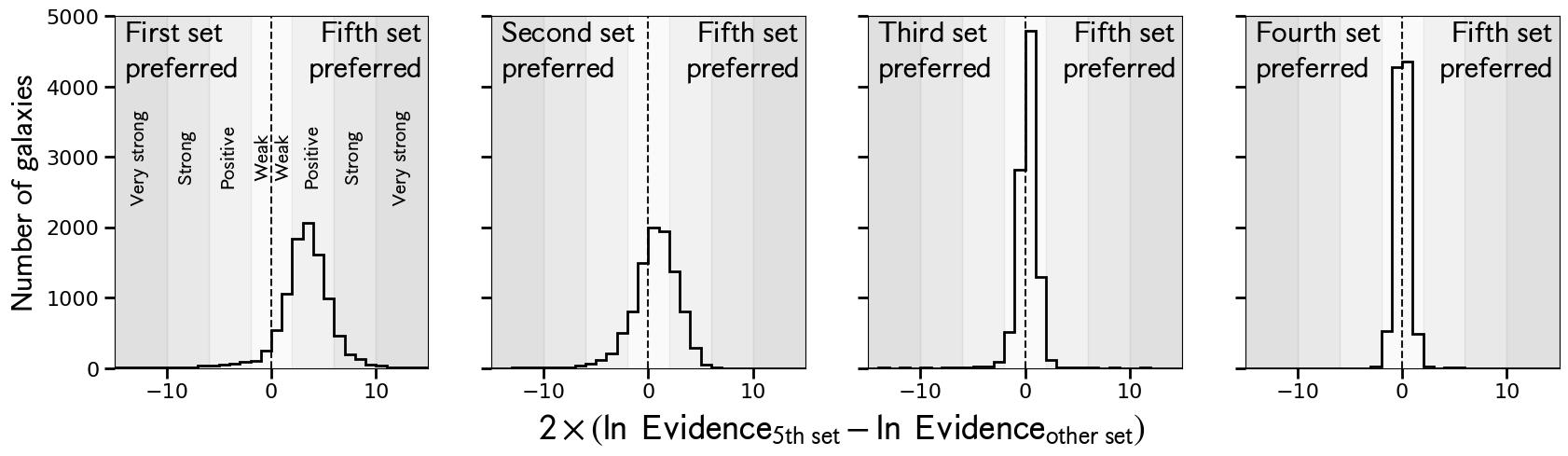}
    \caption{The Bayes factor is used to evaluate whether one model fits the data better than another. In each panel, from left to right, the distribution of the Bayes factor of the fifth set vs. that for the other four sets is shown. In each panel, the distribution of two times the natural log of the Bayes factor, $2 \ln K$ (see Section \ref{sec:bayesian_evidence}), is shown. We adopt the significance criteria developed by \citet{KassRaftery95} in order to interpret these distributions. These are indicated  with the shaded regions in the leftmost panel and labeled accordingly.
    In the leftmost panel, the bulk of the distribution is positive, indicating that the fifth set is favored when comparing it with the first. In the next panel, the fifth set is preferred to the second at a ``weak'' confidence level.  In the following panels, there is no preference between the third and fourth sets and the fifth set.}
    \label{fig:bayesian_evidence}
\end{figure*}

\subsection{Comparison of Profiles Created under the First Set with Studies that Adopt Similar Assumptions}
\label{sec:compare_literature}
\textcolor{black}{We now compare science profiles created under the first set with those from the literature which adopt similar assumptions and datasets. In summary, they are consistent wherever there is overlap in redshift and stellar mass. Below, we walk through each of the studies.}

\noindent\textcolor{black}{\underline{\citet{Nelson16}} measured stellar mass surface density profiles for galaxies on and below the SFMS at $0.7 < z < 1.5$. These are consistent with profiles for star-forming and green valley galaxies in this work.}

\noindent\textcolor{black}{\underline{\citet{Wang17}} measured sSFR and \Av profiles of star-forming galaxies at $0.4 < z < 1.5$. Their \Av profiles agree with those presented here. Their sSFR profiles agree with those for the first set over $10 \leq \log \left(\textrm{M}_{\star} / \textrm{M}_{\odot}\right) < 10.5$. At higher stellar masses, $10.5 \leq \log \left(\textrm{M}_{\star} / \textrm{M}_{\odot}\right) < 11$, sSFR profiles rise with radius, but \citet{Wang17} found shallower profiles than those for the first set here. We note that this work cautions that their sSFR measurements might be significantly affected by oversimplified assumptions adopted in SED fitting.}

\noindent\textcolor{black}{\underline{\citet{Abdurrouf18}} measured sSFR profiles for star-forming, green valley, and quiescent galaxies at $0.8 < z < 1.8$. Their profiles are qualitatively consistent with those for the first set, although normalizations differ for two reasons. First, their sample is selected using spatially integrated SFRs that are measured by combining rest-UV and rest-IR luminosities. Our galaxies are selected using spatially integrated SED fits. Second, we define quiescent galaxies differently. Normalizations differ even if quiescent galaxies are defined in the same manner due to differences in the sample selection.}

\noindent\textcolor{black}{\underline{\citet{Morselli19}} measured sSFR profiles for galaxies above, on, and below the SFMS at $0.2 < z < 1.2$. While their profiles qualitatively agree with those for the first set, normalizations differ for the same two reasons listed in the comparison with \citet{Abdurrouf18}.}

\noindent\textcolor{black}{Lastly, \underline{\citet{Nelson21}} measure sSFR profiles of galaxies on and below the SFMS at $0.7 < z < 1.5$. These agree with profiles of star-forming and green valley galaxies created under the first set.}

\section{Choosing a Set of Assumptions}
\label{sec:choosing_assumption}

In this Section, we determine which set of assumptions in Table \ref{tab:priors} is most appropriate for our galaxy sample; we conclude that the fifth set is. To make this determination, we perform three tests for each of the five sets. First, in Section \ref{sec:bayesian_evidence}, we examine Bayesian evidences. \textcolor{black}{Next, in Section \ref{sec:comparison_sami_manga},} we compare the SFR surface density and dust profiles with those of \textcolor{black}{nearby} galaxies that have these profiles measured from spectroscopy, which are more accurate.
\textcolor{black}{Finally, in Section \ref{sec:dust_inclination}}, we examine dust profiles as a function of galaxy inclination.

\subsection{Comparison of Bayesian Evidences}
\label{sec:bayesian_evidence}

Sets of SED-fitting assumptions may be compared by estimating the Bayesian evidence, which is the product of the likelihood and the prior that is integrated over the entire parameter space. The ratio of the Bayesian evidences of two models, also known as the Bayes factor, is often used to determine whether one model fits the data better than the other \citep[e.g.,][]{HanHan12, HanHan14, HanHan19, Chevallard16, Salmon16, Lawler21, Han23}. Here we compare the Bayesian evidence of the fifth set of assumptions with that of the other sets for SED fits to the {\it spatially integrated fluxes} for the full sample (Section \ref{sec:full_sample}). 

Histograms of the Bayes factors for the fifth versus each of the other four sets are shown in Figure \ref{fig:bayesian_evidence}.
In each panel of Figure \ref{fig:bayesian_evidence}, positive values indicate a preference for the fifth set and negative values denote a preference for the set being compared. A value of zero indicates no preference for either. In order to interpret these distributions and determine the strength of the preference towards one set, the qualitative criteria by \citet{KassRaftery95} are adopted. 
The strength of the preference in favor of one model is categorized as ``weak,'' ``positive,'' ``strong,'' or ``very strong,'' depending on the value of the Bayes factor. These categories are indicated with the shaded regions in Figure \ref{fig:bayesian_evidence}. In the leftmost panel, the histogram is centered at a Bayes factor of about 3. This means that the fifth set is preferred to the first at a ``positive'' confidence level. In the second through fourth panels, the histograms gradually move toward a value of zero. The fifth set is preferred to the second at a ``weak'' confidence level. There is no preference between the third and fifth or fourth and fifth sets. 

\begin{figure*}[t!]
    \centering
    \includegraphics[width=\textwidth]{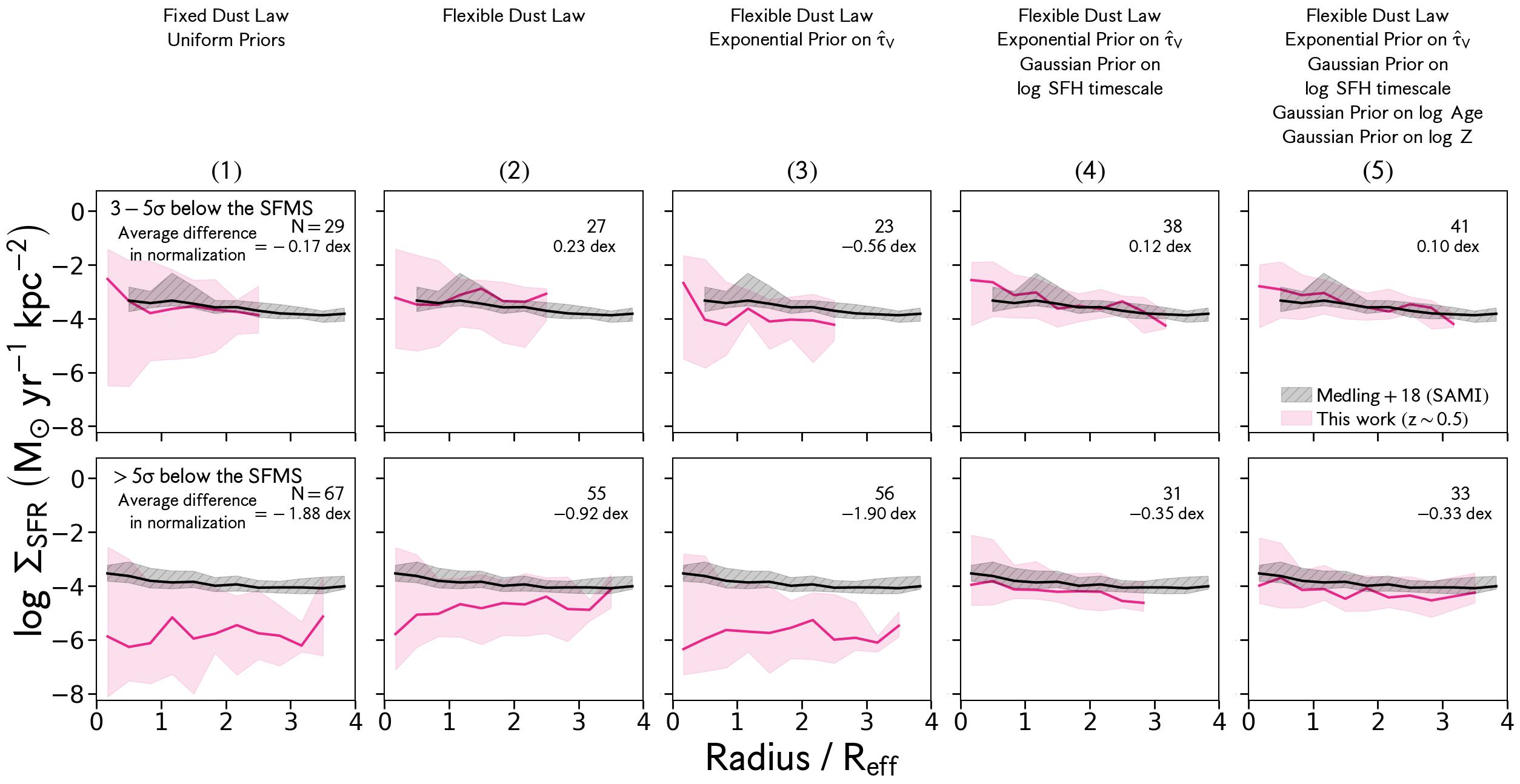}
    \caption{\textcolor{black}{SFR surface density profiles of quiescent galaxies at $z\sim0.5$ (magenta) are compared with those measured from spectroscopy by \citet{Medling18} at $z\sim0$ (black). The 16th--84th percentile ranges are shown as shading.  The average differences in normalization between profiles for each set and those at $z\sim0$ are given. Profiles for the fifth set have the smallest difference. This is because the fifth set assumes Gaussian priors on the log SFH timescale, log age, and log metallicity, with the former having the largest effect.  We compare with galaxies in \citet{Medling18} that overlap ours in stellar mass and SFR.  The top row shows galaxies with  $10.0 < \log \ \left( \textrm{M}_{\star} \ / \ \textrm{M}_{\odot} \right) < 11.0$ and which are $3\sigma-5\sigma$ below the SFM, where $\sigma$ is the scatter in the SFMS.  The bottom row shows the same mass range but for galaxies that are $>5\sigma$ below the SFMS.}}
    \label{fig:sfr_medling}
\end{figure*}

\begin{figure*}[t!]
    \centering
    \includegraphics[width=\textwidth]{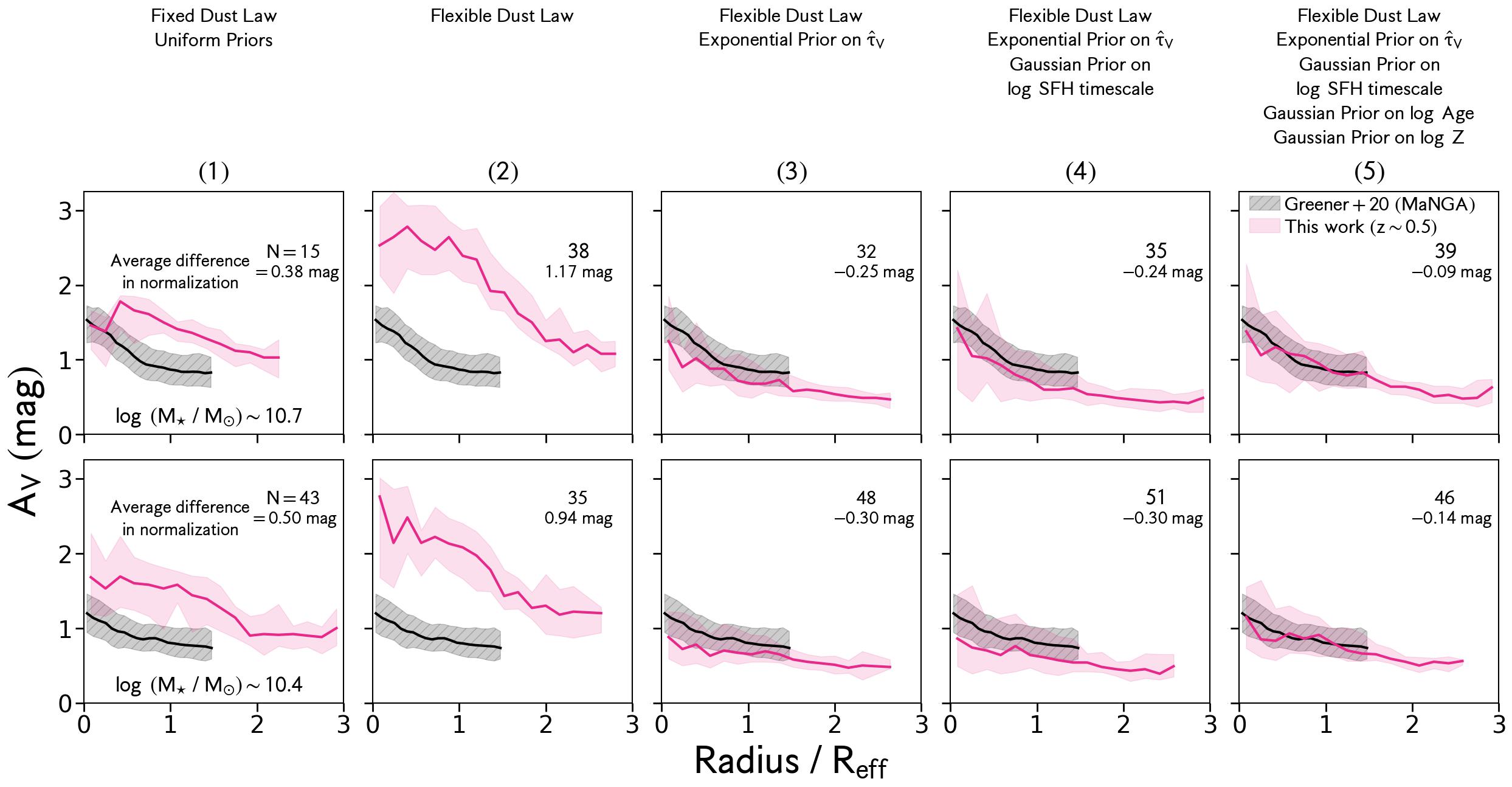}
    \caption{\textcolor{black}{\Av profiles of star-forming galaxies at $z\sim0.5$ (magenta) are compared with those at $z\sim0$ measured from spectroscopy by \citet[][black]{Greener20}. Interquartile ranges are indicated by shading.
    We choose galaxies that overlap two of their bins in stellar mass ($10.58 < \log~\left(\textrm{M}_{\star} / \textrm{M}_{\odot}\right) < 10.86$ and $10.26 < \log~\left(\textrm{M}_{\star} / \textrm{M}_{\odot}\right)<10.58$) and also apply their cut in axis ratio ($>0.35$).
    Average differences in normalization are given, and those for the fifth set are the smallest. This is primarily because the fifth set adopts an exponential prior on the amount of dust.}}
    \label{fig:av_greener}
\end{figure*}

\subsection{Comparison of SFR Surface Density and \Av Profiles with Results from Spectroscopy}
\label{sec:comparison_sami_manga}

\textcolor{black}{We now compare profiles created under each set of assumptions at $z\sim0.5$ with those from the literature that are based on spatially resolved spectroscopic surveys of galaxies at $z\sim0$. These comparisons are meant as first-order sanity checks on the profiles. While we expect the profiles evolve with redshift over $0<z<0.5$, we do not expect this to be significant (by $\lesssim 0.4$ dex in SFR surface density and by $\lesssim0.3$ mag in \AvNoSpace). Further justifications for this are provided in Section \ref{sec:justifications}.}

\textcolor{black}{Comparisons with SFR surface density ($\Sigma_{\textrm{SFR}}$) profiles of quiescent galaxies from SAMI \citep[][A. M. Medling 2025, private communication]{Bryant15, Medling18} are shown in Figure \ref{fig:sfr_medling} and examined in Section \ref{sec:sami}. 
Comparisons with dust profiles of star-forming galaxies from MaNGA \citep[][M. J. Greener 2025, private communication]{Bundy15, Greener20} are shown in Figure \ref{fig:av_greener} and examined in Section \ref{sec:manga}.
In both Figures, profiles for each set are plotted as magenta lines, and those from the literature are shown as black lines.
We measure the average differences in normalization between them. Profiles for the fifth set have the smallest difference and therefore agree the best with those from spectroscopy.}

\textcolor{black}{Profiles from the literature are normalized by the effective radius. Therefore, we do the same. We adopt PSF-corrected effective radii and axis ratios measured in the F160W bandpass from \citet{vdW12}.}

\subsubsection{Comparison with SFR Surface Density Profiles of Quiescent Galaxies}
\label{sec:sami}

\textcolor{black}{To compare with SFR surface density profiles of quiescent galaxies from \citet{Medling18}, we choose a subset of their sample that overlaps ours in stellar mass and SFR (Figure \ref{fig:sfr_medling}). We compare with their profiles over stellar masses $10 < \log\left(\textrm{M}_{\star}/\textrm{M}_{\odot}\right)<11$. Their profiles are compared with ours at high and low SFRs: 3$\sigma$--5$\sigma$ below the SFMS and 5$\sigma$ and greater, where $\sigma$ is the scatter in the SFMS. We assume $\sigma = 0.5$ dex, and our results are not affected by choosing a smaller or larger value.
At high and low SFRs, profiles created under the fifth set have the smallest average difference in normalization ($0.1$ and  $-0.33$ dex, respectively). Therefore, they agree the best. This is because the fifth set assumes Gaussian priors on the log SFH timescale, log age, and log metallicity, with the former having the largest effect. Profiles for the fourth set, which also assume the same prior on log SFH timescale, perform marginally worse. This is because this set assumes uniform priors on log age and log metallicity. For the first three sets, differences in normalization are much larger. This is because they assume a uniform prior on the log SFH timescale.}

\subsubsection{Comparison with \Av Profiles of Star-forming Galaxies}
\label{sec:manga}

\textcolor{black}{To compare with \Av profiles of star-forming galaxies from \citet{Greener20}, we choose a subset of their sample that overlaps ours in stellar mass, SFR, and axis ratio (Figure \ref{fig:av_greener}). Their profiles are measured for galaxies with SFRs that are $>0.71$ dex below the SFMS. We choose galaxies that overlap two of their bins in stellar mass ($10.58 < \log~\left(\textrm{M}_{\star} / \textrm{M}_{\odot}\right) < 10.86$ and $10.26 < \log~\left(\textrm{M}_{\star} / \textrm{M}_{\odot}\right)<10.58$) and also apply their cut in axis ratio (i.e., greater than 0.35). At high and low stellar masses, profiles for the fifth set have the smallest average difference in normalization ($-0.09$ and $-0.14$ mag, respectively). Therefore, they agree the best. This is mainly because the fifth set assumes an exponential prior on the amount of dust, and, to a lesser extent, Gaussian priors on log age and log metallicity. Profiles for the third and fourth sets, which assume the same prior on the amount of dust, perform slightly worse. This is because they assume uniform priors on log age and log metallicity. For the first two sets, differences in normalization are the largest. This is because they assume a uniform prior on the amount of dust.}

\begin{figure*}[t!]
    \centering
    \includegraphics[width=\textwidth]{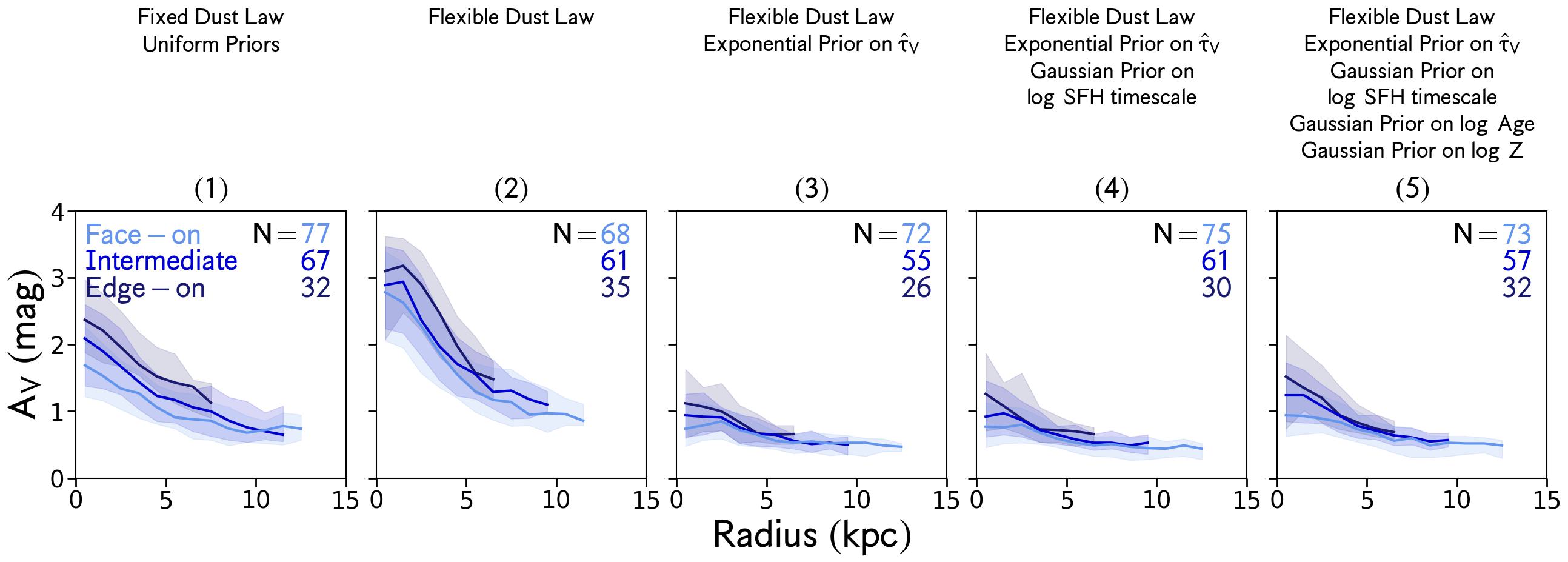}
    \caption{\textcolor{black}{\Av profiles of star-forming galaxies as a function of axis ratio/inclination are qualitatively compared with expectations based on observations of disk galaxies at $z\sim0$. This is because star-forming galaxies at the stellar masses and redshifts considered in this work are generally disk galaxies. As disk galaxies become more inclined, it is expected that the largest increases in the amount of dust are in their centers and the smallest are in their outskirts. Therefore, as inclination increases, \Av profiles should steepen. Profiles for the first and second sets do not display the expected behavior. This is due to the uniform prior on the dust. However, profiles for the third, fourth, and fifth sets do demonstrate the expected behavior. This is due to the exponential prior on the dust. Profiles for each set in one one stellar mass $\left( \log \ \left( \textrm{M}_{\star} \ / \ \textrm{M}_{\odot} \right) \sim 10.3\right)$ and redshift ($z\sim0.9$) bin are shown. Galaxies are divided by axis ratio. Face-on ($b/a > 0.64$), intermediate inclination ($0.34 < b/a < 0.64$), and edge-on galaxies ($b/a < 0.34$) are shown as light, medium, and dark-blue lines, respectively.}}
    \label{fig:av_prof_axis_ratio}
\end{figure*}

\subsubsection{Justifications for Comparing Profiles at $z\sim0$ and $z\sim0.5$}
\label{sec:justifications}

\textcolor{black}{SFR surface density profiles of quiescent galaxies at $z\sim0.5$ are compared with those from spectroscopy at $z\sim0$ as a sanity check. We do not expect perfect agreement, as mild evolution with redshift has been found. Profiles are expected to decrease in normalization by $\sim0.4$ dex with time over $0<z<0.5$ at fixed mass. This is because integrated sSFRs decrease by this amount at fixed mass \citep[][]{Fumagalli14, McDermid15, SalvadorRusinol20}.}

\textcolor{black}{\Av profiles of star-forming galaxies at $z\sim0.5$ are also compared with those from spectroscopy at $z\sim0$ as a sanity check. We do not expect perfect agreement as mild evolution with redshift has been found. Profiles are expected to increase by $\sim0.3$ mag at most with time over $0<z<0.5$ at fixed mass. This is because integrated dust attenuation measured from spectroscopy increases at most by this amount at fixed mass \citep[][]{Dominguez13, Ramraj17, Shapley22}.}

\textcolor{black}{SFR surface density profiles of quiescent galaxies at $z\sim0$ are measured differently from those in this work. However, they are compatible. The former are measured by averaging over regions that are selected to be ionized by young, massive stars. This selection is based on emission-line diagrams. Profiles in this work are measured by averaging over annuli containing rest-UV light, which predominantly arises from young, massive stars, but could arise from other sources, namely X-ray AGNs and old stars. We have removed the first from our sample (Section \ref{sec:xray}) and the second are not observed to be predominant at the stellar masses and redshifts examined in this work \citep{LeCras16, Dantas20}.}

\textcolor{black}{While different SFR tracers are used to measure SFR surface density profiles of quiescent galaxies at $z\sim0$ and those in this work, we find that this does not significantly affect our comparison. The former are measured using H$\alpha$ luminosities corrected for dust. The latter come from dust-corrected rest-UV luminosities. These trace the SFR on different timescales \citep[$\sim10$ Myr for emission lines and $\sim100$ Myr for UV continuum; see, e.g.,][]{Kennicutt98, Kennicutt12} and the calibration from rest-frame luminosity to SFR depends on the burstiness of the SFH \citep[e.g.,][]{Emami19, FloresVelazquez21}.
However, these measurements of the SFR can be reconciled because massive galaxies at both $z\sim0$ and $z\sim0.5$ are not observed to have bursty SFHs \citep[see, e.g.,][]{Guo16b, Hahn17, Trussler20, Bravo23, Weibel23}. This suggests that the SFRs measured from emission lines and UV continuum should be similar for massive galaxies at both $z\sim0$ and $z\sim0.5$.}

\textcolor{black}{\Av profiles of star-forming galaxies at $z\sim0$ and those in this work are measured differently. However, they are compatible. The former are measured by averaging over regions that are selected to be ionized by young, massive stars. The latter are measured differently, and we address this above. \Av is measured from spectroscopy at $z\sim0$ and from photometry in this work. These measurements are comparable. This is because both quantify the total amount of attenuation suffered by young stellar populations, which is due to birth clouds and diffuse dust in the ISM. A caveat is that \Av from spectroscopy is measured over regions with dust from birth clouds and the ISM, whereas \Av from photometry is measured over regions with both types of dust and with dust only from the ISM.}

\subsection{Comparing Dust Profiles of Star-forming Galaxies as a Function of Inclination}
\label{sec:dust_inclination}

We examine dust profiles \textcolor{black}{for each set} as a function of galaxy inclination and \textcolor{black}{qualitatively} compare with expectations based on observations \textcolor{black}{of disk galaxies} at $z\sim0$ (Figure \ref{fig:av_prof_axis_ratio}). \textcolor{black}{Star-forming galaxies at the stellar masses and redshifts examined in this work are disk galaxies in general \citep[e.g.,][]{Kassin12, vdW14, Simons17}}. Since disk galaxies \textcolor{black}{often} have more dust in their centers than in their outskirts, it is expected that when \textcolor{black}{they} are viewed from face-on to edge-on, the largest increases in the amount of dust are seen in their centers and the smallest are seen in the outskirts \citep{Byun94, Giovanelli94, Peletier95, Kuchinski98, Tuffs04, Riello05}. \textcolor{black}
{Therefore, as disk galaxies become more inclined, dust profiles should steepen.}

\textcolor{black}{Profiles are examined at $z\sim0.9$ because there are enough galaxies at this redshift to divide the sample into three bins of inclination with at least 25 galaxies each.  For these three bins in inclination (i.e., axis ratio), we measure \Av profiles for the five sets of assumptions (Figure \ref{fig:av_prof_axis_ratio}). The profiles for the first and second sets do not follow the expected behavior. Instead, as inclination increases, the profiles increase in normalization at all radii. This is due to the uniform prior on the dust. However, the profiles for the third, fourth, and fifth sets demonstrate the expected behavior--as galaxies become more inclined, they steepen. This behavior is due to the exponential prior on the dust.}

\section{sSFR Profiles under the Fifth Set of Assumptions and Comparisons with the First Set}
\label{sec:ssfr_optimal}

\subsection{Descriptions of the sSFR Profiles under the Fifth Set and Fits to Them}

In this Section, we examine median SSFR profiles created under the fifth set of assumptions, our chosen set.  They are shown in the top two rows of Figure \ref{fig:optimal_ssfr}.  The dispersion around these median profiles can be large, indicating that individual profiles can have a variety of shapes. In the bottom \textcolor{black}{two rows of Figure \ref{fig:optimal_ssfr},} linear fits to these profiles are shown. \textcolor{black}{They} are fit out to radii where at least 10 galaxies contribute. Errors on the median profiles \textcolor{black}{are used in the fits} and are shown as shaded areas in Figure \ref{fig:optimal_ssfr}. The errors are the dispersions about the median profiles divided by the square root of the number of galaxies that contribute at each radius. Now we examine SSFR profiles for star-forming, green valley, and quiescent galaxies.

\begin{figure*}[t!]
    \centering
    \includegraphics[width=0.75\textwidth]{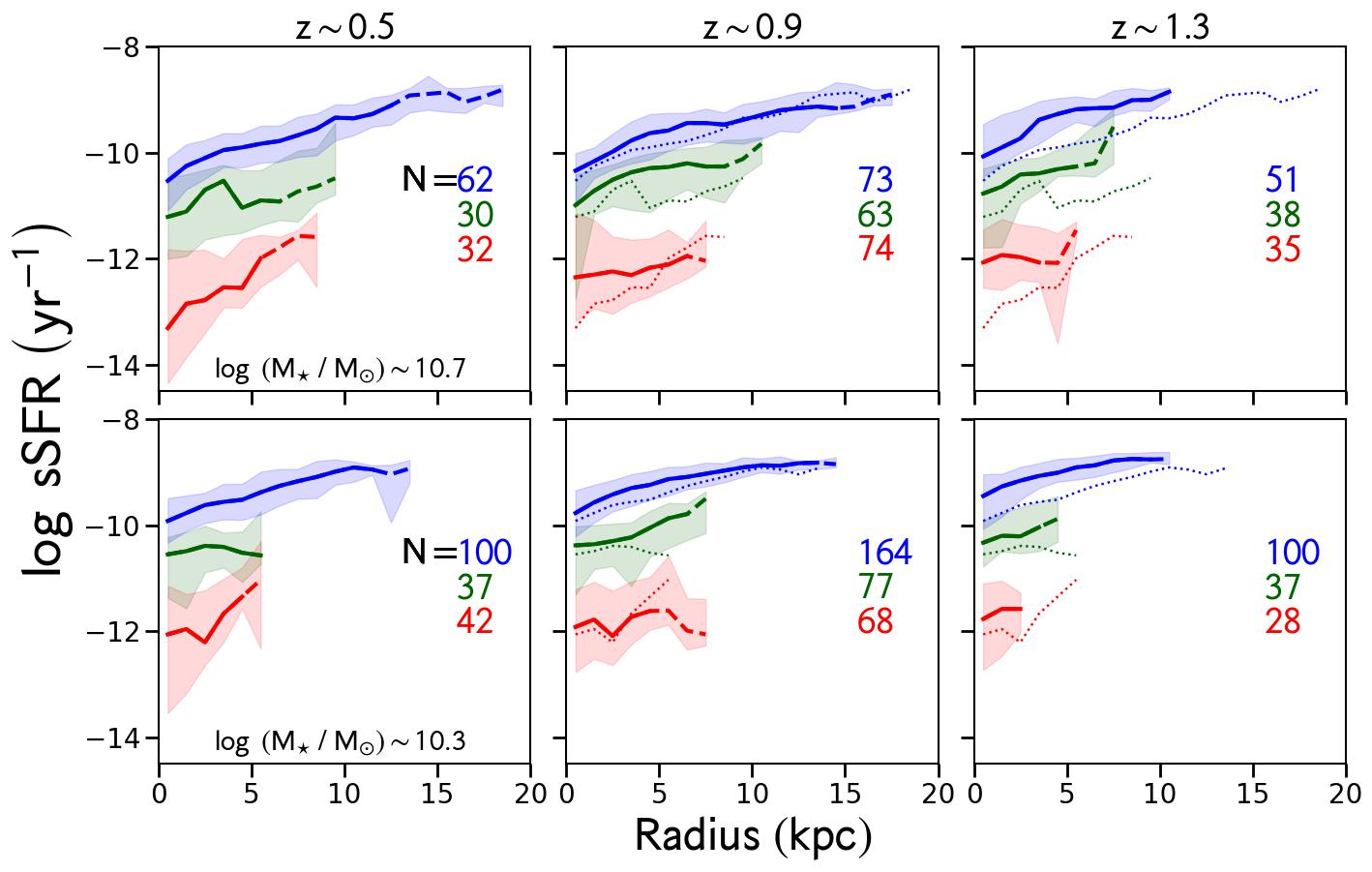}
    \includegraphics[width=0.75\textwidth]{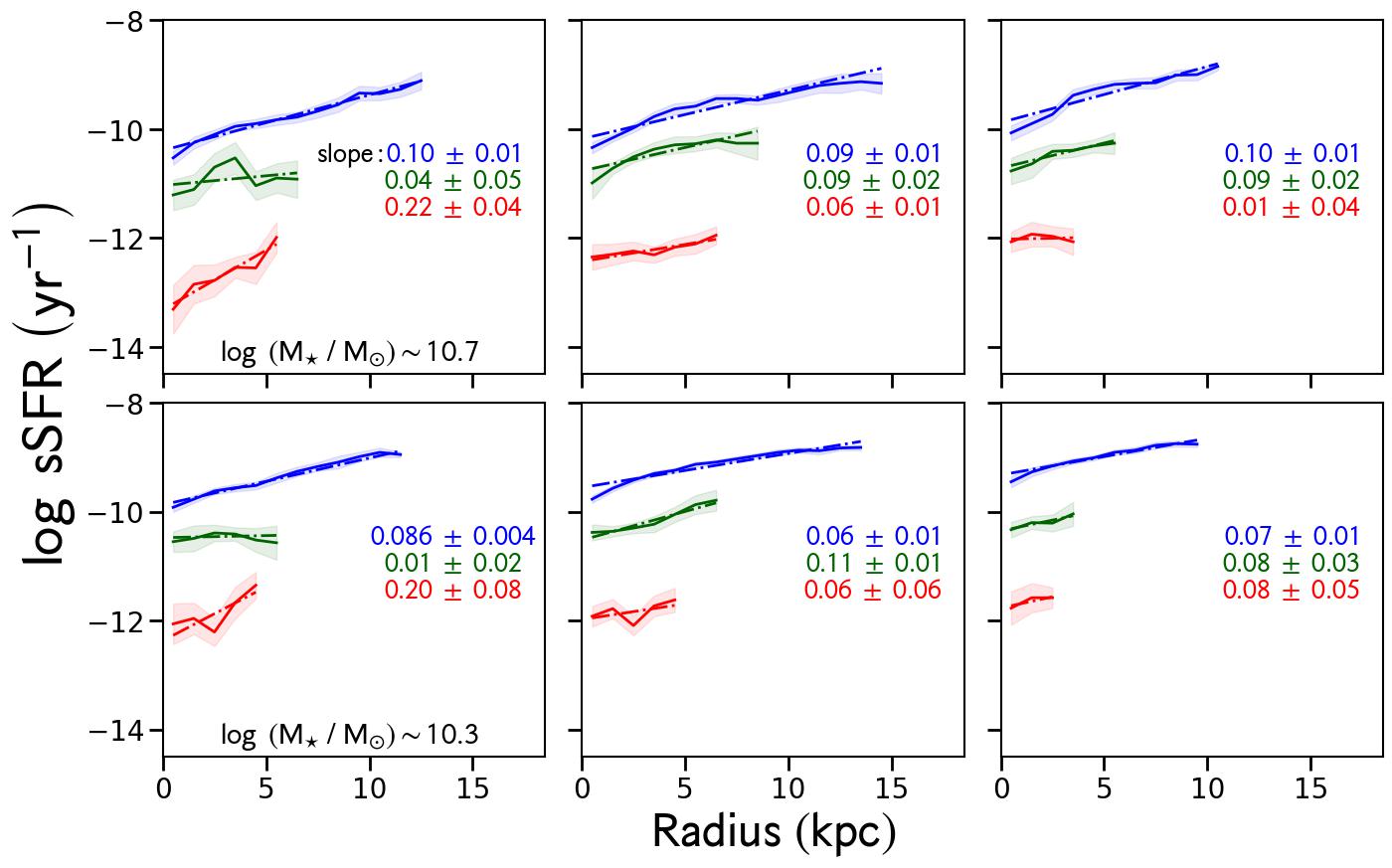}
    \caption{\textcolor{black}{Median specific star-formation rate profiles created under the fifth set of assumptions are shown. In the bottom two rows, fits to the median profiles are shown as dashed-dotted lines. The profiles of green valley galaxies at $z\sim1.3$ have similar slopes to those of quiescent galaxies at $z\sim0.9$, which we expect them to evolve into. This suggests that they quench at all radii at the same time. Profiles of green valley galaxies at $z\sim0.9$ are shallower than those of quiescent galaxies at $z\sim0.5$, which we expect them to evolve into.  This suggests inside-out quenching.
    Median profiles in each mass and redshift bin indicated are shown as thick solid lines and dashed lines wherever at least 10 and at least five galaxies contribute to the profile, respectively. Median profiles are fit at radii where at least 10 galaxies contribute.  Median profiles in the lowest-redshift bin are shown as thin dotted lines in other panels. The 16th--84th percentiles as a function of radius are indicated by the shading.}}
    \label{fig:optimal_ssfr}
    \vspace{1.0cm}
\end{figure*}

\begin{figure*}[t!]
    \centering
    \includegraphics[width=0.75\textwidth]{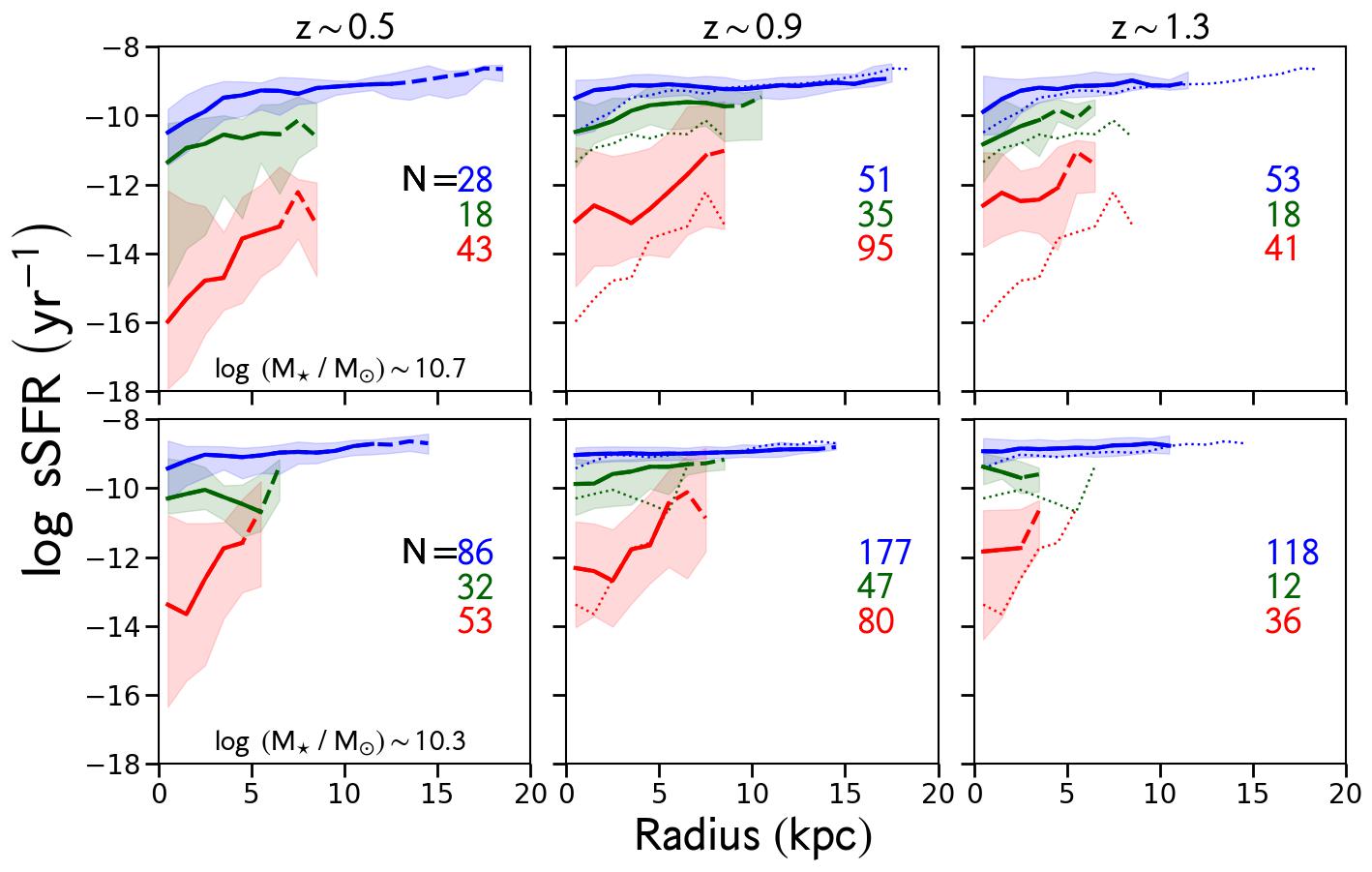}
    \includegraphics[width=0.75\textwidth]{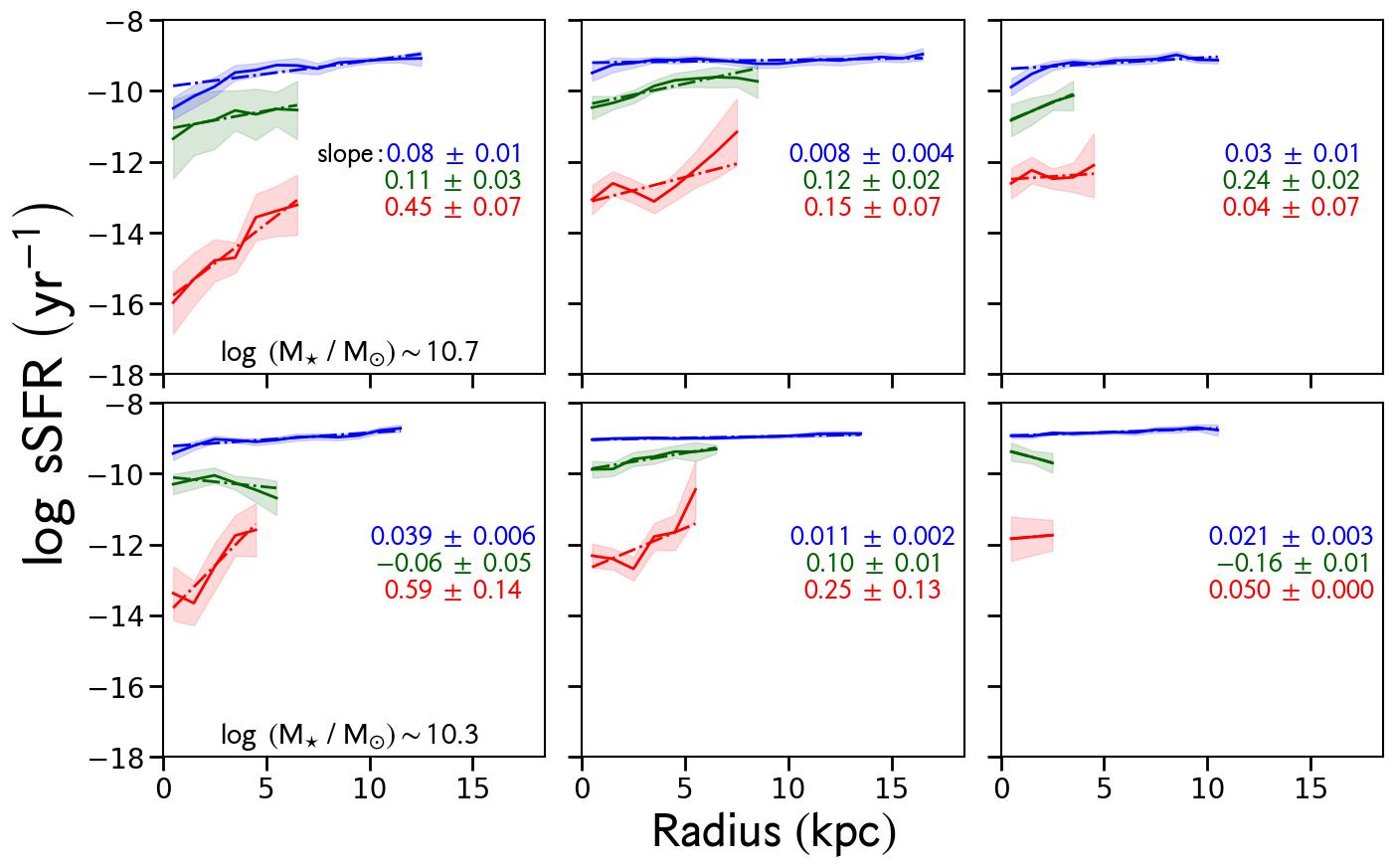}
    \caption{Median specific star-formation rate profiles \textcolor{black}{created} under the first set of assumptions are shown. We expect green valley galaxies at $z\sim 1.3$ to evolve into quiescent galaxies at $z\sim 0.9$.  Comparing these profiles, those of the quiescent galaxies are steeper. This is also the case for green valley galaxies at $z\sim0.9$ and $z\sim0.5$.
    This suggests inside-out quenching. Lines and shading have the same meaning as in Figure \ref{fig:optimal_ssfr}. }
    \label{fig:typical_ssfr}
\end{figure*}

Star-forming galaxies have median sSFR profiles with similar slopes at all redshifts ($0.06\pm0.01$ - $0.1\pm0.01$ dex kpc$^{-1}$). Green valley galaxies have similar slopes to the star-forming  galaxies at $z\sim0.9$ and $z\sim1.3$ ($0.08\pm0.03$ - $0.11\pm0.01$). However, at $z\sim0.5$, they
have flatter slopes than the star-forming galaxies ($0.01\pm0.02$ and $0.04\pm0.05$ versus $0.086\pm0.004$ and $0.10\pm0.01$).  Quiescent galaxies also have similar slopes to those of star-forming and green valley galaxies ($0.01\pm0.04$ -- $0.08\pm0.05$) at $z\sim0.9$ and $z\sim1.3$.  However, at $z\sim0.5$, they are much steeper, $\approx 0.2\pm0.06$.

\textcolor{black}{The measurements of these slopes are robust. This is because we only consider radii of individual profiles where the S/N $>5$ and only calculate a median profile if at least 10 galaxies contribute to it (Section \ref{sec:construct_median_profile}). Furthermore, although the slopes we measure at higher redshift are shallow, as demonstrated in Section \ref{sec:effects_assumptions}, we have the ability to measure steep slopes if present.}

\subsection{Differences between the First and Fifth Sets of Assumptions}
\label{sec:ssfr_differences}

Since there are significant differences between the sSFR profiles for the first and fifth sets of assumptions, they are compared here. \textcolor{black}{A review and explanations of the differences between these sets is given. Since many studies make assumptions similar to those in the first set, findings from these works are also reproduced.} In Figure \ref{fig:typical_ssfr}, the sSFR profiles created under the first set are shown, and they are compared with those from the fifth, which are shown in Figure \ref{fig:optimal_ssfr}. \textcolor{black}{Star-forming galaxies in the first set have significantly flatter profiles than those in the fifth (slopes of approximately $0.010\pm 0.003$ versus $0.1\pm0.01$). Green valley galaxies for the two sets have similar profiles. Profiles of quiescent galaxies have significantly different slopes between the first and fifth sets. At $z\sim0.5$, those in the first set are much steeper: approximately $0.5\pm0.01$ vs. $0.2\pm0.06$. The case is similar at $z\sim0.9$: about $0.2\pm0.01$ vs. $0.06\pm0.03$. The difference goes away at $z\sim1.3$, where the slopes are similar among both sets ($0.04\pm0.07$ and $0.05$ vs. $0.01\pm0.04$ - $0.08\pm0.05$).}

These differences arise from the way observed colors and physical properties relate to each other for the first and fifth sets of assumptions. Under the first set, sSFR does not change much along the color-color profiles of star-forming galaxies (second row in Figure \ref{fig:color_color_tracks}). Therefore, star-forming galaxies have flat sSFR gradients. The sharp transition in the F435W-F775W color between low- and high-sSFR models causes quiescent galaxies to have sSFR profiles that rise quickly with radius. Profiles of green valley galaxies lie between these extremes.
For the fifth set, sSFR varies along the color-color profiles of star-forming galaxies such that sSFR is lower in the centers and is higher in the outskirts. 
The F435W-F775W colors transition more smoothly between low- and high-sSFR models. As a result, quiescent galaxies have shallower profiles. Profiles of green valley galaxies are between these extremes.
}

Studies that adopt assumptions similar to those in the first set find shallow and steep slopes for the sSFR profiles of quiescent galaxies at stellar masses and redshifts similar to those examined in this paper. At $0.5 < z < 1.0$, \citet{Liu18} found that quiescent galaxies have nearly flat sSFR profiles. In contrast, at $0.2 < z < 0.7$, \citet{Morselli19} found steeply rising sSFR profiles for quiescent galaxies (slopes of $\sim0.3 - 0.5$). At intermediate redshifts of $0.7 < z < 1.5$, \citet{Nelson21} found shallow slopes of $\sim0.1$ for the sSFR profiles galaxies below the SFMS. Similar slopes for the profiles of quiescent galaxies are obtained by \citet{Abdurrouf18}, who examined galaxies at $\log \left(\textrm{M}_{\star} / \textrm{M}_{\odot}\right) \geq 10.5$ and at $0.8 < z < 1.8$. However, steep slopes of $\sim0.3--0.5$ are found for the profiles of quiescent galaxies at $0.7 < z < 1.2$ by \citet{Morselli19}. 

\section{Galaxy Evolution under the Fifth Set of Assumptions}
\label{sec:evolution_optimal}

\subsection{Evolution at Fixed Mass}

We now do science with the sSFR profiles created under the fifth set of assumptions (Figure \ref{fig:optimal_ssfr}). We do so at a fixed mass, or under the simple assumption  that galaxies do not gain enough mass with time to leave their respective mass ranges.  We assume that with time, star-forming galaxies become green valley galaxies, and green valley galaxies become quiescent.  Specifically, we assume that the star-forming galaxies at $z\sim 1.3$ and $\sim 0.9$ become green valley galaxies at $z\sim 0.9$ and $z\sim 0.5$, respectively.  We assume that green valley galaxies at $z\sim 1.3$ and $z\sim0.9$ become quiescent galaxies at $z\sim0.9$ and $z\sim0.5$, respectively. Finally, we assume that quiescent galaxies remain quiescent. 

While the assumptions above are oversimplified, we validated some of them. We compared the SFHs of the green valley galaxies with those of quiescent galaxies at $z\sim1.3$ and $z\sim0.9$ and at $z\sim0.9$ and $z\sim0.5$ and find that they are consistent. We also find that the SFHs of quiescent galaxies at $z\sim0.9$ and $z\sim0.5$ overlap with those of quiescent galaxies at $z\sim1.3$ and $z\sim0.9$, respectively. Therefore, our assumptions that these populations are connected at fixed mass are valid.
In Appendix \ref{sec:evolution_optimal_abundance_matching}, we consider bins in mass that evolve with redshift and find quantitatively similar results to those presented here.

The median sSFR profiles of star-forming galaxies at $z\sim1.3$ and those of green valley galaxies at $z\sim0.9$ \textcolor{black}{that we expect them to evolve into} differ most significantly in normalization \textcolor{black}{(by about $0.8$ dex)} but have similar slopes (approximately $0.08\pm 0.01$ and $0.1\pm 0.02 \ \textrm{dex kpc}^{-1}$, respectively). 
At $z\sim0.5$, the median profiles differ in normalization and slope. The normalization decreases from the star-forming to green valley by about $0.8$ dex, and the slopes become shallower by about 0.05.

At all redshifts, \textcolor{black}{profiles of green valley galaxies have higher normalizations than those for quiescent galaxies by around $1$ dex. Green valley galaxy profiles at $z\sim1.3$ have slopes that are similar to those of quiescent galaxies at $z\sim0.9$ (approximately $0.09\pm0.02$ and $0.07\pm0.03$, respectively). Because their slopes are similar and the most significant difference is in their normalizations, this suggests that, as green valley galaxies quench, they decline in sSFR at all radii at the same time. Therefore, if green valley galaxies at $z\sim1.3$ evolve into quiescent galaxies at $z\sim0.9$, they must quench at all radii at the same time.}

Interestingly, at lower redshifts, the sSFR profiles of the quiescent galaxies are considerably steeper than all the other profiles, with slopes of \textcolor{black}{about $0.2\pm0.06$}. Green valley galaxies at $z\sim0.9$ have slopes of \textcolor{black}{about $0.1\pm0.01$}. Therefore, if green valley galaxies at $z\sim0.9$ evolve into quiescent galaxies at $z\sim0.5$, they must reduce star formation in their inner parts first.
This is confirmed by examining the median SFHs of quiescent galaxies at $z\sim0.5$. We find that the median SFH in the centers of these galaxies peaks and declines earlier than that in the outskirts. This is consistent with reducing SFR in their inner parts first.

\subsection{Interpreting Our Findings}
\label{sec:discuss_quenching}

In this Section we compare our results under the fifth sets of assumptions to those from the literature to ours and discuss potential physical mechanisms for quenching galaxies in our sample. In the previous subsection, we find that in the median, quenching occurs at all radii at the same time at $0.7 < z < 1.5$ and from the inside-out at $0.4 < z < 0.7$. 

Quenching mechanisms that are consistent with inside-out quenching include compaction \citep{Zolotov15, Tacchella16b}, quenching due to AGNs (\citealt{Sanchez18} and references therein; \citealt{Nelson19a, Appleby20, Nelson21}), and morphological quenching \citep{Martig09, Genzel14}. A decline in the gas accretion at all radii would be consistent with quenching at all radii at the same time. 

Physical processes that quench galaxies from the outside-in are not consistent with our observations in the median. Such processes include ram pressure stripping \citep[e.g.,][]{Fossati18, Cramer19}, though this process is expected to occur in dense environments, e.g., galaxy clusters. Gas-rich mergers, which are believed to enhance star-formation in the center and suppress star-formation in the outskirts \citep{Mihos96, Snyder11, Barrera-Ballesteros15, Moreno15, Thorp19, Steffen21}, would also be inconsistent with our findings in the median. 

Gas-poor mergers could explain some of the evolution observed under the fifth set of assumptions. Mass-weighted age profiles are observed to progressively flatten as galaxies transition from star-forming to green valley to quiescent, as seen in Figure \ref{fig:all_param_one_bin} and Appendix \ref{sec:science_profiles_all}. Gas-poor mergers are known to flatten age gradients \citep[e.g.,][]{White79, White80, Bekki98, Bekki99, Ogando05, Hopkins09a, Tortora10, Hirschmann15}. Studies that have measured merger rates find that the number density evolution of high-mass $\left(\textrm{M}_{\star} \geq 10^{10.5} \textrm{M}_{\odot}\right)$ quiescent galaxies at $z < 1$ can be accounted for by major mergers of star-forming galaxies (e.g., \citealt{Bell06, Robaina10, Brammer11, Man12, Man16}; though see 
\citealt{Moustakas13}).
However, it is unclear whether there are enough major mergers at low redshift to build up the quiescent galaxy population observed at $z = 0$ at lower masses \citep[e.g.,][]{Weinzirl09, Weigel17, RodriguezMontero19}. 

Another mechanism that can flatten age gradients is one where the orbits of stars in galaxies change in radius with time, regardless of the merger history. In disk galaxies, this process occurs due to perturbations to their galactic potentials and is known as radial migration \citep[e.g.,][]{LyndenBell72, Sellwood02}. While some studies find radial migration significantly influences the radial gradients of stellar populations \citep[e.g.,][]{Roskar08, Schonrich09, Buck20}, others do not \citep[e.g.,][]{Minchev14, Grand15, GrandKawata16, AvilaReese18}. IFU observations of $z\sim0$ spiral galaxies show that radial migration appears to have had minor effects on their growth in size and mass over the last 10 Gyr \citep{Peterken20}. We conclude that radial migration does not contribute significantly to our results.

\section{Summary and Conclusions}
\label{sec:conclusions}

We use multiband HST images from the optical through near-infrared to measure radial flux profiles of 1,440 galaxies with stellar masses $10 \leq \log \left(\textrm{M}_{\star} / \textrm{M}_{\odot}\right) \leq 11.5$ over redshifts $0.4 < z < 1.5$. We fit the resulting broadband SEDs at each radius in these profiles using the {\sc beagle} tool.  We then examine the resulting profiles of sSFR, SFR surface density, stellar mass surface density, attenuation in the rest-frame $V$ band, \AvNoSpace, and mass-weighted age. We do so as function of galaxy stellar mass and redshift. This has been done by other studies \citep[e.g.,][]{Nelson16, Nelson21, Wang17, Abdurrouf18, Liu18, Morselli19} and we revisit it to examine the effects of adopting more physically motivated assumptions in the fitting of the SEDs.

In an ideal world, priors used in SED fits are uninformative, and data are highly constraining.  However, this is rarely the reality.  When this is the case, we suggest an iterative approach.  We start with priors that are minimally informative and compare the models created with these priors with data in color-color space. Then, to optimize the overlap of models with data, priors are edited in a physically meaningful way.  We use this technique to develop five sets of SED-fitting assumptions that build upon each other.

We first consider assumptions that are commonly used in the literature, namely, a dust law with a fixed slope and uniform priors on all free parameters in the SED fits \citep[e.g.,][]{Wuyts12, Hemmati14, Hemmati20, Cibinel15, Morishita15, Mosleh17, Mosleh20, Abdurrouf18, Guo18, Sorba18, Morselli19, Suess19a}. 
Then, we examine the cumulative effects of a number of changes that are based on studies that examine individual components of the SED fits.  These changes are described below. To do so, we gradually incorporate them into the SED fits by defining five sets of SED-fitting assumptions (Section \ref{sec:five_sets_assumptions} and Table \ref{tab:priors}). 

\subsection{The Five Sets of SED-fitting Assumptions}

The first set of assumptions, which is commonly used in the literature, consists of a single dust law and uniform priors on all parameters, including optical depth, log of the star-formation history timescale, log age, and log metallicity.  

For the second set of assumptions, we make a significant change and adopt the flexible dust law developed by \citet{Chevallard13}. A flexible dust law encapsulates the variety of dust laws found for galaxies in the nearby Universe \citep{Wild11, Battisti17, Salim18} and at intermediate redshift \citep{Kriek13, Salmon16, Barisic20}. 

For the third set of assumptions, we add to the second set an exponential prior on the effective optical depth in the rest-frame $V$ band, \tauVeffNoSpace. This is also a significant change.  It ensures that models with little to no dust (\Av $\sim0$ mag) are more probable than models with a lot of dust (\Av $\gtrsim 2$ mag). We make this change because the distribution of dust attenuation values of galaxies in the nearby Universe is approximately exponential \citep{Brinchmann04}. We expect it to be similar at \textcolor{black}{the redshifts studied in this paper because the correlation between mass and dust and the shape of the mass function do not change significantly over $0<z<1.5$ (see \citealt{GarnBest10, Dominguez13, Momcheva13, Ramraj17} and \citealt{Muzzin13, Davidzon17}, respectively).}

For the fourth set, we incorporate a Gaussian prior on the log of the SFH timescale, which allows for extended star-formation histories. \textcolor{black}{This change is made} because studies that adopt nonparametric SFHs find that massive quiescent galaxies at $z\sim1$ peak approximately $1--2$ Gyr after they begin forming stars \citep{Pacifici16}. This change ensures that quiescent galaxies \textcolor{black}{can have star-formation histories that peak about} 1 Gyr after they begin forming stars. When adopting a uniform prior, as done by other studies, \textcolor{black}{star-formation histories tend to peak much earlier, typically} 200--300 Myr after stars begin to form. This change is significant for quiescent galaxies and relevant for our results on quenching. 

For the fifth set of assumptions, broad Gaussian priors to log age and log metallicity \textcolor{black}{are applied}. This is done to be consistent with studies that have determined ages and metallicities of galaxies at $z\sim1$ via spectroscopy \citep{Maiolino08, Hathi09,  Henry13, Gallazzi14, Onodera15, Guo16a, EstradaCarpenter19, Maiolino19}. These changes \textcolor{black}{cause minor} differences \textcolor{black}{in the results between the fourth and fifth sets of assumptions}.

\subsection{Choosing a Set of Assumptions}

\textcolor{black}{We perform five complementary tests to determine which set of assumptions is most appropriate for our sample. In three of the tests, the fifth set performs the best. In two of them, the third, fourth, and fifth sets perform equally well.}

\textcolor{black}{As a first test, model colors are compared with observed integrated colors (Figure \ref{fig:model_obs_colors}). Appropriate priors result in model colors that overlap with the observed colors. This overlap is quantified by calculating the fraction of models that lie within a rectangular region that contains 90\% of the observations (Section \ref{sec:first_set}). For the first set, only about 50\%} of the models lie in this region. For the second set, the fraction rises to 83\%. \textcolor{black}{From the third to the fifth sets, the fraction continues to rise to 93\%. Therefore, the fifth set is the most appropriate for this test.}

Second, we compare the Bayesian evidence of the fifth set of assumptions with that of the other sets for SED fits to the spatially integrated fluxes (see Figure \ref{fig:bayesian_evidence} and Section \ref{sec:bayesian_evidence}). The fifth set is preferred to the first and marginally preferred to the second. It is not preferred over the third or the fourth.

\textcolor{black}{Third, SFR surface density profiles of quiescent galaxies at $z\sim0.5$ are compared with those measured from spectroscopy at $z\sim0$ (see Figure \ref{fig:sfr_medling} and Section \ref{sec:sami}). Justifications for doing this are provided in Section \ref{sec:justifications}. Profiles for the fifth set agree the best, and those for the fourth set perform marginally worse than the fifth.}

\textcolor{black}{Fourth, \Av profiles of star-forming galaxies at $z\sim0.5$ are compared with those measured from spectroscopy at $z\sim0$ (Figure \ref{fig:av_greener} and Section \ref{sec:manga}). Profiles for the fifth set agree the best.}

\textcolor{black}{For the fifth test, the behavior of \Av profiles of star-forming galaxies is examined as a function of galaxy inclination (Figure \ref{fig:av_prof_axis_ratio} and Section \ref{sec:dust_inclination}). Profiles for the third through fifth sets demonstrate the expected behavior, namely, they steepen in their centers as inclination increases, and are therefore preferred for this test.}

\subsection{Evolution of the sSFR Profiles under the Fifth Set of Assumptions}

\textcolor{black}{The median sSFR profiles created under the fifth set are examined to determine their evolution with redshift. This is done by comparing profiles at a fixed mass. The green valley galaxies at $z\sim1.3$ are assumed to evolve into quiescent galaxies at $z\sim0.9$. The median profiles of these green valley and quiescent galaxies each have different normalizations but have similar slopes. Therefore, the sSFR profiles simply decline in normalization with time, quenching at all radii at the same time. The green valley galaxies at $z\sim0.9$ are assumed to evolve into quiescent galaxies at $z\sim0.5$. The median profiles of these green valley and quiescent galaxies each have different normalizations and slopes, with those of the latter being steeper. Therefore, as galaxies quench, sSFR decreases more in their centers than in their outskirts. They quench from the inside-out.}

\textcolor{black}{In summary, we conclude that galaxies quench at all radii at the same time over $z\sim1.3$ to $z\sim0.9$ and from the inside-out over $z\sim0.9$ to $z\sim0.5$.} 

\textcolor{black}{Other studies generally find inside-out quenching at all redshifts considered in this paper. This is because assumptions similar to those in the first set, primarily a fixed dust law, are assumed. Under a fixed dust law, profiles of quiescent galaxies are steep due to a sharp transition in model colors, which results in profiles that change abruptly in the outskirts (Section \ref{sec:effects_assumptions} and Figure \ref{fig:color_color_tracks}). 
Under the fifth set, the transition in color is generally smoother than under the first. However, at $z\sim0.5$, colors transition more sharply than they do at $z\sim0.9$ and at $z\sim1.3$. This results in steeper sSFR profiles of quiescent galaxies in this set at $z\sim0.5$ than at higher redshifts. Consequently, galaxies quench from the inside-out under both the first and fifth sets at $z\sim0.5$.}

\section*{Acknowledgements}

We thank the anonymous referee for providing constructive comments. A.d.l.V., S.A.K., and C.P. would like to acknowledge support from NASA's Astrophysics Data Analysis Program (ADAP) grant Nos. 80NSSC20K0760 and NNX16AF44G and an RSAC grant from the Space Telescope Science Institute. W.W. would like to acknowledge support from ADAP grant No. 80NSSC20K0760 and an RSAC grant. A.d.l.V. gratefully acknowledges support from a Chateaubriand STEM Fellowship and support from the Maryland Space Grant Consortium via NASA grants NNX15AJ21H and 80NSSC20M0049. C.P. was partially supported by the Canadian Space Agency under a contract with NRC Herzberg Astronomy and Astrophysics.

A.d.l.V. and S.A.K. would like to thank S. M. Faber, D. C. Koo, H. Ferguson, and B. Weiner for valuable comments on an earlier draft of this work. A.d.l.V. thanks Michael J. Greener and Anne M. Medling for kindly providing median profiles and galaxy identifiers for their respective samples. 

This work was conducted using computational resources at the Maryland Advanced Research Computing Center (MARCC) and the Advanced Research Computing at Hopkins (ARCH) core facility (\url{rockfish.jhu.edu}), which is supported by the National Science Foundation (NSF) grant No. OAC 1920103. A.d.l.V. and S.A.K. thank Profs. Jaime Combariza and Alex Szalay for directing computational resources at Johns Hopkins University. A.d.l.V. also thanks Prof. Jaime Combariza and Dr. Tanvi Karwal for valuable advice on how to use MARCC. 

All of the HST data used in this paper can be found in MAST: \dataset[10.17909/T94S3X]{http:/dx.doi.org/10.17909/T94S3X}

\software{astropy \citep{2013A&A...558A..33A,2018AJ....156..123A, Astropy2022}, ipython \citep{PER-GRA:2007}, LMFIT \citep{Newville14}, matplotlib \citep{Hunter07}, numpy \citep{harris20}, photutils \citep{photutils22}, scipy \citep{Scipy20}}

\begin{figure*}[t!]
    \includegraphics[width=\textwidth]{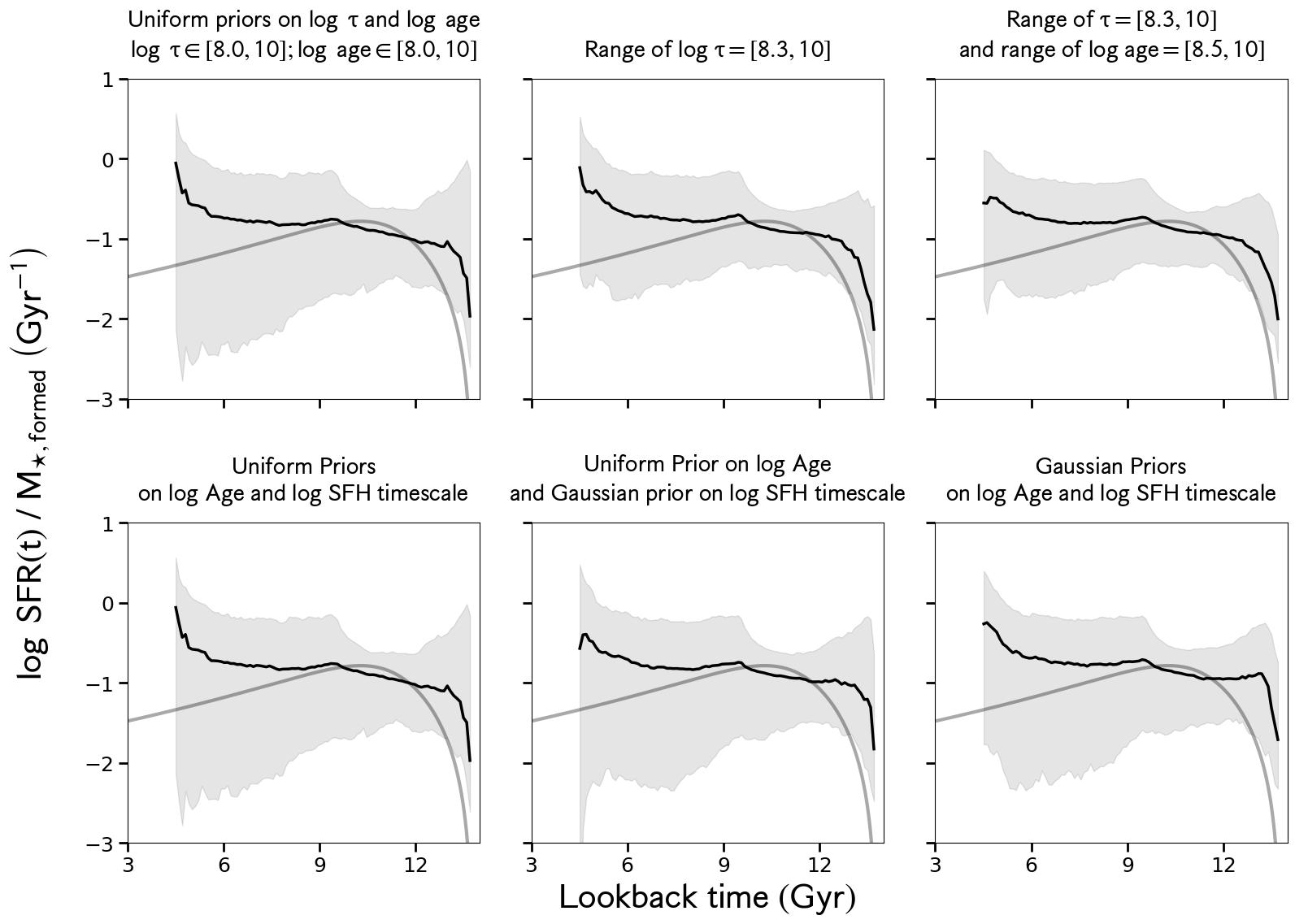}
    \caption{Comparison between the \citet{Madau14} cosmic star-formation rate density (dark-gray line) and median SFHs of model galaxies for two cases: one where only the ranges on the log SFH timescale and log age change (top row) and another where only the priors distributions on these parameters change (bottom row). Median SFHs shown as black solid lines and the 16th--84th percentile ranges are plotted as gray shaded regions. Narrowing the range on $\log~\tau$ has the most significant effect (top-middle panel). Doing so down-weights the probability for models with extreme SFHs, i.e., those that sharply rise or decline with time. Adjusting the range on log age and changing prior distributions from uniform to Gaussian have minor effects. In all cases, model SFHs tend to rise at late times, in contrast with the SFRD. This is because the priors are constructed such that young, star-forming galaxies are abundant at all redshifts. This achieves the highest overlap between observed and model colors and results in model sSFRs that are high in the median at all redshifts.}
    \label{fig:sfhs_sfrd}
\end{figure*}

\appendix

\section{The Effects of Changing the Priors on the Star-formation Histories}
\label{sec:priors_sfh}

Priors on the SFH are adjusted in the fourth and fifth sets following studies based on spectroscopy (Sections \ref{sec:fourth_set} and \ref{sec:fifth_set}). As a result, extreme SFHs, i.e., those that rise or decline sharply with time, occur less often in these sets than in the first. Changing the range on log $\tau$ alone has the most significant effect: it reduces the frequency of extreme SFHs. Changing the range on log age alone has minor effects. Changing the prior distributions from uniform to Gaussian also has minor effects, as the Gaussian priors are broad.

In Figure \ref{fig:sfhs_sfrd}, we show median SFHs and the 16th--84th percentile ranges of 10,000 model SFHs for two cases: one where only the range on $\log~\tau$ and log age is each adjusted, and another where only the prior on each parameter is changed.
Each SFH is randomly drawn from the priors and must end at a redshift between $z=0.4$ and $z=1.5$. The shapes of the model SFHs are compared with that of the cosmic star-formation rate density (SFRD), measured by \citet{Madau14}, which is repeated in all panels of Figure \ref{fig:sfhs_sfrd}. Following \citet{Carnall19}, we show the cosmic SFRD divided by the integral of the cosmic SFRD over time. 

In all cases, the median SFH rises quickly in the first $\sim$Gyr after the Big Bang and rises gradually as lookback time decreases. The median SFH rises at late times, in contrast with the SFRD, because the priors are set such that young, star-forming galaxies are abundant at all redshifts. This is done to achieve the highest overlap between observed and model colors. Therefore, the model SFHs have high sSFRs in the median at all redshifts, which contrasts with that of the SFRD. For the priors under the first set, shown in the top-left panel of Figure \ref{fig:sfhs_sfrd}, the 16th--84th percentile range grows with time such that it spans $\sim3$ dex by $z=0.4$. This is because of the wide range ([8.0, 10.0]) on the log SFH timescale, $\tau$. When narrowing the range to [8.3, 10.0] (top-middle panel), the percentile range is smaller (1.5 dex at $z=0.4$). Narrowing the range on log age from [8.0, 10] to [8.5, 10] (top-right panel) has minor effects. Modifying the priors from uniform to Gaussian (bottom row) also has minor effects. This is because the Gaussian priors are wide.

\section{Color--Color Tracks for Each Set of Assumptions}
\label{sec:color_color_tracks_appendix}

\begin{figure*}[t!]
    \centering
    \includegraphics[width=\textwidth]{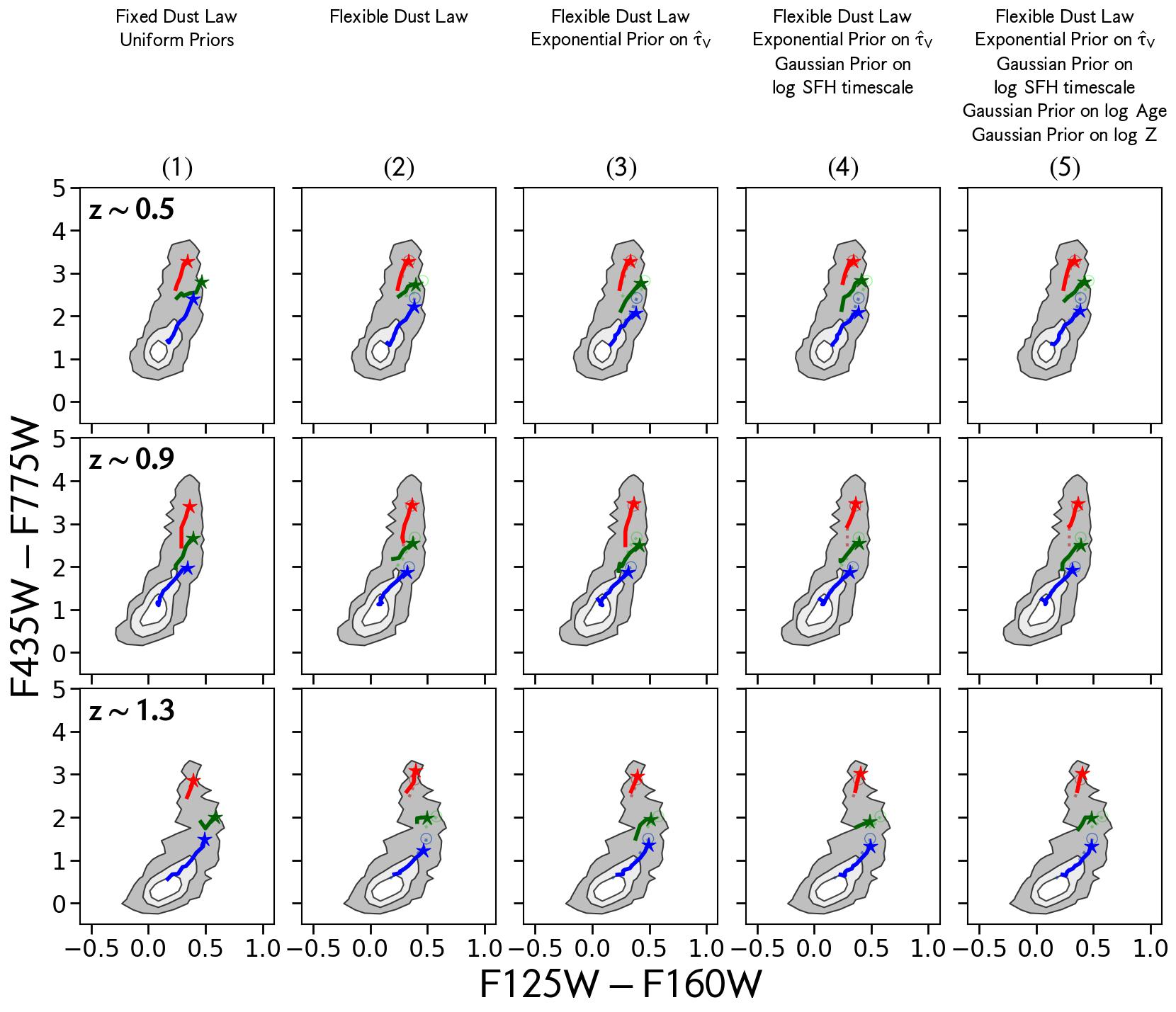}
    \caption{\textcolor{black}{The median color-color tracks for galaxies that have stellar masses $10 < \log\left( \textrm{M}_{\star}/\textrm{M}_{\odot}\right) < 10.5$ are shown here for all sets of assumptions. The definition of star-forming, green valley, and quiescent varies according to the set of assumptions used. Reassuringly, the color-color tracks defined under each set are consistent with each other.} The median central colors are shown as stars. Tracks for galaxies defined under the first set of assumptions are repeated in the second through fifth columns for comparison. They are shown as faint dotted lines, and their central colors are shown as empty circles.}
    \label{fig:color_color_appendix}
\end{figure*}

We show median color-color tracks of star-forming, green valley, and quiescent galaxies defined under each set of assumptions in Figure \ref{fig:color_color_appendix} as blue, green, and red lines, respectively. Color--color tracks in each redshift bin are shown in different rows: $z\sim0.5$ in the top row, $z\sim0.9$ in the middle, and $z\sim1.3$ in the bottom. All galaxies used to create these median profiles have stellar masses \textcolor{black}{$10 < \log\left(\textrm{M}_{\star} / \textrm{M}_{\odot}\right) < 10.5$}. 

The median color-color tracks \textcolor{black}{for all sets of assumptions are in good agreement}. The greatest discrepancy between any set of color-color tracks is that seen for star-forming galaxies at $z\sim0.5$ between the first and third sets, shown as the light-blue circle and dark-blue star, respectively, in the top row, middle column. The central median F435W-F775W colors disagree by 0.3 mag. However, the dispersion about each of these central colors is $\sim0.4$ mag. Therefore, this discrepancy is not significant. 

\section{Science Profiles under All Sets of Assumptions}
\label{sec:science_profiles_all}

\begin{figure*}[t!]
    \centering
    \includegraphics[width=0.75\textwidth]{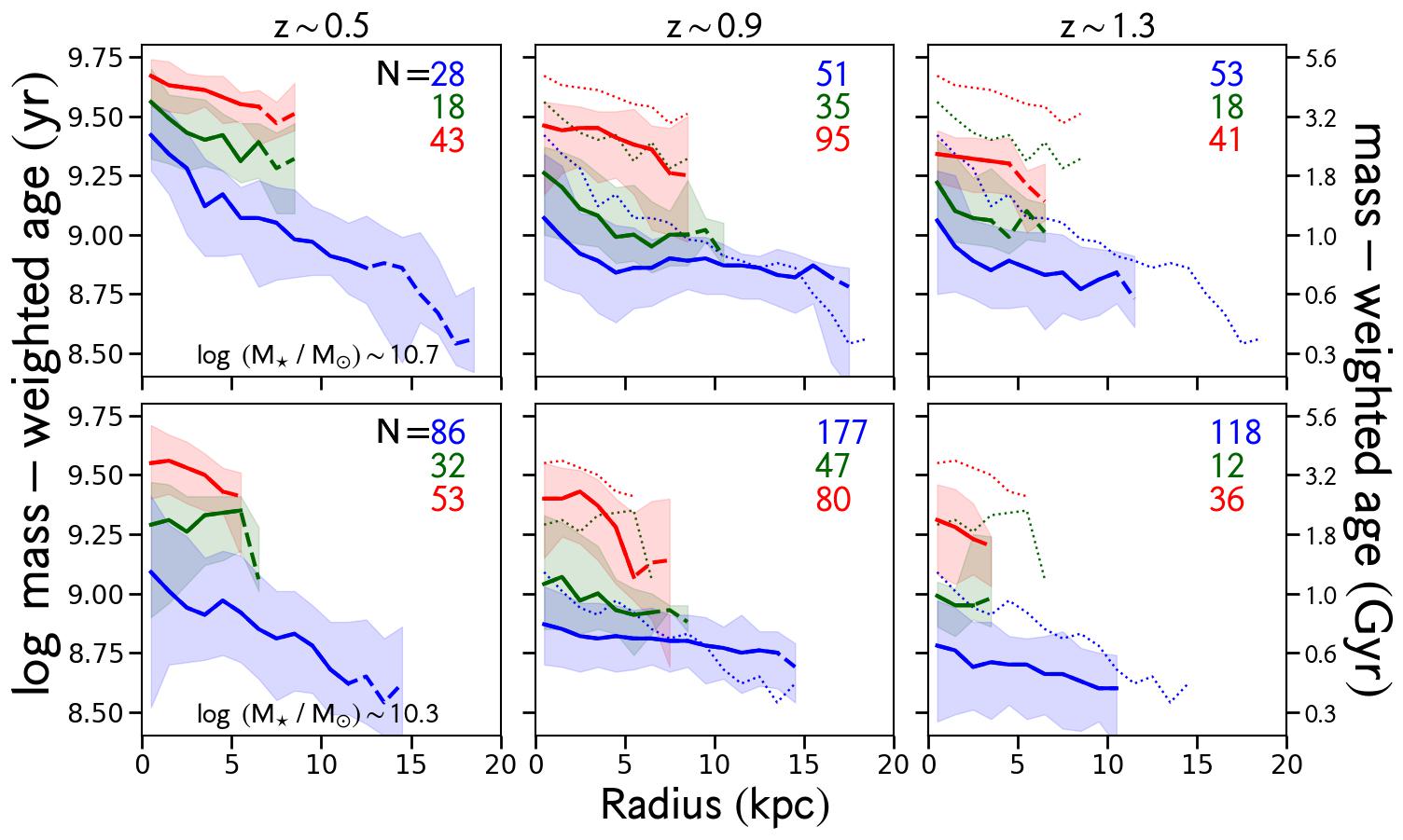}
    \includegraphics[width=0.75\textwidth]{typical_assumptions_log_ssfr_50_r_circularized_all_bins_plot_low_z_nov23.jpg}
    \caption{Median science profiles of mass-weighted age (top) and sSFR (bottom) created under the first set of assumptions for all mass and redshift bins \textcolor{black}{are shown}. In each panel, the meanings of the lines, shadings, and colors are the same as those in the top two rows in Figure \ref{fig:optimal_ssfr}.}
    \label{fig:typical_mw_ssfr}
\end{figure*}

\begin{figure*}[t!]
    \centering
    \includegraphics[width=0.75\textwidth]{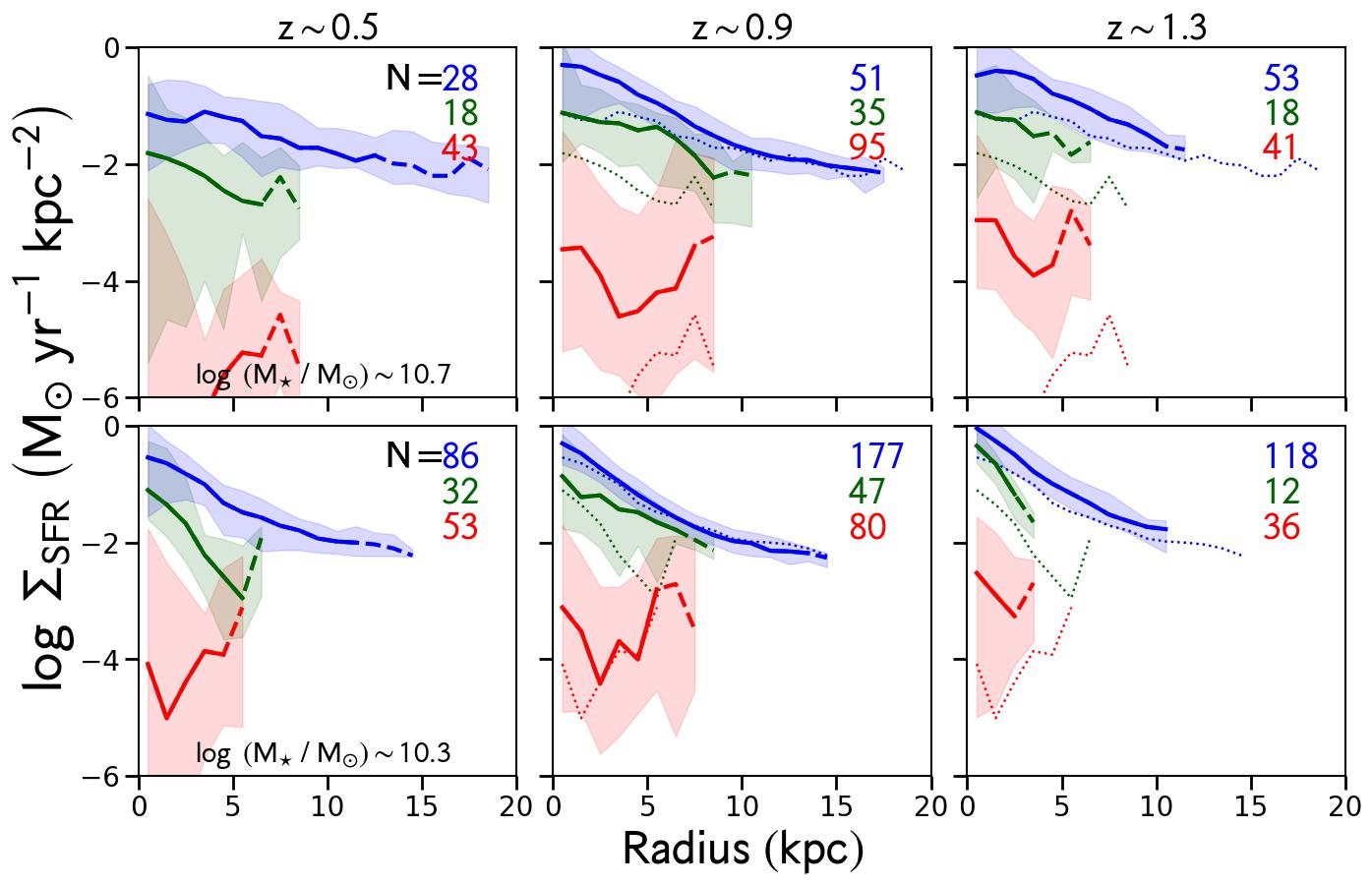}
    \includegraphics[width=0.75\textwidth]{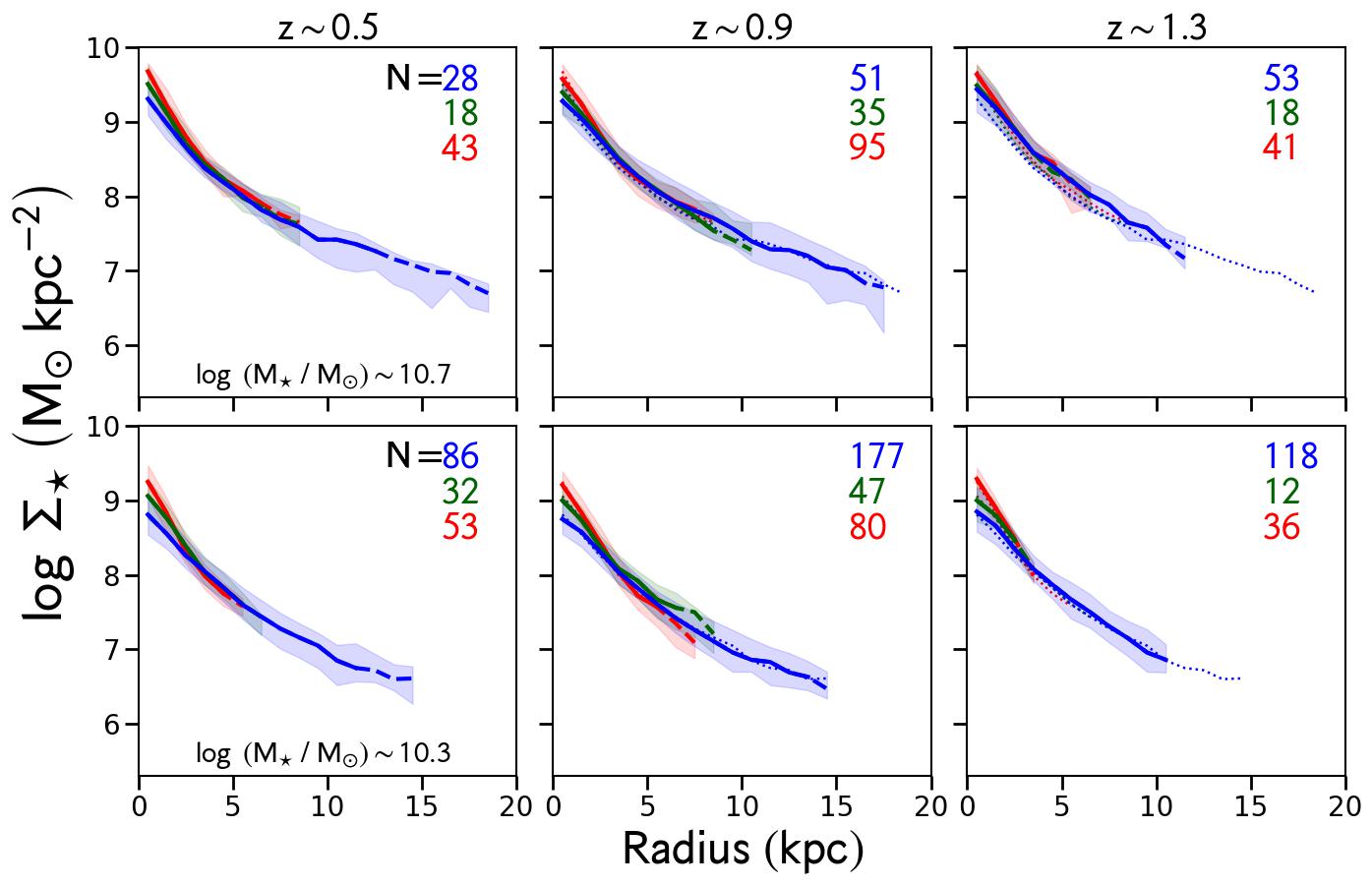}
    \caption{Median science profiles of SFR surface density (top) and stellar mass surface density (bottom) under the first set of assumptions for all mass and redshift bins \textcolor{black}{are shown}. In each panel, the meanings of the lines, shadings, and colors are the same as those in the top two rows in Figure \ref{fig:optimal_ssfr}.}
    \label{fig:typical_sfr_star}
\end{figure*}

\begin{figure*}[t!]
    \centering
    \includegraphics[width=0.75\textwidth]{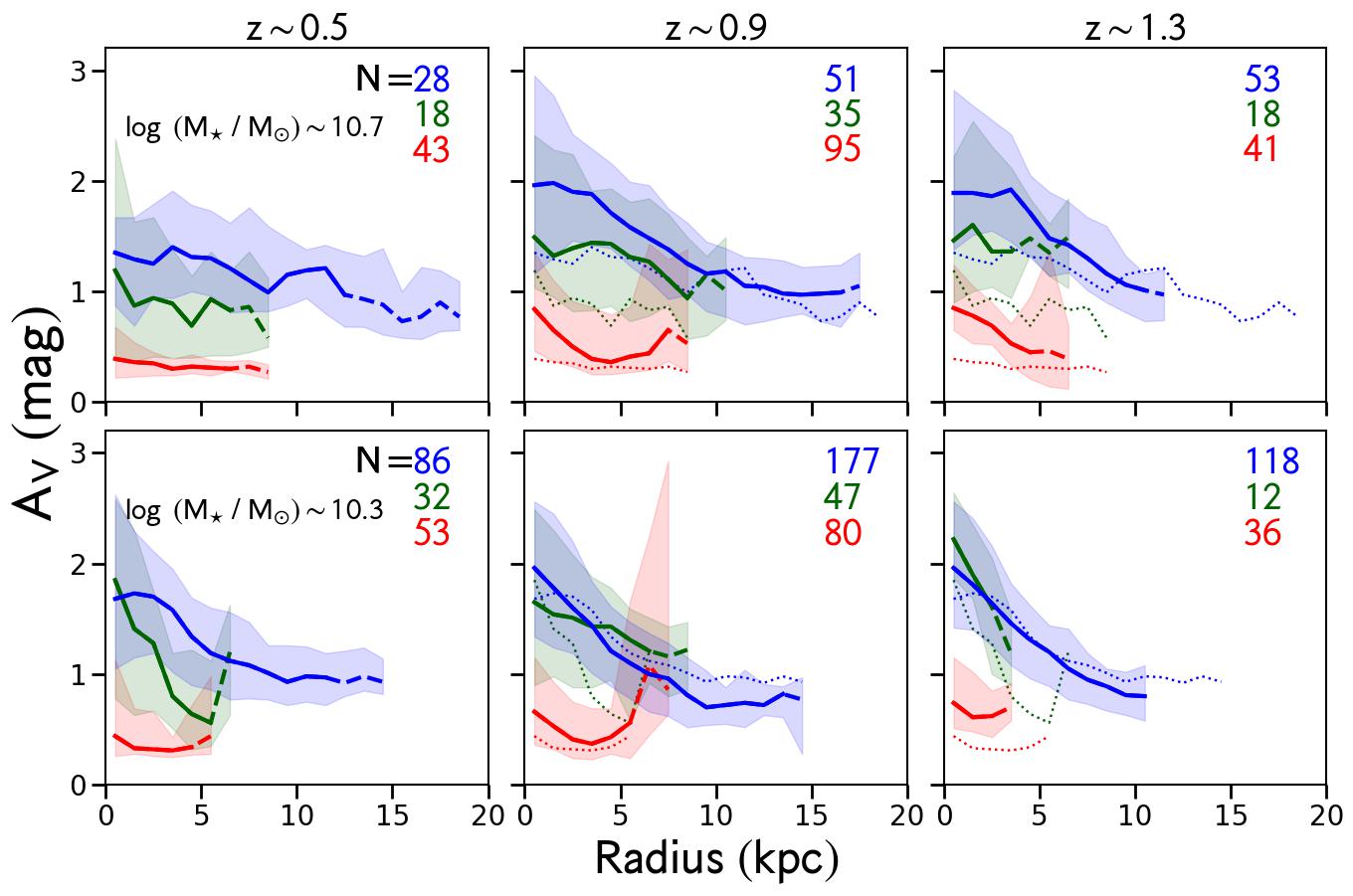}
    \caption{Median science profiles of \Av under the first set of assumptions for all mass and redshift bins \textcolor{black}{are shown}. In each panel, the meanings of the lines, shadings, and colors are the same as those in the top two rows in Figure \ref{fig:optimal_ssfr}.}
    \label{fig:typical_av}
\end{figure*}

\begin{figure*}[t!]
    \centering
    \includegraphics[width=0.75\textwidth]{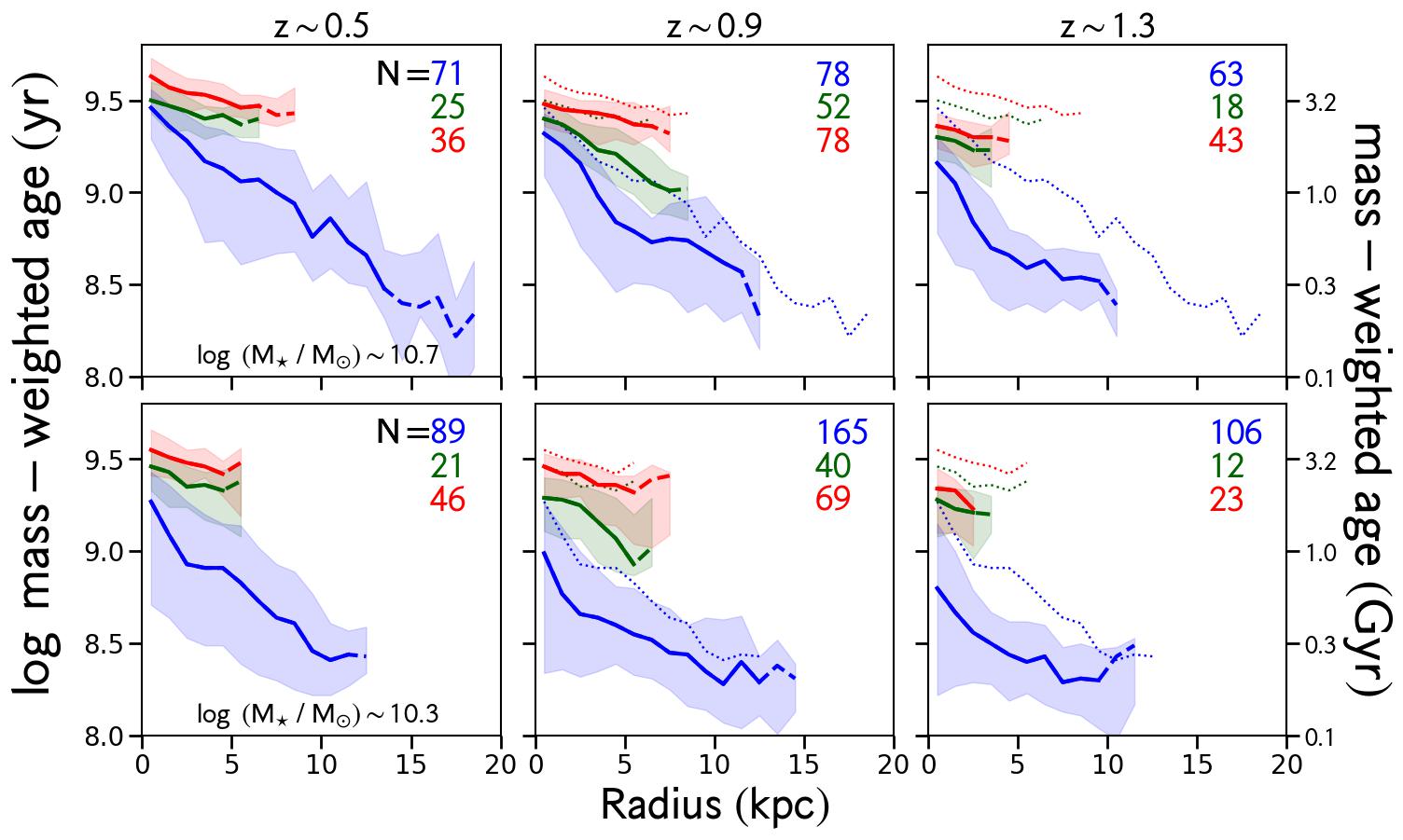}
    \includegraphics[width=0.75\textwidth]{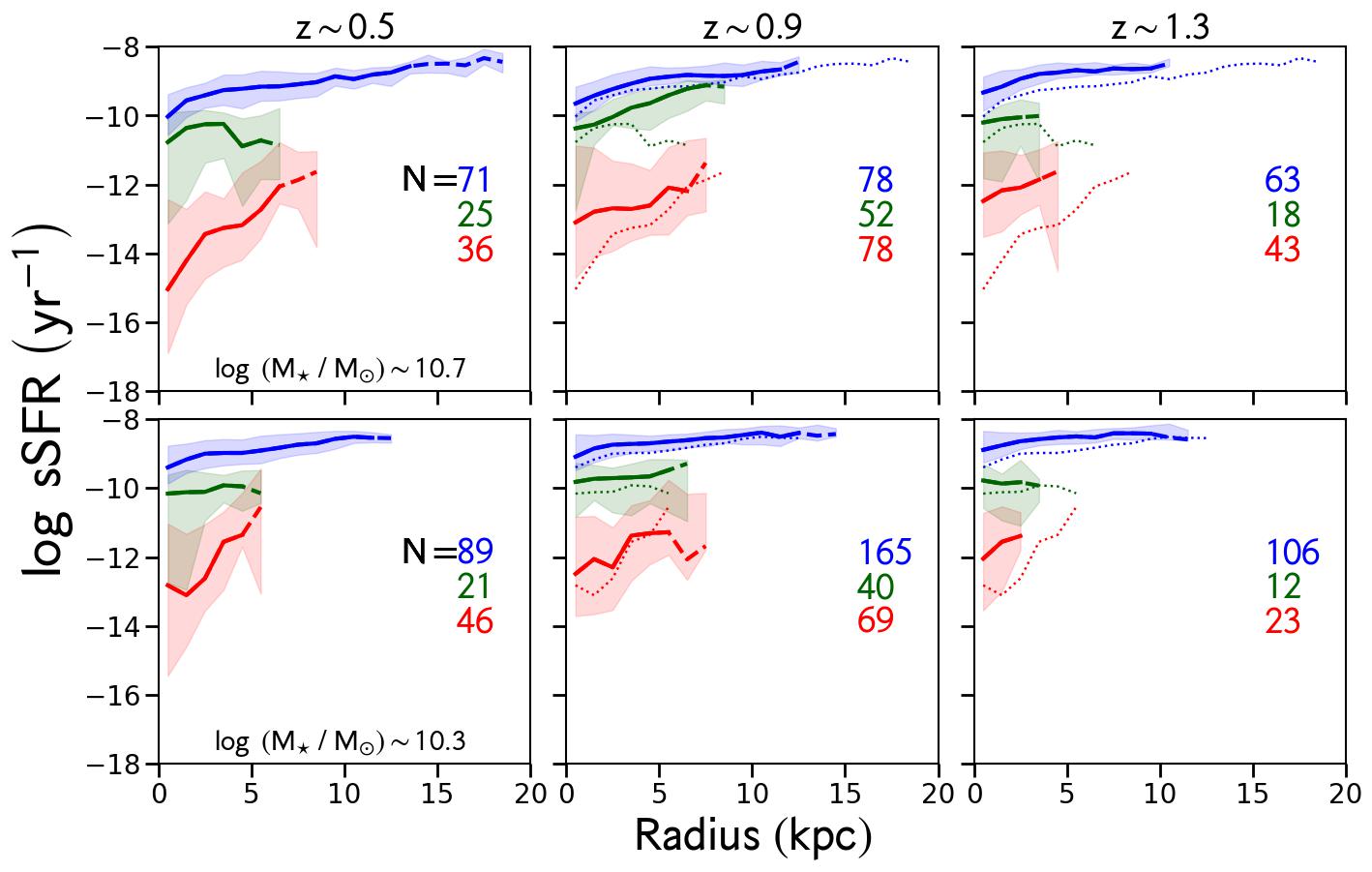}
    \caption{Median science profiles of mass-weighted age (top) and sSFR (bottom) under the second set of assumptions for all mass and redshift bins \textcolor{black}{are shown}. In each panel, the meanings of the lines, shadings, and colors are the same as those in the top two rows in Figure \ref{fig:optimal_ssfr}.}
    \label{fig:intermediate_mw_ssfr}
\end{figure*}

\begin{figure*}[t!]
    \centering
    \includegraphics[width=0.75\textwidth]{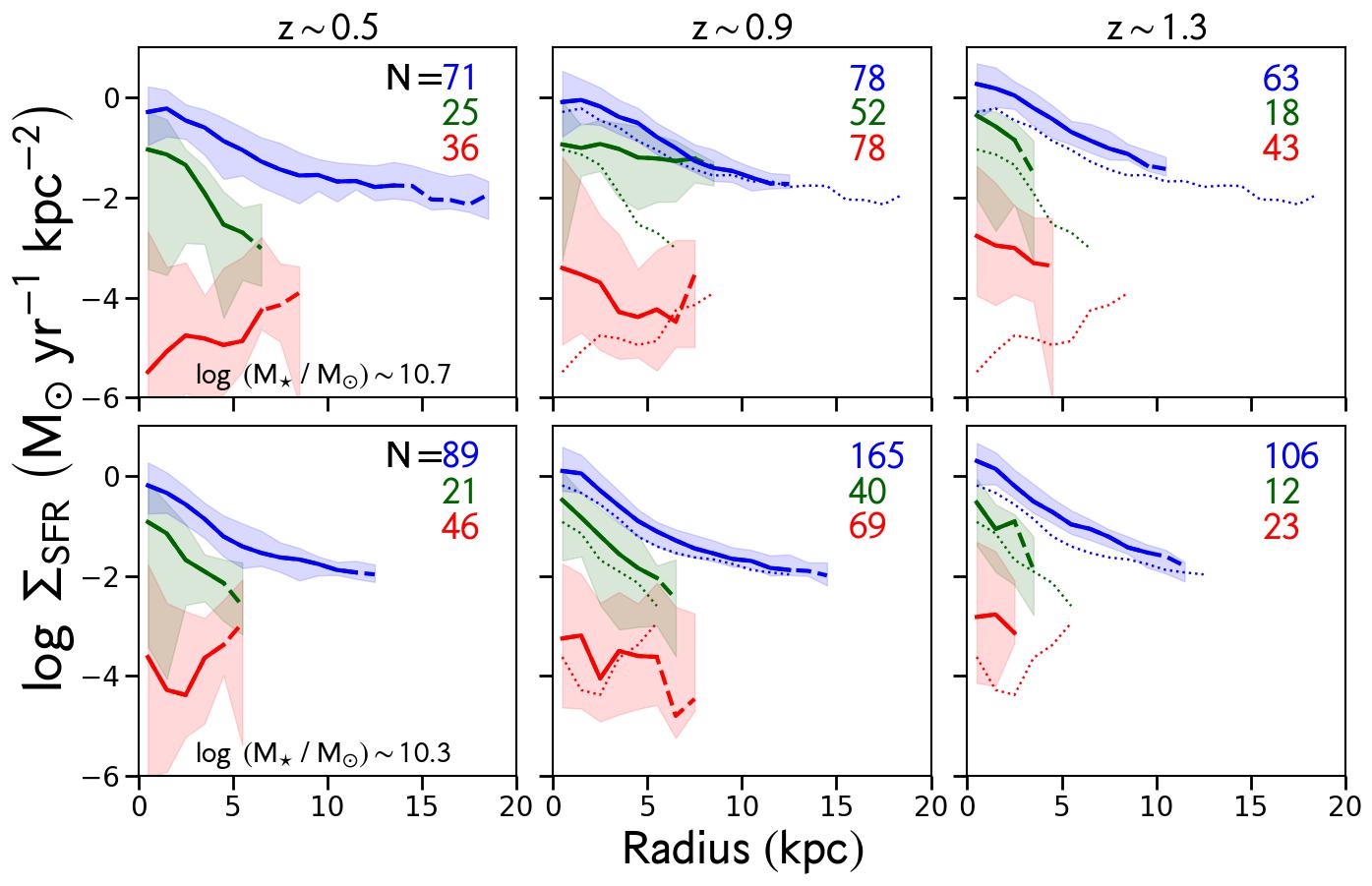}
    \includegraphics[width=0.75\textwidth]{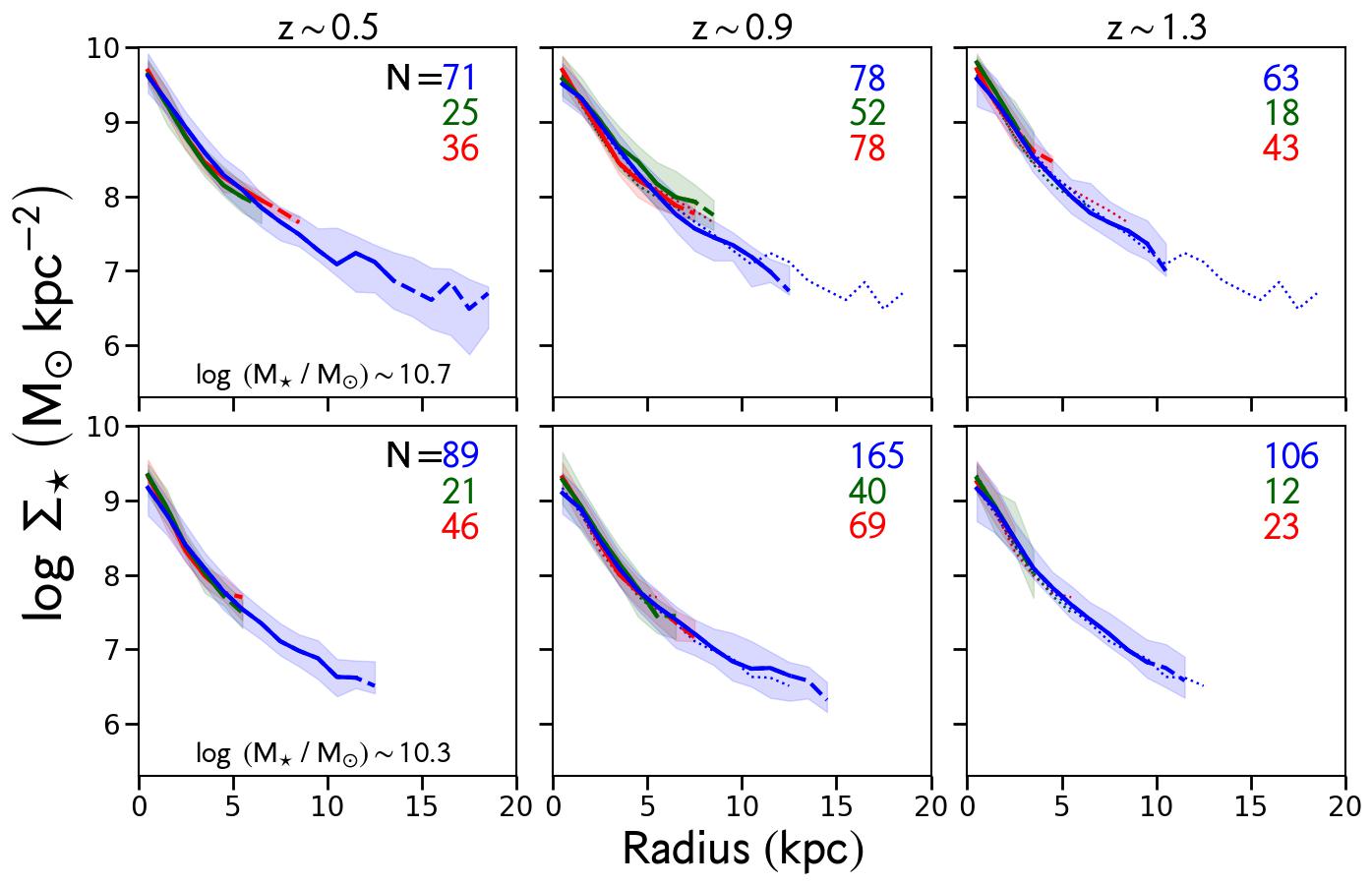}
    \caption{Median science profiles of SFR surface density (top) and stellar mass surface density (bottom) under the second set of assumptions for all mass and redshift bins \textcolor{black}{are shown}. In each panel, the meanings of the lines, shadings, and colors are the same as those in the top two rows in Figure \ref{fig:optimal_ssfr}.}
    \label{fig:intermediate_sfr_star}
\end{figure*}

\begin{figure*}[t!]
    \centering
    \includegraphics[width=0.75\textwidth]{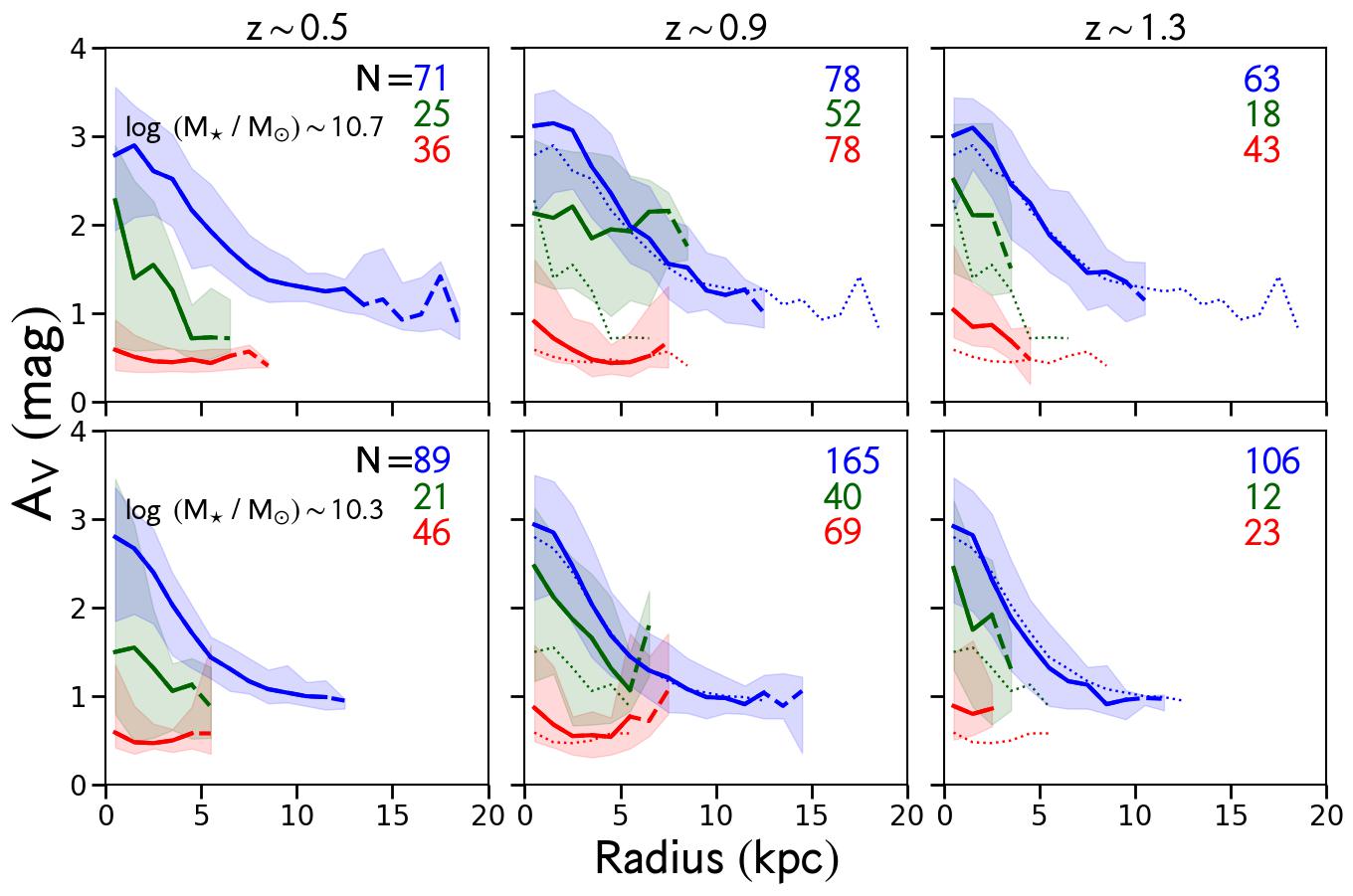}
    \caption{Median science profiles of \Av under the second set of assumptions for all mass and redshift bins \textcolor{black}{are shown}. In each panel, the meanings of the lines, shadings, and colors are the same as those in the top two rows in Figure \ref{fig:optimal_ssfr}.}
    \label{fig:intermediate_av}
\end{figure*}

\begin{figure*}[t!]
    \centering
    \includegraphics[width=0.75\textwidth]{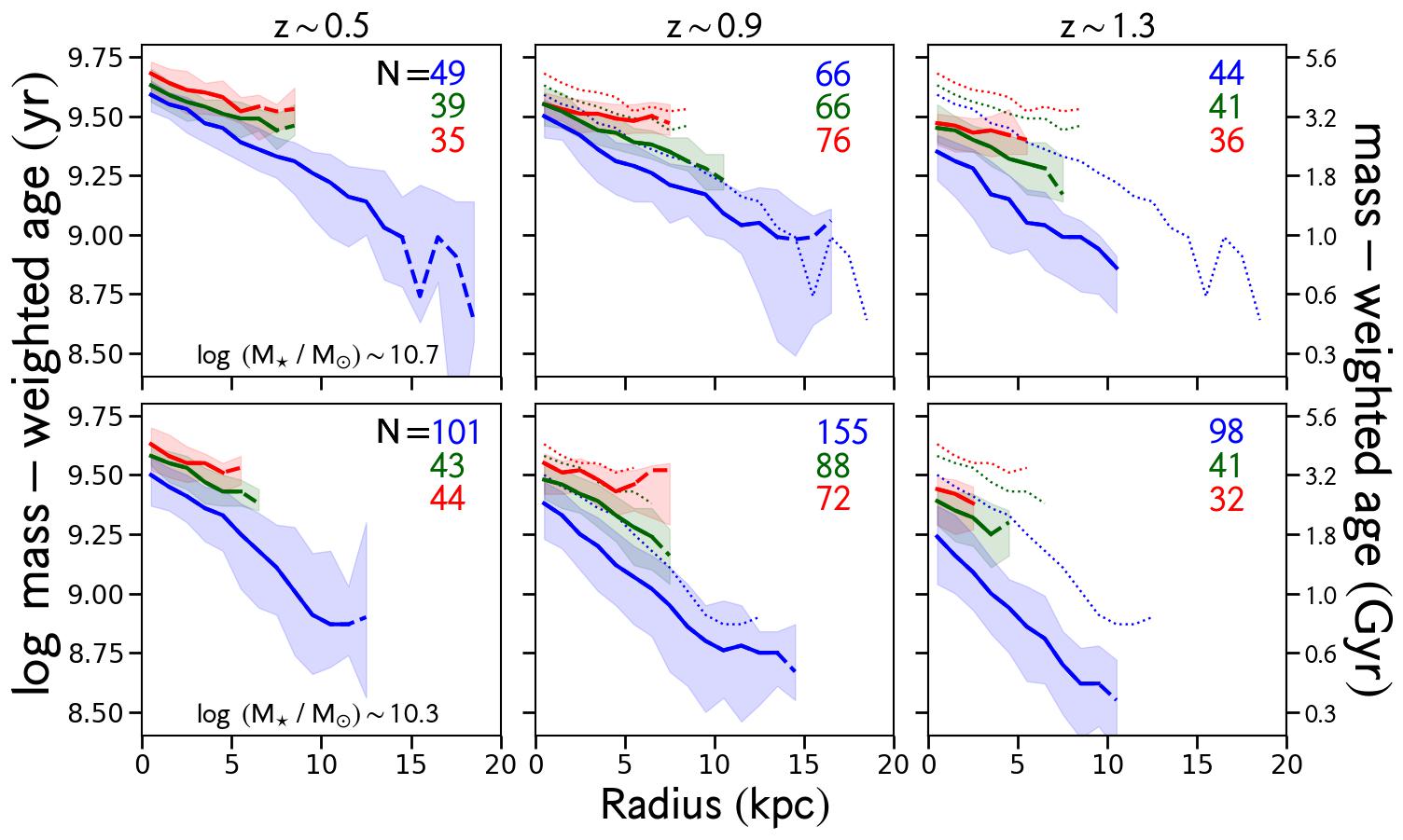}
    \includegraphics[width=0.75\textwidth]{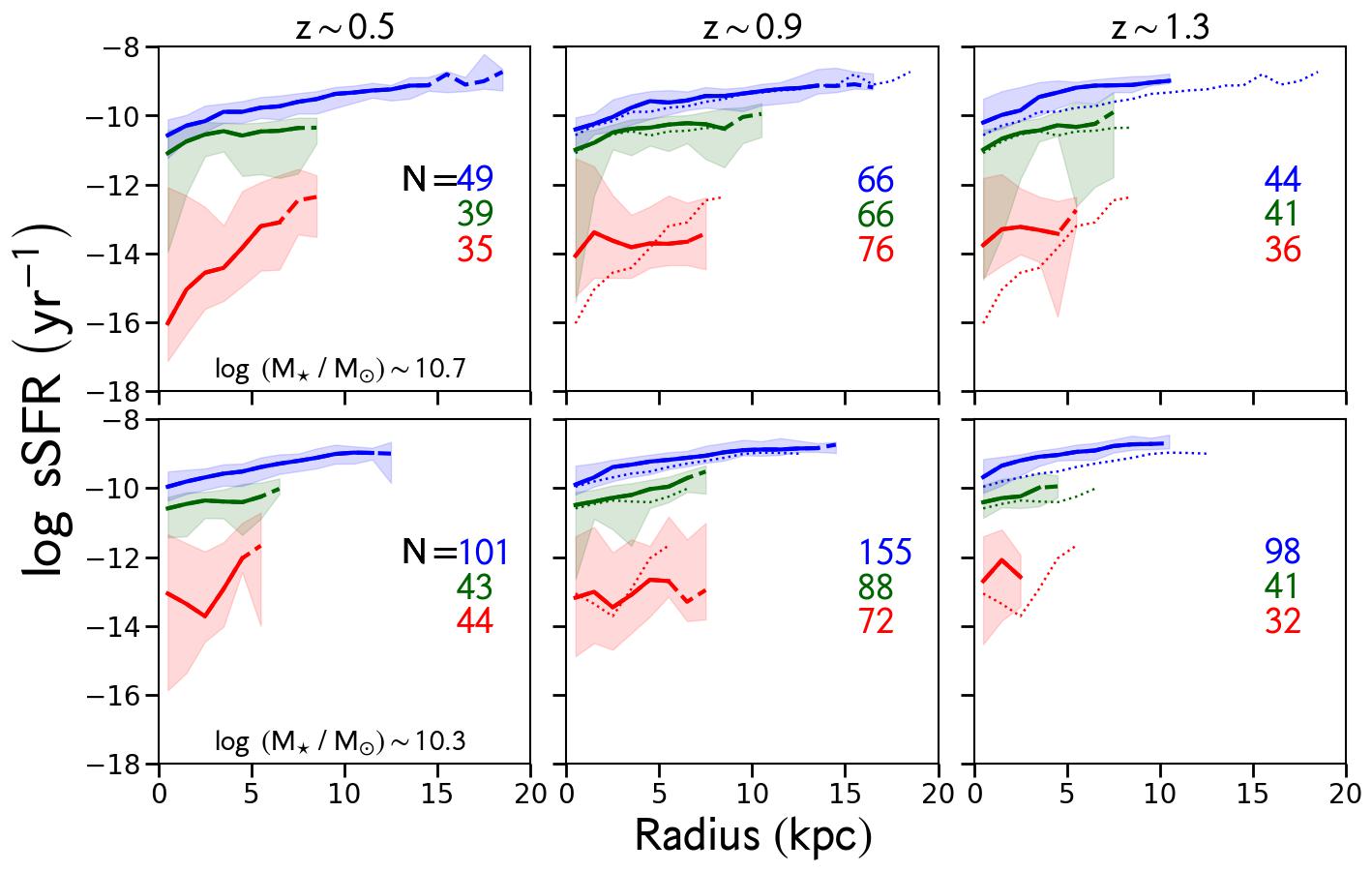}
    \caption{Median science profiles of mass-weighted age (top) and sSFR (bottom) under the third set of assumptions for all mass and redshift bins \textcolor{black}{are shown}. In each panel, the meanings of the lines, shadings, and colors are the same as those in the top two rows in Figure \ref{fig:optimal_ssfr}.}
    \label{fig:intermediate_exp_mw_ssfr}
\end{figure*}

\begin{figure*}[t!]
    \centering
    \includegraphics[width=0.75\textwidth]{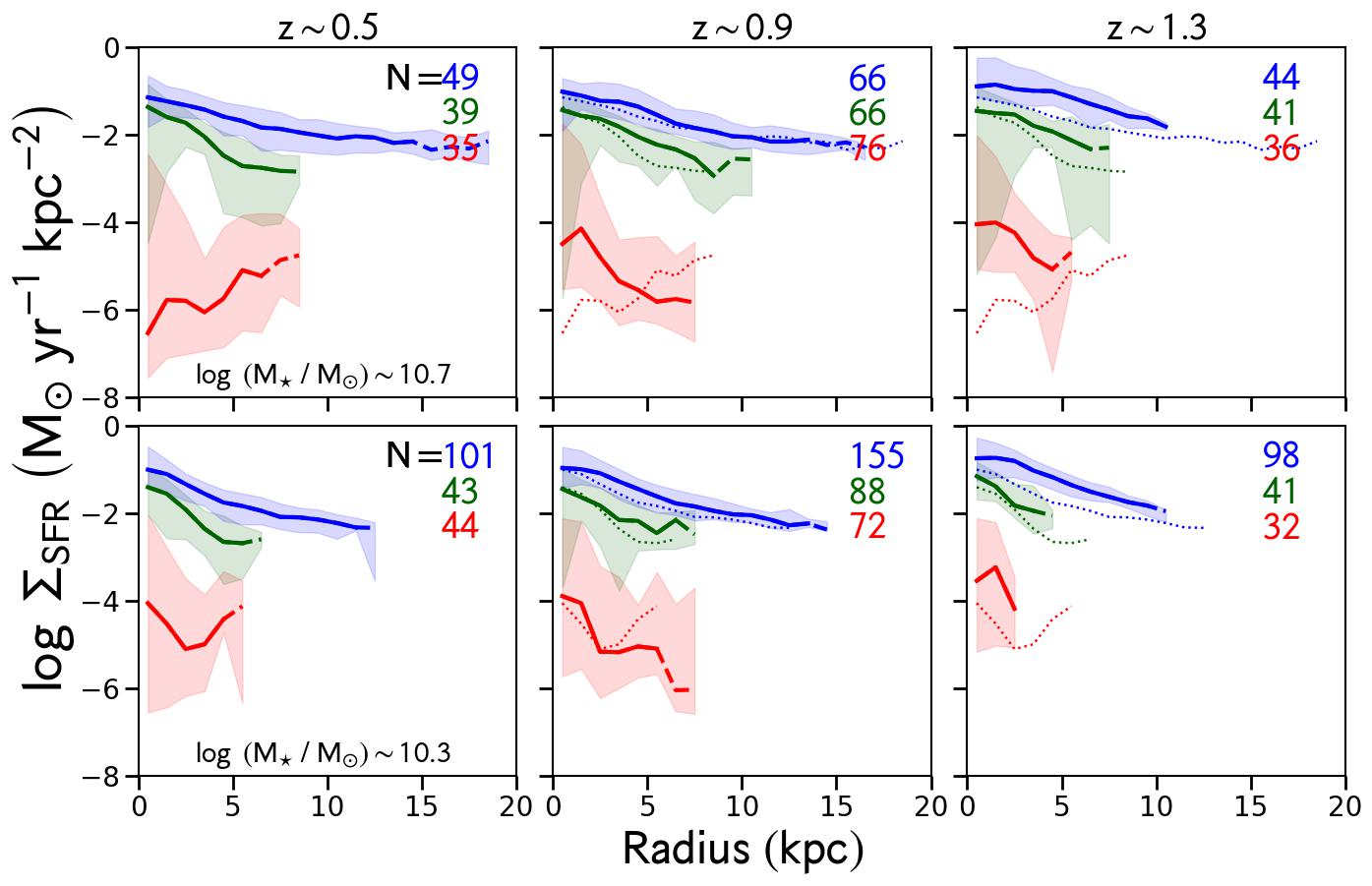}
    \includegraphics[width=0.75\textwidth]{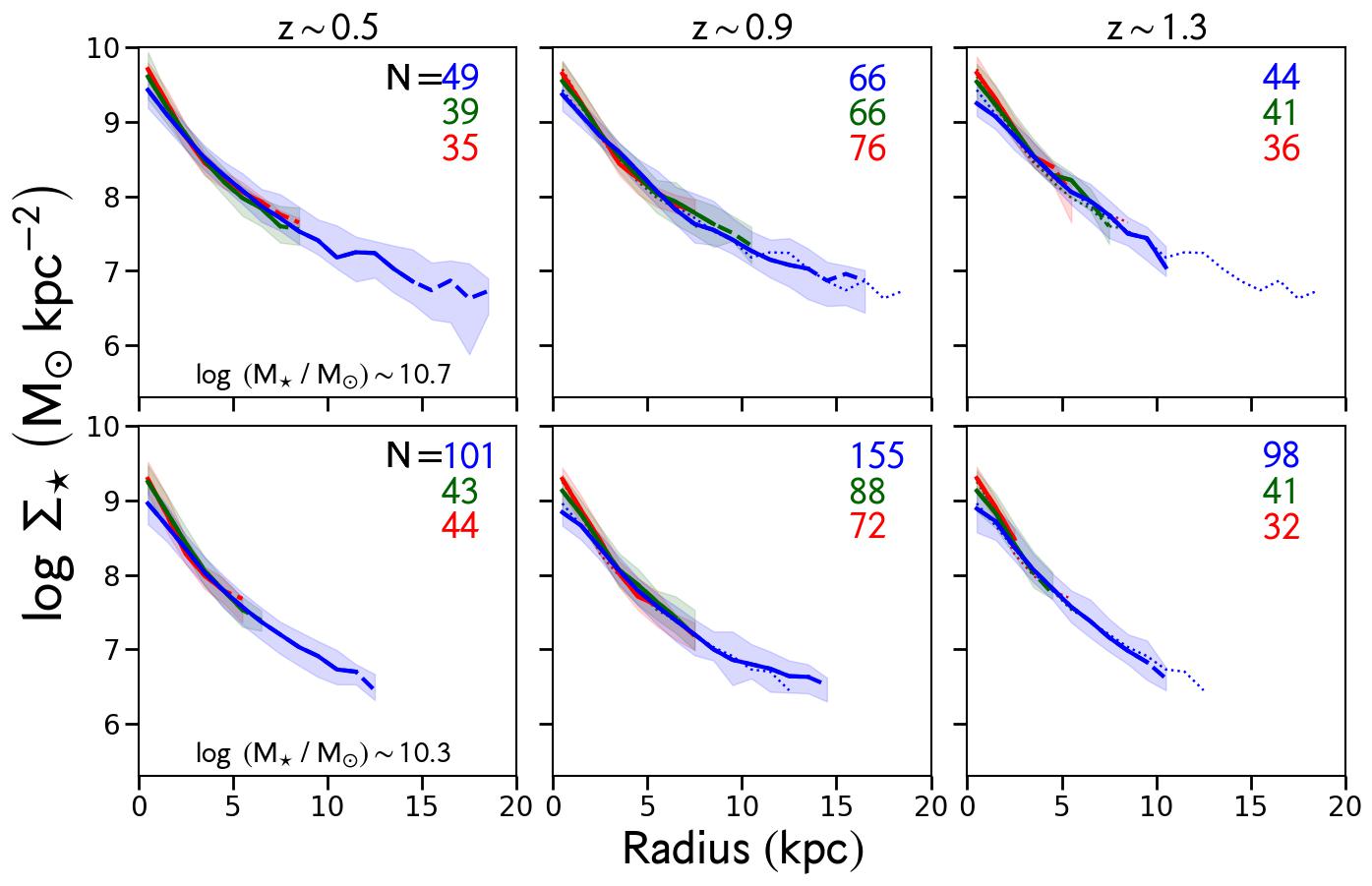} 
    \caption{Median science profiles of SFR surface density (top) and stellar mass surface density (bottom) under the third set of assumptions for all mass and redshift bins \textcolor{black}{are shown}. In each panel, the meanings of the lines, shadings, and colors are the same as those in the top two rows in Figure \ref{fig:optimal_ssfr}.}
    \label{fig:intermediate_exp_sfr_star}
\end{figure*}

\begin{figure*}[t!]
    \centering
    \includegraphics[width=0.75\textwidth]{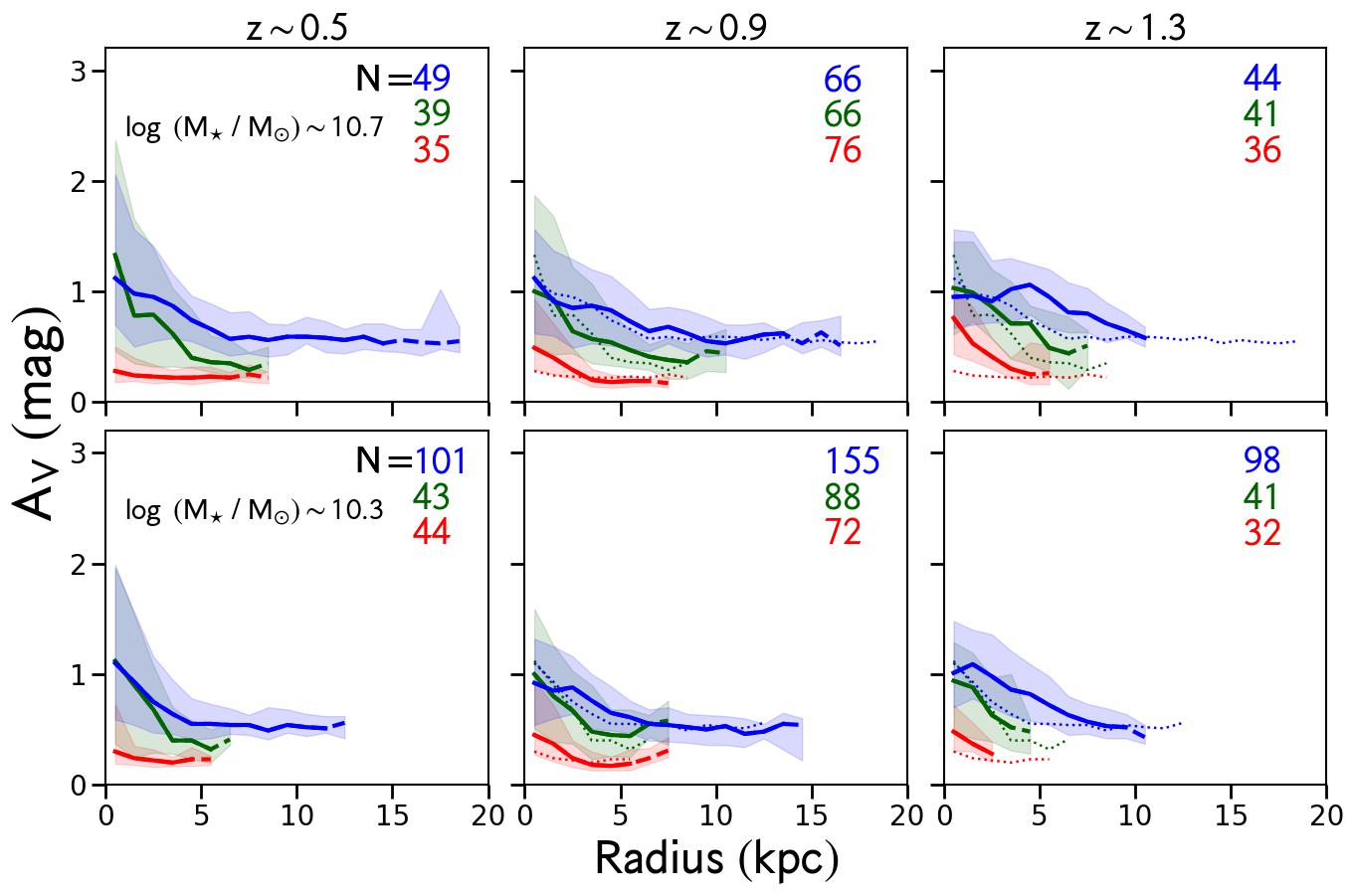}
    \caption{Median science profiles of \Av under the third set of assumptions for all mass and redshift bins \textcolor{black}{are shown}. In each panel, the meanings of the lines, shadings, and colors are the same as those in the top two rows in Figure \ref{fig:optimal_ssfr}.}
    \label{fig:intermediate_exp_av}
\end{figure*}

\begin{figure*}[t!]
    \centering
    \includegraphics[width=0.75\textwidth]{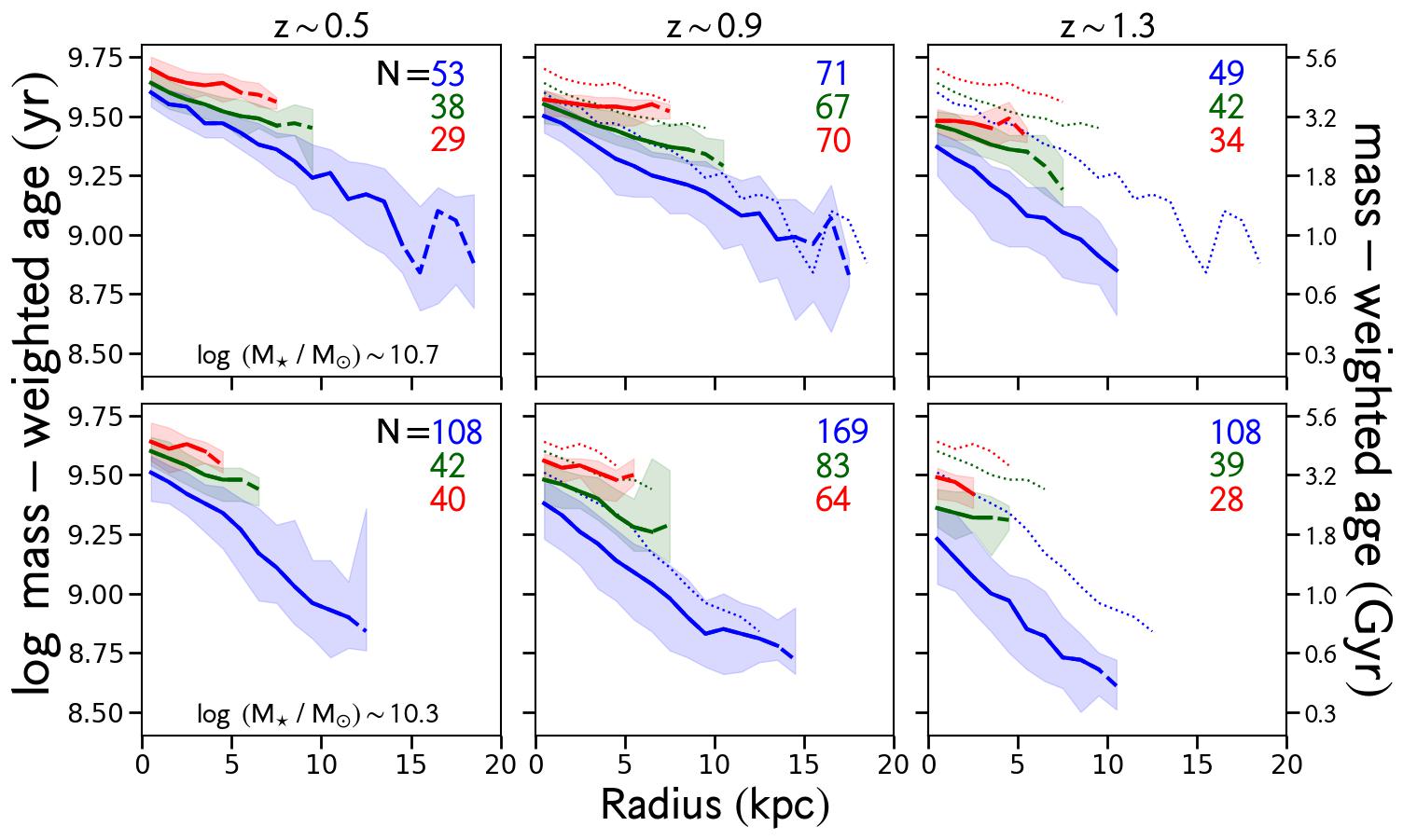}
    \includegraphics[width=0.75\textwidth]{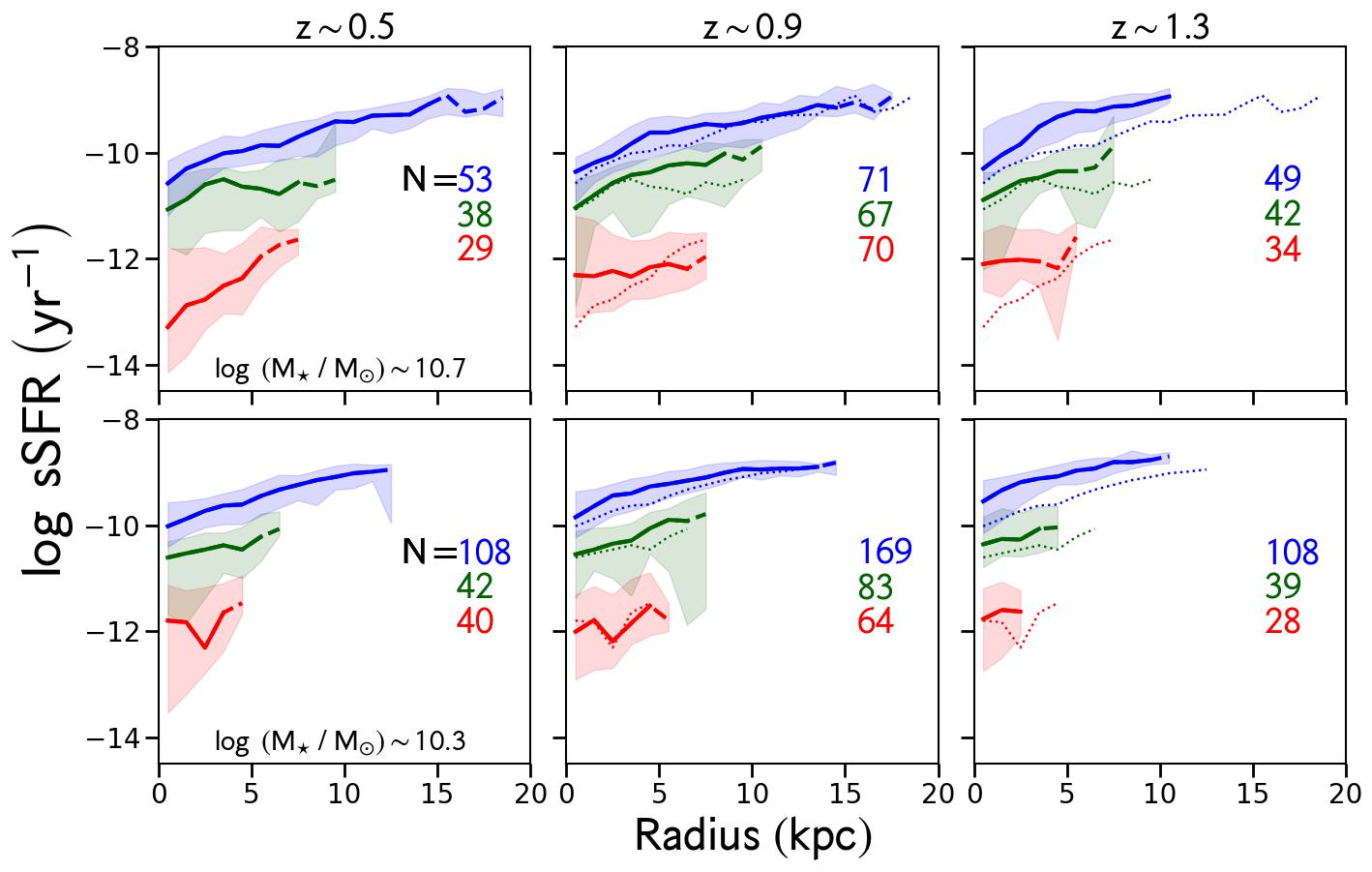}
    \caption{Median science profiles of mass-weighted age (top) and sSFR (bottom) under the fourth set of assumptions for all mass and redshift bins \textcolor{black}{are shown}. In each panel, the meanings of the lines, shadings, and colors are the same as those in the top two rows in Figure \ref{fig:optimal_ssfr}.}
    \label{fig:intermediate_tau_gauss_mw_ssfr}
\end{figure*}

\begin{figure*}[t!]
    \centering
    \includegraphics[width=0.75\textwidth]{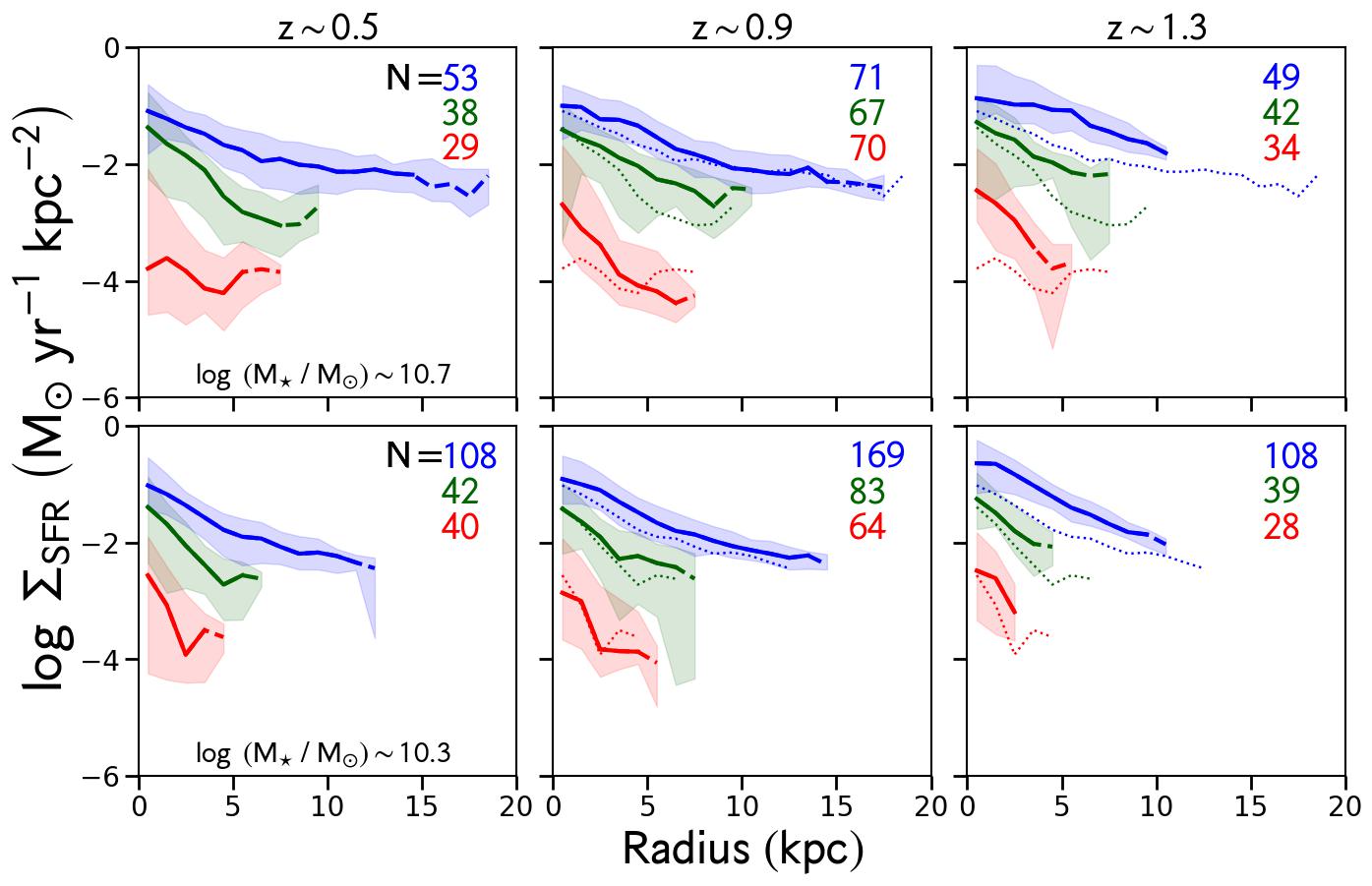}
    \includegraphics[width=0.75\textwidth]{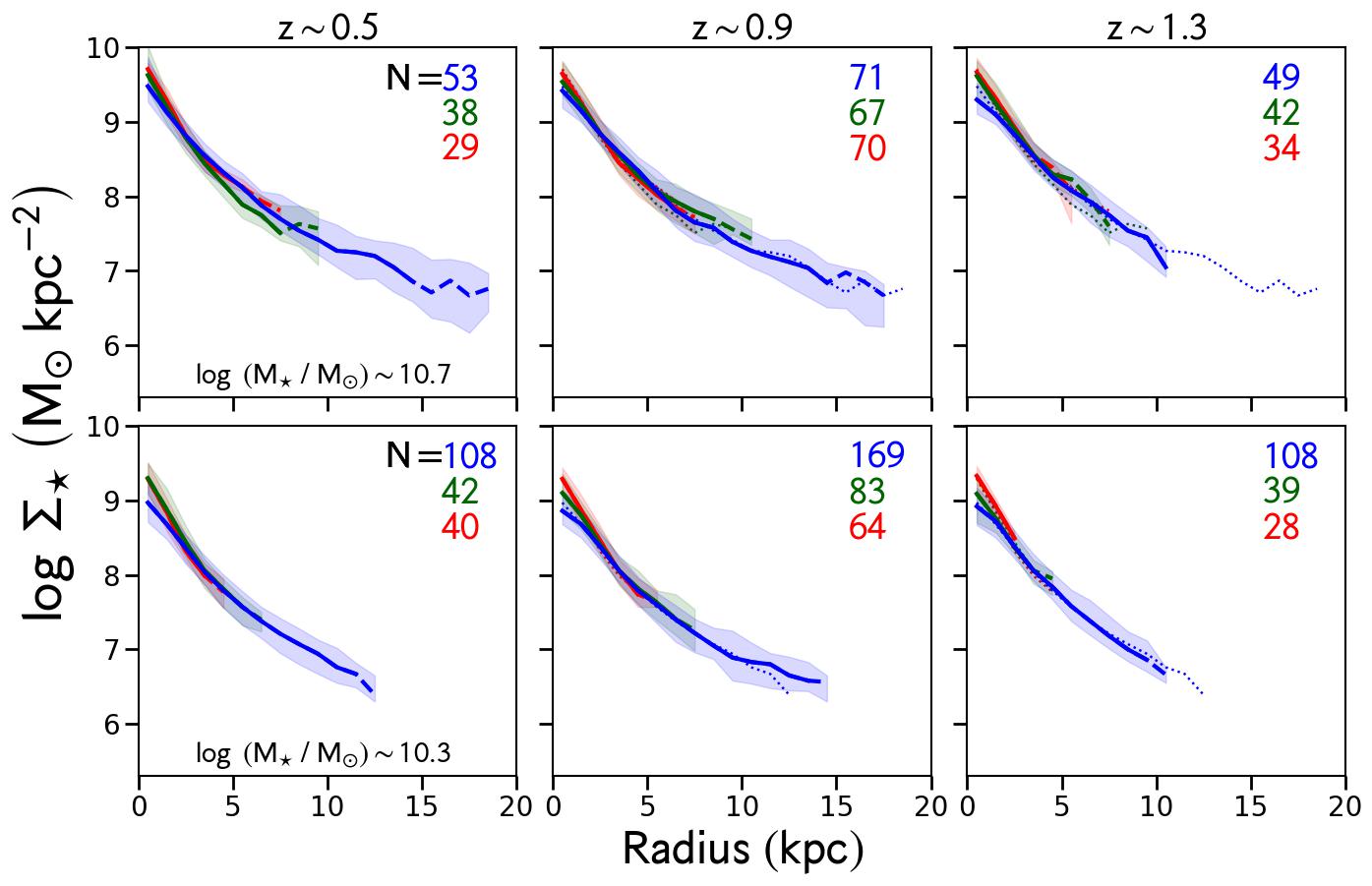}
    \caption{Median science profiles of SFR surface density (top) and stellar mass surface density (bottom) under the fourth set of assumptions for all mass and redshift bins \textcolor{black}{are shown}. In each panel, the meanings of the lines, shadings, and colors are the same as those in the top two rows in Figure \ref{fig:optimal_ssfr}.}
    \label{fig:intermediate_tau_gauss_sfr_star}
\end{figure*}

\begin{figure*}[t!]
    \centering
    \includegraphics[width=0.75\textwidth]{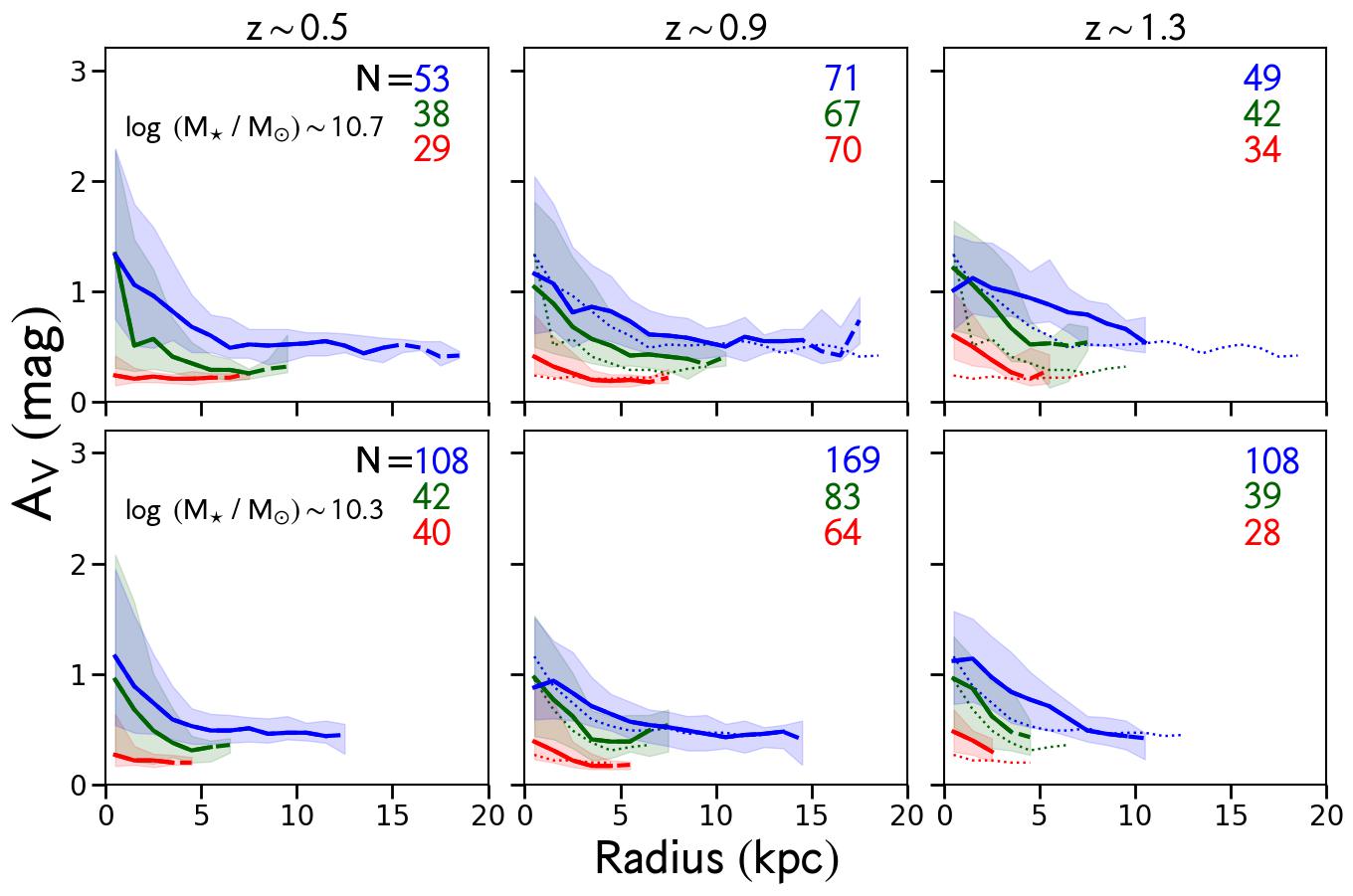}
    \caption{Median science profiles of \Av under the fourth set of assumptions for all mass and redshift bins \textcolor{black}{are shown}. In each panel, the meanings of the lines, shadings, and colors are the same as those in the top two rows in Figure \ref{fig:optimal_ssfr}.}
    \label{fig:intermediate_tau_gauss_av}
\end{figure*}

\begin{figure*}[t!]
    \centering
    \includegraphics[width=0.75\textwidth]{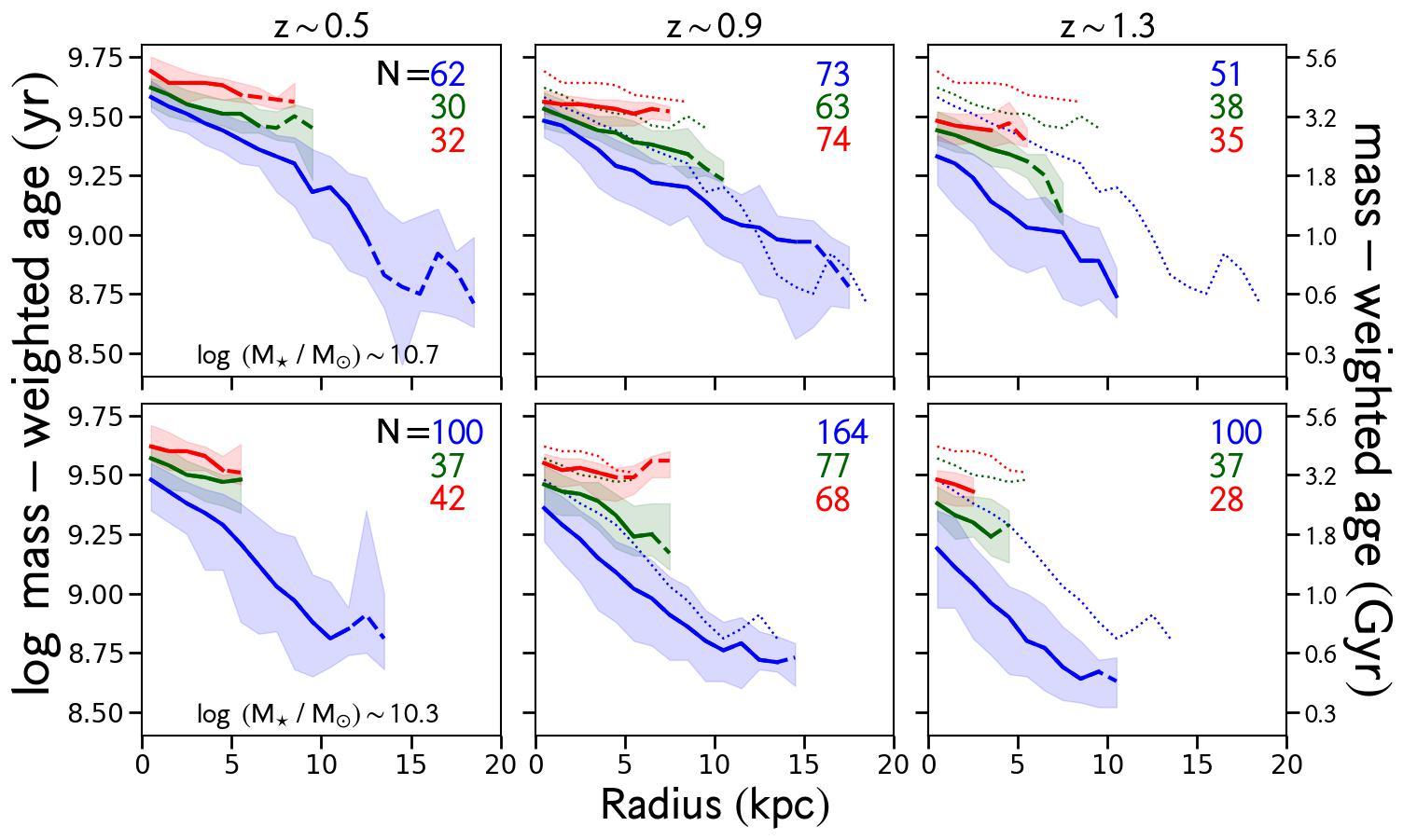}
    \includegraphics[width=0.75\textwidth]{optimal_assumptions_log_ssfr_50_r_circularized_all_bins_plot_low_z_nov23.jpg}
    \caption{Median science profiles of mass-weighted age (top) and sSFR (bottom) under the fifth set of assumptions for all mass and redshift bins \textcolor{black}{are shown}. In each panel, the meanings of the lines, shadings, and colors are the same as those in the top two rows in Figure \ref{fig:optimal_ssfr}.}
    \label{fig:optimal_mw_ssfr}
\end{figure*}

\begin{figure*}[t!]
    \centering
    \includegraphics[width=0.75\textwidth]{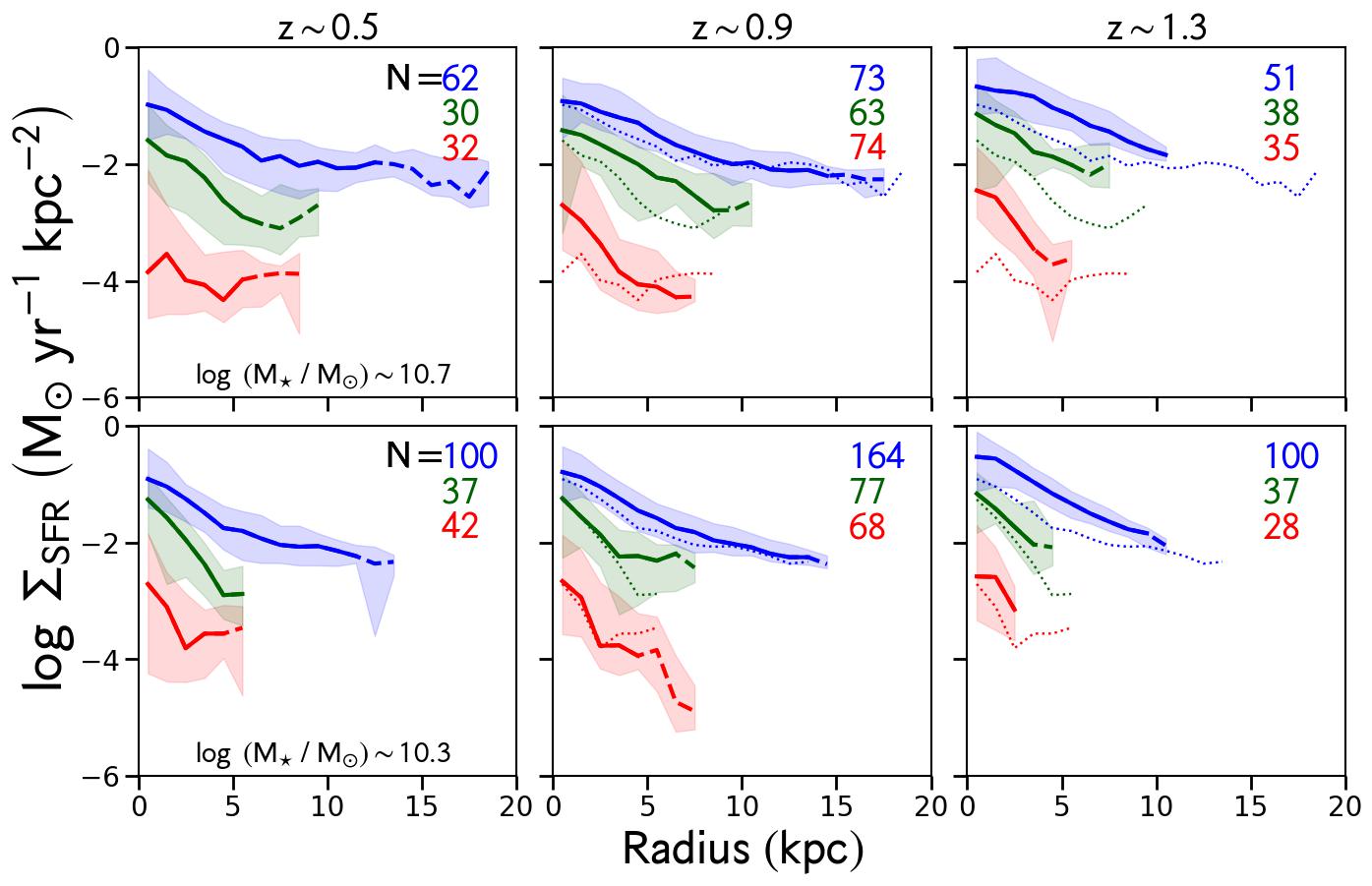}
    \includegraphics[width=0.75\textwidth]{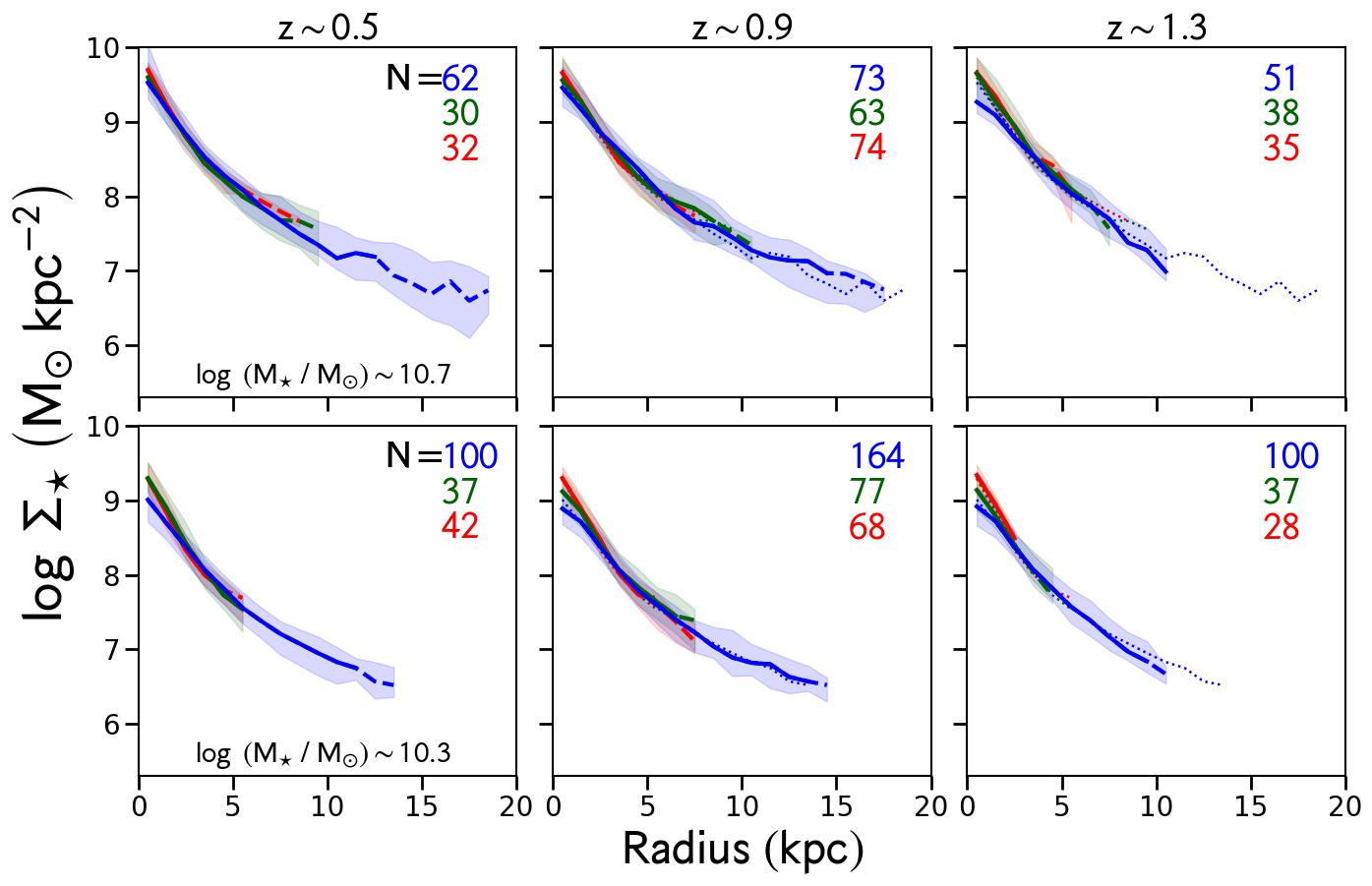}
    \caption{Median science profiles of SFR surface density (top) and stellar mass surface density (bottom) under the fifth set of assumptions for all mass and redshift bins \textcolor{black}{are shown}. In each panel, the meanings of the lines, shadings, and colors are the same as those in the top two rows in Figure \ref{fig:optimal_ssfr}.}
    \label{fig:optimal_sfr_star}
\end{figure*}

\begin{figure*}[t!]
    \centering
    \includegraphics[width=0.75\textwidth]{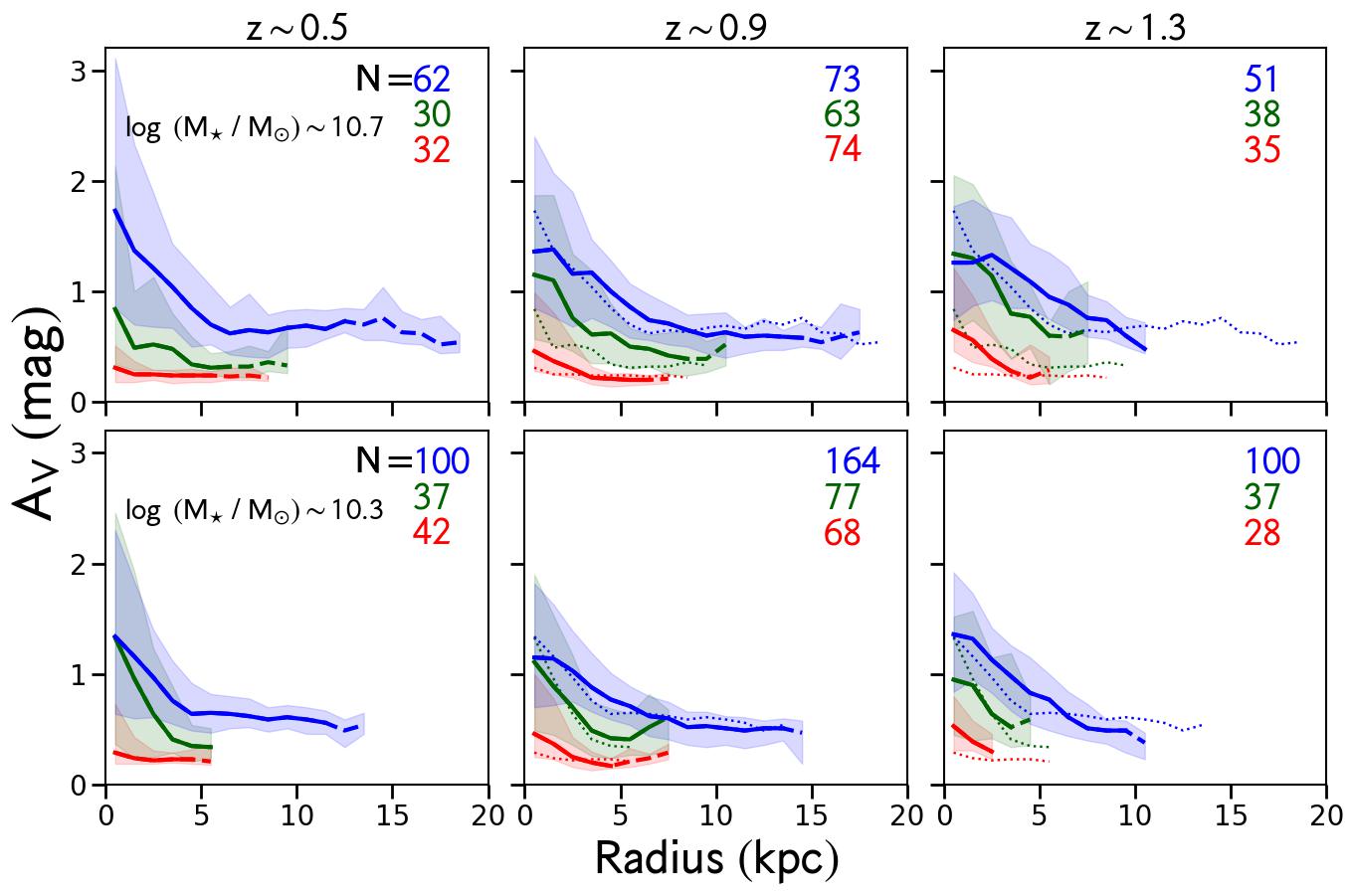}
    \caption{Median science profiles of \Av under the fifth set of assumptions for all mass and redshift bins \textcolor{black}{are shown}. In each panel, the meanings of the lines, shadings, and colors are the same as those in the top two rows in Figure \ref{fig:optimal_ssfr}.}
    \label{fig:optimal_av}
\end{figure*}

In this Appendix, we show science profiles under all sets of assumptions for all mass and redshift bins. For the first set of assumptions, its mass-weighted age and sSFR profiles are shown in Figure \ref{fig:typical_mw_ssfr}, the SFR and stellar mass surface density profiles are shown in Figure \ref{fig:typical_sfr_star}, and \Av profiles are shown in Figure \ref{fig:typical_av}. The same format is repeated for the second, (Figures \ref{fig:intermediate_mw_ssfr} -- \ref{fig:intermediate_av}), third (Figures \ref{fig:intermediate_exp_mw_ssfr} -- \ref{fig:intermediate_exp_av}), fourth (Figures \ref{fig:intermediate_tau_gauss_mw_ssfr} -- \ref{fig:intermediate_tau_gauss_av}), and fifth (Figures \ref{fig:optimal_mw_ssfr} -- \ref{fig:optimal_av}) sets. 



\begin{figure*}
    \centering
    \includegraphics[scale=0.6]{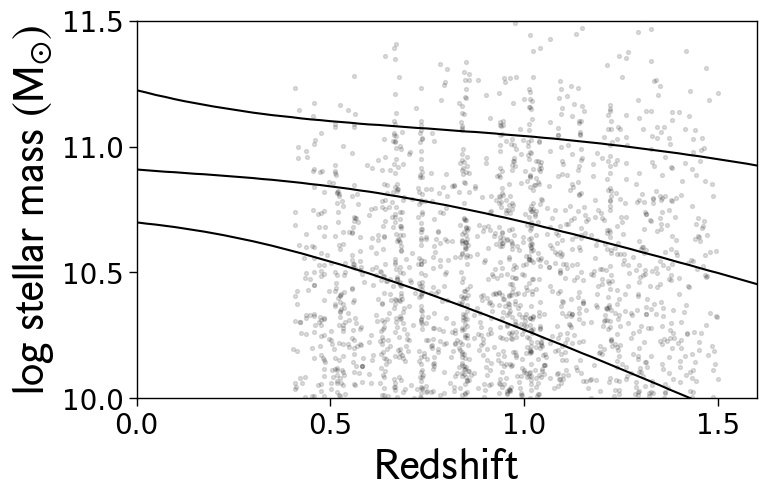}
    \caption{sSFR profiles in mass bins that evolve with redshift are shown here; the bins are indicated with solid lines. They are computed by choosing halo masses at $z=0$ and integrating the fitting functions provided by \citet{Moster13}. Individual galaxies are shown as light-gray dots.}
    \label{fig:abundance_matching_bins}

\end{figure*}

\begin{figure*}[t!]
    \centering
    \includegraphics[width=0.75\textwidth]{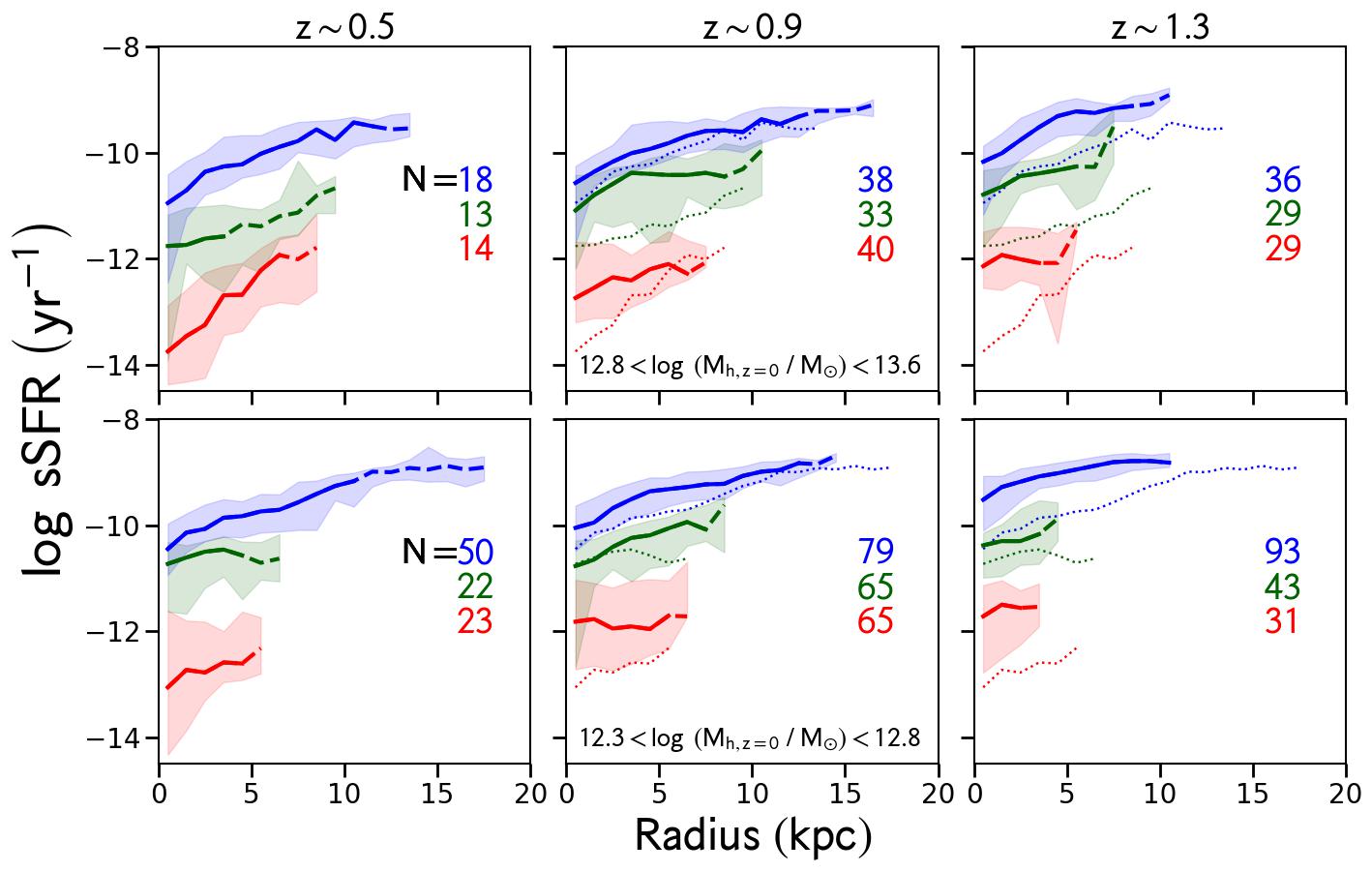}
    \includegraphics[width=0.75\textwidth]{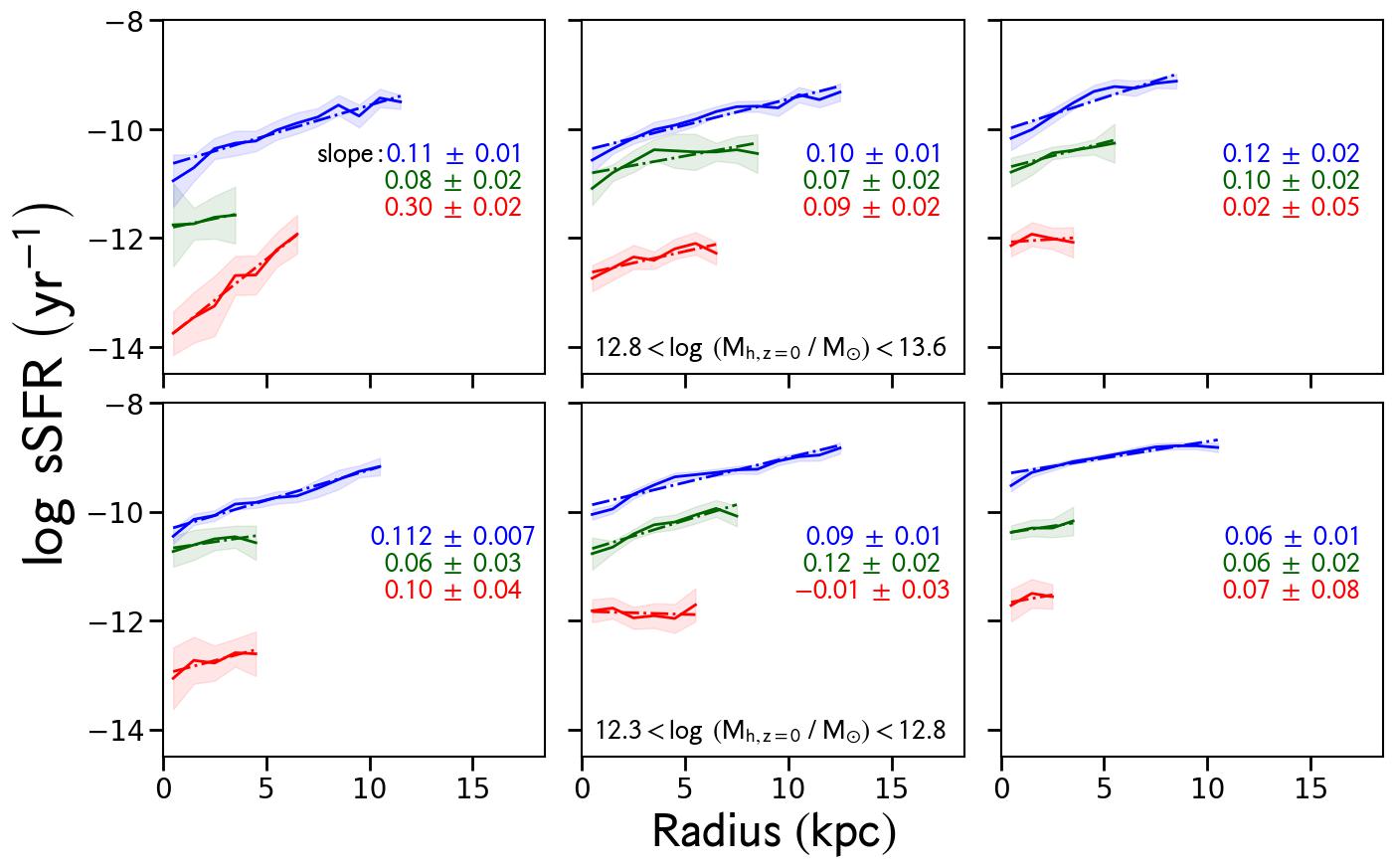}
    \caption{Median science profiles of sSFR under the fifth set of assumptions are measured in mass bins that evolve with redshift, and depicted here. Results from this analysis are similar to those done at fixed mass (see Section \ref{sec:evolution_optimal}). In the top two rows, median profiles are shown, and in the bottom two rows, fits to the median profiles are shown as dashed-dotted lines. In each panel, the meanings of the lines, shadings, and colors are the same as those in the top two rows in Figure \ref{fig:optimal_ssfr}.}
    \label{fig:optimal_ssfr_abundance_matching}
\end{figure*}

\section{Galaxy Evolution under the Fifth Set of Assumptions Considering Evolution in Mass with Redshift}
\label{sec:evolution_optimal_abundance_matching}

In this Appendix, the evolution of sSFR profiles with redshift is examined in bins in mass that also evolve with redshift. The results presented here are quantitatively similar to those in the case of fixed mass bins (see Section \ref{sec:evolution_optimal}). 

To connect galaxies throughout time, we adopt the abundance-matching results by \citet{Moster13}. Instead of two bins of fixed mass, we assume two bins in halo mass at $z=0$ ($\textrm{M}_{h, z=0}$): $12.3 < \log \left(\textrm{M}_{h, z=0}/\textrm{M}_{\odot}\right) < 12.8$ and $12.8 < \log \left(\textrm{M}_{h, z=0}/\textrm{M}_{\odot}\right) < 13.6$. These are shown in Figure \ref{fig:abundance_matching_bins}. We compute curves of the expected evolution in stellar mass with redshift by integrating the fitting functions for the star-formation and mass accretion histories of galaxies provided by \citet{Moster13}. These also take into account mass loss due to stellar evolution. In the lower bin in halo mass, galaxies increase in stellar mass by $\sim0.5$ dex from $z=1.5$ to $z=0.4$, whereas this growth is about 0.2 dex over the same redshift interval in the higher halo mass bin. 

We find similar quantitative results when considering evolution in mass with redshift to what we found in fixed bins in mass (see Section \ref{sec:evolution_optimal}). Median sSFR profiles of galaxies in both bins in halo mass and three redshift bins are shown in the top two rows in Figure \ref{fig:optimal_ssfr_abundance_matching}. Linear fits to the median profiles are shown in the bottom two rows in Figure \ref{fig:optimal_ssfr_abundance_matching}. Star-forming, green valley, and quiescent galaxies in all mass and redshift bins have slopes that are consistent with the slopes measured at fixed mass (Figure \ref{fig:optimal_ssfr}) to within $2\sigma$, with only one exception. Star-forming galaxies in the lower halo mass bin at $z\sim0.5$ have a median profile with a slightly steeper slope than that of star-forming galaxies of stellar mass $\log\left(\textrm{M}_{\star}/\textrm{M}_{\odot}\right) \sim10.3$ at the same redshifts ($0.112 \pm 0.007$ versus $0.086\pm0.004~\textrm{dex kpc}^{-1}$).


\bibliography{sample63}{}

\begin{thebibliography}{}
\expandafter\ifx\csname natexlab\endcsname\relax\def\natexlab#1{#1}\fi
\providecommand{\url}[1]{\href{#1}{#1}}
\providecommand{\dodoi}[1]{doi:~\href{http://doi.org/#1}{\nolinkurl{#1}}}
\providecommand{\doeprint}[1]{\href{http://ascl.net/#1}{\nolinkurl{http://ascl.net/#1}}}
\providecommand{\doarXiv}[1]{\href{https://arxiv.org/abs/#1}{\nolinkurl{https://arxiv.org/abs/#1}}}

\bibitem[{{Abdurro'uf} \& {Akiyama}(2018)}]{Abdurrouf18}
{Abdurro'uf}, \& {Akiyama}, M. 2018, \mnras, 479, 5083,
  \dodoi{10.1093/mnras/sty1771}

\bibitem[{{Appleby} {et~al.}(2020){Appleby}, {Dav{\'e}}, {Kraljic},
  {Angl{\'e}s-Alc{\'a}zar}, \& {Narayanan}}]{Appleby20}
{Appleby}, S., {Dav{\'e}}, R., {Kraljic}, K., {Angl{\'e}s-Alc{\'a}zar}, D., \&
  {Narayanan}, D. 2020, \mnras, 494, 6053, \dodoi{10.1093/mnras/staa1169}

\bibitem[{{Astropy Collaboration} {et~al.}(2013){Astropy Collaboration},
  {Robitaille}, {Tollerud}, {Greenfield}, {Droettboom}, {Bray}, {Aldcroft},
  {Davis}, {Ginsburg}, {Price-Whelan}, {Kerzendorf}, {Conley}, {Crighton},
  {Barbary}, {Muna}, {Ferguson}, {Grollier}, {Parikh}, {Nair}, {Unther},
  {Deil}, {Woillez}, {Conseil}, {Kramer}, {Turner}, {Singer}, {Fox}, {Weaver},
  {Zabalza}, {Edwards}, {Azalee Bostroem}, {Burke}, {Casey}, {Crawford},
  {Dencheva}, {Ely}, {Jenness}, {Labrie}, {Lim}, {Pierfederici}, {Pontzen},
  {Ptak}, {Refsdal}, {Servillat}, \& {Streicher}}]{2013A&A...558A..33A}
{Astropy Collaboration}, {Robitaille}, T.~P., {Tollerud}, E.~J., {et~al.} 2013,
  \aap, 558, A33, \dodoi{10.1051/0004-6361/201322068}

\bibitem[{{Astropy Collaboration} {et~al.}(2018){Astropy Collaboration},
  {Price-Whelan}, {Sip{\H{o}}cz}, {G{\"u}nther}, {Lim}, {Crawford}, {Conseil},
  {Shupe}, {Craig}, {Dencheva}, {Ginsburg}, {VanderPlas}, {Bradley},
  {P{\'e}rez-Su{\'a}rez}, {de Val-Borro}, {Aldcroft}, {Cruz}, {Robitaille},
  {Tollerud}, {Ardelean}, {Babej}, {Bach}, {Bachetti}, {Bakanov}, {Bamford},
  {Barentsen}, {Barmby}, {Baumbach}, {Berry}, {Biscani}, {Boquien}, {Bostroem},
  {Bouma}, {Brammer}, {Bray}, {Breytenbach}, {Buddelmeijer}, {Burke},
  {Calderone}, {Cano Rodr{\'\i}guez}, {Cara}, {Cardoso}, {Cheedella}, {Copin},
  {Corrales}, {Crichton}, {D'Avella}, {Deil}, {Depagne}, {Dietrich}, {Donath},
  {Droettboom}, {Earl}, {Erben}, {Fabbro}, {Ferreira}, {Finethy}, {Fox},
  {Garrison}, {Gibbons}, {Goldstein}, {Gommers}, {Greco}, {Greenfield},
  {Groener}, {Grollier}, {Hagen}, {Hirst}, {Homeier}, {Horton}, {Hosseinzadeh},
  {Hu}, {Hunkeler}, {Ivezi{\'c}}, {Jain}, {Jenness}, {Kanarek}, {Kendrew},
  {Kern}, {Kerzendorf}, {Khvalko}, {King}, {Kirkby}, {Kulkarni}, {Kumar},
  {Lee}, {Lenz}, {Littlefair}, {Ma}, {Macleod}, {Mastropietro}, {McCully},
  {Montagnac}, {Morris}, {Mueller}, {Mumford}, {Muna}, {Murphy}, {Nelson},
  {Nguyen}, {Ninan}, {N{\"o}the}, {Ogaz}, {Oh}, {Parejko}, {Parley}, {Pascual},
  {Patil}, {Patil}, {Plunkett}, {Prochaska}, {Rastogi}, {Reddy Janga},
  {Sabater}, {Sakurikar}, {Seifert}, {Sherbert}, {Sherwood-Taylor}, {Shih},
  {Sick}, {Silbiger}, {Singanamalla}, {Singer}, {Sladen}, {Sooley},
  {Sornarajah}, {Streicher}, {Teuben}, {Thomas}, {Tremblay}, {Turner},
  {Terr{\'o}n}, {van Kerkwijk}, {de la Vega}, {Watkins}, {Weaver}, {Whitmore},
  {Woillez}, {Zabalza}, \& {Astropy Contributors}}]{2018AJ....156..123A}
{Astropy Collaboration}, {Price-Whelan}, A.~M., {Sip{\H{o}}cz}, B.~M., {et~al.}
  2018, \aj, 156, 123, \dodoi{10.3847/1538-3881/aabc4f}

\bibitem[{{Astropy Collaboration} {et~al.}(2022){Astropy Collaboration},
  {Price-Whelan}, {Lim}, {Earl}, {Starkman}, {Bradley}, {Shupe}, {Patil},
  {Corrales}, {Brasseur}, {N{\"o}the}, {Donath}, {Tollerud}, {Morris},
  {Ginsburg}, {Vaher}, {Weaver}, {Tocknell}, {Jamieson}, {van Kerkwijk},
  {Robitaille}, {Merry}, {Bachetti}, {G{\"u}nther}, {Aldcroft},
  {Alvarado-Montes}, {Archibald}, {B{\'o}di}, {Bapat}, {Barentsen},
  {Baz{\'a}n}, {Biswas}, {Boquien}, {Burke}, {Cara}, {Cara}, {Conroy},
  {Conseil}, {Craig}, {Cross}, {Cruz}, {D'Eugenio}, {Dencheva}, {Devillepoix},
  {Dietrich}, {Eigenbrot}, {Erben}, {Ferreira}, {Foreman-Mackey}, {Fox},
  {Freij}, {Garg}, {Geda}, {Glattly}, {Gondhalekar}, {Gordon}, {Grant},
  {Greenfield}, {Groener}, {Guest}, {Gurovich}, {Handberg}, {Hart},
  {Hatfield-Dodds}, {Homeier}, {Hosseinzadeh}, {Jenness}, {Jones}, {Joseph},
  {Kalmbach}, {Karamehmetoglu}, {Ka{\l}uszy{\'n}ski}, {Kelley}, {Kern},
  {Kerzendorf}, {Koch}, {Kulumani}, {Lee}, {Ly}, {Ma}, {MacBride}, {Maljaars},
  {Muna}, {Murphy}, {Norman}, {O'Steen}, {Oman}, {Pacifici}, {Pascual},
  {Pascual-Granado}, {Patil}, {Perren}, {Pickering}, {Rastogi}, {Roulston},
  {Ryan}, {Rykoff}, {Sabater}, {Sakurikar}, {Salgado}, {Sanghi}, {Saunders},
  {Savchenko}, {Schwardt}, {Seifert-Eckert}, {Shih}, {Jain}, {Shukla}, {Sick},
  {Simpson}, {Singanamalla}, {Singer}, {Singhal}, {Sinha}, {Sip{\H{o}}cz},
  {Spitler}, {Stansby}, {Streicher}, {{\v{S}}umak}, {Swinbank}, {Taranu},
  {Tewary}, {Tremblay}, {de Val-Borro}, {Van Kooten}, {Vasovi{\'c}}, {Verma},
  {de Miranda Cardoso}, {Williams}, {Wilson}, {Winkel}, {Wood-Vasey}, {Xue},
  {Yoachim}, {Zhang}, {Zonca}, \& {Astropy Project Contributors}}]{Astropy2022}
{Astropy Collaboration}, {Price-Whelan}, A.~M., {Lim}, P.~L., {et~al.} 2022,
  \apj, 935, 167, \dodoi{10.3847/1538-4357/ac7c74}

\bibitem[{{Avila-Reese} {et~al.}(2018){Avila-Reese}, {Gonz{\'a}lez-Samaniego},
  {Col{\'\i}n}, {Ibarra-Medel}, \& {Rodr{\'\i}guez-Puebla}}]{AvilaReese18}
{Avila-Reese}, V., {Gonz{\'a}lez-Samaniego}, A., {Col{\'\i}n}, P.,
  {Ibarra-Medel}, H., \& {Rodr{\'\i}guez-Puebla}, A. 2018, \apj, 854, 152,
  \dodoi{10.3847/1538-4357/aaab69}

\bibitem[{{Bari{\v{s}}i{\'c}} {et~al.}(2020){Bari{\v{s}}i{\'c}}, {Pacifici},
  {van der Wel}, {Straatman}, {Bell}, {Bezanson}, {Brammer}, {D'Eugenio},
  {Franx}, {van Houdt}, {Maseda}, {Muzzin}, {Sobral}, \& {Wu}}]{Barisic20}
{Bari{\v{s}}i{\'c}}, I., {Pacifici}, C., {van der Wel}, A., {et~al.} 2020,
  \apj, 903, 146, \dodoi{10.3847/1538-4357/abba37}

\bibitem[{{Barrera-Ballesteros} {et~al.}(2015){Barrera-Ballesteros},
  {S{\'a}nchez}, {Garc{\'\i}a-Lorenzo}, {Falc{\'o}n-Barroso}, {Mast},
  {Garc{\'\i}a-Benito}, {Husemann}, {van de Ven}, {Iglesias-P{\'a}ramo},
  {Rosales-Ortega}, {P{\'e}rez-Torres}, {M{\'a}rquez}, {Kehrig}, {Marino},
  {Vilchez}, {Galbany}, {L{\'o}pez-S{\'a}nchez}, {Walcher}, \& {CALIFA
  Collaboration}}]{Barrera-Ballesteros15}
{Barrera-Ballesteros}, J.~K., {S{\'a}nchez}, S.~F., {Garc{\'\i}a-Lorenzo}, B.,
  {et~al.} 2015, \aap, 579, A45, \dodoi{10.1051/0004-6361/201425397}

\bibitem[{{Barro} {et~al.}(2019){Barro}, {P{\'e}rez-Gonz{\'a}lez}, {Cava},
  {Brammer}, {Pandya}, {Eliche Moral}, {Esquej}, {Dom{\'\i}nguez-S{\'a}nchez},
  {Alcalde Pampliega}, {Guo}, {Koekemoer}, {Trump}, {Ashby}, {Cardiel},
  {Castellano}, {Conselice}, {Dickinson}, {Dolch}, {Donley}, {Espino Briones},
  {Faber}, {Fazio}, {Ferguson}, {Finkelstein}, {Fontana}, {Galametz},
  {Gardner}, {Gawiser}, {Giavalisco}, {Grazian}, {Grogin}, {Hathi}, {Hemmati},
  {Hern{\'a}n-Caballero}, {Kocevski}, {Koo}, {Kodra}, {Lee}, {Lin}, {Lucas},
  {Mobasher}, {McGrath}, {Nandra}, {Nayyeri}, {Newman}, {Pforr}, {Peth},
  {Rafelski}, {Rodr{\'\i}guez-Munoz}, {Salvato}, {Stefanon}, {van der Wel},
  {Willner}, {Wiklind}, \& {Wuyts}}]{Barro19}
{Barro}, G., {P{\'e}rez-Gonz{\'a}lez}, P.~G., {Cava}, A., {et~al.} 2019, \apjs,
  243, 22, \dodoi{10.3847/1538-4365/ab23f2}

\bibitem[{{Battisti} {et~al.}(2017){Battisti}, {Calzetti}, \&
  {Chary}}]{Battisti17}
{Battisti}, A.~J., {Calzetti}, D., \& {Chary}, R.~R. 2017, \apj, 851, 90,
  \dodoi{10.3847/1538-4357/aa9a43}

\bibitem[{{Bekki} \& {Shioya}(1998)}]{Bekki98}
{Bekki}, K., \& {Shioya}, Y. 1998, \apj, 497, 108, \dodoi{10.1086/305445}

\bibitem[{{Bekki} \& {Shioya}(1999)}]{Bekki99}
---. 1999, \apj, 513, 108, \dodoi{10.1086/306833}

\bibitem[{{Belfiore} {et~al.}(2018){Belfiore}, {Maiolino}, {Bundy}, {Masters},
  {Bershady}, {Oyarz{\'u}n}, {Lin}, {Cano-Diaz}, {Wake}, {Spindler}, {Thomas},
  {Brownstein}, {Drory}, \& {Yan}}]{Belfiore18}
{Belfiore}, F., {Maiolino}, R., {Bundy}, K., {et~al.} 2018, \mnras, 477, 3014,
  \dodoi{10.1093/mnras/sty768}

\bibitem[{{Bell} \& {de Jong}(2001)}]{Bell01}
{Bell}, E.~F., \& {de Jong}, R.~S. 2001, \apj, 550, 212, \dodoi{10.1086/319728}

\bibitem[{{Bell} {et~al.}(2006){Bell}, {Phleps}, {Somerville}, {Wolf}, {Borch},
  \& {Meisenheimer}}]{Bell06}
{Bell}, E.~F., {Phleps}, S., {Somerville}, R.~S., {et~al.} 2006, \apj, 652,
  270, \dodoi{10.1086/508408}

\bibitem[{{Bell} {et~al.}(2004){Bell}, {Wolf}, {Meisenheimer}, {Rix}, {Borch},
  {Dye}, {Kleinheinrich}, {Wisotzki}, \& {McIntosh}}]{Bell04}
{Bell}, E.~F., {Wolf}, C., {Meisenheimer}, K., {et~al.} 2004, \apj, 608, 752,
  \dodoi{10.1086/420778}

\bibitem[{{Bertin} \& {Arnouts}(1996)}]{Bertin96}
{Bertin}, E., \& {Arnouts}, S. 1996, \aaps, 117, 393,
  \dodoi{10.1051/aas:1996164}

\bibitem[{{Boada} {et~al.}(2015){Boada}, {Tilvi}, {Papovich}, {Quadri},
  {Hilton}, {Finkelstein}, {Guo}, {Bond}, {Conselice}, {Dekel}, {Ferguson},
  {Giavalisco}, {Grogin}, {Kocevski}, {Koekemoer}, \& {Koo}}]{Boada15}
{Boada}, S., {Tilvi}, V., {Papovich}, C., {et~al.} 2015, \apj, 803, 104,
  \dodoi{10.1088/0004-637X/803/2/104}

\bibitem[{Bradley {et~al.}(2020)Bradley, Sipőcz, Robitaille, Tollerud,
  Vinícius, Deil, Barbary, Wilson, Busko, Günther, Cara, Conseil, Bostroem,
  Droettboom, Bray, Bratholm, Lim, Barentsen, Craig, Pascual, Perren, Greco,
  Donath, de~Val-Borro, Kerzendorf, Bach, Weaver, D'Eugenio, Souchereau, \&
  Ferreira}]{Bradley20}
Bradley, L., Sipőcz, B., Robitaille, T., {et~al.} 2020, astropy/photutils:
  1.0.0, 1.0.0,  Zenodo, \dodoi{10.5281/zenodo.4044744}

\bibitem[{Bradley {et~al.}(2022)Bradley, Sipőcz, Robitaille, Tollerud,
  Vinícius, Deil, Barbary, Wilson, Busko, Donath, Günther, Cara, Lim,
  Meßlinger, Conseil, Bostroem, Droettboom, Bray, Bratholm, Barentsen, Craig,
  Rathi, Pascual, Perren, Georgiev, de~Val-Borro, Kerzendorf, Bach, Quint, \&
  Souchereau}]{photutils22}
---. 2022, astropy/photutils: 1.5.0, 1.5.0,  Zenodo,
  \dodoi{10.5281/zenodo.6825092}

\bibitem[{{Brammer} {et~al.}(2011){Brammer}, {Whitaker}, {van Dokkum},
  {Marchesini}, {Franx}, {Kriek}, {Labb{\'e}}, {Lee}, {Muzzin}, {Quadri},
  {Rudnick}, \& {Williams}}]{Brammer11}
{Brammer}, G.~B., {Whitaker}, K.~E., {van Dokkum}, P.~G., {et~al.} 2011, \apj,
  739, 24, \dodoi{10.1088/0004-637X/739/1/24}

\bibitem[{{Bravo} {et~al.}(2023){Bravo}, {Robotham}, {Lagos}, {Davies},
  {Bellstedt}, \& {Thorne}}]{Bravo23}
{Bravo}, M., {Robotham}, A. S.~G., {Lagos}, C. d.~P., {et~al.} 2023, \mnras,
  522, 4481, \dodoi{10.1093/mnras/stad1234}

\bibitem[{{Bressan} {et~al.}(2012){Bressan}, {Marigo}, {Girardi}, {Salasnich},
  {Dal Cero}, {Rubele}, \& {Nanni}}]{Bressan12}
{Bressan}, A., {Marigo}, P., {Girardi}, L., {et~al.} 2012, \mnras, 427, 127,
  \dodoi{10.1111/j.1365-2966.2012.21948.x}

\bibitem[{{Brinchmann} {et~al.}(2013){Brinchmann}, {Charlot}, {Kauffmann},
  {Heckman}, {White}, \& {Tremonti}}]{Brinchmann13}
{Brinchmann}, J., {Charlot}, S., {Kauffmann}, G., {et~al.} 2013, \mnras, 432,
  2112, \dodoi{10.1093/mnras/stt551}

\bibitem[{{Brinchmann} {et~al.}(2004){Brinchmann}, {Charlot}, {White},
  {Tremonti}, {Kauffmann}, {Heckman}, \& {Brinkmann}}]{Brinchmann04}
{Brinchmann}, J., {Charlot}, S., {White}, S.~D.~M., {et~al.} 2004, \mnras, 351,
  1151, \dodoi{10.1111/j.1365-2966.2004.07881.x}

\bibitem[{{Bruzual} \& {Charlot}(2003)}]{BC03}
{Bruzual}, G., \& {Charlot}, S. 2003, \mnras, 344, 1000,
  \dodoi{10.1046/j.1365-8711.2003.06897.x}

\bibitem[{{Bryant} {et~al.}(2015){Bryant}, {Owers}, {Robotham}, {Croom},
  {Driver}, {Drinkwater}, {Lorente}, {Cortese}, {Scott}, {Colless}, {Schaefer},
  {Taylor}, {Konstantopoulos}, {Allen}, {Baldry}, {Barnes}, {Bauer},
  {Bland-Hawthorn}, {Bloom}, {Brooks}, {Brough}, {Cecil}, {Couch}, {Croton},
  {Davies}, {Ellis}, {Fogarty}, {Foster}, {Glazebrook}, {Goodwin}, {Green},
  {Gunawardhana}, {Hampton}, {Ho}, {Hopkins}, {Kewley}, {Lawrence},
  {Leon-Saval}, {Leslie}, {McElroy}, {Lewis}, {Liske}, {L{\'o}pez-S{\'a}nchez},
  {Mahajan}, {Medling}, {Metcalfe}, {Meyer}, {Mould}, {Obreschkow}, {O'Toole},
  {Pracy}, {Richards}, {Shanks}, {Sharp}, {Sweet}, {Thomas}, {Tonini}, \&
  {Walcher}}]{Bryant15}
{Bryant}, J.~J., {Owers}, M.~S., {Robotham}, A.~S.~G., {et~al.} 2015, \mnras,
  447, 2857, \dodoi{10.1093/mnras/stu2635}

\bibitem[{{Buat} {et~al.}(2014){Buat}, {Heinis}, {Boquien}, {Burgarella},
  {Charmandaris}, {Boissier}, {Boselli}, {Le Borgne}, \& {Morrison}}]{Buat14}
{Buat}, V., {Heinis}, S., {Boquien}, M., {et~al.} 2014, \aap, 561, A39,
  \dodoi{10.1051/0004-6361/201322081}

\bibitem[{{Buck}(2020)}]{Buck20}
{Buck}, T. 2020, \mnras, 491, 5435, \dodoi{10.1093/mnras/stz3289}

\bibitem[{{Bundy} {et~al.}(2015){Bundy}, {Bershady}, {Law}, {Yan}, {Drory},
  {MacDonald}, {Wake}, {Cherinka}, {S{\'a}nchez-Gallego}, {Weijmans}, {Thomas},
  {Tremonti}, {Masters}, {Coccato}, {Diamond-Stanic}, {Arag{\'o}n-Salamanca},
  {Avila-Reese}, {Badenes}, {Falc{\'o}n-Barroso}, {Belfiore}, {Bizyaev},
  {Blanc}, {Bland-Hawthorn}, {Blanton}, {Brownstein}, {Byler}, {Cappellari},
  {Conroy}, {Dutton}, {Emsellem}, {Etherington}, {Frinchaboy}, {Fu}, {Gunn},
  {Harding}, {Johnston}, {Kauffmann}, {Kinemuchi}, {Klaene}, {Knapen},
  {Leauthaud}, {Li}, {Lin}, {Maiolino}, {Malanushenko}, {Malanushenko}, {Mao},
  {Maraston}, {McDermid}, {Merrifield}, {Nichol}, {Oravetz}, {Pan}, {Parejko},
  {Sanchez}, {Schlegel}, {Simmons}, {Steele}, {Steinmetz}, {Thanjavur},
  {Thompson}, {Tinker}, {van den Bosch}, {Westfall}, {Wilkinson}, {Wright},
  {Xiao}, \& {Zhang}}]{Bundy15}
{Bundy}, K., {Bershady}, M.~A., {Law}, D.~R., {et~al.} 2015, \apj, 798, 7,
  \dodoi{10.1088/0004-637X/798/1/7}

\bibitem[{{Byun} {et~al.}(1994){Byun}, {Freeman}, \& {Kylafis}}]{Byun94}
{Byun}, Y.~I., {Freeman}, K.~C., \& {Kylafis}, N.~D. 1994, \apj, 432, 114,
  \dodoi{10.1086/174553}

\bibitem[{{Caffau} {et~al.}(2011){Caffau}, {Ludwig}, {Steffen}, {Freytag}, \&
  {Bonifacio}}]{Caffau11}
{Caffau}, E., {Ludwig}, H.~G., {Steffen}, M., {Freytag}, B., \& {Bonifacio}, P.
  2011, \solphys, 268, 255, \dodoi{10.1007/s11207-010-9541-4}

\bibitem[{{Calzetti} {et~al.}(1994){Calzetti}, {Kinney}, \&
  {Storchi-Bergmann}}]{Calzetti94}
{Calzetti}, D., {Kinney}, A.~L., \& {Storchi-Bergmann}, T. 1994, \apj, 429,
  582, \dodoi{10.1086/174346}

\bibitem[{{Carnall} {et~al.}(2019){Carnall}, {Leja}, {Johnson}, {McLure},
  {Dunlop}, \& {Conroy}}]{Carnall19}
{Carnall}, A.~C., {Leja}, J., {Johnson}, B.~D., {et~al.} 2019, \apj, 873, 44,
  \dodoi{10.3847/1538-4357/ab04a2}

\bibitem[{{Carnall} {et~al.}(2018){Carnall}, {McLure}, {Dunlop}, \&
  {Dav{\'e}}}]{Carnall18}
{Carnall}, A.~C., {McLure}, R.~J., {Dunlop}, J.~S., \& {Dav{\'e}}, R. 2018,
  \mnras, 480, 4379, \dodoi{10.1093/mnras/sty2169}

\bibitem[{{Casertano} {et~al.}(2000){Casertano}, {de Mello}, {Dickinson},
  {Ferguson}, {Fruchter}, {Gonzalez-Lopezlira}, {Heyer}, {Hook}, {Levay},
  {Lucas}, {Mack}, {Makidon}, {Mutchler}, {Smith}, {Stiavelli}, {Wiggs}, \&
  {Williams}}]{Casertano00}
{Casertano}, S., {de Mello}, D., {Dickinson}, M., {et~al.} 2000, \aj, 120,
  2747, \dodoi{10.1086/316851}

\bibitem[{{Chabrier}(2003)}]{Chabrier03}
{Chabrier}, G. 2003, \pasp, 115, 763, \dodoi{10.1086/376392}

\bibitem[{{Charlot} \& {Fall}(2000)}]{CF00}
{Charlot}, S., \& {Fall}, S.~M. 2000, \apj, 539, 718, \dodoi{10.1086/309250}

\bibitem[{{Chevallard} \& {Charlot}(2016)}]{Chevallard16}
{Chevallard}, J., \& {Charlot}, S. 2016, \mnras, 462, 1415,
  \dodoi{10.1093/mnras/stw1756}

\bibitem[{{Chevallard} {et~al.}(2013){Chevallard}, {Charlot}, {Wandelt}, \&
  {Wild}}]{Chevallard13}
{Chevallard}, J., {Charlot}, S., {Wandelt}, B., \& {Wild}, V. 2013, \mnras,
  432, 2061, \dodoi{10.1093/mnras/stt523}

\bibitem[{{Cibinel} {et~al.}(2015){Cibinel}, {Le Floc'h}, {Perret}, {Bournaud},
  {Daddi}, {Pannella}, {Elbaz}, {Amram}, \& {Duc}}]{Cibinel15}
{Cibinel}, A., {Le Floc'h}, E., {Perret}, V., {et~al.} 2015, \apj, 805, 181,
  \dodoi{10.1088/0004-637X/805/2/181}

\bibitem[{{Costantin} {et~al.}(2022){Costantin}, {P{\'e}rez-Gonz{\'a}lez},
  {M{\'e}ndez-Abreu}, {Huertas-Company}, {Pampliega}, {Balcells}, {Barro},
  {Ceverino}, {Dimauro}, {S{\'a}nchez}, {Espino-Briones}, \&
  {Koekemoer}}]{Costantin22}
{Costantin}, L., {P{\'e}rez-Gonz{\'a}lez}, P.~G., {M{\'e}ndez-Abreu}, J.,
  {et~al.} 2022, \apj, 929, 121, \dodoi{10.3847/1538-4357/ac5a57}

\bibitem[{{Cramer} {et~al.}(2019){Cramer}, {Kenney}, {Sun}, {Crowl}, {Yagi},
  {J{\'a}chym}, {Roediger}, \& {Waldron}}]{Cramer19}
{Cramer}, W.~J., {Kenney}, J.~D.~P., {Sun}, M., {et~al.} 2019, \apj, 870, 63,
  \dodoi{10.3847/1538-4357/aaefff}

\bibitem[{{Curtis-Lake} {et~al.}(2021){Curtis-Lake}, {Chevallard}, {Charlot},
  \& {Sandles}}]{CurtisLake21}
{Curtis-Lake}, E., {Chevallard}, J., {Charlot}, S., \& {Sandles}, L. 2021,
  \mnras, 503, 4855, \dodoi{10.1093/mnras/stab698}

\bibitem[{{Curtis-Lake} {et~al.}(2013){Curtis-Lake}, {McLure}, {Dunlop},
  {Schenker}, {Rogers}, {Targett}, {Cirasuolo}, {Almaini}, {Ashby}, {Bradshaw},
  {Finkelstein}, {Dickinson}, {Ellis}, {Faber}, {Fazio}, {Ferguson}, {Fontana},
  {Grogin}, {Hartley}, {Kocevski}, {Koekemoer}, {Lai}, {Robertson}, {Vanzella},
  \& {Willner}}]{CurtisLake13}
{Curtis-Lake}, E., {McLure}, R.~J., {Dunlop}, J.~S., {et~al.} 2013, \mnras,
  429, 302, \dodoi{10.1093/mnras/sts338}

\bibitem[{{da Cunha} {et~al.}(2008){da Cunha}, {Charlot}, \&
  {Elbaz}}]{daCunha08}
{da Cunha}, E., {Charlot}, S., \& {Elbaz}, D. 2008, \mnras, 388, 1595,
  \dodoi{10.1111/j.1365-2966.2008.13535.x}

\bibitem[{{Daddi} {et~al.}(2007){Daddi}, {Dickinson}, {Morrison}, {Chary},
  {Cimatti}, {Elbaz}, {Frayer}, {Renzini}, {Pope}, {Alexander}, {Bauer},
  {Giavalisco}, {Huynh}, {Kurk}, \& {Mignoli}}]{Daddi07}
{Daddi}, E., {Dickinson}, M., {Morrison}, G., {et~al.} 2007, \apj, 670, 156,
  \dodoi{10.1086/521818}

\bibitem[{{Dahlen} {et~al.}(2013){Dahlen}, {Mobasher}, {Faber}, {Ferguson},
  {Barro}, {Finkelstein}, {Finlator}, {Fontana}, {Gruetzbauch}, {Johnson},
  {Pforr}, {Salvato}, {Wiklind}, {Wuyts}, {Acquaviva}, {Dickinson}, {Guo},
  {Huang}, {Huang}, {Newman}, {Bell}, {Conselice}, {Galametz}, {Gawiser},
  {Giavalisco}, {Grogin}, {Hathi}, {Kocevski}, {Koekemoer}, {Koo}, {Lee},
  {McGrath}, {Papovich}, {Peth}, {Ryan}, {Somerville}, {Weiner}, \&
  {Wilson}}]{Dahlen13}
{Dahlen}, T., {Mobasher}, B., {Faber}, S.~M., {et~al.} 2013, \apj, 775, 93,
  \dodoi{10.1088/0004-637X/775/2/93}

\bibitem[{{Dantas} {et~al.}(2020){Dantas}, {Coelho}, {de Souza}, \&
  {Gon{\c{c}}alves}}]{Dantas20}
{Dantas}, M.~L.~L., {Coelho}, P.~R.~T., {de Souza}, R.~S., \&
  {Gon{\c{c}}alves}, T.~S. 2020, \mnras, 492, 2996,
  \dodoi{10.1093/mnras/stz3609}

\bibitem[{{Davidzon} {et~al.}(2017){Davidzon}, {Ilbert}, {Laigle}, {Coupon},
  {McCracken}, {Delvecchio}, {Masters}, {Capak}, {Hsieh}, {Le F{\`e}vre},
  {Tresse}, {Bethermin}, {Chang}, {Faisst}, {Le Floc'h}, {Steinhardt}, {Toft},
  {Aussel}, {Dubois}, {Hasinger}, {Salvato}, {Sanders}, {Scoville}, \&
  {Silverman}}]{Davidzon17}
{Davidzon}, I., {Ilbert}, O., {Laigle}, C., {et~al.} 2017, \aap, 605, A70,
  \dodoi{10.1051/0004-6361/201730419}

\bibitem[{{Dom{\'\i}nguez} {et~al.}(2013){Dom{\'\i}nguez}, {Siana}, {Henry},
  {Scarlata}, {Bedregal}, {Malkan}, {Atek}, {Ross}, {Colbert}, {Teplitz},
  {Rafelski}, {McCarthy}, {Bunker}, {Hathi}, {Dressler}, {Martin}, \&
  {Masters}}]{Dominguez13}
{Dom{\'\i}nguez}, A., {Siana}, B., {Henry}, A.~L., {et~al.} 2013, \apj, 763,
  145, \dodoi{10.1088/0004-637X/763/2/145}

\bibitem[{{Elbaz} {et~al.}(2007){Elbaz}, {Daddi}, {Le Borgne}, {Dickinson},
  {Alexander}, {Chary}, {Starck}, {Brandt}, {Kitzbichler}, {MacDonald},
  {Nonino}, {Popesso}, {Stern}, \& {Vanzella}}]{Elbaz07}
{Elbaz}, D., {Daddi}, E., {Le Borgne}, D., {et~al.} 2007, \aap, 468, 33,
  \dodoi{10.1051/0004-6361:20077525}

\bibitem[{{Ellison} {et~al.}(2018){Ellison}, {S{\'a}nchez}, {Ibarra-Medel},
  {Antonio}, {Mendel}, \& {Barrera-Ballesteros}}]{Ellison18}
{Ellison}, S.~L., {S{\'a}nchez}, S.~F., {Ibarra-Medel}, H., {et~al.} 2018,
  \mnras, 474, 2039, \dodoi{10.1093/mnras/stx2882}

\bibitem[{{Emami} {et~al.}(2019){Emami}, {Siana}, {Weisz}, {Johnson}, {Ma}, \&
  {El-Badry}}]{Emami19}
{Emami}, N., {Siana}, B., {Weisz}, D.~R., {et~al.} 2019, \apj, 881, 71,
  \dodoi{10.3847/1538-4357/ab211a}

\bibitem[{{Estrada-Carpenter} {et~al.}(2019){Estrada-Carpenter}, {Papovich},
  {Momcheva}, {Brammer}, {Long}, {Quadri}, {Bridge}, {Dickinson}, {Ferguson},
  {Finkelstein}, {Giavalisco}, {Gosmeyer}, {Lotz}, {Salmon}, {Skelton},
  {Trump}, \& {Weiner}}]{EstradaCarpenter19}
{Estrada-Carpenter}, V., {Papovich}, C., {Momcheva}, I., {et~al.} 2019, \apj,
  870, 133, \dodoi{10.3847/1538-4357/aaf22e}

\bibitem[{{Faber} {et~al.}(2007){Faber}, {Willmer}, {Wolf}, {Koo}, {Weiner},
  {Newman}, {Im}, {Coil}, {Conroy}, {Cooper}, {Davis}, {Finkbeiner}, {Gerke},
  {Gebhardt}, {Groth}, {Guhathakurta}, {Harker}, {Kaiser}, {Kassin},
  {Kleinheinrich}, {Konidaris}, {Kron}, {Lin}, {Luppino}, {Madgwick},
  {Meisenheimer}, {Noeske}, {Phillips}, {Sarajedini}, {Schiavon}, {Simard},
  {Szalay}, {Vogt}, \& {Yan}}]{Faber07}
{Faber}, S.~M., {Willmer}, C.~N.~A., {Wolf}, C., {et~al.} 2007, \apj, 665, 265,
  \dodoi{10.1086/519294}

\bibitem[{{Ferland} {et~al.}(2013){Ferland}, {Porter}, {van Hoof}, {Williams},
  {Abel}, {Lykins}, {Shaw}, {Henney}, \& {Stancil}}]{Ferland13}
{Ferland}, G.~J., {Porter}, R.~L., {van Hoof}, P.~A.~M., {et~al.} 2013, \rmxaa,
  49, 137.
\newblock \doarXiv{1302.4485}

\bibitem[{{Feroz} {et~al.}(2009){Feroz}, {Hobson}, \& {Bridges}}]{Feroz09}
{Feroz}, F., {Hobson}, M.~P., \& {Bridges}, M. 2009, \mnras, 398, 1601,
  \dodoi{10.1111/j.1365-2966.2009.14548.x}

\bibitem[{{Fitzpatrick}(1999)}]{Fitzpatrick99}
{Fitzpatrick}, E.~L. 1999, \pasp, 111, 63, \dodoi{10.1086/316293}

\bibitem[{{Flores Vel{\'a}zquez} {et~al.}(2021){Flores Vel{\'a}zquez},
  {Gurvich}, {Faucher-Gigu{\`e}re}, {Bullock}, {Starkenburg}, {Moreno},
  {Lazar}, {Mercado}, {Stern}, {Sparre}, {Hayward}, {Wetzel}, \&
  {El-Badry}}]{FloresVelazquez21}
{Flores Vel{\'a}zquez}, J.~A., {Gurvich}, A.~B., {Faucher-Gigu{\`e}re}, C.-A.,
  {et~al.} 2021, \mnras, 501, 4812, \dodoi{10.1093/mnras/staa3893}

\bibitem[{{Fossati} {et~al.}(2018){Fossati}, {Mendel}, {Boselli}, {Cuillandre},
  {Vollmer}, {Boissier}, {Consolandi}, {Ferrarese}, {Gwyn}, {Amram}, {Boquien},
  {Buat}, {Burgarella}, {Cortese}, {C{\^o}t{\'e}}, {C{\^o}t{\'e}}, {Durrell},
  {Fumagalli}, {Gavazzi}, {Gomez-Lopez}, {Hensler}, {Koribalski}, {Longobardi},
  {Peng}, {Roediger}, {Sun}, \& {Toloba}}]{Fossati18}
{Fossati}, M., {Mendel}, J.~T., {Boselli}, A., {et~al.} 2018, \aap, 614, A57,
  \dodoi{10.1051/0004-6361/201732373}

\bibitem[{{Fumagalli} {et~al.}(2014){Fumagalli}, {Labb{\'e}}, {Patel}, {Franx},
  {van Dokkum}, {Brammer}, {da Cunha}, {F{\"o}rster Schreiber}, {Kriek},
  {Quadri}, {Rix}, {Wake}, {Whitaker}, {Lundgren}, {Marchesini}, {Maseda},
  {Momcheva}, {Nelson}, {Pacifici}, \& {Skelton}}]{Fumagalli14}
{Fumagalli}, M., {Labb{\'e}}, I., {Patel}, S.~G., {et~al.} 2014, \apj, 796, 35,
  \dodoi{10.1088/0004-637X/796/1/35}

\bibitem[{{Gallazzi} {et~al.}(2014){Gallazzi}, {Bell}, {Zibetti}, {Brinchmann},
  \& {Kelson}}]{Gallazzi14}
{Gallazzi}, A., {Bell}, E.~F., {Zibetti}, S., {Brinchmann}, J., \& {Kelson},
  D.~D. 2014, \apj, 788, 72, \dodoi{10.1088/0004-637X/788/1/72}

\bibitem[{{Gallazzi} {et~al.}(2005){Gallazzi}, {Charlot}, {Brinchmann},
  {White}, \& {Tremonti}}]{Gallazzi05}
{Gallazzi}, A., {Charlot}, S., {Brinchmann}, J., {White}, S. D.~M., \&
  {Tremonti}, C.~A. 2005, \mnras, 362, 41,
  \dodoi{10.1111/j.1365-2966.2005.09321.x}

\bibitem[{{Garn} \& {Best}(2010)}]{GarnBest10}
{Garn}, T., \& {Best}, P.~N. 2010, \mnras, 409, 421,
  \dodoi{10.1111/j.1365-2966.2010.17321.x}

\bibitem[{{Ge} {et~al.}(2021){Ge}, {Mao}, {Lu}, {Cappellari}, {Long}, \&
  {Yan}}]{Ge21}
{Ge}, J., {Mao}, S., {Lu}, Y., {et~al.} 2021, \mnras, 507, 2488,
  \dodoi{10.1093/mnras/stab2341}

\bibitem[{{Genzel} {et~al.}(2014){Genzel}, {F{\"o}rster Schreiber}, {Lang},
  {Tacchella}, {Tacconi}, {Wuyts}, {Bandara}, {Burkert}, {Buschkamp},
  {Carollo}, {Cresci}, {Davies}, {Eisenhauer}, {Hicks}, {Kurk}, {Lilly},
  {Lutz}, {Mancini}, {Naab}, {Newman}, {Peng}, {Renzini}, {Shapiro Griffin},
  {Sternberg}, {Vergani}, {Wisnioski}, {Wuyts}, \& {Zamorani}}]{Genzel14}
{Genzel}, R., {F{\"o}rster Schreiber}, N.~M., {Lang}, P., {et~al.} 2014, \apj,
  785, 75, \dodoi{10.1088/0004-637X/785/1/75}

\bibitem[{{Giavalisco} {et~al.}(2004){Giavalisco}, {Ferguson}, {Koekemoer},
  {Dickinson}, {Alexander}, {Bauer}, {Bergeron}, {Biagetti}, {Brandt},
  {Casertano}, {Cesarsky}, {Chatzichristou}, {Conselice}, {Cristiani}, {Da
  Costa}, {Dahlen}, {de Mello}, {Eisenhardt}, {Erben}, {Fall}, {Fassnacht},
  {Fosbury}, {Fruchter}, {Gardner}, {Grogin}, {Hook}, {Hornschemeier}, {Idzi},
  {Jogee}, {Kretchmer}, {Laidler}, {Lee}, {Livio}, {Lucas}, {Madau},
  {Mobasher}, {Moustakas}, {Nonino}, {Padovani}, {Papovich}, {Park},
  {Ravindranath}, {Renzini}, {Richardson}, {Riess}, {Rosati}, {Schirmer},
  {Schreier}, {Somerville}, {Spinrad}, {Stern}, {Stiavelli}, {Strolger},
  {Urry}, {Vandame}, {Williams}, \& {Wolf}}]{Giavalisco04}
{Giavalisco}, M., {Ferguson}, H.~C., {Koekemoer}, A.~M., {et~al.} 2004, \apjl,
  600, L93, \dodoi{10.1086/379232}

\bibitem[{{Giovanelli} {et~al.}(1994){Giovanelli}, {Haynes}, {Salzer},
  {Wegner}, {da Costa}, \& {Freudling}}]{Giovanelli94}
{Giovanelli}, R., {Haynes}, M.~P., {Salzer}, J.~J., {et~al.} 1994, \aj, 107,
  2036, \dodoi{10.1086/117014}

\bibitem[{{Grand} \& {Kawata}(2016)}]{GrandKawata16}
{Grand}, R.~J.~J., \& {Kawata}, D. 2016, Astronomische Nachrichten, 337, 957,
  \dodoi{10.1002/asna.201612407}

\bibitem[{{Grand} {et~al.}(2015){Grand}, {Kawata}, \& {Cropper}}]{Grand15}
{Grand}, R. J.~J., {Kawata}, D., \& {Cropper}, M. 2015, \mnras, 447, 4018,
  \dodoi{10.1093/mnras/stv016}

\bibitem[{{Greener} {et~al.}(2020){Greener}, {Arag{\'o}n-Salamanca},
  {Merrifield}, {Peterken}, {Fraser-McKelvie}, {Masters}, {Krawczyk},
  {Boardman}, {Boquien}, {Andrews}, {Brinkmann}, \& {Drory}}]{Greener20}
{Greener}, M.~J., {Arag{\'o}n-Salamanca}, A., {Merrifield}, M.~R., {et~al.}
  2020, \mnras, 495, 2305, \dodoi{10.1093/mnras/staa1300}

\bibitem[{{Grogin} {et~al.}(2011){Grogin}, {Kocevski}, {Faber}, {Ferguson},
  {Koekemoer}, {Riess}, {Acquaviva}, {Alexander}, {Almaini}, {Ashby}, {Barden},
  {Bell}, {Bournaud}, {Brown}, {Caputi}, {Casertano}, {Cassata}, {Castellano},
  {Challis}, {Chary}, {Cheung}, {Cirasuolo}, {Conselice}, {Roshan Cooray},
  {Croton}, {Daddi}, {Dahlen}, {Dav{\'e}}, {de Mello}, {Dekel}, {Dickinson},
  {Dolch}, {Donley}, {Dunlop}, {Dutton}, {Elbaz}, {Fazio}, {Filippenko},
  {Finkelstein}, {Fontana}, {Gardner}, {Garnavich}, {Gawiser}, {Giavalisco},
  {Grazian}, {Guo}, {Hathi}, {H{\"a}ussler}, {Hopkins}, {Huang}, {Huang},
  {Jha}, {Kartaltepe}, {Kirshner}, {Koo}, {Lai}, {Lee}, {Li}, {Lotz}, {Lucas},
  {Madau}, {McCarthy}, {McGrath}, {McIntosh}, {McLure}, {Mobasher},
  {Moustakas}, {Mozena}, {Nandra}, {Newman}, {Niemi}, {Noeske}, {Papovich},
  {Pentericci}, {Pope}, {Primack}, {Rajan}, {Ravindranath}, {Reddy}, {Renzini},
  {Rix}, {Robaina}, {Rodney}, {Rosario}, {Rosati}, {Salimbeni}, {Scarlata},
  {Siana}, {Simard}, {Smidt}, {Somerville}, {Spinrad}, {Straughn}, {Strolger},
  {Telford}, {Teplitz}, {Trump}, {van der Wel}, {Villforth}, {Wechsler},
  {Weiner}, {Wiklind}, {Wild}, {Wilson}, {Wuyts}, {Yan}, \& {Yun}}]{Grogin11}
{Grogin}, N.~A., {Kocevski}, D.~D., {Faber}, S.~M., {et~al.} 2011, \apjs, 197,
  35, \dodoi{10.1088/0067-0049/197/2/35}

\bibitem[{{Guo} {et~al.}(2013){Guo}, {Ferguson}, {Giavalisco}, {Barro},
  {Willner}, {Ashby}, {Dahlen}, {Donley}, {Faber}, {Fontana}, {Galametz},
  {Grazian}, {Huang}, {Kocevski}, {Koekemoer}, {Koo}, {McGrath}, {Peth},
  {Salvato}, {Wuyts}, {Castellano}, {Cooray}, {Dickinson}, {Dunlop}, {Fazio},
  {Gardner}, {Gawiser}, {Grogin}, {Hathi}, {Hsu}, {Lee}, {Lucas}, {Mobasher},
  {Nandra}, {Newman}, \& {van der Wel}}]{Guo13}
{Guo}, Y., {Ferguson}, H.~C., {Giavalisco}, M., {et~al.} 2013, \apjs, 207, 24,
  \dodoi{10.1088/0067-0049/207/2/24}

\bibitem[{{Guo} {et~al.}(2016{\natexlab{a}}){Guo}, {Koo}, {Lu}, {Forbes},
  {Rafelski}, {Trump}, {Amor{\'\i}n}, {Barro}, {Dav{\'e}}, {Faber}, {Hathi},
  {Yesuf}, {Cooper}, {Dekel}, {Guhathakurta}, {Kirby}, {Koekemoer},
  {P{\'e}rez-Gonz{\'a}lez}, {Lin}, {Newman}, {Primack}, {Rosario}, {Willmer},
  \& {Yan}}]{Guo16a}
{Guo}, Y., {Koo}, D.~C., {Lu}, Y., {et~al.} 2016{\natexlab{a}}, \apj, 822, 103,
  \dodoi{10.3847/0004-637X/822/2/103}

\bibitem[{{Guo} {et~al.}(2016{\natexlab{b}}){Guo}, {Rafelski}, {Faber}, {Koo},
  {Krumholz}, {Trump}, {Willner}, {Amor{\'\i}n}, {Barro}, {Bell}, {Gardner},
  {Gawiser}, {Hathi}, {Koekemoer}, {Pacifici}, {P{\'e}rez-Gonz{\'a}lez},
  {Ravindranath}, {Reddy}, {Teplitz}, \& {Yesuf}}]{Guo16b}
{Guo}, Y., {Rafelski}, M., {Faber}, S.~M., {et~al.} 2016{\natexlab{b}}, \apj,
  833, 37, \dodoi{10.3847/1538-4357/833/1/37}

\bibitem[{{Guo} {et~al.}(2018){Guo}, {Rafelski}, {Bell}, {Conselice}, {Dekel},
  {Faber}, {Giavalisco}, {Koekemoer}, {Koo}, {Lu}, {Mandelker}, {Primack},
  {Ceverino}, {de Mello}, {Ferguson}, {Hathi}, {Kocevski}, {Lucas},
  {P{\'e}rez-Gonz{\'a}lez}, {Ravindranath}, {Soto}, {Straughn}, \&
  {Wang}}]{Guo18}
{Guo}, Y., {Rafelski}, M., {Bell}, E.~F., {et~al.} 2018, \apj, 853, 108,
  \dodoi{10.3847/1538-4357/aaa018}

\bibitem[{{Gutkin} {et~al.}(2016){Gutkin}, {Charlot}, \& {Bruzual}}]{Gutkin16}
{Gutkin}, J., {Charlot}, S., \& {Bruzual}, G. 2016, \mnras, 462, 1757,
  \dodoi{10.1093/mnras/stw1716}

\bibitem[{{Hahn} {et~al.}(2017){Hahn}, {Tinker}, \& {Wetzel}}]{Hahn17}
{Hahn}, C., {Tinker}, J.~L., \& {Wetzel}, A. 2017, \apj, 841, 6,
  \dodoi{10.3847/1538-4357/aa6d6b}

\bibitem[{{Han} {et~al.}(2023){Han}, {Fan}, {Zheng}, {Bai}, \& {Han}}]{Han23}
{Han}, Y., {Fan}, L., {Zheng}, X.~Z., {Bai}, J.-M., \& {Han}, Z. 2023, \apjs,
  269, 39, \dodoi{10.3847/1538-4365/acfc3a}

\bibitem[{{Han} \& {Han}(2012)}]{HanHan12}
{Han}, Y., \& {Han}, Z. 2012, \apj, 749, 123,
  \dodoi{10.1088/0004-637X/749/2/123}

\bibitem[{{Han} \& {Han}(2014)}]{HanHan14}
---. 2014, \apjs, 215, 2, \dodoi{10.1088/0067-0049/215/1/2}

\bibitem[{{Han} \& {Han}(2019)}]{HanHan19}
---. 2019, \apjs, 240, 3, \dodoi{10.3847/1538-4365/aaeffa}

\bibitem[{Harris {et~al.}(2020)Harris, Millman, van~der Walt, Gommers,
  Virtanen, Cournapeau, Wieser, Taylor, Berg, Smith, Kern, Picus, Hoyer, van
  Kerkwijk, Brett, Haldane, del R{\'{i}}o, Wiebe, Peterson,
  G{\'{e}}rard-Marchant, Sheppard, Reddy, Weckesser, Abbasi, Gohlke, \&
  Oliphant}]{harris20}
Harris, C.~R., Millman, K.~J., van~der Walt, S.~J., {et~al.} 2020, Nature, 585,
  357, \dodoi{10.1038/s41586-020-2649-2}

\bibitem[{{Harvey} {et~al.}(2025){Harvey}, {Conselice}, {Adams}, {Austin},
  {Juod{\v{z}}balis}, {Trussler}, {Li}, {Ormerod}, {Ferreira}, {Lovell},
  {Duan}, {Westcott}, {Harris}, {Bhatawdekar}, {Coe}, {Cohen}, {Caruana},
  {Cheng}, {Driver}, {Frye}, {Furtak}, {Grogin}, {Hathi}, {Holwerda}, {Jansen},
  {Koekemoer}, {Marshall}, {Nonino}, {Vijayan}, {Wilkins}, {Windhorst},
  {Willmer}, {Yan}, \& {Zitrin}}]{Harvey25}
{Harvey}, T., {Conselice}, C.~J., {Adams}, N.~J., {et~al.} 2025, \apj, 978, 89,
  \dodoi{10.3847/1538-4357/ad8c29}

\bibitem[{{Hathi} {et~al.}(2009){Hathi}, {Ferreras}, {Pasquali}, {Malhotra},
  {Rhoads}, {Pirzkal}, {Windhorst}, \& {Xu}}]{Hathi09}
{Hathi}, N.~P., {Ferreras}, I., {Pasquali}, A., {et~al.} 2009, \apj, 690, 1866,
  \dodoi{10.1088/0004-637X/690/2/1866}

\bibitem[{{Hemmati} {et~al.}(2014){Hemmati}, {Miller}, {Mobasher}, {Nayyeri},
  {Ferguson}, {Guo}, {Koekemoer}, {Koo}, \& {Papovich}}]{Hemmati14}
{Hemmati}, S., {Miller}, S.~H., {Mobasher}, B., {et~al.} 2014, \apj, 797, 108,
  \dodoi{10.1088/0004-637X/797/2/108}

\bibitem[{{Hemmati} {et~al.}(2020){Hemmati}, {Mobasher}, {Nayyeri}, {Shahidi},
  {Capak}, {Darvish}, {Chartab}, {Jafariyazani}, \& {Sattari}}]{Hemmati20}
{Hemmati}, S., {Mobasher}, B., {Nayyeri}, H., {et~al.} 2020, \apjl, 896, L17,
  \dodoi{10.3847/2041-8213/ab7243}

\bibitem[{{Henry} {et~al.}(2013){Henry}, {Scarlata}, {Dom{\'\i}nguez},
  {Malkan}, {Martin}, {Siana}, {Atek}, {Bedregal}, {Colbert}, {Rafelski},
  {Ross}, {Teplitz}, {Bunker}, {Dressler}, {Hathi}, {Masters}, {McCarthy}, \&
  {Straughn}}]{Henry13}
{Henry}, A., {Scarlata}, C., {Dom{\'\i}nguez}, A., {et~al.} 2013, \apjl, 776,
  L27, \dodoi{10.1088/2041-8205/776/2/L27}

\bibitem[{{Hirschmann} {et~al.}(2015){Hirschmann}, {Naab}, {Ostriker},
  {Forbes}, {Duc}, {Dav{\'e}}, {Oser}, \& {Karabal}}]{Hirschmann15}
{Hirschmann}, M., {Naab}, T., {Ostriker}, J.~P., {et~al.} 2015, \mnras, 449,
  528, \dodoi{10.1093/mnras/stv274}

\bibitem[{{Hopkins} {et~al.}(2009){Hopkins}, {Cox}, {Younger}, \&
  {Hernquist}}]{Hopkins09a}
{Hopkins}, P.~F., {Cox}, T.~J., {Younger}, J.~D., \& {Hernquist}, L. 2009,
  \apj, 691, 1168, \dodoi{10.1088/0004-637X/691/2/1168}

\bibitem[{Hunter(2007)}]{Hunter07}
Hunter, J.~D. 2007, Computing in Science \& Engineering, 9, 90,
  \dodoi{10.1109/MCSE.2007.55}

\bibitem[{{Iyer} \& {Gawiser}(2017)}]{Iyer17}
{Iyer}, K., \& {Gawiser}, E. 2017, \apj, 838, 127,
  \dodoi{10.3847/1538-4357/aa63f0}

\bibitem[{{Jonsson} {et~al.}(2010){Jonsson}, {Groves}, \& {Cox}}]{Jonsson10}
{Jonsson}, P., {Groves}, B.~A., \& {Cox}, T.~J. 2010, \mnras, 403, 17,
  \dodoi{10.1111/j.1365-2966.2009.16087.x}

\bibitem[{{Karim} {et~al.}(2011){Karim}, {Schinnerer},
  {Mart{\'\i}nez-Sansigre}, {Sargent}, {van der Wel}, {Rix}, {Ilbert},
  {Smol{\v{c}}i{\'c}}, {Carilli}, {Pannella}, {Koekemoer}, {Bell}, \&
  {Salvato}}]{Karim11}
{Karim}, A., {Schinnerer}, E., {Mart{\'\i}nez-Sansigre}, A., {et~al.} 2011,
  \apj, 730, 61, \dodoi{10.1088/0004-637X/730/2/61}

\bibitem[{{Kashino} {et~al.}(2013){Kashino}, {Silverman}, {Rodighiero},
  {Renzini}, {Arimoto}, {Daddi}, {Lilly}, {Sanders}, {Kartaltepe}, {Zahid},
  {Nagao}, {Sugiyama}, {Capak}, {Carollo}, {Chu}, {Hasinger}, {Ilbert},
  {Kajisawa}, {Kewley}, {Koekemoer}, {Kova{\v{c}}}, {Le F{\`e}vre}, {Masters},
  {McCracken}, {Onodera}, {Scoville}, {Strazzullo}, {Symeonidis}, \&
  {Taniguchi}}]{Kashino13}
{Kashino}, D., {Silverman}, J.~D., {Rodighiero}, G., {et~al.} 2013, \apjl, 777,
  L8, \dodoi{10.1088/2041-8205/777/1/L8}

\bibitem[{{Kass} \& {Raftery}(1995)}]{KassRaftery95}
{Kass}, R.~E., \& {Raftery}, A.~E. 1995, J. Am. Stat. Assoc., 90, 773,
  \dodoi{10.1080/01621459.1995.10476572}

\bibitem[{{Kassin} {et~al.}(2012){Kassin}, {Weiner}, {Faber}, {Gardner},
  {Willmer}, {Coil}, {Cooper}, {Devriendt}, {Dutton}, {Guhathakurta}, {Koo},
  {Metevier}, {Noeske}, \& {Primack}}]{Kassin12}
{Kassin}, S.~A., {Weiner}, B.~J., {Faber}, S.~M., {et~al.} 2012, \apj, 758,
  106, \dodoi{10.1088/0004-637X/758/2/106}

\bibitem[{{Kennicutt}(1998)}]{Kennicutt98}
{Kennicutt}, Robert~C., J. 1998, \araa, 36, 189,
  \dodoi{10.1146/annurev.astro.36.1.189}

\bibitem[{{Kennicutt} \& {Evans}(2012)}]{Kennicutt12}
{Kennicutt}, R.~C., \& {Evans}, N.~J. 2012, \araa, 50, 531,
  \dodoi{10.1146/annurev-astro-081811-125610}

\bibitem[{{Kodra} {et~al.}(2023){Kodra}, {Andrews}, {Newman}, {Finkelstein},
  {Fontana}, {Hathi}, {Salvato}, {Wiklind}, {Wuyts}, {Broussard}, {Chartab},
  {Conselice}, {Cooper}, {Dekel}, {Dickinson}, {Ferguson}, {Gawiser}, {Grogin},
  {Iyer}, {Kartaltepe}, {Kassin}, {Koekemoer}, {Koo}, {Lucas}, {Mantha},
  {McIntosh}, {Mobasher}, {Pacifici}, {P{\'e}rez-Gonz{\'a}lez}, \&
  {Santini}}]{Kodra23}
{Kodra}, D., {Andrews}, B.~H., {Newman}, J.~A., {et~al.} 2023, \apj, 942, 36,
  \dodoi{10.3847/1538-4357/ac9f12}

\bibitem[{{Koekemoer} {et~al.}(2011){Koekemoer}, {Faber}, {Ferguson}, {Grogin},
  {Kocevski}, {Koo}, {Lai}, {Lotz}, {Lucas}, {McGrath}, {Ogaz}, {Rajan},
  {Riess}, {Rodney}, {Strolger}, {Casertano}, {Castellano}, {Dahlen},
  {Dickinson}, {Dolch}, {Fontana}, {Giavalisco}, {Grazian}, {Guo}, {Hathi},
  {Huang}, {van der Wel}, {Yan}, {Acquaviva}, {Alexander}, {Almaini}, {Ashby},
  {Barden}, {Bell}, {Bournaud}, {Brown}, {Caputi}, {Cassata}, {Challis},
  {Chary}, {Cheung}, {Cirasuolo}, {Conselice}, {Roshan Cooray}, {Croton},
  {Daddi}, {Dav{\'e}}, {de Mello}, {de Ravel}, {Dekel}, {Donley}, {Dunlop},
  {Dutton}, {Elbaz}, {Fazio}, {Filippenko}, {Finkelstein}, {Frazer}, {Gardner},
  {Garnavich}, {Gawiser}, {Gruetzbauch}, {Hartley}, {H{\"a}ussler},
  {Herrington}, {Hopkins}, {Huang}, {Jha}, {Johnson}, {Kartaltepe},
  {Khostovan}, {Kirshner}, {Lani}, {Lee}, {Li}, {Madau}, {McCarthy},
  {McIntosh}, {McLure}, {McPartland}, {Mobasher}, {Moreira}, {Mortlock},
  {Moustakas}, {Mozena}, {Nandra}, {Newman}, {Nielsen}, {Niemi}, {Noeske},
  {Papovich}, {Pentericci}, {Pope}, {Primack}, {Ravindranath}, {Reddy},
  {Renzini}, {Rix}, {Robaina}, {Rosario}, {Rosati}, {Salimbeni}, {Scarlata},
  {Siana}, {Simard}, {Smidt}, {Snyder}, {Somerville}, {Spinrad}, {Straughn},
  {Telford}, {Teplitz}, {Trump}, {Vargas}, {Villforth}, {Wagner}, {Wandro},
  {Wechsler}, {Weiner}, {Wiklind}, {Wild}, {Wilson}, {Wuyts}, \&
  {Yun}}]{Koekemoer11}
{Koekemoer}, A.~M., {Faber}, S.~M., {Ferguson}, H.~C., {et~al.} 2011, \apjs,
  197, 36, \dodoi{10.1088/0067-0049/197/2/36}

\bibitem[{{Kriek} \& {Conroy}(2013)}]{Kriek13}
{Kriek}, M., \& {Conroy}, C. 2013, \apjl, 775, L16,
  \dodoi{10.1088/2041-8205/775/1/L16}

\bibitem[{{Kriek} {et~al.}(2010){Kriek}, {Labb{\'e}}, {Conroy}, {Whitaker},
  {van Dokkum}, {Brammer}, {Franx}, {Illingworth}, {Marchesini}, {Muzzin},
  {Quadri}, \& {Rudnick}}]{Kriek10}
{Kriek}, M., {Labb{\'e}}, I., {Conroy}, C., {et~al.} 2010, \apjl, 722, L64,
  \dodoi{10.1088/2041-8205/722/1/L64}

\bibitem[{{Kuchinski} {et~al.}(1998){Kuchinski}, {Terndrup}, {Gordon}, \&
  {Witt}}]{Kuchinski98}
{Kuchinski}, L.~E., {Terndrup}, D.~M., {Gordon}, K.~D., \& {Witt}, A.~N. 1998,
  \aj, 115, 1438, \dodoi{10.1086/300287}

\bibitem[{{Labb{\'e}} {et~al.}(2003){Labb{\'e}}, {Franx}, {Rudnick},
  {F{\"o}rster Schreiber}, {Rix}, {Moorwood}, {van Dokkum}, {van der Werf},
  {R{\"o}ttgering}, {van Starkenburg}, {van der Wel}, {Kuijken}, \&
  {Daddi}}]{Labbe03}
{Labb{\'e}}, I., {Franx}, M., {Rudnick}, G., {et~al.} 2003, \aj, 125, 1107,
  \dodoi{10.1086/346140}

\bibitem[{{Lawler} \& {Acquaviva}(2021)}]{Lawler21}
{Lawler}, A.~J., \& {Acquaviva}, V. 2021, \mnras, 502, 3993,
  \dodoi{10.1093/mnras/stab138}

\bibitem[{{Le Cras} {et~al.}(2016){Le Cras}, {Maraston}, {Thomas}, \&
  {York}}]{LeCras16}
{Le Cras}, C., {Maraston}, C., {Thomas}, D., \& {York}, D.~G. 2016, \mnras,
  461, 766, \dodoi{10.1093/mnras/stw1024}

\bibitem[{{Lee} {et~al.}(2015){Lee}, {Sanders}, {Casey}, {Toft}, {Scoville},
  {Hung}, {Le Floc'h}, {Ilbert}, {Zahid}, {Aussel}, {Capak}, {Kartaltepe},
  {Kewley}, {Li}, {Schawinski}, {Sheth}, \& {Xiao}}]{Lee15}
{Lee}, N., {Sanders}, D.~B., {Casey}, C.~M., {et~al.} 2015, \apj, 801, 80,
  \dodoi{10.1088/0004-637X/801/2/80}

\bibitem[{{Leja} {et~al.}(2017){Leja}, {Johnson}, {Conroy}, {van Dokkum}, \&
  {Byler}}]{Leja17}
{Leja}, J., {Johnson}, B.~D., {Conroy}, C., {van Dokkum}, P.~G., \& {Byler}, N.
  2017, \apj, 837, 170, \dodoi{10.3847/1538-4357/aa5ffe}

\bibitem[{{Leja} {et~al.}(2020){Leja}, {Speagle}, {Johnson}, {Conroy}, {van
  Dokkum}, \& {Franx}}]{Leja20}
{Leja}, J., {Speagle}, J.~S., {Johnson}, B.~D., {et~al.} 2020, \apj, 893, 111,
  \dodoi{10.3847/1538-4357/ab7e27}

\bibitem[{{Leja} {et~al.}(2019){Leja}, {Johnson}, {Conroy}, {van Dokkum},
  {Speagle}, {Brammer}, {Momcheva}, {Skelton}, {Whitaker}, {Franx}, \&
  {Nelson}}]{Leja19}
{Leja}, J., {Johnson}, B.~D., {Conroy}, C., {et~al.} 2019, \apj, 877, 140,
  \dodoi{10.3847/1538-4357/ab1d5a}

\bibitem[{{Li} {et~al.}(2018){Li}, {Lelli}, {McGaugh}, \& {Schombert}}]{Li18}
{Li}, P., {Lelli}, F., {McGaugh}, S., \& {Schombert}, J. 2018, \aap, 615, A3,
  \dodoi{10.1051/0004-6361/201732547}

\bibitem[{{Li} {et~al.}(2021){Li}, {Lelli}, {McGaugh}, {Schombert}, \&
  {Chae}}]{Li21}
{Li}, P., {Lelli}, F., {McGaugh}, S., {Schombert}, J., \& {Chae}, K.-H. 2021,
  \aap, 646, L13, \dodoi{10.1051/0004-6361/202040101}

\bibitem[{{Lin} {et~al.}(2019){Lin}, {Hsieh}, {Pan}, {Rembold}, {S{\'a}nchez},
  {Argudo-Fern{\'a}ndez}, {Rowlands}, {Belfiore}, {Bizyaev}, {Lacerna},
  {Riffel}, {Rong}, {Yuan}, {Drory}, {Maiolino}, \& {Wilcots}}]{Lin19}
{Lin}, L., {Hsieh}, B.-C., {Pan}, H.-A., {et~al.} 2019, \apj, 872, 50,
  \dodoi{10.3847/1538-4357/aafa84}

\bibitem[{{Liu} {et~al.}(2018){Liu}, {Jia}, {Yesuf}, {Faber}, {Koo}, {Guo},
  {Bell}, {Jiang}, {Wang}, {Koekemoer}, {Zheng}, {Fang}, {Barro},
  {P{\'e}rez-Gonz{\'a}lez}, {Dekel}, {Kocevski}, {Hathi}, {Croton},
  {Huertas-Company}, {Meng}, {Tong}, \& {Liu}}]{Liu18}
{Liu}, F.~S., {Jia}, M., {Yesuf}, H.~M., {et~al.} 2018, \apj, 860, 60,
  \dodoi{10.3847/1538-4357/aac20d}

\bibitem[{{Lo Faro} {et~al.}(2017){Lo Faro}, {Buat}, {Roehlly},
  {Alvarez-Marquez}, {Burgarella}, {Silva}, \& {Efstathiou}}]{LoFaro17}
{Lo Faro}, B., {Buat}, V., {Roehlly}, Y., {et~al.} 2017, \mnras, 472, 1372,
  \dodoi{10.1093/mnras/stx1901}

\bibitem[{{Lower} {et~al.}(2020){Lower}, {Narayanan}, {Leja}, {Johnson},
  {Conroy}, \& {Dav{\'e}}}]{Lower20}
{Lower}, S., {Narayanan}, D., {Leja}, J., {et~al.} 2020, \apj, 904, 33,
  \dodoi{10.3847/1538-4357/abbfa7}

\bibitem[{{Lu} {et~al.}(2024){Lu}, {Daddi}, {Maraston}, {Dickinson}, {Haro},
  {Gobat}, {Renzini}, {Giavalisco}, {Bagley}, {Calabr{\`o}}, {Cheng}, {de la
  Vega}, {D'Eugenio}, {Elbaz}, {Finkelstein}, {G{\'o}mez-Guijarro}, {Gu},
  {Hathi}, {Huertas-Company}, {Kartaltepe}, {Koekemoer}, {Henry}, {Lyu},
  {Magnelli}, {Mobasher}, {Papovich}, {Pirzkal}, {Rich}, {Tacchella}, \&
  {Yung}}]{Lu24}
{Lu}, S., {Daddi}, E., {Maraston}, C., {et~al.} 2024, Nature Astronomy,
  \dodoi{10.1038/s41550-024-02391-9}

\bibitem[{{Luo} {et~al.}(2017){Luo}, {Brandt}, {Xue}, {Lehmer}, {Alexander},
  {Bauer}, {Vito}, {Yang}, {Basu-Zych}, {Comastri}, {Gilli}, {Gu},
  {Hornschemeier}, {Koekemoer}, {Liu}, {Mainieri}, {Paolillo}, {Ranalli},
  {Rosati}, {Schneider}, {Shemmer}, {Smail}, {Sun}, {Tozzi}, {Vignali}, \&
  {Wang}}]{Luo17}
{Luo}, B., {Brandt}, W.~N., {Xue}, Y.~Q., {et~al.} 2017, \apjs, 228, 2,
  \dodoi{10.3847/1538-4365/228/1/2}

\bibitem[{{Ly} {et~al.}(2012){Ly}, {Malkan}, {Kashikawa}, {Ota}, {Shimasaku},
  {Iye}, \& {Currie}}]{Ly12}
{Ly}, C., {Malkan}, M.~A., {Kashikawa}, N., {et~al.} 2012, \apjl, 747, L16,
  \dodoi{10.1088/2041-8205/747/1/L16}

\bibitem[{{Lynden-Bell} \& {Kalnajs}(1972)}]{LyndenBell72}
{Lynden-Bell}, D., \& {Kalnajs}, A.~J. 1972, \mnras, 157, 1,
  \dodoi{10.1093/mnras/157.1.1}

\bibitem[{{Madau} \& {Dickinson}(2014)}]{Madau14}
{Madau}, P., \& {Dickinson}, M. 2014, \araa, 52, 415,
  \dodoi{10.1146/annurev-astro-081811-125615}

\bibitem[{{Maiolino} \& {Mannucci}(2019)}]{Maiolino19}
{Maiolino}, R., \& {Mannucci}, F. 2019, \aapr, 27, 3,
  \dodoi{10.1007/s00159-018-0112-2}

\bibitem[{{Maiolino} {et~al.}(2008){Maiolino}, {Nagao}, {Grazian}, {Cocchia},
  {Marconi}, {Mannucci}, {Cimatti}, {Pipino}, {Ballero}, {Calura}, {Chiappini},
  {Fontana}, {Granato}, {Matteucci}, {Pastorini}, {Pentericci}, {Risaliti},
  {Salvati}, \& {Silva}}]{Maiolino08}
{Maiolino}, R., {Nagao}, T., {Grazian}, A., {et~al.} 2008, \aap, 488, 463,
  \dodoi{10.1051/0004-6361:200809678}

\bibitem[{{Man} \& {Belli}(2018)}]{Man18}
{Man}, A., \& {Belli}, S. 2018, Nature Astronomy, 2, 695,
  \dodoi{10.1038/s41550-018-0558-1}

\bibitem[{{Man} {et~al.}(2012){Man}, {Toft}, {Zirm}, {Wuyts}, \& {van der
  Wel}}]{Man12}
{Man}, A. W.~S., {Toft}, S., {Zirm}, A.~W., {Wuyts}, S., \& {van der Wel}, A.
  2012, \apj, 744, 85, \dodoi{10.1088/0004-637X/744/2/85}

\bibitem[{{Man} {et~al.}(2016){Man}, {Zirm}, \& {Toft}}]{Man16}
{Man}, A. W.~S., {Zirm}, A.~W., \& {Toft}, S. 2016, \apj, 830, 89,
  \dodoi{10.3847/0004-637X/830/2/89}

\bibitem[{{Maraston}(2005)}]{Maraston05}
{Maraston}, C. 2005, \mnras, 362, 799, \dodoi{10.1111/j.1365-2966.2005.09270.x}

\bibitem[{{Martig} {et~al.}(2009){Martig}, {Bournaud}, {Teyssier}, \&
  {Dekel}}]{Martig09}
{Martig}, M., {Bournaud}, F., {Teyssier}, R., \& {Dekel}, A. 2009, \apj, 707,
  250, \dodoi{10.1088/0004-637X/707/1/250}

\bibitem[{{McDermid} {et~al.}(2015){McDermid}, {Alatalo}, {Blitz}, {Bournaud},
  {Bureau}, {Cappellari}, {Crocker}, {Davies}, {Davis}, {de Zeeuw}, {Duc},
  {Emsellem}, {Khochfar}, {Krajnovi{\'c}}, {Kuntschner}, {Morganti}, {Naab},
  {Oosterloo}, {Sarzi}, {Scott}, {Serra}, {Weijmans}, \& {Young}}]{McDermid15}
{McDermid}, R.~M., {Alatalo}, K., {Blitz}, L., {et~al.} 2015, \mnras, 448,
  3484, \dodoi{10.1093/mnras/stv105}

\bibitem[{{Medling} {et~al.}(2018){Medling}, {Cortese}, {Croom}, {Green},
  {Groves}, {Hampton}, {Ho}, {Davies}, {Kewley}, {Moffett}, {Schaefer},
  {Taylor}, {Zafar}, {Bekki}, {Bland-Hawthorn}, {Bloom}, {Brough}, {Bryant},
  {Catinella}, {Cecil}, {Colless}, {Couch}, {Drinkwater}, {Driver},
  {Federrath}, {Foster}, {Goldstein}, {Goodwin}, {Hopkins}, {Lawrence},
  {Leslie}, {Lewis}, {Lorente}, {Owers}, {McDermid}, {Richards}, {Sharp},
  {Scott}, {Sweet}, {Taranu}, {Tescari}, {Tonini}, {van de Sande}, {Walcher},
  \& {Wright}}]{Medling18}
{Medling}, A.~M., {Cortese}, L., {Croom}, S.~M., {et~al.} 2018, \mnras, 475,
  5194, \dodoi{10.1093/mnras/sty127}

\bibitem[{{Mihos} \& {Hernquist}(1996)}]{Mihos96}
{Mihos}, J.~C., \& {Hernquist}, L. 1996, \apj, 464, 641, \dodoi{10.1086/177353}

\bibitem[{{Minchev} {et~al.}(2014){Minchev}, {Chiappini}, \&
  {Martig}}]{Minchev14}
{Minchev}, I., {Chiappini}, C., \& {Martig}, M. 2014, \aap, 572, A92,
  \dodoi{10.1051/0004-6361/201423487}

\bibitem[{{Momcheva} {et~al.}(2013){Momcheva}, {Lee}, {Ly}, {Salim}, {Dale},
  {Ouchi}, {Finn}, \& {Ono}}]{Momcheva13}
{Momcheva}, I.~G., {Lee}, J.~C., {Ly}, C., {et~al.} 2013, \aj, 145, 47,
  \dodoi{10.1088/0004-6256/145/2/47}

\bibitem[{{Momcheva} {et~al.}(2016){Momcheva}, {Brammer}, {van Dokkum},
  {Skelton}, {Whitaker}, {Nelson}, {Fumagalli}, {Maseda}, {Leja}, {Franx},
  {Rix}, {Bezanson}, {Da Cunha}, {Dickey}, {F{\"o}rster Schreiber},
  {Illingworth}, {Kriek}, {Labb{\'e}}, {Ulf Lange}, {Lundgren}, {Magee},
  {Marchesini}, {Oesch}, {Pacifici}, {Patel}, {Price}, {Tal}, {Wake}, {van der
  Wel}, \& {Wuyts}}]{Momcheva16}
{Momcheva}, I.~G., {Brammer}, G.~B., {van Dokkum}, P.~G., {et~al.} 2016, \apjs,
  225, 27, \dodoi{10.3847/0067-0049/225/2/27}

\bibitem[{{Moreno} {et~al.}(2015){Moreno}, {Torrey}, {Ellison}, {Patton},
  {Bluck}, {Bansal}, \& {Hernquist}}]{Moreno15}
{Moreno}, J., {Torrey}, P., {Ellison}, S.~L., {et~al.} 2015, \mnras, 448, 1107,
  \dodoi{10.1093/mnras/stv094}

\bibitem[{{Morishita} {et~al.}(2015){Morishita}, {Ichikawa}, {Noguchi},
  {Akiyama}, {Patel}, {Kajisawa}, \& {Obata}}]{Morishita15}
{Morishita}, T., {Ichikawa}, T., {Noguchi}, M., {et~al.} 2015, \apj, 805, 34,
  \dodoi{10.1088/0004-637X/805/1/34}

\bibitem[{{Morselli} {et~al.}(2019){Morselli}, {Popesso}, {Cibinel}, {Oesch},
  {Montes}, {Atek}, {Illingworth}, \& {Holden}}]{Morselli19}
{Morselli}, L., {Popesso}, P., {Cibinel}, A., {et~al.} 2019, \aap, 626, A61,
  \dodoi{10.1051/0004-6361/201834559}

\bibitem[{{Mosleh} {et~al.}(2020){Mosleh}, {Hosseinnejad},
  {Hosseini-ShahiSavandi}, \& {Tacchella}}]{Mosleh20}
{Mosleh}, M., {Hosseinnejad}, S., {Hosseini-ShahiSavandi}, S.~Z., \&
  {Tacchella}, S. 2020, \apj, 905, 170, \dodoi{10.3847/1538-4357/abc7cc}

\bibitem[{{Mosleh} {et~al.}(2017){Mosleh}, {Tacchella}, {Renzini}, {Carollo},
  {Molaeinezhad}, {Onodera}, {Khosroshahi}, \& {Lilly}}]{Mosleh17}
{Mosleh}, M., {Tacchella}, S., {Renzini}, A., {et~al.} 2017, \apj, 837, 2,
  \dodoi{10.3847/1538-4357/aa5f14}

\bibitem[{{Moster} {et~al.}(2013){Moster}, {Naab}, \& {White}}]{Moster13}
{Moster}, B.~P., {Naab}, T., \& {White}, S. D.~M. 2013, \mnras, 428, 3121,
  \dodoi{10.1093/mnras/sts261}

\bibitem[{{Moustakas} {et~al.}(2013){Moustakas}, {Coil}, {Aird}, {Blanton},
  {Cool}, {Eisenstein}, {Mendez}, {Wong}, {Zhu}, \& {Arnouts}}]{Moustakas13}
{Moustakas}, J., {Coil}, A.~L., {Aird}, J., {et~al.} 2013, \apj, 767, 50,
  \dodoi{10.1088/0004-637X/767/1/50}

\bibitem[{{Muzzin} {et~al.}(2013){Muzzin}, {Marchesini}, {Stefanon}, {Franx},
  {McCracken}, {Milvang-Jensen}, {Dunlop}, {Fynbo}, {Brammer}, {Labb{\'e}}, \&
  {van Dokkum}}]{Muzzin13}
{Muzzin}, A., {Marchesini}, D., {Stefanon}, M., {et~al.} 2013, \apj, 777, 18,
  \dodoi{10.1088/0004-637X/777/1/18}

\bibitem[{{Nagaraj} {et~al.}(2022){Nagaraj}, {Forbes}, {Leja},
  {Foreman-Mackey}, \& {Hayward}}]{Nagaraj22}
{Nagaraj}, G., {Forbes}, J.~C., {Leja}, J., {Foreman-Mackey}, D., \& {Hayward},
  C.~C. 2022, \apj, 932, 54, \dodoi{10.3847/1538-4357/ac6c80}

\bibitem[{{Nelson} {et~al.}(2018){Nelson}, {Pillepich}, {Springel},
  {Weinberger}, {Hernquist}, {Pakmor}, {Genel}, {Torrey}, {Vogelsberger},
  {Kauffmann}, {Marinacci}, \& {Naiman}}]{Nelson18}
{Nelson}, D., {Pillepich}, A., {Springel}, V., {et~al.} 2018, \mnras, 475, 624,
  \dodoi{10.1093/mnras/stx3040}

\bibitem[{{Nelson} {et~al.}(2019){Nelson}, {Springel}, {Pillepich},
  {Rodriguez-Gomez}, {Torrey}, {Genel}, {Vogelsberger}, {Pakmor}, {Marinacci},
  {Weinberger}, {Kelley}, {Lovell}, {Diemer}, \& {Hernquist}}]{Nelson19a}
{Nelson}, D., {Springel}, V., {Pillepich}, A., {et~al.} 2019, Computational
  Astrophysics and Cosmology, 6, 2, \dodoi{10.1186/s40668-019-0028-x}

\bibitem[{{Nelson} {et~al.}(2016){Nelson}, {van Dokkum}, {F{\"o}rster
  Schreiber}, {Franx}, {Brammer}, {Momcheva}, {Wuyts}, {Whitaker}, {Skelton},
  {Fumagalli}, {Hayward}, {Kriek}, {Labb{\'e}}, {Leja}, {Rix}, {Tacconi}, {van
  der Wel}, {van den Bosch}, {Oesch}, {Dickey}, \& {Ulf Lange}}]{Nelson16}
{Nelson}, E.~J., {van Dokkum}, P.~G., {F{\"o}rster Schreiber}, N.~M., {et~al.}
  2016, \apj, 828, 27, \dodoi{10.3847/0004-637X/828/1/27}

\bibitem[{{Nelson} {et~al.}(2021){Nelson}, {Tacchella}, {Diemer}, {Leja},
  {Hernquist}, {Whitaker}, {Weinberger}, {Pillepich}, {Nelson}, {Terrazas},
  {Nevin}, {Brammer}, {Burkhart}, {Cochrane}, {van Dokkum}, {Johnson},
  {Marinacci}, {Mowla}, {Pakmor}, {Skelton}, {Speagle}, {Springel}, {Torrey},
  {Vogelsberger}, \& {Wuyts}}]{Nelson21}
{Nelson}, E.~J., {Tacchella}, S., {Diemer}, B., {et~al.} 2021, \mnras, 508,
  219, \dodoi{10.1093/mnras/stab2131}

\bibitem[{{Newville} {et~al.}(2014){Newville}, {Stensitzki}, {Allen}, \&
  {Ingargiola}}]{Newville14}
{Newville}, M., {Stensitzki}, T., {Allen}, D.~B., \& {Ingargiola}, A. 2014,
  {LMFIT: Non-Linear Least-Square Minimization and Curve-Fitting for Python},
  0.8.0, Zenodo,  Zenodo, \dodoi{10.5281/zenodo.11813}

\bibitem[{{Noeske} {et~al.}(2007){Noeske}, {Weiner}, {Faber}, {Papovich},
  {Koo}, {Somerville}, {Bundy}, {Conselice}, {Newman}, {Schiminovich}, {Le
  Floc'h}, {Coil}, {Rieke}, {Lotz}, {Primack}, {Barmby}, {Cooper}, {Davis},
  {Ellis}, {Fazio}, {Guhathakurta}, {Huang}, {Kassin}, {Martin}, {Phillips},
  {Rich}, {Small}, {Willmer}, \& {Wilson}}]{Noeske07}
{Noeske}, K.~G., {Weiner}, B.~J., {Faber}, S.~M., {et~al.} 2007, \apjl, 660,
  L43, \dodoi{10.1086/517926}

\bibitem[{{Ogando} {et~al.}(2005){Ogando}, {Maia}, {Chiappini}, {Pellegrini},
  {Schiavon}, \& {da Costa}}]{Ogando05}
{Ogando}, R. L.~C., {Maia}, M. A.~G., {Chiappini}, C., {et~al.} 2005, \apjl,
  632, L61, \dodoi{10.1086/497824}

\bibitem[{{Oke} \& {Gunn}(1983)}]{Oke83}
{Oke}, J.~B., \& {Gunn}, J.~E. 1983, \apj, 266, 713, \dodoi{10.1086/160817}

\bibitem[{{Onodera} {et~al.}(2015){Onodera}, {Carollo}, {Renzini},
  {Cappellari}, {Mancini}, {Arimoto}, {Daddi}, {Gobat}, {Strazzullo},
  {Tacchella}, \& {Yamada}}]{Onodera15}
{Onodera}, M., {Carollo}, C.~M., {Renzini}, A., {et~al.} 2015, \apj, 808, 161,
  \dodoi{10.1088/0004-637X/808/2/161}

\bibitem[{{Pacifici} {et~al.}(2012){Pacifici}, {Charlot}, {Blaizot}, \&
  {Brinchmann}}]{Pacifici12}
{Pacifici}, C., {Charlot}, S., {Blaizot}, J., \& {Brinchmann}, J. 2012, \mnras,
  421, 2002, \dodoi{10.1111/j.1365-2966.2012.20431.x}

\bibitem[{{Pacifici} {et~al.}(2015){Pacifici}, {da Cunha}, {Charlot}, {Rix},
  {Fumagalli}, {Wel}, {Franx}, {Maseda}, {van Dokkum}, {Brammer}, {Momcheva},
  {Skelton}, {Whitaker}, {Leja}, {Lundgren}, {Kassin}, \& {Yi}}]{Pacifici15}
{Pacifici}, C., {da Cunha}, E., {Charlot}, S., {et~al.} 2015, \mnras, 447, 786,
  \dodoi{10.1093/mnras/stu2447}

\bibitem[{{Pacifici} {et~al.}(2016){Pacifici}, {Kassin}, {Weiner}, {Holden},
  {Gardner}, {Faber}, {Ferguson}, {Koo}, {Primack}, {Bell}, {Dekel}, {Gawiser},
  {Giavalisco}, {Rafelski}, {Simons}, {Barro}, {Croton}, {Dav{\'e}}, {Fontana},
  {Grogin}, {Koekemoer}, {Lee}, {Salmon}, {Somerville}, \&
  {Behroozi}}]{Pacifici16}
{Pacifici}, C., {Kassin}, S.~A., {Weiner}, B.~J., {et~al.} 2016, \apj, 832, 79,
  \dodoi{10.3847/0004-637X/832/1/79}

\bibitem[{{Peletier} {et~al.}(1995){Peletier}, {Valentijn}, {Moorwood},
  {Freudling}, {Knapen}, \& {Beckman}}]{Peletier95}
{Peletier}, R.~F., {Valentijn}, E.~A., {Moorwood}, A.~F.~M., {et~al.} 1995,
  \aap, 300, L1.
\newblock \doarXiv{astro-ph/9506108}

\bibitem[{{Peng} {et~al.}(2010{\natexlab{a}}){Peng}, {Ho}, {Impey}, \&
  {Rix}}]{Peng10}
{Peng}, C.~Y., {Ho}, L.~C., {Impey}, C.~D., \& {Rix}, H.-W. 2010{\natexlab{a}},
  \aj, 139, 2097, \dodoi{10.1088/0004-6256/139/6/2097}

\bibitem[{{Peng} {et~al.}(2010{\natexlab{b}}){Peng}, {Lilly}, {Kova{\v{c}}},
  {Bolzonella}, {Pozzetti}, {Renzini}, {Zamorani}, {Ilbert}, {Knobel},
  {Iovino}, {Maier}, {Cucciati}, {Tasca}, {Carollo}, {Silverman}, {Kampczyk},
  {de Ravel}, {Sanders}, {Scoville}, {Contini}, {Mainieri}, {Scodeggio},
  {Kneib}, {Le F{\`e}vre}, {Bardelli}, {Bongiorno}, {Caputi}, {Coppa}, {de la
  Torre}, {Franzetti}, {Garilli}, {Lamareille}, {Le Borgne}, {Le Brun},
  {Mignoli}, {Perez Montero}, {Pello}, {Ricciardelli}, {Tanaka}, {Tresse},
  {Vergani}, {Welikala}, {Zucca}, {Oesch}, {Abbas}, {Barnes}, {Bordoloi},
  {Bottini}, {Cappi}, {Cassata}, {Cimatti}, {Fumana}, {Hasinger}, {Koekemoer},
  {Leauthaud}, {Maccagni}, {Marinoni}, {McCracken}, {Memeo}, {Meneux}, {Nair},
  {Porciani}, {Presotto}, \& {Scaramella}}]{YPeng10}
{Peng}, Y.-j., {Lilly}, S.~J., {Kova{\v{c}}}, K., {et~al.} 2010{\natexlab{b}},
  \apj, 721, 193, \dodoi{10.1088/0004-637X/721/1/193}

\bibitem[{P\'erez \& Granger(2007)}]{PER-GRA:2007}
P\'erez, F., \& Granger, B.~E. 2007, Computing in Science and Engineering, 9,
  21, \dodoi{10.1109/MCSE.2007.53}

\bibitem[{{Peterken} {et~al.}(2020){Peterken}, {Merrifield},
  {Arag{\'o}n-Salamanca}, {Fraser-McKelvie}, {Avila-Reese}, {Riffel}, {Knapen},
  \& {Drory}}]{Peterken20}
{Peterken}, T., {Merrifield}, M., {Arag{\'o}n-Salamanca}, A., {et~al.} 2020,
  \mnras, 495, 3387, \dodoi{10.1093/mnras/staa1303}

\bibitem[{{Pforr} {et~al.}(2012){Pforr}, {Maraston}, \& {Tonini}}]{Pforr12}
{Pforr}, J., {Maraston}, C., \& {Tonini}, C. 2012, \mnras, 422, 3285,
  \dodoi{10.1111/j.1365-2966.2012.20848.x}

\bibitem[{{Pierini} {et~al.}(2004){Pierini}, {Gordon}, {Witt}, \&
  {Madsen}}]{Pierini04}
{Pierini}, D., {Gordon}, K.~D., {Witt}, A.~N., \& {Madsen}, G.~J. 2004, \apj,
  617, 1022, \dodoi{10.1086/425651}

\bibitem[{{Pillepich} {et~al.}(2018){Pillepich}, {Nelson}, {Hernquist},
  {Springel}, {Pakmor}, {Torrey}, {Weinberger}, {Genel}, {Naiman}, {Marinacci},
  \& {Vogelsberger}}]{Pillepich18}
{Pillepich}, A., {Nelson}, D., {Hernquist}, L., {et~al.} 2018, \mnras, 475,
  648, \dodoi{10.1093/mnras/stx3112}

\bibitem[{{Planck Collaboration} {et~al.}(2016){Planck Collaboration}, {Ade},
  {Aghanim}, {Arnaud}, {Ashdown}, {Aumont}, {Baccigalupi}, {Banday},
  {Barreiro}, {Bartlett}, {Bartolo}, {Battaner}, {Battye}, {Benabed},
  {Beno{\^\i}t}, {Benoit-L{\'e}vy}, {Bernard}, {Bersanelli}, {Bielewicz},
  {Bock}, {Bonaldi}, {Bonavera}, {Bond}, {Borrill}, {Bouchet}, {Boulanger},
  {Bucher}, {Burigana}, {Butler}, {Calabrese}, {Cardoso}, {Catalano},
  {Challinor}, {Chamballu}, {Chary}, {Chiang}, {Chluba}, {Christensen},
  {Church}, {Clements}, {Colombi}, {Colombo}, {Combet}, {Coulais}, {Crill},
  {Curto}, {Cuttaia}, {Danese}, {Davies}, {Davis}, {de Bernardis}, {de Rosa},
  {de Zotti}, {Delabrouille}, {D{\'e}sert}, {Di Valentino}, {Dickinson},
  {Diego}, {Dolag}, {Dole}, {Donzelli}, {Dor{\'e}}, {Douspis}, {Ducout},
  {Dunkley}, {Dupac}, {Efstathiou}, {Elsner}, {En{\ss}lin}, {Eriksen},
  {Farhang}, {Fergusson}, {Finelli}, {Forni}, {Frailis}, {Fraisse},
  {Franceschi}, {Frejsel}, {Galeotta}, {Galli}, {Ganga}, {Gauthier}, {Gerbino},
  {Ghosh}, {Giard}, {Giraud-H{\'e}raud}, {Giusarma}, {Gjerl{\o}w},
  {Gonz{\'a}lez-Nuevo}, {G{\'o}rski}, {Gratton}, {Gregorio}, {Gruppuso},
  {Gudmundsson}, {Hamann}, {Hansen}, {Hanson}, {Harrison}, {Helou},
  {Henrot-Versill{\'e}}, {Hern{\'a}ndez-Monteagudo}, {Herranz}, {Hildebrandt},
  {Hivon}, {Hobson}, {Holmes}, {Hornstrup}, {Hovest}, {Huang}, {Huffenberger},
  {Hurier}, {Jaffe}, {Jaffe}, {Jones}, {Juvela}, {Keih{\"a}nen}, {Keskitalo},
  {Kisner}, {Kneissl}, {Knoche}, {Knox}, {Kunz}, {Kurki-Suonio}, {Lagache},
  {L{\"a}hteenm{\"a}ki}, {Lamarre}, {Lasenby}, {Lattanzi}, {Lawrence}, {Leahy},
  {Leonardi}, {Lesgourgues}, {Levrier}, {Lewis}, {Liguori}, {Lilje},
  {Linden-V{\o}rnle}, {L{\'o}pez-Caniego}, {Lubin}, {Mac{\'\i}as-P{\'e}rez},
  {Maggio}, {Maino}, {Mandolesi}, {Mangilli}, {Marchini}, {Maris}, {Martin},
  {Martinelli}, {Mart{\'\i}nez-Gonz{\'a}lez}, {Masi}, {Matarrese}, {McGehee},
  {Meinhold}, {Melchiorri}, {Melin}, {Mendes}, {Mennella}, {Migliaccio},
  {Millea}, {Mitra}, {Miville-Desch{\^e}nes}, {Moneti}, {Montier}, {Morgante},
  {Mortlock}, {Moss}, {Munshi}, {Murphy}, {Naselsky}, {Nati}, {Natoli},
  {Netterfield}, {N{\o}rgaard-Nielsen}, {Noviello}, {Novikov}, {Novikov},
  {Oxborrow}, {Paci}, {Pagano}, {Pajot}, {Paladini}, {Paoletti}, {Partridge},
  {Pasian}, {Patanchon}, {Pearson}, {Perdereau}, {Perotto}, {Perrotta},
  {Pettorino}, {Piacentini}, {Piat}, {Pierpaoli}, {Pietrobon}, {Plaszczynski},
  {Pointecouteau}, {Polenta}, {Popa}, {Pratt}, {Pr{\'e}zeau}, {Prunet},
  {Puget}, {Rachen}, {Reach}, {Rebolo}, {Reinecke}, {Remazeilles}, {Renault},
  {Renzi}, {Ristorcelli}, {Rocha}, {Rosset}, {Rossetti}, {Roudier},
  {Rouill{\'e} d'Orfeuil}, {Rowan-Robinson}, {Rubi{\~n}o-Mart{\'\i}n},
  {Rusholme}, {Said}, {Salvatelli}, {Salvati}, {Sandri}, {Santos},
  {Savelainen}, {Savini}, {Scott}, {Seiffert}, {Serra}, {Shellard}, {Spencer},
  {Spinelli}, {Stolyarov}, {Stompor}, {Sudiwala}, {Sunyaev}, {Sutton},
  {Suur-Uski}, {Sygnet}, {Tauber}, {Terenzi}, {Toffolatti}, {Tomasi},
  {Tristram}, {Trombetti}, {Tucci}, {Tuovinen}, {T{\"u}rler}, {Umana},
  {Valenziano}, {Valiviita}, {Van Tent}, {Vielva}, {Villa}, {Wade}, {Wandelt},
  {Wehus}, {White}, {White}, {Wilkinson}, {Yvon}, {Zacchei}, \&
  {Zonca}}]{Planck16}
{Planck Collaboration}, {Ade}, P.~A.~R., {Aghanim}, N., {et~al.} 2016, \aap,
  594, A13, \dodoi{10.1051/0004-6361/201525830}

\bibitem[{{Price} {et~al.}(2014){Price}, {Kriek}, {Brammer}, {Conroy},
  {F{\"o}rster Schreiber}, {Franx}, {Fumagalli}, {Lundgren}, {Momcheva},
  {Nelson}, {Skelton}, {van Dokkum}, {Whitaker}, \& {Wuyts}}]{Price14}
{Price}, S.~H., {Kriek}, M., {Brammer}, G.~B., {et~al.} 2014, \apj, 788, 86,
  \dodoi{10.1088/0004-637X/788/1/86}

\bibitem[{{Puglisi} {et~al.}(2016){Puglisi}, {Rodighiero}, {Franceschini},
  {Talia}, {Cimatti}, {Baronchelli}, {Daddi}, {Renzini}, {Schawinski},
  {Mancini}, {Silverman}, {Gruppioni}, {Lutz}, {Berta}, \&
  {Oliver}}]{Puglisi16}
{Puglisi}, A., {Rodighiero}, G., {Franceschini}, A., {et~al.} 2016, \aap, 586,
  A83, \dodoi{10.1051/0004-6361/201526782}

\bibitem[{{Qin} {et~al.}(2022){Qin}, {Zheng}, {Fang}, {Pan}, {Wuyts}, {Shi},
  {Peng}, {Gonzalez}, {Bian}, {Huang}, {Gu}, {Liu}, {Tan}, {Shi}, {Ren},
  {Zhang}, {Qiao}, {Wen}, \& {Liu}}]{Qin22}
{Qin}, J., {Zheng}, X.~Z., {Fang}, M., {et~al.} 2022, \mnras, 511, 765,
  \dodoi{10.1093/mnras/stac132}

\bibitem[{{Ramraj} {et~al.}(2017){Ramraj}, {Gilbank}, {Blyth}, {Skelton},
  {Glazebrook}, {Bower}, \& {Balogh}}]{Ramraj17}
{Ramraj}, R., {Gilbank}, D.~G., {Blyth}, S.-L., {et~al.} 2017, \mnras, 466,
  3143, \dodoi{10.1093/mnras/stw3262}

\bibitem[{{Riello} \& {Patat}(2005)}]{Riello05}
{Riello}, M., \& {Patat}, F. 2005, \mnras, 362, 671,
  \dodoi{10.1111/j.1365-2966.2005.09348.x}

\bibitem[{{Robaina} {et~al.}(2010){Robaina}, {Bell}, {van der Wel},
  {Somerville}, {Skelton}, {McIntosh}, {Meisenheimer}, \& {Wolf}}]{Robaina10}
{Robaina}, A.~R., {Bell}, E.~F., {van der Wel}, A., {et~al.} 2010, \apj, 719,
  844, \dodoi{10.1088/0004-637X/719/1/844}

\bibitem[{{Rodighiero} {et~al.}(2010){Rodighiero}, {Cimatti}, {Gruppioni},
  {Popesso}, {Andreani}, {Altieri}, {Aussel}, {Berta}, {Bongiovanni},
  {Brisbin}, {Cava}, {Cepa}, {Daddi}, {Dominguez-Sanchez}, {Elbaz}, {Fontana},
  {F{\"o}rster Schreiber}, {Franceschini}, {Genzel}, {Grazian}, {Lutz},
  {Magdis}, {Magliocchetti}, {Magnelli}, {Maiolino}, {Mancini}, {Nordon},
  {Perez Garcia}, {Poglitsch}, {Santini}, {Sanchez-Portal}, {Pozzi},
  {Riguccini}, {Saintonge}, {Shao}, {Sturm}, {Tacconi}, {Valtchanov},
  {Wetzstein}, \& {Wieprecht}}]{Rodighiero10}
{Rodighiero}, G., {Cimatti}, A., {Gruppioni}, C., {et~al.} 2010, \aap, 518,
  L25, \dodoi{10.1051/0004-6361/201014624}

\bibitem[{{Rodr{\'\i}guez Montero} {et~al.}(2019){Rodr{\'\i}guez Montero},
  {Dav{\'e}}, {Wild}, {Angl{\'e}s-Alc{\'a}zar}, \&
  {Narayanan}}]{RodriguezMontero19}
{Rodr{\'\i}guez Montero}, F., {Dav{\'e}}, R., {Wild}, V.,
  {Angl{\'e}s-Alc{\'a}zar}, D., \& {Narayanan}, D. 2019, \mnras, 490, 2139,
  \dodoi{10.1093/mnras/stz2580}

\bibitem[{{Ro{\v{s}}kar} {et~al.}(2008){Ro{\v{s}}kar}, {Debattista}, {Quinn},
  {Stinson}, \& {Wadsley}}]{Roskar08}
{Ro{\v{s}}kar}, R., {Debattista}, V.~P., {Quinn}, T.~R., {Stinson}, G.~S., \&
  {Wadsley}, J. 2008, \apjl, 684, L79, \dodoi{10.1086/592231}

\bibitem[{{Salim} {et~al.}(2018){Salim}, {Boquien}, \& {Lee}}]{Salim18}
{Salim}, S., {Boquien}, M., \& {Lee}, J.~C. 2018, \apj, 859, 11,
  \dodoi{10.3847/1538-4357/aabf3c}

\bibitem[{{Salim} \& {Narayanan}(2020)}]{Salim20}
{Salim}, S., \& {Narayanan}, D. 2020, \araa, 58, 529,
  \dodoi{10.1146/annurev-astro-032620-021933}

\bibitem[{{Salmon} {et~al.}(2016){Salmon}, {Papovich}, {Long}, {Willner},
  {Finkelstein}, {Ferguson}, {Dickinson}, {Duncan}, {Faber}, {Hathi},
  {Koekemoer}, {Kurczynski}, {Newman}, {Pacifici}, {P{\'e}rez-Gonz{\'a}lez}, \&
  {Pforr}}]{Salmon16}
{Salmon}, B., {Papovich}, C., {Long}, J., {et~al.} 2016, \apj, 827, 20,
  \dodoi{10.3847/0004-637X/827/1/20}

\bibitem[{{Salvador-Rusi{\~n}ol} {et~al.}(2020){Salvador-Rusi{\~n}ol},
  {Vazdekis}, {La Barbera}, {Beasley}, {Ferreras}, {Negri}, \& {Dalla
  Vecchia}}]{SalvadorRusinol20}
{Salvador-Rusi{\~n}ol}, N., {Vazdekis}, A., {La Barbera}, F., {et~al.} 2020,
  Nature Astronomy, 4, 252, \dodoi{10.1038/s41550-019-0955-0}

\bibitem[{{S{\'a}nchez} {et~al.}(2018){S{\'a}nchez}, {Avila-Reese},
  {Hernandez-Toledo}, {Cortes-Su{\'a}rez}, {Rodr{\'\i}guez-Puebla},
  {Ibarra-Medel}, {Cano-D{\'\i}az}, {Barrera-Ballesteros}, {Negrete},
  {Calette}, {de Lorenzo-C{\'a}ceres}, {Ortega-Minakata}, {Aquino},
  {Valenzuela}, {Clemente}, {Storchi-Bergmann}, {Riffel}, {Schimoia}, {Riffel},
  {Rembold}, {Brownstein}, {Pan}, {Yates}, {Mallmann}, \&
  {Bitsakis}}]{Sanchez18}
{S{\'a}nchez}, S.~F., {Avila-Reese}, V., {Hernandez-Toledo}, H., {et~al.} 2018,
  \rmxaa, 54, 217.
\newblock \doarXiv{1709.05438}

\bibitem[{{Santini} {et~al.}(2015){Santini}, {Ferguson}, {Fontana}, {Mobasher},
  {Barro}, {Castellano}, {Finkelstein}, {Grazian}, {Hsu}, {Lee}, {Lee},
  {Pforr}, {Salvato}, {Wiklind}, {Wuyts}, {Almaini}, {Cooper}, {Galametz},
  {Weiner}, {Amorin}, {Boutsia}, {Conselice}, {Dahlen}, {Dickinson},
  {Giavalisco}, {Grogin}, {Guo}, {Hathi}, {Kocevski}, {Koekemoer},
  {Kurczynski}, {Merlin}, {Mortlock}, {Newman}, {Paris}, {Pentericci},
  {Simons}, \& {Willner}}]{Santini15}
{Santini}, P., {Ferguson}, H.~C., {Fontana}, A., {et~al.} 2015, \apj, 801, 97,
  \dodoi{10.1088/0004-637X/801/2/97}

\bibitem[{{Schlafly} \& {Finkbeiner}(2011)}]{Schlafly11}
{Schlafly}, E.~F., \& {Finkbeiner}, D.~P. 2011, \apj, 737, 103,
  \dodoi{10.1088/0004-637X/737/2/103}

\bibitem[{{Schlegel} {et~al.}(1998){Schlegel}, {Finkbeiner}, \&
  {Davis}}]{Schlegel98}
{Schlegel}, D.~J., {Finkbeiner}, D.~P., \& {Davis}, M. 1998, \apj, 500, 525,
  \dodoi{10.1086/305772}

\bibitem[{{Sch{\"o}nrich} \& {Binney}(2009)}]{Schonrich09}
{Sch{\"o}nrich}, R., \& {Binney}, J. 2009, \mnras, 396, 203,
  \dodoi{10.1111/j.1365-2966.2009.14750.x}

\bibitem[{{Schreiber} {et~al.}(2015){Schreiber}, {Pannella}, {Elbaz},
  {B{\'e}thermin}, {Inami}, {Dickinson}, {Magnelli}, {Wang}, {Aussel}, {Daddi},
  {Juneau}, {Shu}, {Sargent}, {Buat}, {Faber}, {Ferguson}, {Giavalisco},
  {Koekemoer}, {Magdis}, {Morrison}, {Papovich}, {Santini}, \&
  {Scott}}]{Schreiber15}
{Schreiber}, C., {Pannella}, M., {Elbaz}, D., {et~al.} 2015, \aap, 575, A74,
  \dodoi{10.1051/0004-6361/201425017}

\bibitem[{{Sellwood} \& {Binney}(2002)}]{Sellwood02}
{Sellwood}, J.~A., \& {Binney}, J.~J. 2002, \mnras, 336, 785,
  \dodoi{10.1046/j.1365-8711.2002.05806.x}

\bibitem[{{Shanks} {et~al.}(2021){Shanks}, {Ansarinejad}, {Bielby}, {Heywood},
  {Metcalfe}, \& {Wang}}]{Shanks21}
{Shanks}, T., {Ansarinejad}, B., {Bielby}, R.~M., {et~al.} 2021, \mnras, 505,
  1509, \dodoi{10.1093/mnras/stab1226}

\bibitem[{{Shapley} {et~al.}(2023){Shapley}, {Sanders}, {Reddy}, {Topping}, \&
  {Brammer}}]{Shapley23}
{Shapley}, A.~E., {Sanders}, R.~L., {Reddy}, N.~A., {Topping}, M.~W., \&
  {Brammer}, G.~B. 2023, \apj, 954, 157, \dodoi{10.3847/1538-4357/acea5a}

\bibitem[{{Shapley} {et~al.}(2022){Shapley}, {Sanders}, {Salim}, {Reddy},
  {Kriek}, {Mobasher}, {Coil}, {Siana}, {Price}, {Shivaei}, {Dunlop}, {McLure},
  \& {Cullen}}]{Shapley22}
{Shapley}, A.~E., {Sanders}, R.~L., {Salim}, S., {et~al.} 2022, \apj, 926, 145,
  \dodoi{10.3847/1538-4357/ac4742}

\bibitem[{{Silva} {et~al.}(1998){Silva}, {Granato}, {Bressan}, \&
  {Danese}}]{Silva98}
{Silva}, L., {Granato}, G.~L., {Bressan}, A., \& {Danese}, L. 1998, \apj, 509,
  103, \dodoi{10.1086/306476}

\bibitem[{{Simons} {et~al.}(2017){Simons}, {Kassin}, {Weiner}, {Faber},
  {Trump}, {Heckman}, {Koo}, {Pacifici}, {Primack}, {Snyder}, \& {de la
  Vega}}]{Simons17}
{Simons}, R.~C., {Kassin}, S.~A., {Weiner}, B.~J., {et~al.} 2017, \apj, 843,
  46, \dodoi{10.3847/1538-4357/aa740c}

\bibitem[{{Skelton} {et~al.}(2014){Skelton}, {Whitaker}, {Momcheva}, {Brammer},
  {van Dokkum}, {Labb{\'e}}, {Franx}, {van der Wel}, {Bezanson}, {Da Cunha},
  {Fumagalli}, {F{\"o}rster Schreiber}, {Kriek}, {Leja}, {Lundgren}, {Magee},
  {Marchesini}, {Maseda}, {Nelson}, {Oesch}, {Pacifici}, {Patel}, {Price},
  {Rix}, {Tal}, {Wake}, \& {Wuyts}}]{Skelton14}
{Skelton}, R.~E., {Whitaker}, K.~E., {Momcheva}, I.~G., {et~al.} 2014, \apjs,
  214, 24, \dodoi{10.1088/0067-0049/214/2/24}

\bibitem[{{Snyder} {et~al.}(2011){Snyder}, {Cox}, {Hayward}, {Hernquist}, \&
  {Jonsson}}]{Snyder11}
{Snyder}, G.~F., {Cox}, T.~J., {Hayward}, C.~C., {Hernquist}, L., \& {Jonsson},
  P. 2011, \apj, 741, 77, \dodoi{10.1088/0004-637X/741/2/77}

\bibitem[{{Sorba} \& {Sawicki}(2018)}]{Sorba18}
{Sorba}, R., \& {Sawicki}, M. 2018, \mnras, 476, 1532,
  \dodoi{10.1093/mnras/sty186}

\bibitem[{{Speagle} {et~al.}(2014){Speagle}, {Steinhardt}, {Capak}, \&
  {Silverman}}]{Speagle14}
{Speagle}, J.~S., {Steinhardt}, C.~L., {Capak}, P.~L., \& {Silverman}, J.~D.
  2014, \apjs, 214, 15, \dodoi{10.1088/0067-0049/214/2/15}

\bibitem[{{Steffen} {et~al.}(2021){Steffen}, {Fu}, {Comerford}, {Dai}, {Feng},
  {Gross}, \& {Xue}}]{Steffen21}
{Steffen}, J.~L., {Fu}, H., {Comerford}, J.~M., {et~al.} 2021, \apj, 909, 120,
  \dodoi{10.3847/1538-4357/abe2a5}

\bibitem[{{Suess} {et~al.}(2019){Suess}, {Kriek}, {Price}, \&
  {Barro}}]{Suess19a}
{Suess}, K.~A., {Kriek}, M., {Price}, S.~H., \& {Barro}, G. 2019, \apj, 877,
  103, \dodoi{10.3847/1538-4357/ab1bda}

\bibitem[{{Tacchella} {et~al.}(2016){Tacchella}, {Dekel}, {Carollo},
  {Ceverino}, {DeGraf}, {Lapiner}, {Mandelker}, \& {Primack}}]{Tacchella16b}
{Tacchella}, S., {Dekel}, A., {Carollo}, C.~M., {et~al.} 2016, \mnras, 458,
  242, \dodoi{10.1093/mnras/stw303}

\bibitem[{{Tacchella} {et~al.}(2018){Tacchella}, {Carollo}, {F{\"o}rster
  Schreiber}, {Renzini}, {Dekel}, {Genzel}, {Lang}, {Lilly}, {Mancini},
  {Onodera}, {Tacconi}, {Wuyts}, \& {Zamorani}}]{Tacchella18}
{Tacchella}, S., {Carollo}, C.~M., {F{\"o}rster Schreiber}, N.~M., {et~al.}
  2018, \apj, 859, 56, \dodoi{10.3847/1538-4357/aabf8b}

\bibitem[{{Taylor} {et~al.}(2011){Taylor}, {Hopkins}, {Baldry}, {Brown},
  {Driver}, {Kelvin}, {Hill}, {Robotham}, {Bland-Hawthorn}, {Jones}, {Sharp},
  {Thomas}, {Liske}, {Loveday}, {Norberg}, {Peacock}, {Bamford}, {Brough},
  {Colless}, {Cameron}, {Conselice}, {Croom}, {Frenk}, {Gunawardhana},
  {Kuijken}, {Nichol}, {Parkinson}, {Phillipps}, {Pimbblet}, {Popescu},
  {Prescott}, {Sutherland}, {Tuffs}, {van Kampen}, \& {Wijesinghe}}]{Taylor11}
{Taylor}, E.~N., {Hopkins}, A.~M., {Baldry}, I.~K., {et~al.} 2011, \mnras, 418,
  1587, \dodoi{10.1111/j.1365-2966.2011.19536.x}

\bibitem[{{Thorp} {et~al.}(2019){Thorp}, {Ellison}, {Simard}, {S{\'a}nchez}, \&
  {Antonio}}]{Thorp19}
{Thorp}, M.~D., {Ellison}, S.~L., {Simard}, L., {S{\'a}nchez}, S.~F., \&
  {Antonio}, B. 2019, \mnras, 482, L55, \dodoi{10.1093/mnrasl/sly185}

\bibitem[{{Tomczak} {et~al.}(2016){Tomczak}, {Quadri}, {Tran}, {Labb{\'e}},
  {Straatman}, {Papovich}, {Glazebrook}, {Allen}, {Brammer}, {Cowley},
  {Dickinson}, {Elbaz}, {Inami}, {Kacprzak}, {Morrison}, {Nanayakkara},
  {Persson}, {Rees}, {Salmon}, {Schreiber}, {Spitler}, \&
  {Whitaker}}]{Tomczak16}
{Tomczak}, A.~R., {Quadri}, R.~F., {Tran}, K.-V.~H., {et~al.} 2016, \apj, 817,
  118, \dodoi{10.3847/0004-637X/817/2/118}

\bibitem[{{Tortora} {et~al.}(2010){Tortora}, {Napolitano}, {Cardone},
  {Capaccioli}, {Jetzer}, \& {Molinaro}}]{Tortora10}
{Tortora}, C., {Napolitano}, N.~R., {Cardone}, V.~F., {et~al.} 2010, \mnras,
  407, 144, \dodoi{10.1111/j.1365-2966.2010.16938.x}

\bibitem[{{Tress} {et~al.}(2019){Tress}, {Ferreras}, {P{\'e}rez-Gonz{\'a}lez},
  {Bressan}, {Barro}, {Dom{\'\i}nguez-S{\'a}nchez}, \&
  {Eliche-Moral}}]{Tress19}
{Tress}, M., {Ferreras}, I., {P{\'e}rez-Gonz{\'a}lez}, P.~G., {et~al.} 2019,
  \mnras, 488, 2301, \dodoi{10.1093/mnras/stz1851}

\bibitem[{{Tress} {et~al.}(2018){Tress}, {M{\'a}rmol-Queralt{\'o}}, {Ferreras},
  {P{\'e}rez-Gonz{\'a}lez}, {Barro}, {Pampliega}, {Cava},
  {Dom{\'\i}nguez-S{\'a}nchez}, {Eliche-Moral}, {Espino-Briones}, {Esquej},
  {Hern{\'a}n-Caballero}, {Rodighiero}, \& {Rodriguez-Mu{\~n}oz}}]{Tress18}
{Tress}, M., {M{\'a}rmol-Queralt{\'o}}, E., {Ferreras}, I., {et~al.} 2018,
  \mnras, 475, 2363, \dodoi{10.1093/mnras/stx3334}

\bibitem[{{Trussler} {et~al.}(2020){Trussler}, {Maiolino}, {Maraston}, {Peng},
  {Thomas}, {Goddard}, \& {Lian}}]{Trussler20}
{Trussler}, J., {Maiolino}, R., {Maraston}, C., {et~al.} 2020, \mnras, 491,
  5406, \dodoi{10.1093/mnras/stz3286}

\bibitem[{{Tuffs} {et~al.}(2004){Tuffs}, {Popescu}, {V{\"o}lk}, {Kylafis}, \&
  {Dopita}}]{Tuffs04}
{Tuffs}, R.~J., {Popescu}, C.~C., {V{\"o}lk}, H.~J., {Kylafis}, N.~D., \&
  {Dopita}, M.~A. 2004, \aap, 419, 821, \dodoi{10.1051/0004-6361:20035689}

\bibitem[{{van der Wel} {et~al.}(2012){van der Wel}, {Bell}, {H{\"a}ussler},
  {McGrath}, {Chang}, {Guo}, {McIntosh}, {Rix}, {Barden}, {Cheung}, {Faber},
  {Ferguson}, {Galametz}, {Grogin}, {Hartley}, {Kartaltepe}, {Kocevski},
  {Koekemoer}, {Lotz}, {Mozena}, {Peth}, \& {Peng}}]{vdW12}
{van der Wel}, A., {Bell}, E.~F., {H{\"a}ussler}, B., {et~al.} 2012, \apjs,
  203, 24, \dodoi{10.1088/0067-0049/203/2/24}

\bibitem[{{van der Wel} {et~al.}(2014){van der Wel}, {Franx}, {van Dokkum},
  {Skelton}, {Momcheva}, {Whitaker}, {Brammer}, {Bell}, {Rix}, {Wuyts},
  {Ferguson}, {Holden}, {Barro}, {Koekemoer}, {Chang}, {McGrath},
  {H{\"a}ussler}, {Dekel}, {Behroozi}, {Fumagalli}, {Leja}, {Lundgren},
  {Maseda}, {Nelson}, {Wake}, {Patel}, {Labb{\'e}}, {Faber}, {Grogin}, \&
  {Kocevski}}]{vdW14}
{van der Wel}, A., {Franx}, M., {van Dokkum}, P.~G., {et~al.} 2014, \apj, 788,
  28, \dodoi{10.1088/0004-637X/788/1/28}

\bibitem[{{Vidal-Garc{\'\i}a} {et~al.}(2017){Vidal-Garc{\'\i}a}, {Charlot},
  {Bruzual}, \& {Hubeny}}]{VidalGarcia2017}
{Vidal-Garc{\'\i}a}, A., {Charlot}, S., {Bruzual}, G., \& {Hubeny}, I. 2017,
  \mnras, 470, 3532, \dodoi{10.1093/mnras/stx1324}

\bibitem[{{Villaume} {et~al.}(2015){Villaume}, {Conroy}, \&
  {Johnson}}]{Villaume15}
{Villaume}, A., {Conroy}, C., \& {Johnson}, B.~D. 2015, \apj, 806, 82,
  \dodoi{10.1088/0004-637X/806/1/82}

\bibitem[{Virtanen {et~al.}(2020)Virtanen, Gommers, Oliphant, Haberland, Reddy,
  Cournapeau, Burovski, Peterson, Weckesser, Bright, {van der Walt}, Brett,
  Wilson, Millman, Mayorov, Nelson, Jones, Kern, Larson, Carey, Polat, Feng,
  Moore, {VanderPlas}, Laxalde, Perktold, Cimrman, Henriksen, Quintero, Harris,
  Archibald, Ribeiro, Pedregosa, {van Mulbregt}, \& {SciPy 1.0
  Contributors}}]{Scipy20}
Virtanen, P., Gommers, R., Oliphant, T.~E., {et~al.} 2020, Nature Methods, 17,
  261, \dodoi{10.1038/s41592-019-0686-2}

\bibitem[{{Walcher} {et~al.}(2008){Walcher}, {Lamareille}, {Vergani},
  {Arnouts}, {Buat}, {Charlot}, {Tresse}, {Le F{\`e}vre}, {Bolzonella},
  {Brinchmann}, {Pozzetti}, {Zamorani}, {Bottini}, {Garilli}, {Le Brun},
  {Maccagni}, {Milliard}, {Scaramella}, {Scodeggio}, {Vettolani}, {Zanichelli},
  {Adami}, {Bardelli}, {Cappi}, {Ciliegi}, {Contini}, {Franzetti}, {Foucaud},
  {Gavignaud}, {Guzzo}, {Ilbert}, {Iovino}, {McCracken}, {Marano}, {Marinoni},
  {Mazure}, {Meneux}, {Merighi}, {Paltani}, {Pell{\`o}}, {Pollo}, {Radovich},
  {Zucca}, {Lonsdale}, \& {Martin}}]{Walcher08}
{Walcher}, C.~J., {Lamareille}, F., {Vergani}, D., {et~al.} 2008, \aap, 491,
  713, \dodoi{10.1051/0004-6361:200810704}

\bibitem[{{Wang} {et~al.}(2023){Wang}, {Leja}, {Bezanson}, {Johnson},
  {Khullar}, {Labb{\'e}}, {Price}, {Weaver}, \& {Whitaker}}]{Wang23}
{Wang}, B., {Leja}, J., {Bezanson}, R., {et~al.} 2023, \apjl, 944, L58,
  \dodoi{10.3847/2041-8213/acba99}

\bibitem[{{Wang} {et~al.}(2017){Wang}, {Faber}, {Liu}, {Guo}, {Pacifici},
  {Koo}, {Kassin}, {Mao}, {Fang}, {Chen}, {Koekemoer}, {Kocevski}, \&
  {Ashby}}]{Wang17}
{Wang}, W., {Faber}, S.~M., {Liu}, F.~S., {et~al.} 2017, \mnras, 469, 4063,
  \dodoi{10.1093/mnras/stx1148}

\bibitem[{{Weibel} {et~al.}(2023){Weibel}, {Wang}, \& {Lilly}}]{Weibel23}
{Weibel}, A., {Wang}, E., \& {Lilly}, S.~J. 2023, \apj, 950, 102,
  \dodoi{10.3847/1538-4357/accffc}

\bibitem[{{Weigel} {et~al.}(2017){Weigel}, {Schawinski}, {Caplar}, {Carpineti},
  {Hart}, {Kaviraj}, {Keel}, {Kruk}, {Lintott}, {Nichol}, {Simmons}, \&
  {Smethurst}}]{Weigel17}
{Weigel}, A.~K., {Schawinski}, K., {Caplar}, N., {et~al.} 2017, \apj, 845, 145,
  \dodoi{10.3847/1538-4357/aa8097}

\bibitem[{{Weinzirl} {et~al.}(2009){Weinzirl}, {Jogee}, {Khochfar}, {Burkert},
  \& {Kormendy}}]{Weinzirl09}
{Weinzirl}, T., {Jogee}, S., {Khochfar}, S., {Burkert}, A., \& {Kormendy}, J.
  2009, \apj, 696, 411, \dodoi{10.1088/0004-637X/696/1/411}

\bibitem[{{Whitaker} {et~al.}(2012){Whitaker}, {van Dokkum}, {Brammer}, \&
  {Franx}}]{Whitaker12}
{Whitaker}, K.~E., {van Dokkum}, P.~G., {Brammer}, G., \& {Franx}, M. 2012,
  \apjl, 754, L29, \dodoi{10.1088/2041-8205/754/2/L29}

\bibitem[{{Whitaker} {et~al.}(2011){Whitaker}, {Labb{\'e}}, {van Dokkum},
  {Brammer}, {Kriek}, {Marchesini}, {Quadri}, {Franx}, {Muzzin}, {Williams},
  {Bezanson}, {Illingworth}, {Lee}, {Lundgren}, {Nelson}, {Rudnick}, {Tal}, \&
  {Wake}}]{Whitaker11}
{Whitaker}, K.~E., {Labb{\'e}}, I., {van Dokkum}, P.~G., {et~al.} 2011, \apj,
  735, 86, \dodoi{10.1088/0004-637X/735/2/86}

\bibitem[{{Whitaker} {et~al.}(2014){Whitaker}, {Franx}, {Leja}, {van Dokkum},
  {Henry}, {Skelton}, {Fumagalli}, {Momcheva}, {Brammer}, {Labb{\'e}},
  {Nelson}, \& {Rigby}}]{Whitaker14}
{Whitaker}, K.~E., {Franx}, M., {Leja}, J., {et~al.} 2014, \apj, 795, 104,
  \dodoi{10.1088/0004-637X/795/2/104}

\bibitem[{{White}(1979)}]{White79}
{White}, S.~D.~M. 1979, \mnras, 189, 831, \dodoi{10.1093/mnras/189.4.831}

\bibitem[{{White}(1980)}]{White80}
---. 1980, \mnras, 191, 1P, \dodoi{10.1093/mnras/191.1.1P}

\bibitem[{{Wild} {et~al.}(2011){Wild}, {Charlot}, {Brinchmann}, {Heckman},
  {Vince}, {Pacifici}, \& {Chevallard}}]{Wild11}
{Wild}, V., {Charlot}, S., {Brinchmann}, J., {et~al.} 2011, \mnras, 417, 1760,
  \dodoi{10.1111/j.1365-2966.2011.19367.x}

\bibitem[{{Wu} {et~al.}(2018){Wu}, {van der Wel}, {Gallazzi}, {Bezanson},
  {Pacifici}, {Straatman}, {Franx}, {Bari{\v{s}}i{\'c}}, {Bell}, {Brammer},
  {Calhau}, {Chauke}, {van Houdt}, {Maseda}, {Muzzin}, {Rix}, {Sobral},
  {Spilker}, {van de Sande}, {van Dokkum}, \& {Wild}}]{Wu18}
{Wu}, P.-F., {van der Wel}, A., {Gallazzi}, A., {et~al.} 2018, \apj, 855, 85,
  \dodoi{10.3847/1538-4357/aab0a6}

\bibitem[{{Wuyts} {et~al.}(2008){Wuyts}, {Labb{\'e}}, {F{\"o}rster Schreiber},
  {Franx}, {Rudnick}, {Brammer}, \& {van Dokkum}}]{Wuyts08}
{Wuyts}, S., {Labb{\'e}}, I., {F{\"o}rster Schreiber}, N.~M., {et~al.} 2008,
  \apj, 682, 985, \dodoi{10.1086/588749}

\bibitem[{{Wuyts} {et~al.}(2011){Wuyts}, {F{\"o}rster Schreiber}, {Lutz},
  {Nordon}, {Berta}, {Altieri}, {Andreani}, {Aussel}, {Bongiovanni}, {Cepa},
  {Cimatti}, {Daddi}, {Elbaz}, {Genzel}, {Koekemoer}, {Magnelli}, {Maiolino},
  {McGrath}, {P{\'e}rez Garc{\'\i}a}, {Poglitsch}, {Popesso}, {Pozzi},
  {Sanchez-Portal}, {Sturm}, {Tacconi}, \& {Valtchanov}}]{Wuyts11}
{Wuyts}, S., {F{\"o}rster Schreiber}, N.~M., {Lutz}, D., {et~al.} 2011, \apj,
  738, 106, \dodoi{10.1088/0004-637X/738/1/106}

\bibitem[{{Wuyts} {et~al.}(2012){Wuyts}, {F{\"o}rster Schreiber}, {Genzel},
  {Guo}, {Barro}, {Bell}, {Dekel}, {Faber}, {Ferguson}, {Giavalisco}, {Grogin},
  {Hathi}, {Huang}, {Kocevski}, {Koekemoer}, {Koo}, {Lotz}, {Lutz}, {McGrath},
  {Newman}, {Rosario}, {Saintonge}, {Tacconi}, {Weiner}, \& {van der
  Wel}}]{Wuyts12}
{Wuyts}, S., {F{\"o}rster Schreiber}, N.~M., {Genzel}, R., {et~al.} 2012, \apj,
  753, 114, \dodoi{10.1088/0004-637X/753/2/114}

\bibitem[{{Xue} {et~al.}(2016){Xue}, {Luo}, {Brandt}, {Alexander}, {Bauer},
  {Lehmer}, \& {Yang}}]{Xue16}
{Xue}, Y.~Q., {Luo}, B., {Brandt}, W.~N., {et~al.} 2016, \apjs, 224, 15,
  \dodoi{10.3847/0067-0049/224/2/15}

\bibitem[{{Zibetti} {et~al.}(2013){Zibetti}, {Gallazzi}, {Charlot}, {Pierini},
  \& {Pasquali}}]{Zibetti13}
{Zibetti}, S., {Gallazzi}, A., {Charlot}, S., {Pierini}, D., \& {Pasquali}, A.
  2013, \mnras, 428, 1479, \dodoi{10.1093/mnras/sts126}

\bibitem[{{Zolotov} {et~al.}(2015){Zolotov}, {Dekel}, {Mandelker}, {Tweed},
  {Inoue}, {DeGraf}, {Ceverino}, {Primack}, {Barro}, \& {Faber}}]{Zolotov15}
{Zolotov}, A., {Dekel}, A., {Mandelker}, N., {et~al.} 2015, \mnras, 450, 2327,
  \dodoi{10.1093/mnras/stv740}

\end{thebibliography}
\bibliographystyle{aasjournal}

\end{document}